\documentclass{aastex}
\usepackage{spr-astr-addons}
\usepackage{url}

\usepackage{booktabs}
\usepackage{subfigure}
\usepackage{graphicx}
\usepackage{subfloat}
\usepackage{hyperref}

\RequirePackage{color}
\renewcommand{\arraystretch}{1.3}
\renewcommand\sec{\ \mathrm{s}}
\newcommand\eiso{$E_\mathrm{iso}$}
\newcommand\epeak{$E_\mathrm{peak}$}
\newcommand\Tqvd{$T_\mathrm{90}\ $}
\newcommand\Gpccc{$\mathrm{Gpc^3}$}
\newcommand\Msun{$\mathrm{M_\odot}$}
\newcommand\fluence{\mathrm{erg\ cm^{-2}}}
\newcommand\phflux{\mathrm{ph\ cm^{-2}\ s^{-1}}}
\newcommand\Nall{41}
\newcommand\Nlong{24}
\newcommand\Nshort{10}
\newcommand\Nsgr{7}

\begin{document}

\title{Detection of short high-energy transients in the local universe with SVOM/ECLAIRs}
\slugcomment{Not to appear in Nonlearned J., 45.}
\shorttitle{Short HE transients with SVOM/ECLAIRs}
\shortauthors{B. Arcier et al.}

\author{B. Arcier} \and \author{JL. Atteia} \and \author{O. Godet} \and \author{S. Mate}\and \author{S. Guillot}
\affil{IRAP, Universite de Toulouse, CNES, CNRS, UPS, Toulouse, France\\ \textit{Corresponding author: Benjamin.Arcier@irap.omp.eu}}
\and
\author{N. Dagoneau} \and \author{J. Rodriguez}\and \author{D. Gotz} \and \author{S. Schanne}
\affil{Lab AIM, CEA/CNRS/Universite Paris-Saclay, Universite de Paris, F-91191 Gif-sur-Yvette, France}
\and
\author{M. G. Bernardini\altaffilmark{1}}
\affil{INAF – Osservatorio Astronomico di Brera, via Bianchi 46, 23807 Merate (LC), Italy}

\altaffiltext{1}{LUPM – Laboratoire Univers et Particules Montpellier, Place Eugene Bataillon, 34090 Montpellier, France}

\begin{abstract}

   The coincidental detection of the gravitational wave event GW~170817 and the associated gamma-ray burst GRB~170817A marked the advent of multi-messenger astronomy and represented a milestone in the study of GRBs. Significant progress in this field is expected in the coming years with the increased sensitivity of gravitational waves detectors and the launch of new facilities for the high-energy survey of the sky. In this context, the launch of \textit{SVOM} in mid-2022, with its two wide-field high-energy instruments ECLAIRs and GRM, will foster the possibilities of coincidental transient detection with gravitational waves and gamma-rays events.
   The purpose of this paper is to assess the ability of \textit{SVOM}/ECLAIRs to detect and quickly characterize high-energy transients in the local Universe (\mbox{z $\leq$ 0.3}), and to discuss the contribution of this instrument to multi-messenger astronomy and to gamma-ray burst (GRB) astrophysics in the 2020's.
   A list of local HE transients, along with their main characteristics, is constructed through an extensive literature survey. This list includes \Nall\ transients: \Nlong\ long GRBs, \Nshort\ short GRBs and \Nsgr\ SGR Giant Flares.
   The detectability of these transients with ECLAIRs is assessed with detailed simulations using tools developed for the \textit{SVOM} mission, including a GEANT4 simulation of the energy response and a simulated trigger algorithm representative of the onboard trigger algorithm.
   \textit{SVOM}/ECLAIRs would have been able to detect 88\% of the short high-energy transients in our list: 22 out of \Nlong\ long GRBs, 8 out of \Nshort\ short GRBs and 6 out of \Nsgr\ SGR Giant Flares. The SNR for almost all detections will be sufficiently high to allow the on-board ECLAIRs trigger algorithm to derive the localisation of the transient, transmitting it to the \textit{SVOM} satellite and ground-based instruments. Coupled with the anti-solar pointing strategy of \textit{SVOM}, this will enable an optimal follow-up of the events, allowing the observation of their afterglows, supernovae/kilonovae counterparts, and host galaxies.
   We conclude the paper with a discussion of the unique contribution expected from \textit{SVOM} and of the possibility of simultaneous GW detection for each type of transient in our sample.

\end{abstract}


\section{Introduction}
\label{sec:intro}

High-energy transients in the local universe are privileged targets for the \textit{Space-based multi-band astronomical Variable Objects Monitor} (\textit{SVOM}) especially in the nascent multi-messenger astrophysics context.
The potential of this new field has been beautifully illustrated by the coincidental detection of GRB~170817A and GW~170817, which confirmed the link between short gamma-ray bursts and neutron star mergers \citep{Abbott2017b}, and the subsequent detection of the kilonova AT~2017gfo \citep{Tanvir2017_170817A}, whose optical spectrum evidenced the production of r-process elements \citep{Pian2017}. 

The launch of \textit{SVOM} in mid-2022 will add new resources for multi-messenger astrophysics as described by \cite{Wei2016}. 
We discuss here the performance of ECLAIRs, the hard X-ray imager of \textit{SVOM} \citep{Godet2014}, for the detection and near real-time identification of various types of short high-energy transients in the local universe: short GRBs (SGRBs), long GRBs (LGRBs) and soft gamma-ray repeaters giant flares (SGR GFs), in the perspective of studying their multi-messenger emission.
The motivation for this study is twofold: first, to verify the capability of ECLAIRs to detect these events and second, to check whether ECLAIRs data are sufficient to recognize the various types of events, especially their likelihood to be associated with transient Gravitational Wave (GW) signals. 
Being able to detect and quickly follow-up these events could shed some light on some important GRBs topics, such as the nature of long GRBs without a supernova (SN) \citep{Gehrels2006_060614, Yang2015_060614}, the rate of SGR GFs in our local Universe \citep{Hurley2011_SGR}, the differences between short GRBs with and without extended emission \citep{Barthelmy2005_050724, Jin2016_050709} and the connection between X-Ray Flashes (XRFs) and classical long GRBs \citep{Sakamoto2005, Sakamoto2008}.
Our study encompasses \Nall\ short extra-galactic transients closer than \mbox{$z = 0.3$}, a volume chosen to contain a sufficient number of events of each type.
While we have tried to compile the most complete list of high-energy transients in the local universe, we cannot guarantee that we have not missed some of them.
This is, however, not a problem in the context of this paper, since we seek to get a representative sample of events to study their appearance and detectability by ECLAIRs, and not a complete sample.

The construction of the sample is discussed in Sect. \ref{sec:sample}. 
The ECLAIRs instrument is presented in Sect. \ref{sec:eclairs},  while its performance for the detection and classification of events in the local universe is discussed in Sect. \ref{sec:snr}.
Section \ref{sec:discussion} places 
these results in their astrophysical context.

This work relies on a complete simulation package involving realistic response matrices, the detailed simulation of the instrument background \citep{Mate2019}, and a computation of high-energy transients Signal-to-Noise Ratio (SNR) taking into account the main features of the on-board trigger algorithm. 
In all this paper, we use the cosmological parameters measured by the Planck collaboration: $H_\mathrm{0} = 67.4~\mathrm{km~s^{-1}}~\mathrm{Mpc}^{-1}$ and $\Omega_\mathrm{m} = 0.315$ \citep{PlanckCollaboration2018}. The errors quoted in this paper are given at the $1\sigma$ confidence level.

\section{The local GRB sample}
\label{sec:sample}

\subsection{Construction of the sample} 
\label{sub:sample_construct}

The construction of our local GRB sample starts with the selection of 33 GRBs with a redshift smaller than $z=0.3$ in the public GRB table made available by J.~Greiner\footnote{\url{http://www.mpe.mpg.de/~jcg/grbgen.html}} (up to the end of 2019). The redshift $z=0.3$ has been chosen so that the number of short GRBs included in the sample is approximately $10$, giving a statistically meaningful sample of local events.
We limited ourselves to GRBs with a secured redshift measured either on their host galaxy or with the GRB afterglow spectroscopy.\footnote{GRB~150424A, whose redshift is not secure \citep{Tanvir2015_150424A}, has been excluded} After a rapid survey of the literature, we have added GRB~040701 at $z = 0.2146$ \citep{Kelson2004_040701, Soderberg2005_040701}, reaching a sample with 34 GRBs.
We note that this GRB sample is by no means exhaustive.

The sample has been divided into \Nlong\ long GRBs (LGRBs) and \Nshort\ short GRBs (SGRBs).
The list of events in each class can be found in Tables \ref{tab:detection_long} (for LGRBs) and \ref{tab:detection_short} (for SGRBs). 
It is important to note that the short/long classification of GRBs is based on the literature, and not only on their duration and spectral hardness.

To this list we have added 3 GFs from SGRs located in our galaxy or the Large Magellanic Cloud, and 3 giant flare candidates located in nearby galaxies, all taken from \cite{Hurley2011}, and another candidate GF~200415A that has been recently discovered \citep{Svinkin2020_200415A}. 
They are listed in Table \ref{tab:detection_gflare}.
Confirmed giant flares are associated with a magnetar detected before the giant flare (\mbox{GF 790305, GF 980827, GF 041227}), the other events are considered as candidate giant flares.
Giant flares from soft gamma repeaters typically consist of a short hard pulse followed by an oscillating tail several orders of magnitude weaker \citep{Mazets2005_041227, Tanaka2007b_980827}.
In all cases, we consider only the properties of first short pulse here, as this is the only component that can be seen at distances greater than a few hundreds of kiloparsecs.

GRBs will be referred by GRB~YYMMDD in the remaining of this paper,
while giant flares will be designated with the same date-based denomination preceded by GF, irrespective of their classification as secure or candidate giant flares. For example, the giant flare from SGR 1806-20 will be designated as GF~041227. The correspondence between the SGR origin of the giant flares and the date-based denomination can be found in Table \ref{tab:intrinsic_gflare}. 

Within these categories, some sub-categories have been created to emphasize specific characteristics for some of the bursts.

GRB~150518A, GRB~100316D and GRB~060218 have their high-energy emission recorded for thousands of seconds and they have therefore been placed in the ultra-long GRBs group (ulGRB) \citep{Campana2006_060218, Starling2011_100316D, Gendre2013, Levan2014, Sakamoto2015_150518A, Dagoneau2020}. 
Some other GRBs are classified as XRFs because of their soft prompt emission, such as GRB~020903 \citep{Sakamoto2005} and GRB~040701 \citep{Barraud2004_040701,Pelangeon2008_040701}. GRB~031203 is also sometimes referred to as an XRF \citep{Watson2004_031203}. However, since its soft X-ray emission cannot be unambiguously associated with the prompt emission \citep{Watson2006_031203} and its peak energy measured with INTEGRAL is $> 100$~keV \citep{Sazonov2004_031203, Ulanov2005_031203}, we stay conservative and do not include it in our list of XRFs. GRB~060218 and GRB~100316D are also sometimes referred as XRFs \citep{Soderberg2006_060218_nature, Cano2011_100316D}. However, we consider here that their main characteristic is their duration, and not the softness of their spectrum, and we class them in the ulGRB category. Finally, for some GRBs, the short/long paradigm based on the duration does not exactly apply, and they have been classified according to the consensus found in the literature:

\begin{itemize}

\item GRB~050709 displays a short hard spike followed by a softer extended emission, so that \Tqvd$ = 160 \sec$ \citep{Villasenor2005_050709},\footnote{According to \cite{BATcat2016}, \Tqvd refers to the duration of the burst over which $90\%$ of its photon fluence has been emitted.} an unusually long duration for a short GRB. However, as the spectral properties of the first spike are typical of short GRBs, and since \mbox{GRB~050709} was not associated with a surpernova \citep{Fox2005_050709}, but with a possible kilonova \citep{Jin2016_050709}, 
we classify it as a short GRB. In this paper, it will be considered as a representative member of the sub-category of ``short GRBs with extended emission'' (eeSGRB).
\item GRB~050724 has also been assigned to the eeSGRBs sub-category, because of its extended emission. This is further justified by the fact that the fluence of the tail represents only 10\% of the total burst fluence \citep{Barthelmy2005_050724}, explaining why the \Tqvd remains equal to $\approx 3 \sec$.
\item GRB~060614 can be considered as a long-duration burst with \Tqvd $ \approx 109 \sec$ and a first peak lasting $\sim 5\sec$, but it has a temporal lag and peak flux close to those of short GRBs, in addition to the lack of associated supernova \citep{Zhang2006_060614}. While its origin remains uncertain \citep{Xu2009_060505, GalYam2006_060614, Gehrels2006_060614}, it will be considered as a long GRB in the rest of this paper.

\end{itemize}

Beyond the classification in duration, a clear demarcation in energy is visible in Fig. \ref{fig:Eiso_all}, where three low-energy GRBs are detected at distances that are typically ten times closer than the bulk of the ``classical'' population. With isotropic energies \eiso\ below $10^{48}~\mathrm{erg}$, GRB~170817A (short), GRB~980425 (long) and GRB~111005A (long) are much fainter than the bulk of the long or short GRB population. Although the criterion is based on the energetic rather than luminosity of the bursts, they have been placed in a "low-luminosity" group (llSGRBs and llLGRBs) to remain consistent with the literature appellation. 
In summary, three groups of high-energy transients have been considered in the local Universe, which are the long GRBs, the short GRBs and the SGR giant flares. 
The long GRBs are divided into 4 sub-categories: 17 classical long GRBs (LGRBs), 3 ultra-long GRBs (ulGRBs), 2 X-Ray Flashes (XRFs) and 2 low-luminosity long GRBs (llLGRBS). Similarly, the short GRBs are divided into 3 categories: 7 classical short GRBs (SGRBs), 2 short GRBs with an extended emission (eeSGRBs) and 1 low-luminosity short GRBs (llSGRBs).  
This classification into long GRBs, short GRBs and giant flares will be used throughout the article.

\subsection{A deeper look into the local sample} 
\label{sub:sample_insight}

The main detected features of the \Nall\ transients in our sample are summarized in Tables \ref{tab:detection_long}, \ref{tab:detection_short} and \ref{tab:detection_gflare} (in appendices A, B, C), for the long GRBs, short GRBs and SGR giant flares respectively. For each event, the tables indicate the satellite(s) that have detected it, the event sub-category, the presence/absence of an afterglow or supernova and references for the host galaxy and the burst duration. Overall $30$ of the \Nall\ GRBs of our sample have been detected by \textit{Swift}/BAT \citep{Barthelmy2005}, $15$ by \textit{Wind}/KONUS \citep{Aptekar1995_konus}, $8$ by \textit{Fermi}/GBM \citep{Meegan2009_gbm}, $7$ by \textit{INTEGRAL} \citep{Winkler2003}, $4$ by \textit{HETE-2} \citep{Atteia2003_fregate, Shirasaki2003}, $3$ by \textit{CGRO}/BATSE, and $1$ by either \textit{BeppoSAX} \citep{Boella1997_beppo}, \textit{RHESSI} \citep{Lin2002_rhessi} or \textit{MAXI} \citep{Matsuoka2009_maxi}.

The time scale for which the peak flux has been given depends on the category of transient: long GRBs have their peak flux measured over a 1s duration, whereas the peak flux of short GRBs and SGR giant flares is measured over a duration of 64ms. The rationale for this choice lies in the fact that the peak flux of short GRBs and SGR giant flares is underestimated by taking a one-second interval, their total duration being usually below one second.

For \textit{Swift}/BAT, the peak flux was only available in the 1s timescale in the \textit{Swift}/BAT catalog \citep{BATcat2016}. Unless specified, the 64ms peak flux used in Table \ref{tab:spectrum_short} has been calculated from the 64ms light curves, normalized to recover the 1s peak flux on the 1s timescale.

A dedicated column $P_\mathrm{flux, norm}$ has been created in Tables \ref{tab:spectrum_long}, \ref{tab:spectrum_short} and \ref{tab:spectrum_gflare}, containing the peak flux information in the common $15$--$150~\mathrm{keV}$ energy range, to compare the high-energy transients in Fig. \ref{fig:T90_Pflux}. This peak flux is computed from the other peak flux column and the corresponding transient spectrum from Tables \ref{tab:spectrum_long}, \ref{tab:spectrum_short} and \ref{tab:spectrum_gflare}.

GRB~050709, GRB~060614, GRB~180728A and \newline GRB~190829A exhibit a clear decomposition into two distinct parts, with distinct spectral models. The variability in their spectrum is so significant that fitting a spectral model for the total duration of the burst (time-integrated spectrum) gives poor results compared to a time-resolved fit. Thus, for these GRBs, the spectral properties of each of the two parts is given separately.

The peak energy $E_\mathrm{peak}$ presented in Tables \ref{tab:spectrum_long}, \ref{tab:spectrum_short} and \ref{tab:spectrum_gflare} comes from the best spectral model of the time-integrated spectrum, which can be the Band function \citep{Band1993}, a cutoff power law (CPL, sometimes also called Comptonized model) or in the case of SGR GF an optically-thin thermal bremsstrahlung (OTTB) model. Few particular cases need to be noted:
\begin{itemize}
\item For \mbox{GRB~191019A}, the \textit{Swift}/BAT time-integrated spectral model extracted from the catalog has no peak energy $E_\mathrm{peak}$. However, the peak flux spectral model has one, this value of $E_\mathrm{peak}$ has been considerFed instead. 
\item Similarly, the \textit{Fermi}/GBM catalog does not contain any model with a peak energy for GRB~150101B. However, the work of \cite{Burns2018_150101B} indicates that the first 16 ms of the burst can be fitted by Comptonized spectral model with a peak energy $E_\mathrm{peak} = 550 \pm 190~\mathrm{keV}$. Because the burst has a short duration (\Tqvd $= 0.23 \sec$) and the tail only represents $10$\% of the global fluence \citep{Burns2018_150101B}, the peak flux spectrum and the time-integrated spectrum should be rather similar. The peak energy of the time-integrated spectrum of the burst has therefore been chosen to be $E_\mathrm{peak} = 550 \pm 190~\mathrm{keV}$, equal to the peak energy of the peak flux spectrum.
\item For GRB~031203, the spectral model used is a power law but the work of \cite{Sazonov2004_031203} identifies a lower limit for the peak energy. The peak energy lower limit has been kept, even if the spectral model indicated in the table remains a power law.
\item \mbox{GRB~120422A}, \mbox{GRB 040701} and \mbox{GRB 020903} only have upper limits on their peak energies. Their time-integrated spectral models remain power-laws, but the upper limit on \epeak\ is indicated in the spectral properties.
\end{itemize}

Tables \ref{tab:intrinsic_long}, \ref{tab:intrinsic_short} and \ref{tab:intrinsic_gflare} summarize the intrinsic properties of the local transients.
The intrinsic peak energy $E_\mathrm{peak,i}$ is computed from the peak energy and the redshift, with the classical formula $E_\mathrm{peak,i} = E_\mathrm{peak}\times\left( 1+z \right)$.
The isotropic energy \eiso\ is directly taken from the literature. However, when the information was not available or when the assumptions taken in the paper implied the use of a different time-integrated spectral model than the one mentioned in Tables \ref{tab:spectrum_long}, \ref{tab:spectrum_short} and \ref{tab:spectrum_gflare}, the isotropic energy has been calculated with the following equations. 

\begin{eqnarray}
    \label{equ:eiso}
    E_\mathrm{iso} = S_\mathrm{bol} \times \frac{4\pi\ D_\mathrm{l}(z)^2}{1+z}\\
    \label{equ:sbol}
    \mathrm{with}\ S_{\mathrm{bol}} = S_{E_{\mathrm{min}}\rightarrow E_{\mathrm{max}}}\times \frac{\int_{\frac{1}{1+z}}^{\frac{10^4}{1+z}}E\ N(E)\ dE}{\int_{E_{min}}^{E_{max}}E\ N(E)\ dE}
\end{eqnarray}
Equ. \ref{equ:eiso} gives the isotropic energy that has been emitted by the transient in its reference frame, from an initial fluence $S_{E_{\mathrm{min}}\rightarrow E_{\mathrm{max}}}$ measured in the energy range $\left[E_\mathrm{min};\ E_\mathrm{max}\right]$, a spectral model $\mathrm{N(E)}$ and a redshift measurement $z$ giving a luminosity distance $D_\mathrm{l}(z)$. In this equation, the term $S_\mathrm{bol}$ represents the fluence extrapolated to a common energy range from $1~\mathrm{keV}$ to $10~\mathrm{MeV}$ in the source frame. The term $1+z$ below each energy appears because the spectral model $N(E)$ is defined in the observer frame, so a redshift correction has to be taken into account for the energy released in the transient reference frame. 

When there is no peak energy in the spectral model used for the calculation (power-law model), the isotropic energy calculated with this formula tends to overestimate the low- and/or high-energy contributions, these cases are discussed below. 
\begin{itemize}
    \item For GRB~031203, a power-law spectrum is given with a lower limit on the peak energy. To calculate the isotropic energy given in Table \ref{tab:intrinsic_long}, a CPL with $\alpha=-1.63$ and $E_\mathrm{peak}=300~\mathrm{keV}$ has been used, giving $E_\mathrm{iso}=1.4\times 10^{50}~\mathrm{erg}$. Assuming the same model but with $E_\mathrm{peak}$ = 200~keV and 500~keV, the isotropic energy becomes $E_\mathrm{iso}=1.3\times 10^{50}~\mathrm{erg}$ and $E_\mathrm{iso}=1.7\times 10^{50}~\mathrm{erg}$, respectively. Therefore the value of $E_\mathrm{peak}$ has little impact on the calculated isotropic energy.
    \item For transients with a power-law index $\Gamma>-1.6$, $E_\mathrm{iso}$ has not been calculated. Without any peak energy, a unique power-law with such an index will overestimate high- energy contributions, jeopardising the isotropic energy calculation. GRB~150518A ($\Gamma=-1.3$), GRB~080905A ($\Gamma=-1.33$), GRB~060502B ($\Gamma=-0.98$), GRB~050826 ($\Gamma=-1.16$) and GRB~050509B ($\Gamma=-1.57$) are also included.
    \item For transients with a power-law index $\Gamma\leq-1.6$, $E_\mathrm{iso}$ has been calculated. We consider that the high-energy contribution is not overestimated with such an index. It includes GRB~191019A ($\Gamma=-2.25$), GRB~130702A ($\Gamma=-2.44$); GRB~111225A ($\Gamma=-1.7$), GRB~051109B ($\Gamma=-1.97$), GF~050906 ($\Gamma=-1.66$) and GRB~050724 ($\Gamma=-1.89$).
\end{itemize}

\begin{figure*}[ht]
\subfigure[]{
\label{fig:Eiso_all}
\centering
\epsscale{0.98}
\plotone{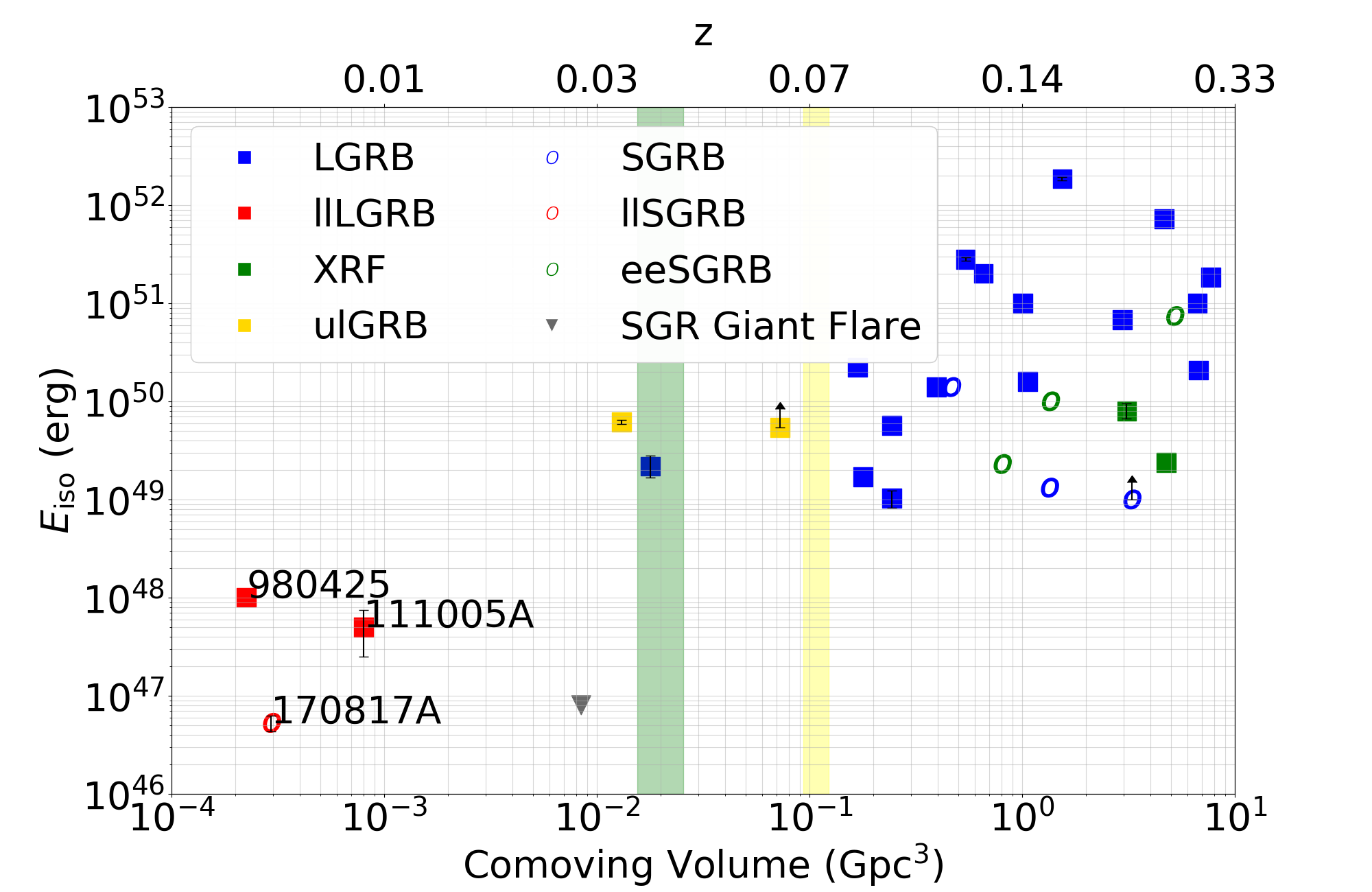}
}
\subfigure[]{
\centering

\plotone{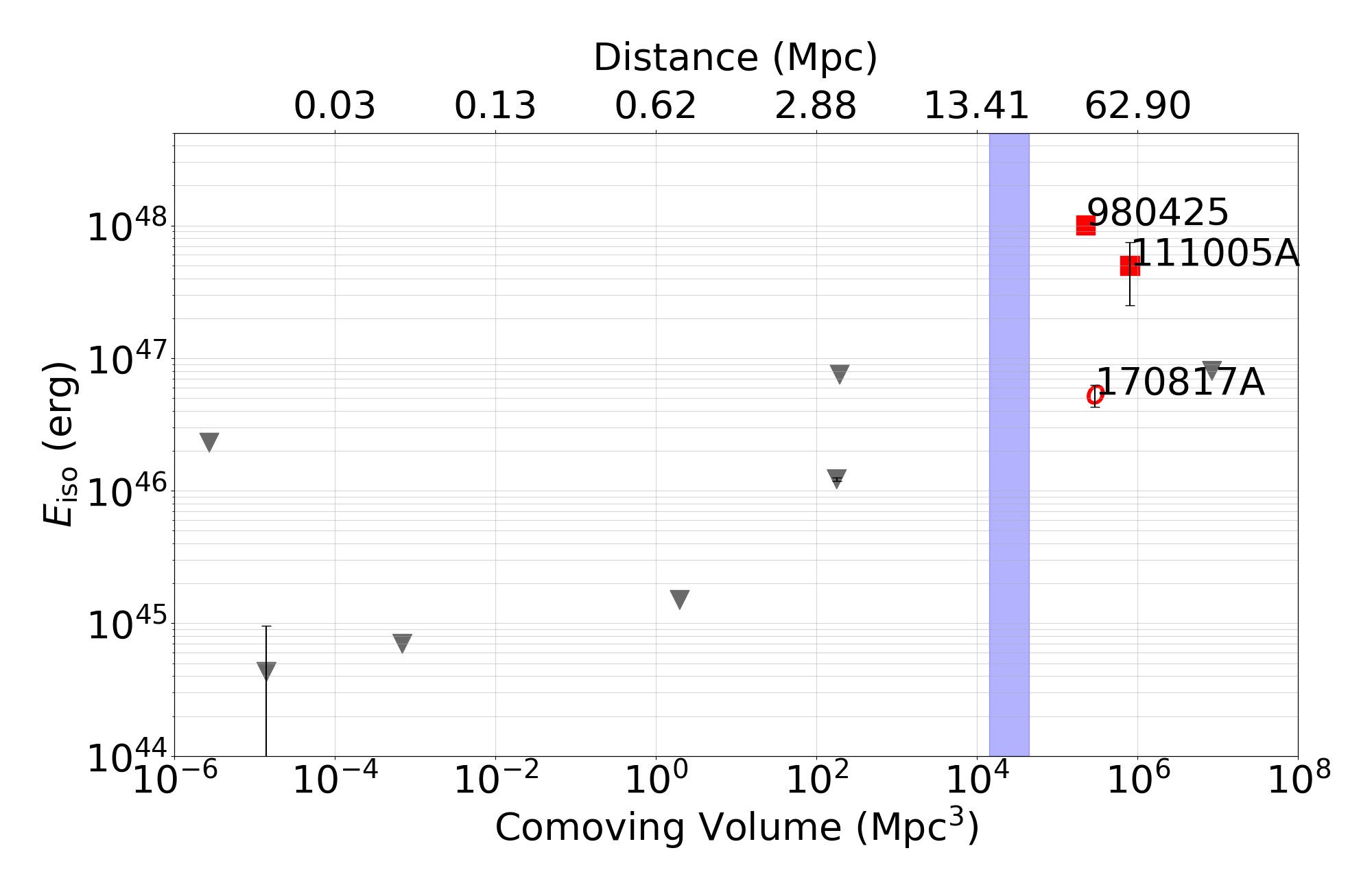}
\label{fig:Eiso_gflare}}
\caption{Intrinsic properties of GRBs in our sample: \eiso\ as a function of the co-moving volume for high-energy transients. The symbols represent the three main categories of GRBs: full squares for long GRBs, empty circles for short GRBs and triangles for SGR giant flares. The colors describe the sub-categories defined in Table \ref{tab:detection_long} and Table \ref{tab:detection_short}. (a) GRBs with $\mathrm{z}<0.3$. The green and yellow bands represent the O4 LIGO sensitivity limits for respectively a NS-NS and a BH-NS merger. (b) SGR giant flares and low-luminosity GRBs. The blue band represents the approximate distance of the Virgo Cluster. }
\end{figure*}

Figures \ref{fig:Eiso_all} and \ref{fig:Eiso_gflare} display selected parameters of Tables \ref{tab:intrinsic_long}, \ref{tab:intrinsic_short} and \ref{tab:intrinsic_gflare}. Two figures are needed to ensure the readability while covering the full distance range of local transients. On Fig. \ref{fig:Eiso_all}, low-luminosity events such as GRB~170817A stand clearly apart. Indeed, except for GF~050906 (which is not a GRB), no GRB has been measured from a co-moving volume ranging from $10^{-3}$ to $\sim 5\times 10^{-2}~\mathrm{Gpc^3}$.
We also note that the most energetic events in our sample have \eiso $\sim 10^{52}$ erg, well below the maximum isotropic energy measured for GRBs at redshifts z $\geq 1$, \eiso $\sim 3 \times 10^{54}$ erg \citep{Atteia2017}.

\begin{figure}[ht]
\centering
\epsscale{1}
\plotone{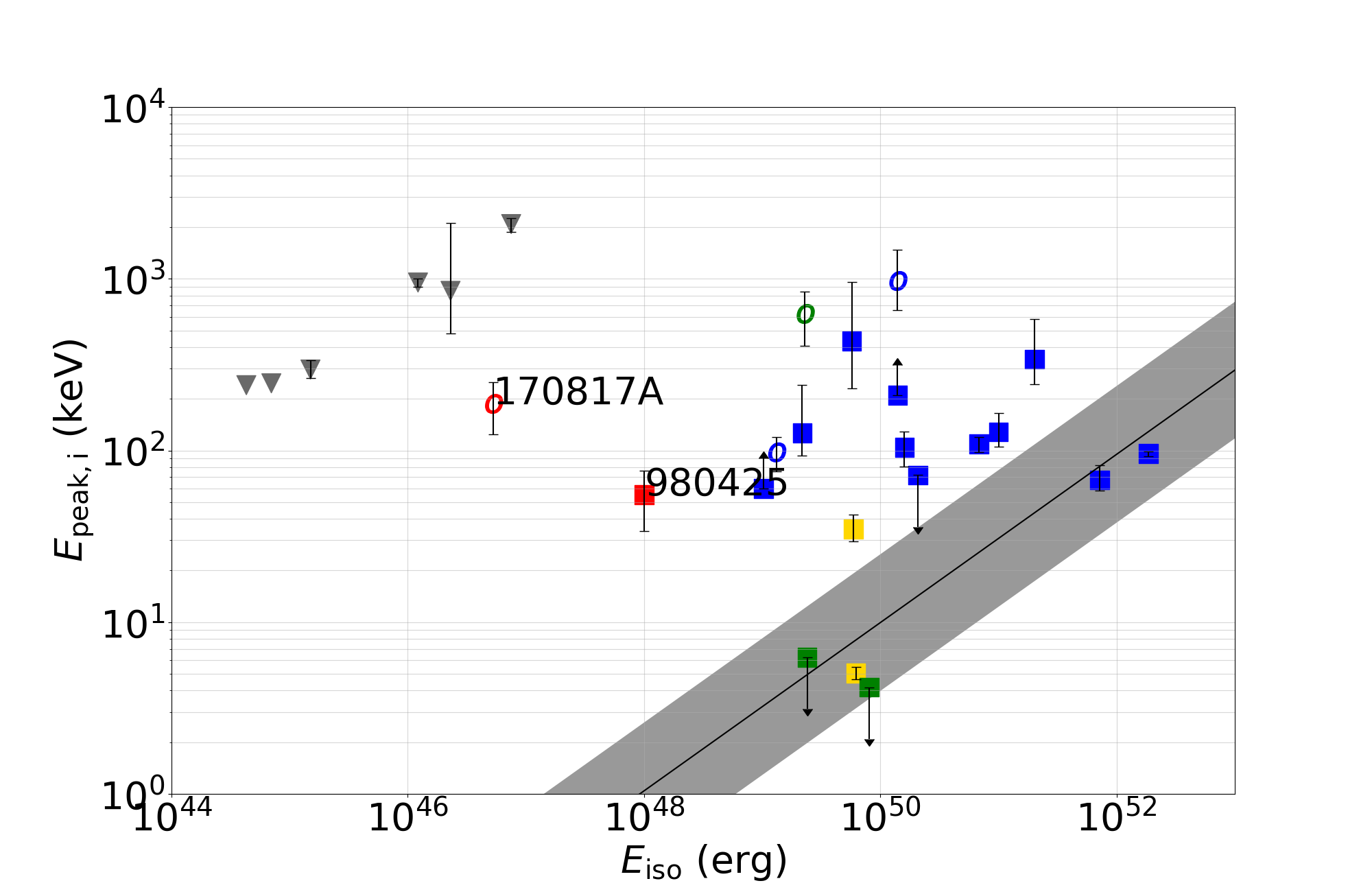}
\caption{Intrinsic properties of GRBs in our sample: the Amati relation. The black line has been plotted according to \mbox{\cite{Amati2006}}, the grey area representing a vertical logarithmic deviation of 0.4 compared to the best-fitting power law values for the Amati relation ${E_\mathrm{peak,i}} = 95\times E_\mathrm{iso}^{0.49}$. Color-coding and shape-coding are the same as in Fig. \ref{fig:Eiso_all}.}
\label{fig:Amati}
\end{figure}

\begin{figure}[ht]
\centering
\plotone{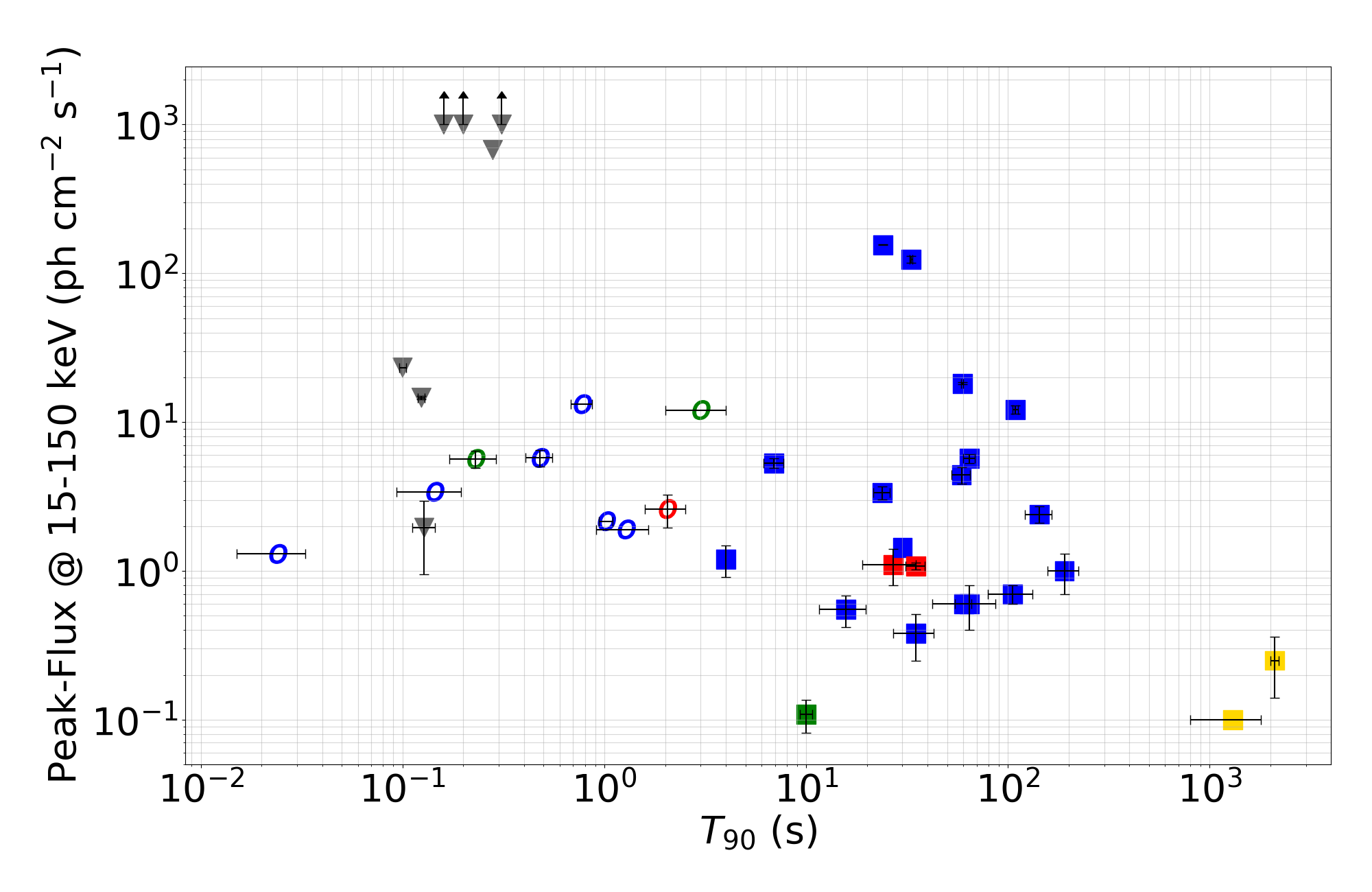}
\caption{Observed properties of GRBs in our sample: \Tqvd at $15$--$350~\mathrm{keV}$ vs Peak Flux ($15$--$150~\mathrm{keV}$). The SGR Giant Flare peak flux and the short GRBs peak flux are measured on a 64ms long time interval and the long GRBs one on a 1s long time interval. The transients whose peak flux has not been measured in the $15$--$150$~keV energy band have been translated thanks to the fluence model provided in Tables \ref{tab:spectrum_long}, \ref{tab:spectrum_short} and \ref{tab:spectrum_gflare}. Color-coding and shape-coding are the same as in Fig. \ref{fig:Eiso_all}.}
\label{fig:T90_Pflux}
\end{figure}

Figure \ref{fig:Amati} represents the GRB distribution in the \epeak --\eiso\ plane for GRBs with measured peak energy. Some GRBs such as GRB~060218 and GRB~030329 follow the Amati relation \citep{Amati2006} down to low values of \eiso . However, other long GRBs such as GRB~98042, GRB~161219B or GRB~171205A are outliers of this relation. At $\mathrm{z}<0.3$, it is possible to detect much fainter GRBs, allowing the observation of a population with more variety than at higher redshift. This diversity includes GRBs with lower luminosities that are outliers of the Amati relation, in agreement with the work of \cite{Heussaff2013}. SGR GFs seem to follow a relation between \eiso and $E_\mathrm{peak,i}$ reminiscent of the Amati relationship (see \cite{Zhang2020_amati_gflare} for a discussion about the possible origin of such correlation).

In Fig. \ref{fig:T90_Pflux}, the usual bi-modal population of long GRBs /short GRBs is less clear, as most of the $T_\mathrm{90}$-values come from the \textit{Swift}/BAT instrument. Indeed, by extending its flux sensitivity as well as its low-energy threshold, BAT can see parts of the prompt emission that would have been buried into the noise for BATSE and previous GRB missions. With its low energy threshold, ECLAIRs might also be able to observe longer prompt emission for short GRBs.  

\section{The \textit{SVOM}/ECLAIRs instrument}
\label{sec:eclairs}
The \textit{SVOM} mission is a Sino-French mission dedicated to the observation of GRBs and other high-energy transients, which will be launched in mid-$2022$.
It will operate on a LEO orbit ($625~\mathrm{km}$, an orbital period of $96~\mathrm{min}$) with a $30$ degree inclination and quasi anti-solar pointing strategy. The \textit{SVOM} spacecraft will be operated in a similar manner as \textit{Swift}, encompassing two wide-field gamma-ray detectors: ECLAIRs \citep{Godet2014} and the Gamma-Ray Burst Monitor GRM \citep{Zhao2012}, and two narrow-field instruments: the Microchannel X-ray Telescope MXT \citep{Gotz2014_MXT} and the Visible Telescope VT \citep{Wu2012_VT}, for the rapid follow-up of GRBs. In addition to the satellite, the mission will benefit from a dedicated ground segment with the Ground-based Wide Angle Camera GWAC \citep{Turpin2020}, the COLIBRI telescope (Catching OpticaL and Infrared BRIght transients, \citealt{Fuentes2020_COLIBRI}) and the Chinese-Ground Follow-up Telescope C-GFT.

\begin{figure}[t]
\centering
\plotone{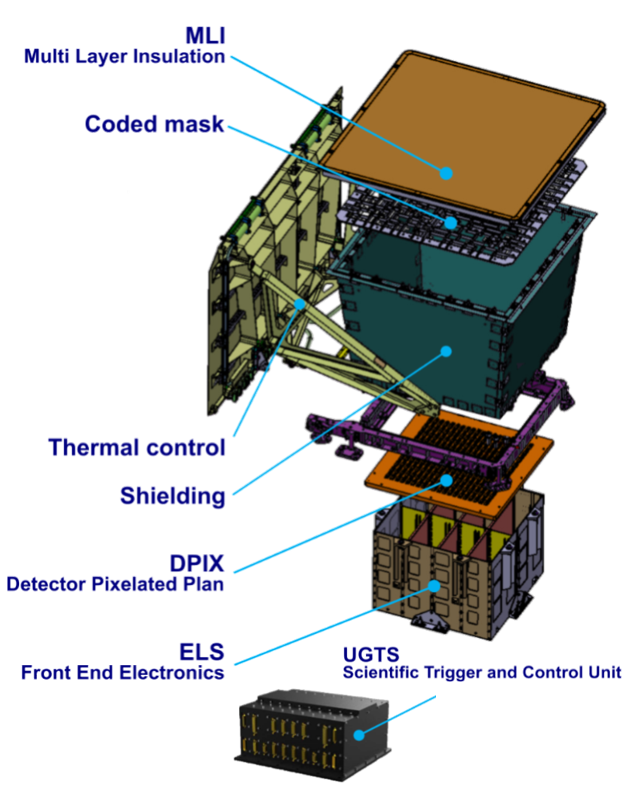}
\caption{Schematic view of the ECLAIRs instrument.}
\label{fig:eclairs}
\end{figure}

ECLAIRs is the wide-field hard X-ray imager of \textit{SVOM}, in charge of the autonomous detection and localisation of GRB prompt emission in near real-time. It can be considered as a smaller analog of the \textit{Swift}/BAT (Burst Alert Telescope, \citealt{Barthelmy2005}). The main features of the instrument and its performances have been described in \citet{Godet2014, Schanne2019, Mate2019, Dagoneau2020}, and they are summarized in Table \ref{tab:eclairs}. In a few words, ECLAIRs encompasses a $\sim 1000~\mathrm{cm}^{2}$ detection plane with its readout electronics, which looks at the sky through a coded-mask (see Fig. \ref{fig:eclairs}). Its energy range extends from $4$ to $150~\mathrm{keV}$, and the instrument is operated in photon counting mode. A passive lateral Pb/Al/Cu shield blocks the hard X-ray radiation originating from outside the \mbox{2 sr} field of view and provides fluorescence lines useful to monitor the energy scale of the detectors. 
A digital processing unit (UGTS)\footnote{UGTS is a French acronym for the ECLAIRs data processing unit: ``Unit\'e de Gestion et de Traitement Scientifique''.} controls the detection plane and analyses in near real-time the detected events to look for high-energy transients. 

The detection plane is made of 6400 pixels of CdTe ($4 \times 4$~mm$^2$ and 1~mm thick, \citealt{Remoue2010}). It is located \mbox{46 cm} below a $54 \times 54$~cm$^2$ coded mask, made of 4 quadrants with $23 \times 23$ elements each, of which 40\% are open.
This geometry guarantees a point spread function with a full width at half maximum (FWHM) of 52~arcminutes and a point source localization error (PSLE) better than  13~arcminute  (90\%  confidence  level)  for sources at the detection limit ($SNR\sim 7\sigma$). ECLAIRs is thus a compact instrument, which has been optimized considering the limited resources (mass, power, volume) available on the \textit{SVOM} spacecraft. 

The UGTS configures the instrument and searches for new transient sources within reconstructed sky images of the field of view (FoV). When a new source is detected, it alerts immediately the satellite about its location. The position is then sent to the ground through the VHF antenna network to alert the community that a transient has been detected and to start the follow-up. Hard X-ray transients are identified as count-rate excesses detected on the full detection plane or just a fraction of it. Count-rate excesses are monitored in 4 energy bands on time scales ranging from 10~ms to 20~s, the whole process aiming at detecting short transients. When a count-rate excess is detected, a sky image is constructed in the same time and energy intervals, and the trigger is validated if a significant excess is found in the sky image. Longer transients are directly searched in sky images constructed cyclically on time scales ranging from 20~s to 20~minutes. A detailed description of the ECLAIRs trigger can be found in \citet{Schanne2019}. Finally, all recorded events are sent to the ground, allowing delayed data analysis on the ground for the accurate calibration of the detectors and offline searches of faint transient sources undetected on-board.

Two major challenges of the instrument are the 4~keV energy threshold and the on-board detection of transients on top of a strongly varying background modulated by the transit of Earth in the field of view.
The 4~keV energy threshold is imposed by the requirement to be sensitive to X-Ray Flashes and highly redshifted GRBs. 
This requirement puts significant constraints on the coded mask structure, whose transparent elements must be fully open, and on the detectors, which must have very low intrinsic noise. The mask requirements have been solved with a sandwich structure made of a punched plate of tantalum inserted between two complex titanium pieces.
For the detectors, ECLAIRs uses Schottky CdTe detectors with intrinsically low leakage current operated at $- 20^\circ$~C, which are powered and read out by a low noise ASIC \citep{Gevin2009, Lacombe2018}. 
The detection of GRBs on top of a highly variable background is based on the use of timing and image information as described in \citet{Schanne2019}. 
Detailed simulations based on realistic instrument background and performance including both the real trigger software and several GRB catalogs predict the detection of about $60~\mathrm{GRB\ yr ^{-1}}$ \citep{Wei2016}, several non-GRB extra-galactic transients, dozens of AGNs and hundreds of galactic X-ray transients and persistent sources (this last number is strongly dependent on the pointing strategy because X-ray transients are concentrated in the galactic plane).

\section{Signal-to-noise ratio computation with \textit{SVOM}/ECLAIRs}
\label{sec:snr}

The purpose of this section is to assess the detectability by \textit{SVOM}/ECLAIRs of each high-energy transient in our local sample, by computing its count and image SNRs for a detection in the fully-coded FoV, hereafter called "best SNRs". The fully coded FoV represents the fraction of the sky in which a source will be able to completely illuminate the detection plane of ECLAIRs, maximising the number of photons received by the instrument. On \textit{SVOM}/ECLAIRs, the fully-coded FoV is a square with a side of $22~\deg$. In addition to the best SNR, the fraction of the ECLAIRs FoV in which the GRBs is detectable with a SNR $\geq 6.5$ has also been computed.

\subsection{Methodology of the SNR calculation}
\label{sub:methodology}

The SNR calculation proceeds along the following steps:
\begin{itemize}
\item First, the expected background in space is estimated with simulations based on to the Particle Interaction Recycling Approach PIRA \citep{Mate2019}.
\item Then the photon events are simulated using the transient spectral properties given in Tables \ref{tab:spectrum_long}, \ref{tab:spectrum_short} and \ref{tab:spectrum_gflare}.
\item The SNR is then computed, taking into account the event counts on the detector and the background generated in the previous steps.
\item A sky image is created in the energy range and time interval for which the count SNR has the maximum value. The SNR of the transient source in this image is then computed, giving the image SNR of the transient.
\end{itemize}

This operation can be performed for transients placed at different locations in the ECLAIRS FoV, allowing us to estimate in which fraction of the FoV the high-energy transient will be detected.

\subsubsection{Background simulation}
\label{sub:bkg_simulation}

\begin{figure}[t]
\centering
\plotone{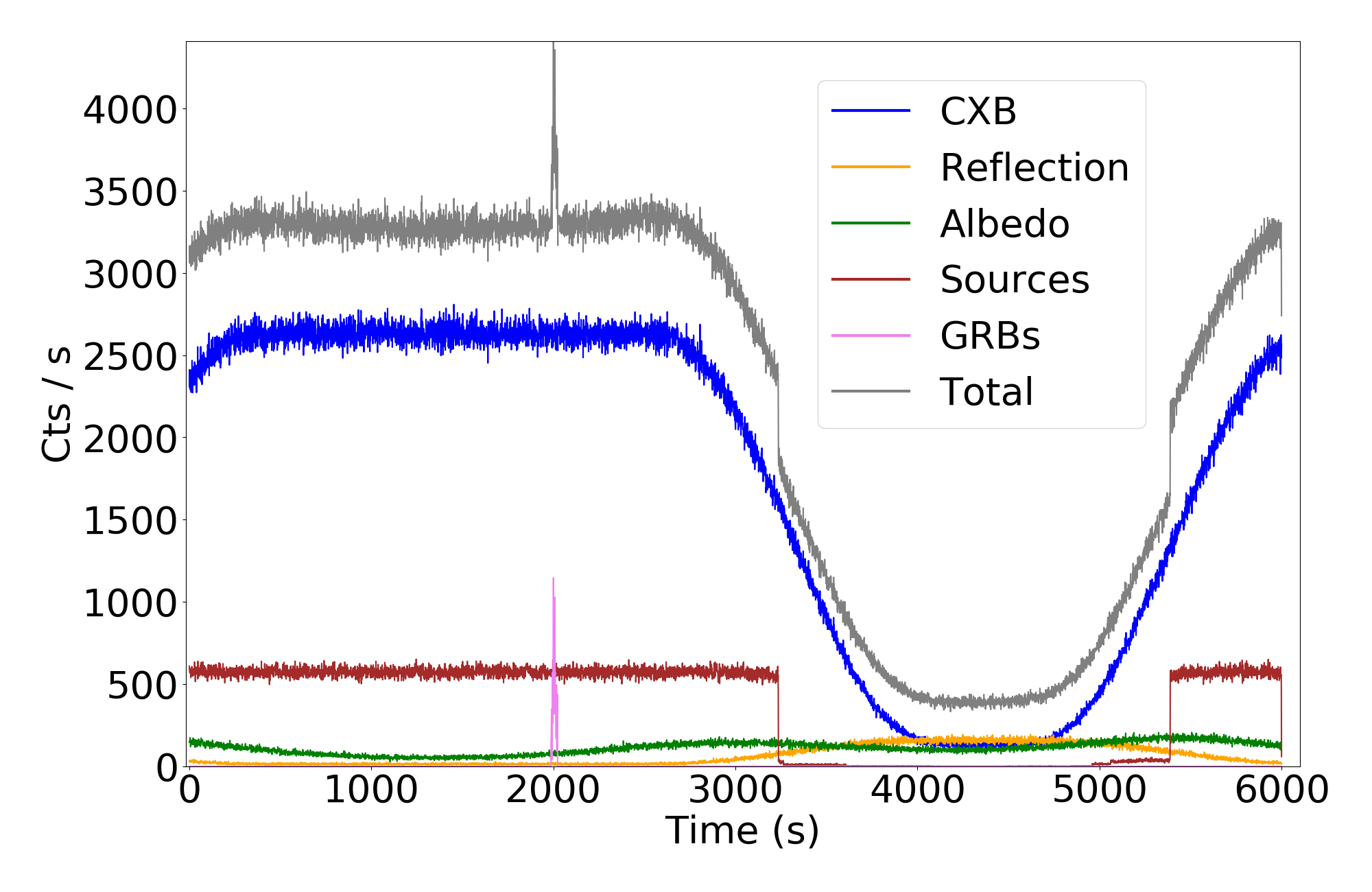}
\caption{Example of PIRA background light curve (grey curve), with GRB~111005A (pink curve) superimposed on it. The total background is the addition of several components: the Cosmic X-ray Background, which is modulated by Earth transits in the FoV (blue curve); Albedo and reflection (yellow and green curves), which are also modulated by Earth transits, and the Crab nebula, which is hiding and rising behind the Earth (brown curve). }
\label{fig:PIRA_lc}
\end{figure}

The sensitivity of ECLAIRs is limited by the background counts, which are dominated by the cosmic X-ray background CXB \citep{Churazov2007, Ajello2008}, with additional components from the CXB reflection on the Earth atmosphere \citep{Churazov2007} and from the interaction of cosmic rays with the Earth's atmosphere, called albedo \citep{Sazonov2007, Ajello2008}. These background events are all simulated using PIRA \citep{Mate2019}, a software that relies on a pre-computed database of particle-instrument interactions simulated using the GEANT4 toolkit \citep{Allison2016_GEANT4} to estimate the dynamical evolution of the background along the orbit. In the following, the high-energy transient is placed in the portion of the orbit when the Earth is completely out of the FoV, thus maximizing the number of events generated by the CXB. An example of generated orbit can be seen in Fig. \ref{fig:PIRA_lc}.

\subsubsection{Other X-ray/gamma-ray sources}
\label{sub:xsources}
Once the background is generated, it is possible to add photons coming from known X-ray sources \citep{Dagoneau2020_xsources} by using ray-tracing algorithms.
Some X-ray sources have a significant contribution, reaching for example $\sim 600~\mathrm{cts/s}$ in the 4--150$~\mathrm{keV}$ energy range for the Crab pulsar and its nebula. The simulation conditions chosen here represent a standard orbit with an extra-galactic pointing law \citep{Wei2016}, which means that the instrument will avoid observing the Galactic plane throughout this orbit. In these conditions, there is no strong X-ray source in the field of view of ECLAIRs. The performance could be different for transients located close to some bright galactic sources such as the Crab or Sco-X1. 

\subsubsection{High-energy transient photon list generation}
\label{sub:movegrb}
\begin{figure}[t]
\centering
\plotone{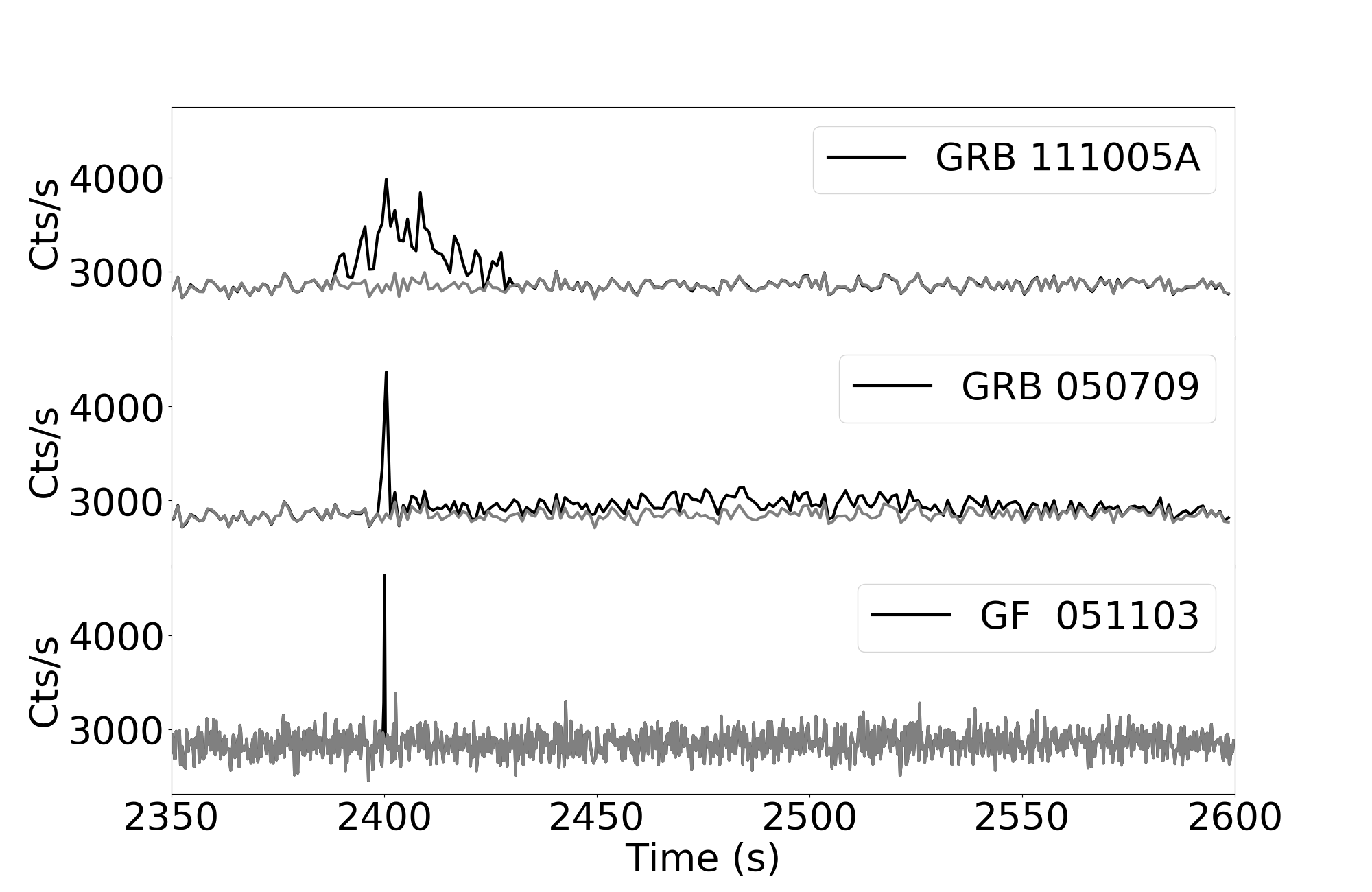}
\caption{Lightcurves of GRB~111005A, GRB~050709 and GF~051103 obtained with the \textit{movegrb} algorithm. The top and middle curves have a bin timescale of $1\sec$, while the bottom one has a bin timescale of $0.2\sec$. The grey lightcurves represent the background simulated without the GRB.}
\label{fig:GRBs_lc}
\end{figure}

The first step of the transient simulation consists in translating the source spectrum from its observed energy bands into \textit{SVOM}/ECLAIRs energy bands. The tool used, called \textit{movegrb} \citep{Antier2016}, allows to generate a list of photons in the ECLAIRs energy range, from a transient with a known spectrum and light curve. The spectral properties of the HE transients come from Tables \ref{tab:spectrum_long}, \ref{tab:spectrum_short} and \ref{tab:spectrum_gflare}. For the light curves, there are several sources:
\begin{itemize}
\item To keep the consistency of the light curve morphology with respect to the morphology that would be observed by \textit{SVOM}/ECLAIRs, we preferentially use the light curves measured by \textit{Swift}/BAT, which has an energy range similar to ECLAIRs. The light curves are taken from the \textit{Swift}/BAT light curve catalog \citep{BATcat2016}, using the interactive BAT light curves\footnote{BATlightcurves, Sakamoto T., Barthelmy S., see \url{https://swift.gsfc.nasa.gov/results/batgrbcat/}}  available over \mbox{1~s} and \mbox{64~ms} timescales. Transients with \Tqvd larger than  10~s have been sampled with 1~s bins, while the transients with shorter durations have been sampled with 64~ms bins. Exceptions are made for GRB~150101B and GRB~160821B whose morphology was much better sampled at 16 ms because of their small duration.
\item The light curves from transients that have not been seen by \textit{Swift}/BAT are extracted from the instrument which displays a well-sampled light curve. For GRB~130702A the \textit{Fermi}/GBM light curve has been taken directly from \url{https://heasarc.gsfc.nasa.gov/FTP/Fermi/data/gbm/bursts/}.
\item For some transients, the light curves have been directly extracted from the associated paper. GRB~980827 \citep{Tanaka2007b_980827}, GRB~980425 \citep{Pian2000_980425}, GRB~020903 \citep{Sakamoto2004_020903}, GRB~031203 \citep{Sazonov2004_031203}, GRB~040701 \citep{Barraud2004_040701}, GRB~041227 \citep{Mazets2005_041227} and also GRB~050709 \citep{Villasenor2005_050709}, GF~051103 \citep{Hurley2010_051103}, GF~070201 \citep{Mazets2008}, GRB
~170817A \citep{Goldstein2017_170817A} and GF~200415A \citep{Frederiks2020_200415A}.
\end{itemize}
The lightcurves used can be found in the appendix in Fig. \ref{fig:LC1}.1.
Light curves in counts are normalized by the \textit{movegrb} simulator to recover the fluence measured in the ECLAIRs energy range. The second step is then to superimpose the simulated HE transients onto the generated orbit, with the same method as for X-ray sources. An example of a long GRB, a short GRB and a SGR GF superimposed on the background can be found in Fig. \ref{fig:GRBs_lc}.

\subsubsection{Computation of the Signal-to-Noise ratio and the FoV detectability fraction}
\label{sub:SNR_computation}

The count and image SNRs are finally computed by applying a simulated \textit{SVOM}/ECLAIRs count trigger algorithm \citep{Schanne2019} to the data. As it is the case for the flight algorithm, the trigger simulation computes the SNR in 4 energy bands (a possible configuration is $4$--$120$, $4$--$25$, $15$--$50$ and $25$--$120~\mathrm{keV}$), 12 timescales (logarithmically spaced from 10 ms to 20.48 s) and 9 zones on the detection plane. 
The count SNR is computed as $SNR=C_\mathrm{GRB}/\sqrt{C_\mathrm{other}}$, where $C_\mathrm{GRB}$ is the number of events generated by the GRB, and $C_\mathrm{other}$ is the number of background events (including the X-ray sources inside the FoV). 
The difference with the on-board count-rate trigger algorithm is that there is no background estimation involved for this work. As the origin of the events created on the detection plane is tracked, the background is computed by selecting the appropriate events.
The SNRs calculated here can thus be considered as best case SNRs.
However, since the background count-rate is flat on the portion of the orbit where the transient is placed (the Earth is completely out of the FoV), the background estimation used by the on-board count trigger (based on the fit of a 1D quadratic function) is close to $C_\mathrm{other}$.

When the count rate SNR exceeds a count threshold $\eta_c$, a sky image of the excess (characterized by its energy band, timescale and zone) is reconstructed with a deconvolution algorithm \citep{Caroli1987}. If this image displays an unknown source with a significance exceeding a preset image threshold $\eta_i$, the GRB is considered as detected and localized by ECLAIRs, which will send a slew request to the satellite. The count and image SNR obtained are listed in Tables \ref{tab:SNR_long}, \ref{tab:SNR_short} and \ref{tab:SNR_gflare}. In the present work, the count and image thresholds have been set to the same value: $\eta_c = 6.5$ and $\eta_i = 6.5$.

The fraction of the FoV in which a transient is detectable has also been evaluated by measuring the count and image SNRs at different sky locations. The FoV of ECLAIRs is divided into $199 \times 199$ sky pixels: as the ECLAIRs FoV is a square projected on the sky, pixels on the edge have a larger angular size than pixels in the center. For practical reasons and to save computation time, the SNR has been computed on a $19\times 19$ pixels grid (at pixel locations $-99, -88, ... 0 ... 88, 99$) and has been interpolated to get an estimation of the SNR for each pixel. For each pixel location, the high-energy transient is simulated and the count SNR estimated. The image SNR is then obtained by creating an image whose parameters (energy band, timescale) maximize the count SNR. 

\subsection{Results}
\label{sub:results}

The SNRs obtained for all the high-energy transients can be found in Tables \ref{tab:SNR_long}, \ref{tab:SNR_short} and \ref{tab:SNR_gflare}. As can be seen, most of the transients that have been detected in the local Universe by previous gamma-ray instruments will be detectable by ECLAIRs.

\begin{figure*}[t]
\subfigure[]{
\centering
\epsscale{0.98}
\plotone{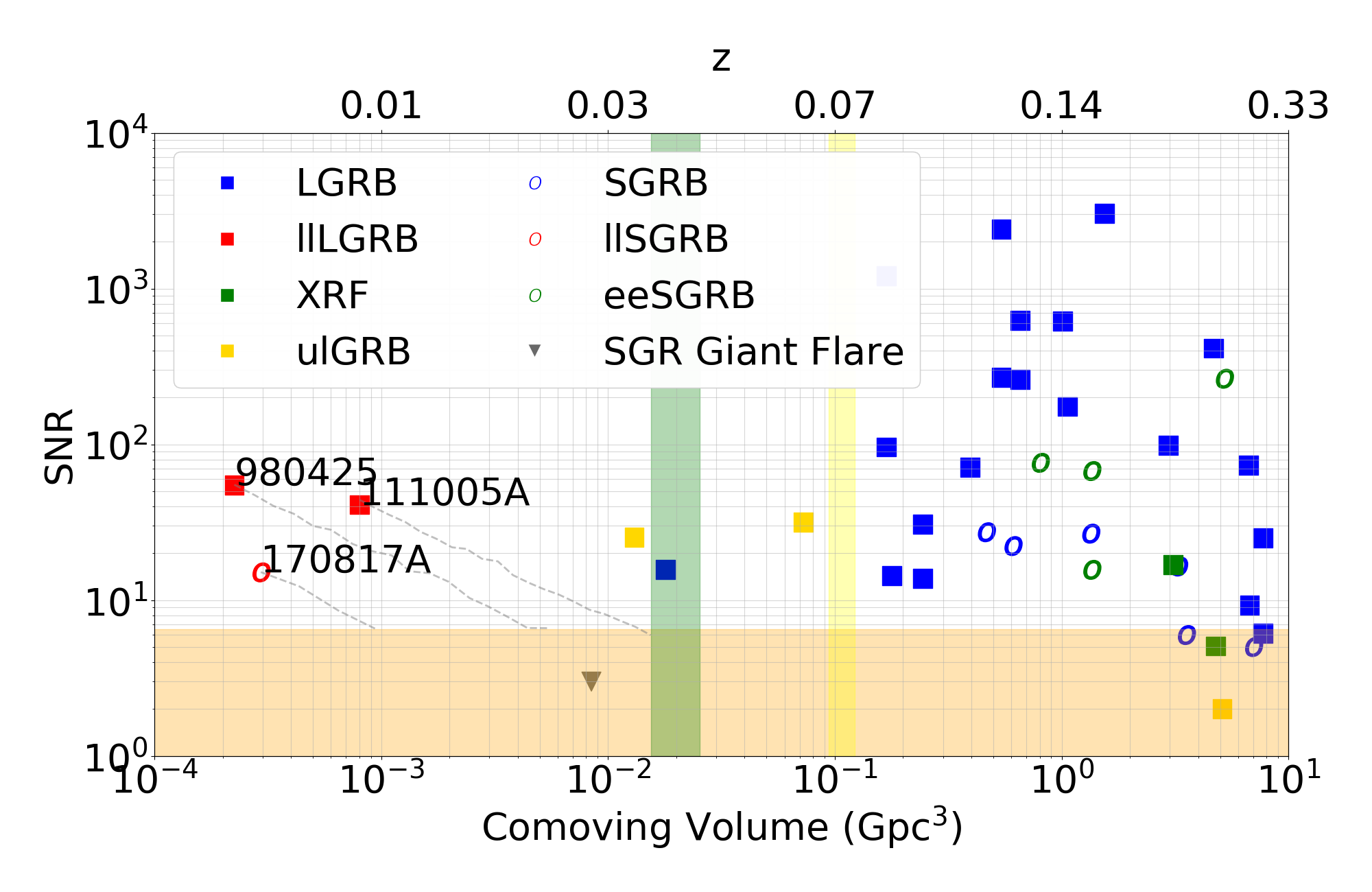}
\label{fig:SNR_all}
}
\subfigure[]{
\centering
\plotone{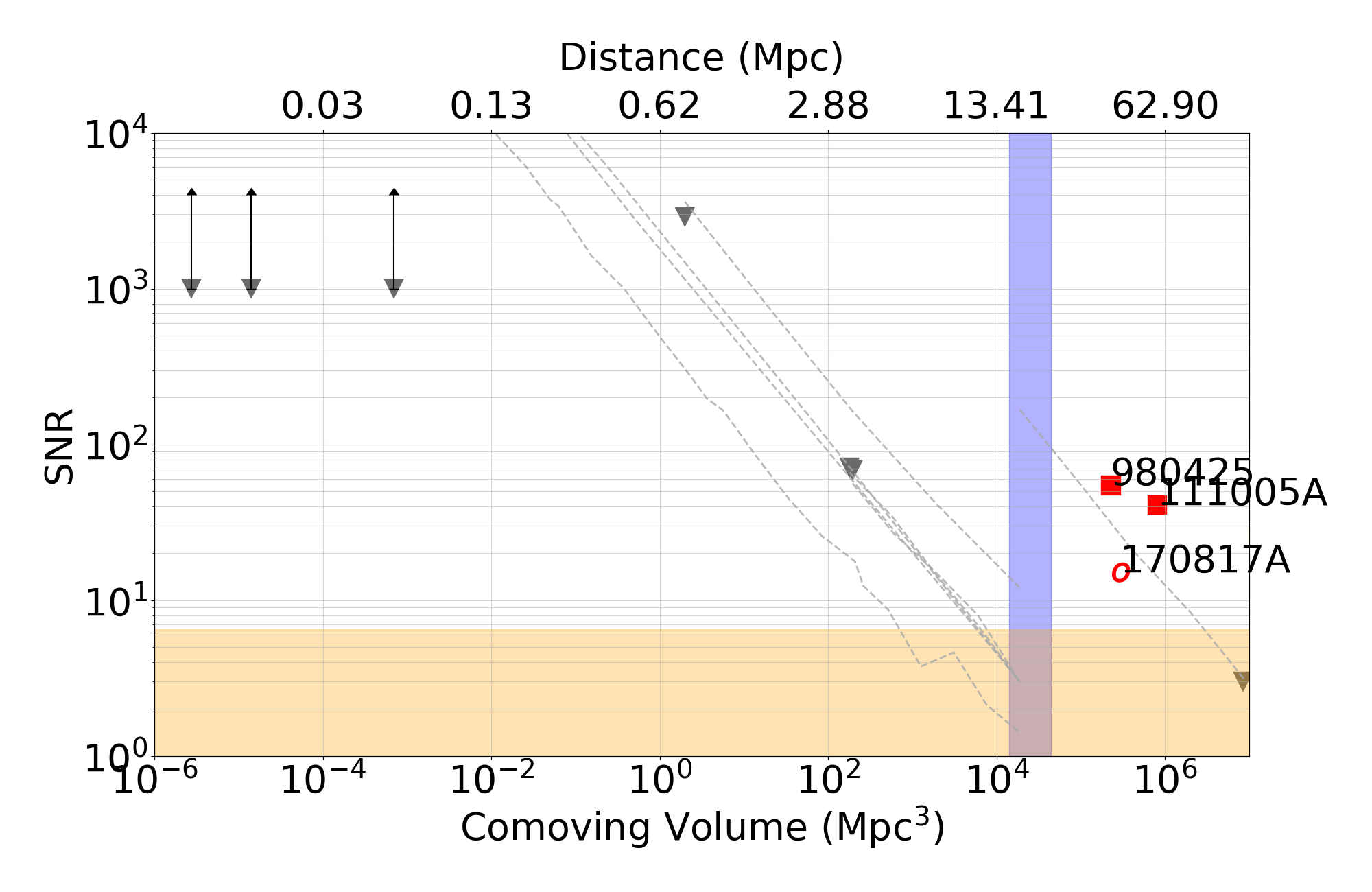}
\label{fig:SNR_gflare}
}
\caption{ECLAIRs on-axis count SNR for transients in our sample. The orange horizontal band represents the detection limit of ECLAIRs at $\mathrm{SNR} = 6.5$. (a) On-axis count SNR obtained for the local GRBs. The green and yellow bands represent the O4 LIGO distance sensitivity limits for NS-NS and BH-NS mergers respectively. The light grey trails represent the evolution of the on-axis count SNR with the redshift.  (b) On-axis SNR calculated for the SGR Giant Flares. The blue band represents the approximate distance from the Virgo Cluster. GF~790205, GF~980827 and GF~041227 on the top left have SNR over $10^4$ and would most likely saturate the instrument, we show them with arrows to improve the readability of the graph.}
\end{figure*}

The count and image SNRs given in the tables represent the median value of the SNR calculated for the sky pixels in the fully coded field of view. The energy range and timescale given represent the configuration for which these 
count and image SNRs are obtained. The fraction of the FoV represents the portion of the $2.0~\mathrm{sr}$ ECLAIRs FoV where both the count SNR and image SNR are larger than $6.5$, triggering a slew request from ECLAIRs to the platform of the \textit{SVOM} satellite. These SNRs have also been plotted for each transient relatively to the co-moving volume in Fig. \ref{fig:SNR_all} and \ref{fig:SNR_gflare}.

For LGRBs, $21/\Nlong\ (88\%)$ have both a count SNR and the resulting image SNR above 6.5. For SGRBs, $8/\Nshort\ (82\%)$ and for SGR giant flares $6/7\ (86\%)$. It can be noticed that the image SNR is systematically lower than the count SNR for SGRBs and Giant Flares (except for GRB~050709-p2, the extended emission of GRB~050709). The reason is that the timescale on which the detection has been made is minute, and therefore only a small number of events are available to reconstruct the sky image. This degrades significantly the performance of the reconstruction. 
It is always possible that by looking at longer timescales, more photons would be collected (both from the transient and the background) that could increase the image SNR and consequently the fraction of the FoV in which a slew request could be made. This possibility is not considered here, in order to stay as close as possible to the on-board trigger behavior.

Similarly, for LGRBs whose maximum count SNR is detected over timescales of several seconds and not limited by the number of photons, the image SNR can be significantly lower than the count-rate SNR for very high values of the count-rate SNR (GRB~030329 and GRB~180728A for example). This effect results from the difference in calculation between the count SNR and image SNR. The count SNR divides the signal counts by a background estimation, so that the count SNR scales as the signal counts. Whereas the image SNR is obtained through deconvolution of the shadowgram, which mixes signal and background counts. As a consequence, both the signal and the background contribute to the noise on the image and when the signal dominates over the background, the image SNR scales as the square root of the signal.

\subsection{Events identification with ECLAIRs}

\begin{figure}[t]
\centering
\epsscale{1}
\plotone{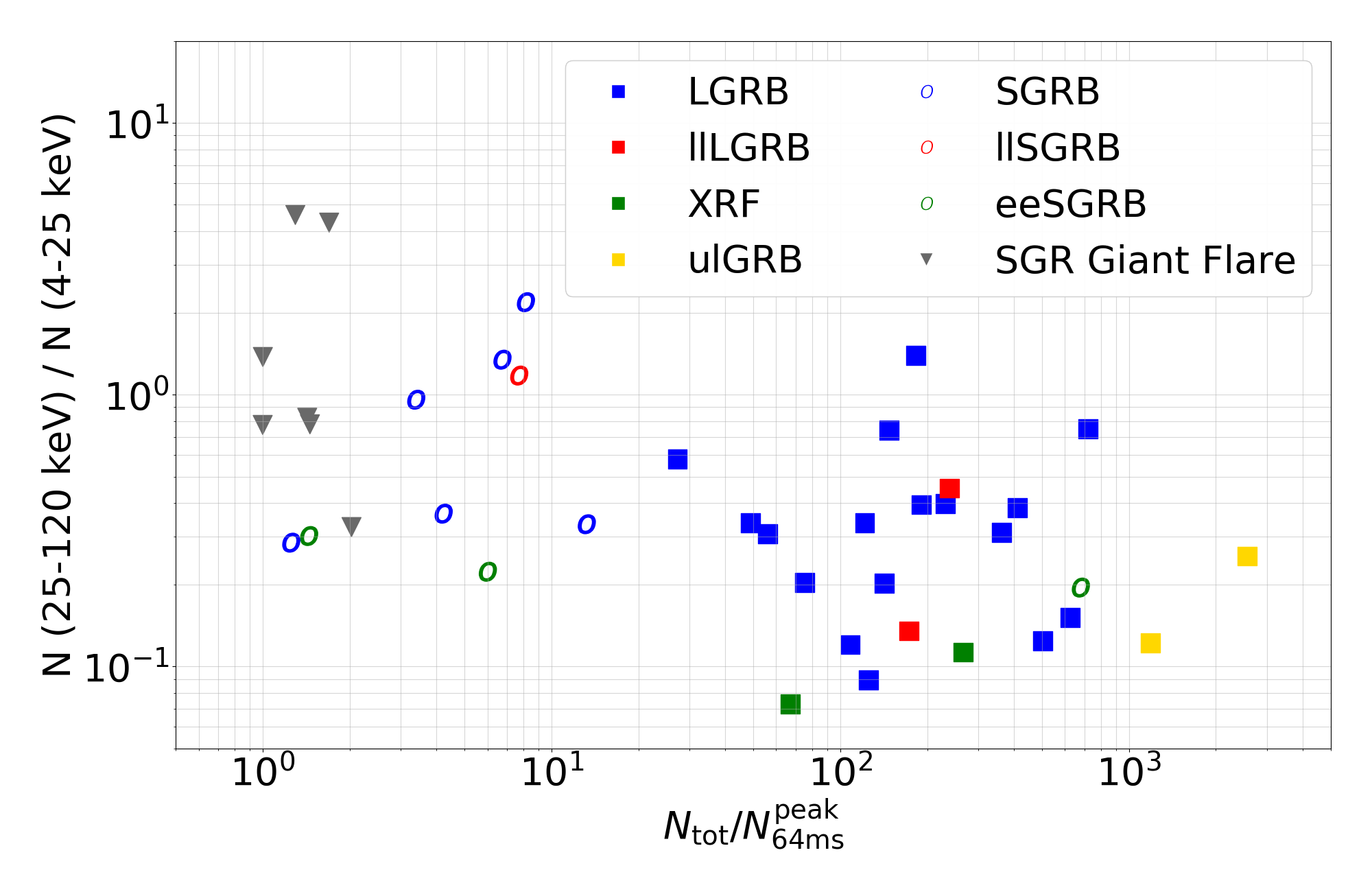}
\caption{Hardness ratio N(25--120 keV)/N(4--25 keV) as a function of the ratio N$_\mathrm{tot}$/N$^\mathrm{peak}_\mathrm{64ms}$ in the $4$--$120~\mathrm{keV}$ energy range.}
\label{fig:hardness_pfluxNtot}
\end{figure}

Once an event is detected by \textit{SVOM}/ECLAIRs and/or GRM, it is important to rapidly infer its nature. At first, the only data available relate to the prompt emission measured by the gamma-ray instruments. The simulation results in Sect. \ref{sub:SNR_computation} can be used to find some criteria for assessing the source type/nature. Fig. \ref{fig:hardness_pfluxNtot} shows that the three classes of events in our sample are relatively well separated in a plane showing the hardness of the transients as a function of their ``flatness''. The hardness is defined as the ratio of the counts in the energy range $25-120~\mathrm{keV}$ to the counts in the energy range $4-25~\mathrm{keV}$. The flatness is defined by the ratio of the total counts over the peak counts (on the 64ms timescale), both measured in the $4-120~\mathrm{keV}$ energy range. 
As shown in Fig. \ref{fig:hardness_pfluxNtot}, these two parameters enable to separate the three classes of transients in our sample. This classification cleanly separates the long GRBs from the short GRBs and SGR Giant Flares classes. The ultra-long category also seems to be easily identified in this graph. The lightcurve shape might be used to separate eeSGRBs (with a first short peak followed by an extended emission) from classical long GRBs. 

The ECLAIRs instrument covers an energy band extending from $4~\mathrm{keV}$ to $120~\mathrm{keV}$ that is small compared to other instruments such as \textit{Fermi}/GBM (from $\sim8~\mathrm{keV}$ to $\sim40~\mathrm{MeV}$, \citealt{Meegan2009_gbm}). The problem of spectral identification with ECLAIRS only is thus similar to \textit{Swift}/BAT, whose narrow energy band often prevents the measure of the peak energy. Nonetheless, \textit{SVOM} has another gamma-ray instrument called GRM, which operates in the 15~keV -- 5~MeV energy range. Since GRM covers a field of view that encompasses the ECLAIRs one, most ECLAIRS transients, except the softest ones, will also be detected by the GRM, offering a joint energy range of 4~keV -- 5~MeV for spectral analysis. In this range, \textit{SVOM} will provide spectral information on the prompt emission with an accuracy comparable to that of \textit{Fermi}/GBM \citep{Bernardini2017}. This will help identifying more precisely the nature of the high-energy transients detected by \textit{SVOM}, with the measure of their peak energy and broadband spectral model.

\section{Discussion}
\label{sec:discussion}

\subsection{Long GRBs with and without supernova}
\label{sub:snless_grbs}

After the first detection of GRB~980425 \citep{Soffitta1998_980425}, a long GRB associated with a supernova SN1998bw \citep{Tinney1998_980425, Galama1998_980425}, the origin of long GRBs seemed intrinsically linked to the core-collapse of massive stars (so-called collapsars). In this way, a supernova is expected to rise few days after the detection of nearby long GRBs (z $\leq 0.1$), although the collapsar model does not systematically predict its appearance \citep{Tominaga2007}. However, detecting the supernova associated with a long GRB requires an active follow-up with sensitive instruments, explaining in some cases why several nearby long GRBs have not been associated with supernova despite their proximity:
\begin{itemize}
\item GRB~050219A ($\mathrm{z}=0.211$) had no optical observations sufficiently deep to detect a supernova \citep{UgartePostigo2005_050219A, Berger2005_050219A}. A SN was not searched because the redshift of the host galaxy was measured  several years after the GRB \citep{Rossi2014_050219A}.
\item For GRB~050826 ($\mathrm{z}=0.297$), no optical searches for an optical transient have been performed. The redshift has been measured in $2014$ by \cite{Rossi2014_050219A}.
\item For GRB~051109B ($\mathrm{z}=0.080$), no observations have been performed to find a supernova. The redshift has been measured in July 2006 by \cite{Perley2006_051109B}.
\item GRB~080517 ($\mathrm{z}=0.089$) was satisfying the possible high-z criteria from \cite{Ukawatta2008}, which might explain why no SN search has been performed.
\item For GRB~111225A ($\mathrm{z}=0.297$), the GRB redshift has been discovered more than 3 years after the GRB \citep{Thoene2014_111225A}, which explains why no supernova search has been conducted.
\item For GRB~191019A ($\mathrm{z}=0.248$), an optical source thought to be the afterglow has been found, but no further search has been documented for the moment.

\end{itemize}

These examples show the importance of having a fast redshift determination that will trigger deeper and longer follow-ups of nearby GRBs in order to seek for a potential supernova association. 
However, there are also some GRBs where deep searches for an associated supernova have been made, and no evidence of such a signal has been found. The limiting magnitude of the observations has been compared to well-known supernovae associated with long GRBs such as SN 2003dh / GRB~030329 \citep{Stanek2003_030329, Hjorth2003_030329} or SN 1998bw / GRB~980425 \citep{Tinney1998_980425}. 
\begin{itemize}
    \item The absence of supernova for GRB~060614 and GRB~060505 has been demonstrated down to limits hundreds of times fainter than typical LGRB/SN associations \citep{Fynbo2006_060614}.
    \item Using Hubble Space Telescope observations, \cite{Soderberg2005_040701} have demonstrated with high confidence that GRB~040701 lacked an associated supernova, even if one takes into account a possible high extinction coming from its host.
    \item For the low-luminosity GRB~111005A \citep{Michalowski2018_111005A}, the near infrared and mid-infrared threshold was $\sim$20 times fainter than usual SNe associated with GRBs, which leads to the conclusion that the origin might be different from the usual collapsar model \citep{Tanga2018_111005A}.
 \end{itemize}
 
For this reason, in Table \ref{tab:detection_long} the long GRBs in the first category do not have a flag for the presence or absence of supernova, whereas GRBs in the second category have been flagged with \textbf{NO} supernova associated.

\begin{figure}[t]
\centering
\plotone{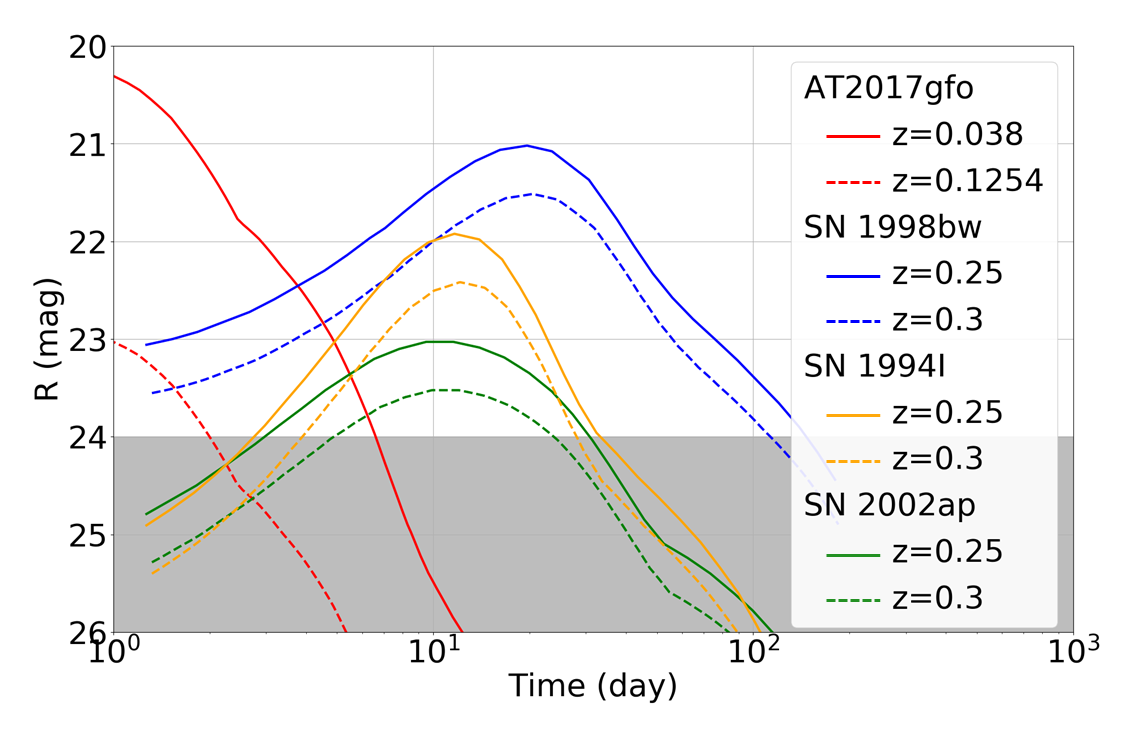}
\caption{Expected lightcurves in the R-band from three Type Ic supernovae and the kilonova AT2017gfo compared to the VT limiting magnitude. Lightcurves at $z=0.25$ are from \cite{Soderberg2005_040701}. The kilonova lightcurve is extracted from \cite{Coperthwaite2017}, originally at the GRB~170817A redshift ($z=0.0093$). The grey band represents the V-band limiting magnitude of VT for an exposure time of $4,800\ \sec$, corresponding to $\sim 2$ orbits of \textit{SVOM}.}
\label{fig:SN_lc}
\end{figure}

The systematic follow-up of \textit{SVOM}/ECLAIRs GRBs with the VT (with R-band limiting magnitudes $m\mathrm{_R} \sim 22.5$ for $\mathrm{SNR} \geq 5$ in a $300 \sec$ long exposure, and $m\mathrm{_R} \sim 24$ for a $4,800 \sec$ long exposure) and/or with the GFTs (with an R-band limiting magnitude $m\mathrm{_R} \sim 22.8$ for $\mathrm{SNR} \geq 5$ in a $300 \sec$ long exposure) will permit detecting the appearance of the associated supernova. Figure \ref{fig:SN_lc} illustrates how powerful could be the VT to detect Type Ic supernovae within the local Universe ($z \leq 0.3$). The Fig. \ref{fig:SN_lc} also shows that within the range of the O4 BNS detection limit ($z\sim 0.038$), the red component of a kilonova similar to AT2017gfo would be easily detectable by the VT. In fact, such kilonova would be detectable up to a redshift of $0.15$. The detection of an associated kilonova in debated cases would solve the issue of classification for SN-less GRBs.

Therefore having the VT optical telescope on-board \textit{SVOM} (more powerful than the \textit{Swift}/UVOT) will enable to put stringent constraints on the presence or absence of a supernova associated to long GRBs. The possibility to reliably identify long GRBs with and without a supernova will allow to better characterize the prompt emission and afterglow of both types of GRBs, using the full instrument suite of \textit{SVOM}. This will lead to a better understanding of the origin of long GRBs without SN whose nature is still debated \citep{Gehrels2006_060614, Yang2015_060614}.

\subsection{Events with gravitational waves counterparts}
\label{sub:gws}

The coincidental detection of GW~170817, a transient signal of gravitational waves and GRB~170817A, a short gamma-ray burst \citep{Abbott2017a, Abbott2017b, Goldstein2017_170817A, Savchenko2017_170817A} marked an essential milestone in our understanding of GRB physics, confirming the link between short GRBs and binary neutron star (BNS) mergers.
Based on the results presented in Table \ref{tab:SNR_short}, GRB~170817A would have been detected in 63\% of the ECLAIRs FoV with a maximum $\mathrm{SNR}_0 \approx 16.4$ in the centre of the FoV (see also the internal \textit{SVOM} study on GRB~170817A by S. Schane\footnote{ \url{http://www.svom.fr/en/portfolio/svom-in-the-era-of-gravitational-waves/} The results are summarized in Fig.~2, which also shows that GRB~170817A would have been detected by ECLAIRs up to $35\deg$ off-axis and by GRM up to an off-axis angle of $\sim 50\deg$.}). The ECLAIRs detection would have triggered an automatic slew of the satellite, allowing the Visible Telescope VT and ground-based telescopes associated with \textit{SVOM} (the French/Mexican telescope COLIBRI and the Chinese telescope C-GFT) to observe the kilonova associated with GW~170817.

The larger horizon of the O4 LIGO/Virgo campaign and the advent of the Kamioka Gravitational Wave Detector KAGRA \citep{Somiya2012, Aso2013} in 2022-2023 \citep{Abbott2018_LIGO} will increase to $160$--$190~\mathrm{Mpc}$ the distance up to which gravitational waves from BNS mergers could be detected, expanding by a factor three the explored volume compared to the O3 campaign. However, even if the increase in sensitivity of GW detectors for O4 run will likely result in a larger number of BNS detections, it is not clear if this will enhance the detection of coincidental EM high-energy counterparts. For example, While 6 BNS merger candidates have been Searched for EM counterparts during the O3 run of LIGO/Virgo, none has been found. This shows how challenging is the search for EM counterpart.
Indeed, since the GRB jet axis is in general likely to be pointed away from Earth, the GRB luminosity drops very quickly with the off-axis angle. 
This is clearly illustrated in the case of the off-axis GRB 170817A since the prompt emission from this source would be hardly detectable by ECLAIRs if located at a distance larger than 50 Mpc (see Fig. \ref{fig:SNR_all}). 
On the other hand, on-axis GRBs are too rare to be detected at distances smaller than \mbox{480 Mpc} ($z=0.1$), too far for the detection of gravitational waves from BNS mergers (but not for the third generation of GW detectors, \citealt{Punturo2010_einsteintelescope, Reitz2019_cosmis_explorer}).

NS-BH mergers may also produce GRBs \citep{Ciolfi2018} that would be associated with strong GW signals. The eeSGRBs sub-population could be associated with NS-BH merger progenitors according to \citet{Troja2008} and \citet{Gompertz2020}. According to \cite{Abbott2018_LIGO}, the volume sampled by O4 for NS-BH detection, reaching a distance limit of \mbox{330 Mpc} ($\sim 0.1$ \Gpccc), might permit the detection of few ($1^{+91}_{-1}~\mathrm{yr}^{-1}$) NS-BH mergers. The joint detection of a GW signal and a GRB from 200 Mpc to 330 Mpc would probably be the manifestation of such an event. 
Long GRBs without SN could represent a possible third source of transient GWs. While the origin of these events remains mysterious, their lack of SN would be naturally explained if they are due to mergers.
Our sample contains four events of this type (XRF~040701, GRB~060505, GRB~060614 and GRB~111005A) at redshifts ranging from \mbox{z = 0.013} to \mbox{z = 0.215}. In this context, GRB~111005A at a distance of only \mbox{57 Mpc} appears especially interesting, because the horizon for its detection with ECLAIRs (170~Mpc) is comparable to the horizon of GW detectors for BNS detection in the O4 run (Fig. \ref{fig:SNR_all}). The detection of an event like GRB~111005A within 170~Mpc with ECLAIRs during the LVC O4 run would thus permit confirming or discarding a merger origin for GRBs without SN. This method has already been used in the past to strengthen a SGR origin for SGR~070201 in M31. A SGR origin was favored considering that a merger would have produced a burst of gravitational waves, which was not observed by LIGO at that time \citep{Abbott2008_070201}. 

Even if the coincidental detection of a short GRB and a BNS merger during O4 remains speculative, they are not the only candidates for coincidental GRB and GW emission. Considering the youth of the field, it will be crucial to systematically look for coincidental transient signals detected by ECLAIRs and the GW detectors during O4, in order to assess their origin. This is also true for the GRM, which has a larger field of view than ECLAIRs for the detection of short GRBs (nearly 6~sr), albeit with much lower localization accuracy.

\subsection{The nature of X-ray flashes}
\label{sub:xray-flash}

XRFs are high-energy transients whose emission is dominated by low energy photons, with most of their fluence below \mbox{30~keV}. Their origin is still highly debated.

According to \cite{Barraud2005}, XRFs could be less energetic events with a contrast of Lorentz factor between internal shells (between 1 and 2) which is smaller than that of classical GRBs. This low contrast could lead to smaller energy dissipation inside the jet, leading to the characteristic XRF spectral energy distribution peaking between few keV and few tens of keV, with little or no emission above $50~\mathrm{keV}$ \citep{Heise2001, Kippen2003, Barraud2003}. 

Other explanations invoke extrinsic parameters, like different viewing angles, XRFs being GRBs seen from the side \citep{Yamazaki2004, Lamb2005}. However, according to \cite{Sakamoto2008} XRFs and classical GRBs have different X-ray afterglow light-curves, which cannot be explained only by the jet orientation.

The quasi absence of high-energy photons \citep{Kippen2003} in these events requires an instrument particularly sensitive to low energies, for their detection.
From February 2001 to September 2003, HETE-2 has detected 16 XRFs and 19 XRRs (X-Ray Rich GRBs) \citep{Sakamoto2005}, mostly thanks to the low low-energy threshold of two of its instruments:

\begin{itemize}
\item The Wield-field X-ray Monitor WXM \citep{Shirasaki2003} sensitive in the $2-25~\mathrm{keV}$ energy range, with an effective area of $85.4~\mathrm{cm^2}$ at $8.3~\mathrm{keV}$.
\item The FREnch GAmma TElescope FREGATE \citep{Atteia2003_fregate} sensitive in the $6-400~\mathrm{keV}$ energy range, with an effective area larger than $120~\mathrm{cm^2}$ in the range $6-200~\mathrm{keV}$.
\end{itemize} 
Thanks to its low energy threshold ($4~\mathrm{keV}$) and its effective area of $\sim400~\mathrm{cm^2}$ at $20~\mathrm{keV}$, \textit{SVOM}/ECLAIRs should be able to detect at least as many XRFs as HETE-2, providing the unique opportunity to test the nature of their differences with classical GRBs. This will take place through the detailed study of their X-ray and visible afterglows, host galaxies and associated supernovae, if any. 
The ability of \textit{SVOM} to slew towards ECLAIRs detected XRFs in few minutes will improve our knowledge of their optical and X-ray afterglow emission. Coupled with the prompt emission characteristics such as $E_\mathrm{peak}^\mathrm{obs}$, it will allow to put further constraints on the geometrical jet models discussed before \citep{Zhang2004, Yamazaki2004, Lamb2005}, but also to derive the jet structure and micro-physics.
Finally, this new XRF sample will also provide better constraints on the XRF and GRB luminosity functions. 

\subsection{The nature of ultra-long GRBs}
\label{sub:ul-grb}

The potentialities brought by \textit{SVOM}/ECLAIRs in this field have been presented in a dedicated paper by \citet{Dagoneau2020}, and will not be discussed here.
Their main conclusion is that ECLAIRs may detect ulGRBs at a rate comparable to \textit{Swift}/BAT, taking into account the differences in sensitivity, duty cycle and field of view of both instruments. The longer duration of \textit{SVOM} pointings (up to 20 hours) combined with the capability to send all the counts recorded to the ground could also allow the ground-based detection of these GRBs.
These observations will shed new light on the nature of the ulGRB phenomenon. 
Finally, we note that the ulGRB class is heterogeneous, including genuine ultra-long GRBs, like GRB~111209A (at z=0.677) and nearby low-energy transients, which have been attributed to SN shock breakout, like GRB~060218 or GRB~100316D. Considering their small distances, the detection of such events with ECLAIRs will permit to constrain their gravitational wave emission with GW interferometers during the O4 run, and confirm their nature.

\subsection{Detection of SGR Giant Flares in the Virgo cluster}
\label{sub:sgr_virgo}

SGR Giants Flares are much more luminous than classical SGR bursts, and can be detected up to a few Mpc (GRB~200415A and GF~051103, for example, have been detected respectively at distances of 3.5 and \mbox{3.6 Mpc}), and even up to \mbox{130 Mpc} for GF~050906 (but see below for a discussion on the nature of this event). Based on the work of \citet{Cline1982_790305, Crider2006_970110, Levan2007_050906, Hurley2010_051103, Ofek2008_070201, Svinkin2020_200415A}, most of the detected Giant Flares seem to come from our direct neighborhood.

Figure \ref{fig:SNR_gflare} shows with light grey trails the expected SNR for Giant Flares as a function of distance. Only GF~070201 and GF~050906 seem to be sufficiently luminous so that their count SNR is above 6.5 at the distance of the Virgo Cluster ($\sim$16.5 Mpc), with the flare in the fully coded FoV. The cluster spans a radius of approximately $8\deg$ on the sky, which means that the whole cluster would fit inside the \textit{SVOM}/ECLAIRs fully coded field of view (a square of about $22 \times 22 \deg^2$).
Our simulations show that GF~070201 would have SNR$_\mathrm{0}=8$ on the $10~\mathrm{ms}$ timescale (SNR$_\mathrm{0}$ representing the SNR obtained in the fully coded FoV). This corresponds to $\sim 45$ photons on the detector, which is not enough to produce a sky image with a SNR greater than $6.5$ (SNR$\sim 5.9\sigma$, using only the photons from the Giant Flare). This giant flare will therefore not be localized in real time by \textit{SVOM}, which means that there will be no follow-up of such an event. However, since \textit{SVOM} transmits all detected photons to the ground, it will be possible to look for short spikes when ECLAIRs points at the Virgo cluster, during pre-planned large programs (for instance, the search for relativistic Tidal Disruption Events in the Virgo Cluster with MXT, see \citealt{Wei2016}). Such analyses may lead to the detection of an excess of very short transients due to giant flares in the Virgo Cluster. Depending on the luminosity function of giant flares, the brightest of them could lead to rapid Target of Opportunity observations with \textit{SVOM} or other space/ground facilities, that may allow their detailed follow-up. 

M31, the putative host of GF~070201, has an estimated star formation rate (SFR) of $\sim 0.7$ \Msun$~\mathrm{yr^{-1}}$ \citep{Lewis2015}.
This SFR can be compared to the star formation rate of all the galaxies in the Virgo Cluster which amounts to $50-100$ \Msun $\mathrm{yr}^{-1}$  \citep{Boselli2016}. Assuming that the rate of SGR giant flares is directly correlated to neutron stars creation and therefore to the massive star formation rate, the event rate from the Virgo Cluster could be up to two orders of magnitude greater than for large spiral galaxies like M31 or the Milky Way.

The second giant flare candidate that could be detectable in the Virgo Cluster is GF~050906 (SGR~0331-1439), detected by \textit{Swift}/BAT \citep{Parsons2005_050906, Levan2007_050906}. This event is however significantly different from the other giant flares in our sample. First, it is  associated with a local galaxy IC328 at distance of $\approx 130$ Mpc, more than twenty times further away than GF~970110, the second most distant giant flare, detected in NGC6946 at a distance of $5.9~\mathrm{Mpc}$ \citep{Crider2006_970110}.
Second, the spectrum of this flare is much softer than other giant flares (GRB~200415A, GF~051103), while having approximately the same energy as GF~051103 (\eiso $\sim 7.5\times 10^{46}~\mathrm{erg}$, see Table \ref{tab:intrinsic_gflare}). This softness results in a much larger number of photons, making it easily detectable in the Virgo cluster with a count SNR value of $\sim 150$ in the fully coded FoV.

The significant differences between GF~050906 and the other SGR giant flares (spectrum and volumetric rate) questions the nature of this event. A recent work by \cite{Dichiara2020} shows that this event might instead belong to a category of low-luminosity short GRBs with typical energies around $10^{46-47}~\mathrm{erg}$, close to the energy of GRB~170817A.

\begin{figure}[t]
\centering
\plotone{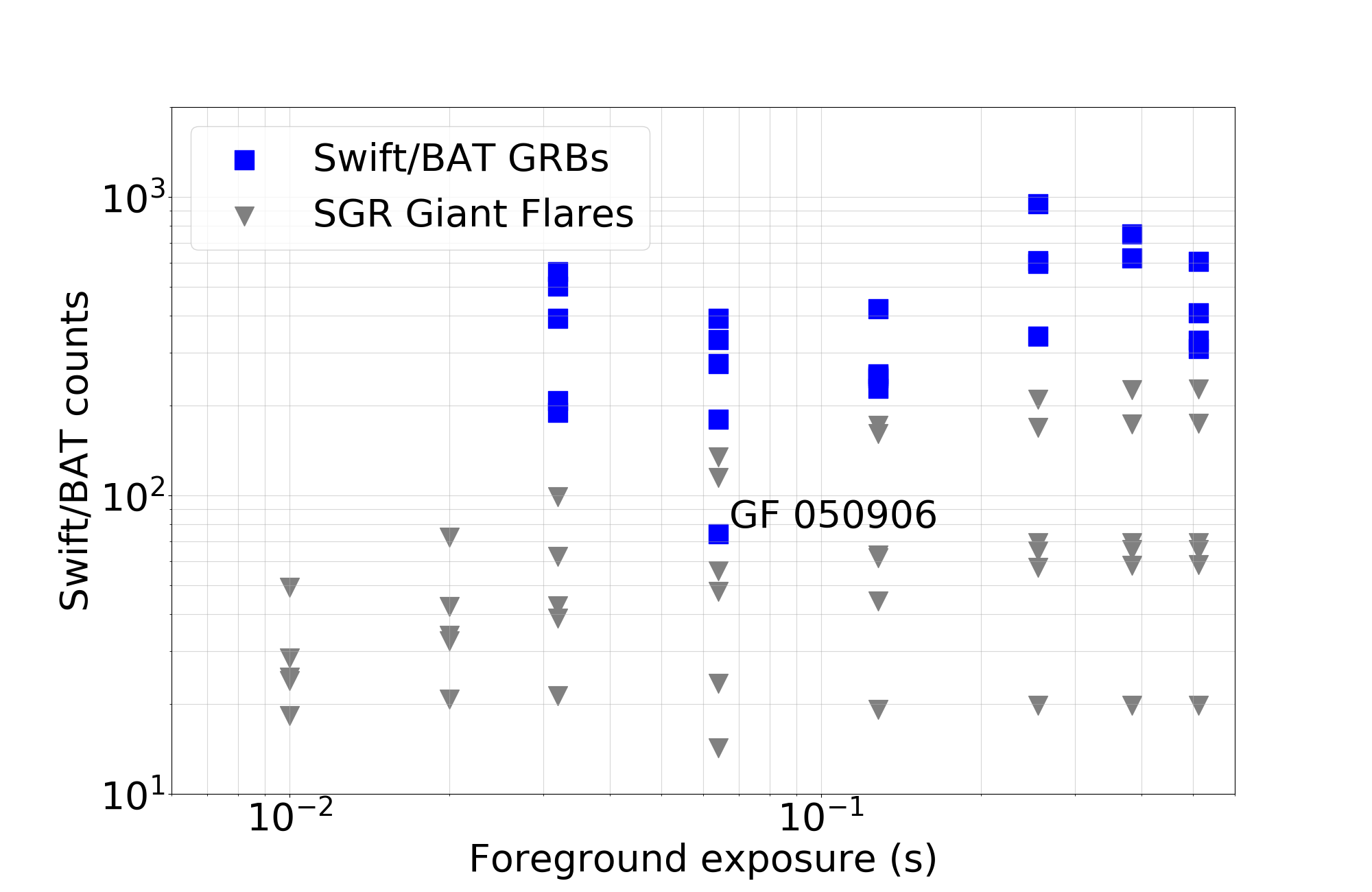}
\caption{Comparison between the faintest \textit{Swift}/BAT GRBs (blue squares) and SGR Giant Flares simulated in the Virgo Cluster (grey triangles), showing that SGR Giant Flares in the Virgo Cluster are fainter than the faintest GRBs detected with \textit{Swift}/BAT. GF~050906 is the faintest event detected by \textit{Swift}/BAT.}
\label{fig:swift_gf_comp}
\end{figure}

We have made a comparison between the dimmest GRBs observed by \textit{Swift}/BAT and simulated Giant Flares in the Virgo Cluster seen by this same instrument in Fig. \ref{fig:swift_gf_comp}. This comparison has been made on the smallest timescales for which \textit{Swift}/BAT has triggered ($32~\mathrm{ms}$, $64~\mathrm{ms}$, $128~\mathrm{ms}$, $384~\mathrm{ms}$, $512~\mathrm{ms}$), using the recorded GRB counts from the GCN. For all timescales, the only event from \textit{Swift}/BAT dimmer than some of the Giant Flares is GF~050906. It has been detected on a $64~\mathrm{ms}$ with a count-level $\sim 5$ times lower than the second faintest detection by \textit{Swift}/BAT on this timescale. Nonetheless, the other faint events with a number of counts similar to GF~050906 such as GRB~070406, GRB~080702B and GRB~100216A have been detected or confirmed only with the help of ground analysis. The in-flight detection of GF~050906 seems to be a "once in the \textit{Swift}/BAT lifetime" opportunity. 
To conclude, the fact that \textit{Swift}/BAT has not detected any Giant Flare when pointing towards the Virgo cluster seems to be acceptable.

Additional simulations have been performed to determine the count SNR registered by GRM for Giant Flare events in the Virgo Cluster. Thanks to the wide energy range and the sensitivity to high-energy photons, GRM will also be able to detect these events, allowing coincident searches between the two \textit{SVOM} gamma-ray instruments. 

\section{Conclusion}
\label{sec:conclusion}
We have shown that \textit{SVOM}/ECLAIRs will be able to detect the large diversity of short high-energy transients known in the local Universe. Once an event is detected, \textit{SVOM} will be able to obtain unique diagnostics thanks to its slewing capability, the on-board narrow field instruments and the follow-up by ground telescopes. Furthermore, the photons recorded by ECLAIRs are entirely sent to the ground: exotic transients not detected by the on-board trigger could be found by the offline trigger on ground and later-on followed. Finally, ECLAIRs low-energy threshold of $4\ \mathrm{keV}$ will be an asset to detect soft transients in the hard X-ray range, and its combination with GRM might allow an even better spectral characterization at low energies than what is currently achieved by \textit{Fermi}/GBM.

Various local high-energy transients are also within the range of the GW detectors LIGO and VIRGO, which permitted for example detecting the well-known GRB~170817A coincident with GW~170817 at $z=0.0093$. The multi-messenger astronomy and the detection or lack of gravitational waves could shed crucial light on the origin of long GRBs without SN such as GRB~111005A and GRB~060614. The VT, C-GFT and Colibri telescopes and their systematic follow-up strategy of long GRBs will put strong constraints on associated supernovae, if the redshift is measured sufficiently early to allow the adequate follow-up of selected nearby GRBs. GW observations might also put strict limits on the GW energy liberated by the SGR Giant Flares progenitors, an opportunity to derive rates for the existence of such objects. 

However, one condition for this success is a long lasting lifetime of \textit{SVOM}/ECLAIRs, as it is the case for \textit{Fermi}/GBM and \textit{Swift}/BAT for example. Low-z high-energy transients are scarcely detected, even if all the instruments detections are added: the \Nlong\ long GRBs and \Nshort\ short GRBs detected and listed in this paper span over two decades, from May 1998 to November 2019 ($\sim 1.7\ \mathrm{yr^{-1}}$). The SGR Giant Flares are even rarer, only \Nsgr\ of them have been detected from March 1979 to April 2020 ($\sim 0.2\ \mathrm{yr^{-1}}$). \textit{SVOM} will have all the assets to detect and characterise these events, their progenitors and their hosts, if they happen in the lifespan of the mission...

\acknowledgments
ECLAIRs is a cooperation between CNES, CEA and CNRS, with CNES acting as prime contractor.
This work is supported by CNES and the Paul Sabatier University.
J\'er\^ome Rodriguez acknowledges partial funding from the French Space Agency (CNES) and the 
French Programme National des Hautes-Energies (PNHE)
We would like to thank Fr\'ed\'eric Daigne and Maxime Bocquier from IAP, Paris, for providing the tools to transport GRB light curves to different redshifts, developed during the thesis of Sarah Antier at CEA Paris-Saclay. 
This article made use of the GRB table maintained by J. Greiner (see footnote $1$ of Sect. \ref{sub:sample_construct}).

\nocite{*}
\bibliographystyle{spr-mp-nameyear-cnd}
\bibliography{sample}

\begin{thebibliography}{334}
\ifx \bisbn   \undefined \def \bisbn  #1{ISBN #1}\fi
\ifx \binits  \undefined \def \binits#1{#1} \fi
\ifx \bauthor  \undefined \def \bauthor#1{#1} \fi
\ifx \batitle  \undefined \def \batitle#1{#1} \fi
\ifx \bjtitle  \undefined \def \bjtitle#1{#1}\fi
\ifx \bvolume  \undefined \def \bvolume#1{\textbf{#1}}\fi
\ifx \byear  \undefined \def \byear#1{#1} \fi
\ifx \bissue  \undefined \def \bissue#1{#1} \fi
\ifx \bfpage  \undefined \def \bfpage#1{#1} \fi
\ifx \blpage  \undefined \def \blpage #1{#1} \fi
\ifx \burl  \undefined \def \burl#1{\textsf{#1}} \fi
\ifx \doiurl  \undefined \def \doiurl#1{\textsf{#1}} \fi
\ifx \betal  \undefined \def \betal{\textit{et al.}} \fi
\ifx \binstitute  \undefined \def \binstitute#1{#1} \fi
\ifx \binstitutionaled  \undefined \def \binstitutionaled#1{#1} \fi
\ifx \bctitle  \undefined \def \bctitle#1{#1} \fi
\ifx \beditor  \undefined \def \beditor#1{#1} \fi
\ifx \bpublisher  \undefined \def \bpublisher#1{#1} \fi
\ifx \bbtitle  \undefined \def \bbtitle#1{#1} \fi
\ifx \bedition  \undefined \def \bedition#1{#1} \fi
\ifx \bseriesno  \undefined \def \bseriesno#1{#1} \fi
\ifx \blocation  \undefined \def \blocation#1{#1} \fi
\ifx \bsertitle  \undefined \def \bsertitle#1{#1} \fi
\ifx \bsnm \undefined \def \bsnm#1{#1} \fi
\ifx \bsuffix \undefined \def \bsuffix#1{#1} \fi
\ifx \bparticle \undefined \def \bparticle#1{#1} \fi
\ifx \barticle \undefined \def \barticle#1{#1} \fi
\ifx \bconfdate \undefined \def \bconfdate #1{#1} \fi
\ifx \botherref \undefined \def \botherref #1{#1} \fi
\ifx \url \undefined \def \url#1{\textsf{#1}} \fi
\ifx \bchapter \undefined \def \bchapter#1{#1} \fi
\ifx \bbook \undefined \def \bbook#1{#1} \fi
\ifx \bcomment \undefined \def \bcomment#1{#1} \fi
\ifx \oauthor \undefined \def \oauthor#1{#1} \fi
\ifx \citeauthoryear \undefined \def \citeauthoryear#1{#1} \fi
\ifx \endbibitem  \undefined \def \endbibitem {}\fi
\ifx \bconflocation  \undefined \def \bconflocation#1{#1} \fi
\ifx \arxivurl  \undefined \def \arxivurl#1{\textsf{#1}} \fi

\bibitem[\protect\citeauthoryear{Abbott et~al.}{2017}]{Abbott2017b}
\begin{barticle}
\bauthor{\bsnm{Abbott}, \binits{B.P.}},
\bauthor{\bsnm{Abbott}, \binits{R.}},
\bauthor{\bsnm{Abbott}, \binits{T.D.}},
\bauthor{\bsnm{Acernese}, \binits{F.}},
\bauthor{\bsnm{Ackley}, \binits{K.}},
\bauthor{\bsnm{Adams}, \binits{C.}},
\bauthor{\bsnm{Adams}, \binits{T.}},
\bauthor{\bsnm{Addesso}, \binits{P.}},
\bauthor{\bsnm{Adhikari}, \binits{R.X.}},
\bauthor{\bsnm{Adya}, \binits{V.B.}},
\bauthor{\bsnm{Affeldt}, \binits{C.}},
\bauthor{\bsnm{Afrough}, \binits{M.}},
\bauthor{\bsnm{Agarwal}, \binits{B.}},
\bauthor{\bsnm{Agathos}, \binits{M.}},
\bauthor{\bsnm{Agatsuma}, \binits{K.}},
\bauthor{\bsnm{Aggarwal}, \binits{N.}},
\bauthor{\bsnm{Aguiar}, \binits{O.D.}},
\bauthor{\bsnm{Aiello}, \binits{L.}},
\bauthor{\bsnm{Ain}, \binits{A.}},
\bauthor{\bsnm{Ajith}, \binits{P.}},
\bauthor{\bsnm{Allen}, \binits{B.}},
\bauthor{\bsnm{Allen}, \binits{G.}},
\bauthor{\bsnm{Allocca}, \binits{A.}},
\bauthor{\bsnm{Aloy}, \binits{M.A.}},
\bauthor{\bsnm{Altin}, \binits{P.A.}},
\bauthor{\bsnm{Amato}, \binits{A.}},
\bauthor{\bsnm{Ananyeva}, \binits{A.}},
\bauthor{\bsnm{Anderson}, \binits{S.B.}},
\bauthor{\bsnm{Anderson}, \binits{W.G.}},
\bauthor{\bsnm{Angelova}, \binits{S.V.}},
\bauthor{\bsnm{Antier}, \binits{S.}},
\bauthor{\bsnm{Appert}, \binits{S.}},
\bauthor{\bsnm{Arai}, \binits{K.}},
\bauthor{\bsnm{Araya}, \binits{M.C.}},
\bauthor{\bsnm{Areeda}, \binits{J.S.}},
\bauthor{\bsnm{Arnaud}, \binits{N.}},
\bauthor{\bsnm{Arun}, \binits{K.G.}},
\bauthor{\bsnm{Ascenzi}, \binits{S.}},
\bauthor{\bsnm{Ashton}, \binits{G.}},
\bauthor{\bsnm{Ast}, \binits{M.}},
\bauthor{\bsnm{Aston}, \binits{S.M.}},
\bauthor{\bsnm{Astone}, \binits{P.}},
\bauthor{\bsnm{Atallah}, \binits{D.V.}},
\bauthor{\bsnm{Aufmuth}, \binits{P.}},
\bauthor{\bsnm{Aulbert}, \binits{C.}},
\bauthor{\bsnm{AultONeal}, \binits{K.}},
\bauthor{\bsnm{Austin}, \binits{C.}},
\bauthor{\bsnm{Avila-Alvarez}, \binits{A.}},
\bauthor{\bsnm{Babak}, \binits{S.}},
\bauthor{\bsnm{Bacon}, \binits{P.}},
\bauthor{\bsnm{Bader}, \binits{M.K.M.}},
\bauthor{\bsnm{Bae}, \binits{S.}},
\bauthor{\bsnm{Baker}, \binits{P.T.}},
\bauthor{\bsnm{Baldaccini}, \binits{F.}},
\bauthor{\bsnm{Ballardin}, \binits{G.}},
\bauthor{\bsnm{Ballmer}, \binits{S.W.}},
\bauthor{\bsnm{Banagiri}, \binits{S.}},
\bauthor{\bsnm{Barayoga}, \binits{J.C.}},
\bauthor{\bsnm{Barclay}, \binits{S.E.}},
\bauthor{\bsnm{Barish}, \binits{B.C.}},
\bauthor{\bsnm{Barker}, \binits{D.}},
\bauthor{\bsnm{Barkett}, \binits{K.}},
\bauthor{\bsnm{Barone}, \binits{F.}},
\bauthor{\bsnm{Barr}, \binits{B.}},
\bauthor{\bsnm{Barsotti}, \binits{L.}},
\bauthor{\bsnm{Barsuglia}, \binits{M.}},
\bauthor{\bsnm{Barta}, \binits{D.}},
\bauthor{\bsnm{Bartlett}, \binits{J.}},
\bauthor{\bsnm{Bartos}, \binits{I.}},
\bauthor{\bsnm{Bassiri}, \binits{R.}},
\bauthor{\bsnm{Basti}, \binits{A.}},
\bauthor{\bsnm{Batch}, \binits{J.C.}}:
\bjtitle{\apj}
\bvolume{848},
\bfpage{13}
(\byear{2017}).
doi:\doiurl{10.3847/2041-8213/aa920c}
\end{barticle}
\endbibitem

\bibitem[\protect\citeauthoryear{{Abbott} et~al.}{2017}]{Abbott2017a}
\begin{barticle}
\bauthor{\bsnm{{Abbott}}, \binits{B.P.}},
\bauthor{\bsnm{{Abbott}}, \binits{R.}},
\bauthor{\bsnm{{Abbott}}, \binits{T.D.}},
\bauthor{\bsnm{{Acernese}}, \binits{F.}},
\bauthor{\bsnm{{Ackley}}, \binits{K.}},
\bauthor{\bsnm{{Adams}}, \binits{C.}},
\bauthor{\bsnm{{Adams}}, \binits{T.}},
\bauthor{\bsnm{{Addesso}}, \binits{P.}},
\bauthor{\bsnm{{Adhikari}}, \binits{R.X.}},
\bauthor{\bsnm{{Adya}}, \binits{V.B.}},
\bauthor{\bsnm{{Affeldt}}, \binits{C.}},
\bauthor{\bsnm{{Afrough}}, \binits{M.}},
\bauthor{\bsnm{{Agarwal}}, \binits{B.}},
\bauthor{\bsnm{{Agathos}}, \binits{M.}},
\bauthor{\bsnm{{Agatsuma}}, \binits{K.}},
\bauthor{\bsnm{{Aggarwal}}, \binits{N.}},
\bauthor{\bsnm{{Aguiar}}, \binits{O.D.}},
\bauthor{\bsnm{{Aiello}}, \binits{L.}},
\bauthor{\bsnm{{Ain}}, \binits{A.}},
\bauthor{\bsnm{{Ajith}}, \binits{P.}},
\bauthor{\bsnm{{Allen}}, \binits{B.}},
\bauthor{\bsnm{{Allen}}, \binits{G.}},
\bauthor{\bsnm{{Allocca}}, \binits{A.}},
\bauthor{\bsnm{{Altin}}, \binits{P.A.}},
\bauthor{\bsnm{{Amato}}, \binits{A.}},
\bauthor{\bsnm{{Ananyeva}}, \binits{A.}},
\bauthor{\bsnm{{Anderson}}, \binits{S.B.}},
\bauthor{\bsnm{{Anderson}}, \binits{W.G.}},
\bauthor{\bsnm{{Angelova}}, \binits{S.V.}},
\bauthor{\bsnm{{Antier}}, \binits{S.}},
\bauthor{\bsnm{{Appert}}, \binits{S.}},
\bauthor{\bsnm{{Arai}}, \binits{K.}},
\bauthor{\bsnm{{Araya}}, \binits{M.C.}},
\bauthor{\bsnm{{Areeda}}, \binits{J.S.}},
\bauthor{\bsnm{{Arnaud}}, \binits{N.}},
\bauthor{\bsnm{{Arun}}, \binits{K.G.}},
\bauthor{\bsnm{{Ascenzi}}, \binits{S.}},
\bauthor{\bsnm{{Ashton}}, \binits{G.}},
\bauthor{\bsnm{{Ast}}, \binits{M.}},
\bauthor{\bsnm{{Aston}}, \binits{S.M.}},
\bauthor{\bsnm{{Astone}}, \binits{P.}},
\bauthor{\bsnm{{Atallah}}, \binits{D.V.}},
\bauthor{\bsnm{{Aufmuth}}, \binits{P.}},
\bauthor{\bsnm{{Aulbert}}, \binits{C.}},
\bauthor{\bsnm{{AultONeal}}, \binits{K.}},
\bauthor{\bsnm{{Austin}}, \binits{C.}},
\bauthor{\bsnm{{Avila-Alvarez}}, \binits{A.}},
\bauthor{\bsnm{{Babak}}, \binits{S.}},
\bauthor{\bsnm{{Bacon}}, \binits{P.}},
\bauthor{\bsnm{{Bader}}, \binits{M.K.M.}},
\bauthor{\bsnm{{Bae}}, \binits{S.}},
\bauthor{\bsnm{{Baker}}, \binits{P.T.}},
\bauthor{\bsnm{{Baldaccini}}, \binits{F.}},
\bauthor{\bsnm{{Ballardin}}, \binits{G.}},
\bauthor{\bsnm{{Ballmer}}, \binits{S.W.}},
\bauthor{\bsnm{{Banagiri}}, \binits{S.}},
\bauthor{\bsnm{{Barayoga}}, \binits{J.C.}},
\bauthor{\bsnm{{Barclay}}, \binits{S.E.}},
\bauthor{\bsnm{{Barish}}, \binits{B.C.}},
\bauthor{\bsnm{{Barker}}, \binits{D.}},
\bauthor{\bsnm{{Barkett}}, \binits{K.}},
\bauthor{\bsnm{{Barone}}, \binits{F.}},
\bauthor{\bsnm{{Barr}}, \binits{B.}},
\bauthor{\bsnm{{Barsotti}}, \binits{L.}},
\bauthor{\bsnm{{Barsuglia}}, \binits{M.}},
\bauthor{\bsnm{{Barta}}, \binits{D.}},
\bauthor{\bsnm{{Barthelmy}}, \binits{S.D.}},
\bauthor{\bsnm{{Bartlett}}, \binits{J.}},
\bauthor{\bsnm{{Bartos}}, \binits{I.}},
\bauthor{\bsnm{{Bassiri}}, \binits{R.}},
\bauthor{\bsnm{{Basti}}, \binits{A.}},
\bauthor{\bsnm{{Batch}}, \binits{J.C.}},
\bauthor{\bsnm{{Bawaj}}, \binits{M.}},
\bauthor{\bsnm{{Bayley}}, \binits{J.C.}},
\bauthor{\bsnm{{Bazzan}}, \binits{M.}},
\bauthor{\bsnm{{B{\'e}csy}}, \binits{B.}},
\bauthor{\bsnm{{Beer}}, \binits{C.}},
\bauthor{\bsnm{{Bejger}}, \binits{M.}},
\bauthor{\bsnm{{Belahcene}}, \binits{I.}},
\bauthor{\bsnm{{Bell}}, \binits{A.S.}},
\bauthor{\bsnm{{Berger}}, \binits{B.K.}},
\bauthor{\bsnm{{Bergmann}}, \binits{G.}},
\bauthor{\bsnm{{Bero}}, \binits{J.J.}},
\bauthor{\bsnm{{Berry}}, \binits{C.P.L.}},
\bauthor{\bsnm{{Bersanetti}}, \binits{D.}},
\bauthor{\bsnm{{Bertolini}}, \binits{A.}},
\bauthor{\bsnm{{Betzwieser}}, \binits{J.}},
\bauthor{\bsnm{{Bhagwat}}, \binits{S.}},
\bauthor{\bsnm{{Bhandare}}, \binits{R.}},
\bauthor{\bsnm{{Bilenko}}, \binits{I.A.}},
\bauthor{\bsnm{{Billingsley}}, \binits{G.}},
\bauthor{\bsnm{{Billman}}, \binits{C.R.}},
\bauthor{\bsnm{{Birch}}, \binits{J.}},
\bauthor{\bsnm{{Birney}}, \binits{R.}},
\bauthor{\bsnm{{Birnholtz}}, \binits{O.}},
\bauthor{\bsnm{{Biscans}}, \binits{S.}},
\bauthor{\bsnm{{Biscoveanu}}, \binits{S.}},
\bauthor{\bsnm{{Bisht}}, \binits{A.}},
\bauthor{\bsnm{{Bitossi}}, \binits{M.}},
\bauthor{\bsnm{{Biwer}}, \binits{C.}},
\bauthor{\bsnm{{Bizouard}}, \binits{M.A.}},
\bauthor{\bsnm{{Blackburn}}, \binits{J.K.}},
\bauthor{\bsnm{{Blackman}}, \binits{J.}},
\bauthor{\bsnm{{Blair}}, \binits{C.D.}},
\bauthor{\bsnm{{Blair}}, \binits{D.G.}},
\bauthor{\bsnm{{Blair}}, \binits{R.M.}},
\bauthor{\bsnm{{Bloemen}}, \binits{S.}},
\bauthor{\bsnm{{Bock}}, \binits{O.}},
\bauthor{\bsnm{{Bode}}, \binits{N.}},
\bauthor{\bsnm{{Boer}}, \binits{M.}},
\bauthor{\bsnm{{Bogaert}}, \binits{G.}},
\bauthor{\bsnm{{Bohe}}, \binits{A.}},
\bauthor{\bsnm{{Bondu}}, \binits{F.}},
\bauthor{\bsnm{{Bonilla}}, \binits{E.}},
\bauthor{\bsnm{{Bonnand}}, \binits{R.}},
\bauthor{\bsnm{{Boom}}, \binits{B.A.}},
\bauthor{\bsnm{{Bork}}, \binits{R.}},
\bauthor{\bsnm{{Boschi}}, \binits{V.}},
\bauthor{\bsnm{{Bose}}, \binits{S.}},
\bauthor{\bsnm{{Bossie}}, \binits{K.}},
\bauthor{\bsnm{{Bouffanais}}, \binits{Y.}},
\bauthor{\bsnm{{Bozzi}}, \binits{A.}},
\bauthor{\bsnm{{Bradaschia}}, \binits{C.}},
\bauthor{\bsnm{{Brady}}, \binits{P.R.}},
\bauthor{\bsnm{{Branchesi}}, \binits{M.}},
\bauthor{\bsnm{{Brau}}, \binits{J.E.}},
\bauthor{\bsnm{{Briant}}, \binits{T.}},
\bauthor{\bsnm{{Brillet}}, \binits{A.}},
\bauthor{\bsnm{{Brinkmann}}, \binits{M.}},
\bauthor{\bsnm{{Brisson}}, \binits{V.}},
\bauthor{\bsnm{{Brockill}}, \binits{P.}},
\bauthor{\bsnm{{Broida}}, \binits{J.E.}},
\bauthor{\bsnm{{Brooks}}, \binits{A.F.}},
\bauthor{\bsnm{{Brown}}, \binits{D.A.}},
\bauthor{\bsnm{{Brown}}, \binits{D.D.}},
\bauthor{\bsnm{{Brunett}}, \binits{S.}},
\bauthor{\bsnm{{Buchanan}}, \binits{C.C.}},
\bauthor{\bsnm{{Buikema}}, \binits{A.}},
\bauthor{\bsnm{{Bulik}}, \binits{T.}},
\bauthor{\bsnm{{Bulten}}, \binits{H.J.}},
\bauthor{\bsnm{{Buonanno}}, \binits{A.}},
\bauthor{\bsnm{{Buskulic}}, \binits{D.}},
\bauthor{\bsnm{{Buy}}, \binits{C.}},
\bauthor{\bsnm{{Byer}}, \binits{R.L.}}:
\bjtitle{\apjl}
\bvolume{848}(\bissue{2}),
\bfpage{12}
(\byear{2017}).
\arxivurl{1710.05833}.
doi:\doiurl{10.3847/2041-8213/aa91c9}
\end{barticle}
\endbibitem

\bibitem[\protect\citeauthoryear{{Abbott} et~al.}{2018a}]{Abbott2018_LIGO}
\begin{barticle}
\bauthor{\bsnm{{Abbott}}, \binits{B.P.}},
\bauthor{\bsnm{{Abbott}}, \binits{R.}},
\bauthor{\bsnm{{Abbott}}, \binits{T.D.}},
\bauthor{\bsnm{{Abernathy}}, \binits{M.R.}},
\bauthor{\bsnm{{Acernese}}, \binits{F.}},
\bauthor{\bsnm{{Ackley}}, \binits{K.}},
\bauthor{\bsnm{{Adams}}, \binits{C.}},
\bauthor{\bsnm{{Adams}}, \binits{T.}},
\bauthor{\bsnm{{Addesso}}, \binits{P.}},
\bauthor{\bsnm{{Adhikari}}, \binits{R.X.}},
\bauthor{\bsnm{{Adya}}, \binits{V.B.}},
\bauthor{\bsnm{{Affeldt}}, \binits{C.}},
\bauthor{\bsnm{{Agathos}}, \binits{M.}},
\bauthor{\bsnm{{Agatsuma}}, \binits{K.}},
\bauthor{\bsnm{{Aggarwal}}, \binits{N.}},
\bauthor{\bsnm{{Aguiar}}, \binits{O.D.}},
\bauthor{\bsnm{{Aiello}}, \binits{L.}},
\bauthor{\bsnm{{Ain}}, \binits{A.}},
\bauthor{\bsnm{{Ajith}}, \binits{P.}},
\bauthor{\bsnm{{Akutsu}}, \binits{T.}},
\bauthor{\bsnm{{Allen}}, \binits{B.}},
\bauthor{\bsnm{{Allocca}}, \binits{A.}},
\bauthor{\bsnm{{Altin}}, \binits{P.A.}},
\bauthor{\bsnm{{Ananyeva}}, \binits{A.}},
\bauthor{\bsnm{{Anderson}}, \binits{S.B.}},
\bauthor{\bsnm{{Anderson}}, \binits{W.G.}},
\bauthor{\bsnm{{Ando}}, \binits{M.}},
\bauthor{\bsnm{{Appert}}, \binits{S.}},
\bauthor{\bsnm{{Arai}}, \binits{K.}},
\bauthor{\bsnm{{Araya}}, \binits{A.}},
\bauthor{\bsnm{{Araya}}, \binits{M.C.}},
\bauthor{\bsnm{{Areeda}}, \binits{J.S.}},
\bauthor{\bsnm{{Arnaud}}, \binits{N.}},
\bauthor{\bsnm{{Arun}}, \binits{K.G.}},
\bauthor{\bsnm{{Asada}}, \binits{H.}},
\bauthor{\bsnm{{Ascenzi}}, \binits{S.}},
\bauthor{\bsnm{{Ashton}}, \binits{G.}},
\bauthor{\bsnm{{Aso}}, \binits{Y.}},
\bauthor{\bsnm{{Ast}}, \binits{M.}},
\bauthor{\bsnm{{Aston}}, \binits{S.M.}},
\bauthor{\bsnm{{Astone}}, \binits{P.}},
\bauthor{\bsnm{{Atsuta}}, \binits{S.}},
\bauthor{\bsnm{{Aufmuth}}, \binits{P.}},
\bauthor{\bsnm{{Aulbert}}, \binits{C.}},
\bauthor{\bsnm{{Avila-Alvarez}}, \binits{A.}},
\bauthor{\bsnm{{Awai}}, \binits{K.}},
\bauthor{\bsnm{{Babak}}, \binits{S.}},
\bauthor{\bsnm{{Bacon}}, \binits{P.}},
\bauthor{\bsnm{{Bader}}, \binits{M.K.M.}},
\bauthor{\bsnm{{Baiotti}}, \binits{L.}},
\bauthor{\bsnm{{Baker}}, \binits{P.T.}},
\bauthor{\bsnm{{Baldaccini}}, \binits{F.}},
\bauthor{\bsnm{{Ballardin}}, \binits{G.}},
\bauthor{\bsnm{{Ballmer}}, \binits{S.W.}},
\bauthor{\bsnm{{Barayoga}}, \binits{J.C.}},
\bauthor{\bsnm{{Barclay}}, \binits{S.E.}},
\bauthor{\bsnm{{Barish}}, \binits{B.C.}},
\bauthor{\bsnm{{Barker}}, \binits{D.}},
\bauthor{\bsnm{{Barone}}, \binits{F.}},
\bauthor{\bsnm{{Barr}}, \binits{B.}},
\bauthor{\bsnm{{Barsotti}}, \binits{L.}},
\bauthor{\bsnm{{Barsuglia}}, \binits{M.}},
\bauthor{\bsnm{{Barta}}, \binits{D.}},
\bauthor{\bsnm{{Bartlett}}, \binits{J.}},
\bauthor{\bsnm{{Barton}}, \binits{M.A.}},
\bauthor{\bsnm{{Bartos}}, \binits{I.}},
\bauthor{\bsnm{{Bassiri}}, \binits{R.}},
\bauthor{\bsnm{{Basti}}, \binits{A.}},
\bauthor{\bsnm{{Batch}}, \binits{J.C.}},
\bauthor{\bsnm{{Baune}}, \binits{C.}},
\bauthor{\bsnm{{Bavigadda}}, \binits{V.}},
\bauthor{\bsnm{{Bazzan}}, \binits{M.}},
\bauthor{\bsnm{{B{\'e}csy}}, \binits{B.}},
\bauthor{\bsnm{{Beer}}, \binits{C.}},
\bauthor{\bsnm{{Bejger}}, \binits{M.}},
\bauthor{\bsnm{{Belahcene}}, \binits{I.}},
\bauthor{\bsnm{{Belgin}}, \binits{M.}},
\bauthor{\bsnm{{Bell}}, \binits{A.S.}},
\bauthor{\bsnm{{Berger}}, \binits{B.K.}},
\bauthor{\bsnm{{Bergmann}}, \binits{G.}},
\bauthor{\bsnm{{Berry}}, \binits{C.P.L.}},
\bauthor{\bsnm{{Bersanetti}}, \binits{D.}},
\bauthor{\bsnm{{Bertolini}}, \binits{A.}},
\bauthor{\bsnm{{Betzwieser}}, \binits{J.}},
\bauthor{\bsnm{{Bhagwat}}, \binits{S.}},
\bauthor{\bsnm{{Bhandare}}, \binits{R.}},
\bauthor{\bsnm{{Bilenko}}, \binits{I.A.}},
\bauthor{\bsnm{{Billingsley}}, \binits{G.}},
\bauthor{\bsnm{{Billman}}, \binits{C.R.}},
\bauthor{\bsnm{{Birch}}, \binits{J.}},
\bauthor{\bsnm{{Birney}}, \binits{R.}},
\bauthor{\bsnm{{Birnholtz}}, \binits{O.}},
\bauthor{\bsnm{{Biscans}}, \binits{S.}},
\bauthor{\bsnm{{Bisht}}, \binits{A.}},
\bauthor{\bsnm{{Bitossi}}, \binits{M.}},
\bauthor{\bsnm{{Biwer}}, \binits{C.}},
\bauthor{\bsnm{{Bizouard}}, \binits{M.A.}},
\bauthor{\bsnm{{Blackburn}}, \binits{J.K.}},
\bauthor{\bsnm{{Blackman}}, \binits{J.}},
\bauthor{\bsnm{{Blair}}, \binits{C.D.}},
\bauthor{\bsnm{{Blair}}, \binits{D.G.}},
\bauthor{\bsnm{{Blair}}, \binits{R.M.}},
\bauthor{\bsnm{{Bloemen}}, \binits{S.}},
\bauthor{\bsnm{{Bock}}, \binits{O.}},
\bauthor{\bsnm{{Boer}}, \binits{M.}},
\bauthor{\bsnm{{Bogaert}}, \binits{G.}},
\bauthor{\bsnm{{Bohe}}, \binits{A.}},
\bauthor{\bsnm{{Bondu}}, \binits{F.}},
\bauthor{\bsnm{{Bonnand }}, \binits{R.}},
\bauthor{\bsnm{{Boom}}, \binits{B.A.}},
\bauthor{\bsnm{{Bork}}, \binits{R.}},
\bauthor{\bsnm{{Boschi}}, \binits{V.}},
\bauthor{\bsnm{{Bose}}, \binits{S.}},
\bauthor{\bsnm{{Bouffanais}}, \binits{Y.}},
\bauthor{\bsnm{{Bozzi}}, \binits{A.}},
\bauthor{\bsnm{{Bradaschia}}, \binits{C.}},
\bauthor{\bsnm{{Brady}}, \binits{P.R.}},
\bauthor{\bsnm{{Braginsky}}, \binits{V.B.}},
\bauthor{\bsnm{{Branchesi}}, \binits{M.}},
\bauthor{\bsnm{{Brau}}, \binits{J.E.}},
\bauthor{\bsnm{{Briant}}, \binits{T.}},
\bauthor{\bsnm{{Brillet}}, \binits{A.}},
\bauthor{\bsnm{{Brinkmann}}, \binits{M.}},
\bauthor{\bsnm{{Brisson}}, \binits{V.}},
\bauthor{\bsnm{{Brockill}}, \binits{P.}},
\bauthor{\bsnm{{Broida}}, \binits{J.E.}},
\bauthor{\bsnm{{Brooks}}, \binits{A.F.}},
\bauthor{\bsnm{{Brown}}, \binits{D.A.}},
\bauthor{\bsnm{{Brown}}, \binits{D.D.}},
\bauthor{\bsnm{{Brown}}, \binits{N.M.}},
\bauthor{\bsnm{{Brunett}}, \binits{S.}},
\bauthor{\bsnm{{Buchanan}}, \binits{C.C.}},
\bauthor{\bsnm{{Buikema}}, \binits{A.}},
\bauthor{\bsnm{{Bulik}}, \binits{T.}},
\bauthor{\bsnm{{Bulten}}, \binits{H.J.}},
\bauthor{\bsnm{{Buonanno}}, \binits{A.}},
\bauthor{\bsnm{{Buskulic}}, \binits{D.}},
\bauthor{\bsnm{{Buy}}, \binits{C.}},
\bauthor{\bsnm{{Byer}}, \binits{R.L.}},
\bauthor{\bsnm{{Cabero}}, \binits{M.}},
\bauthor{\bsnm{{Cadonati}}, \binits{L.}},
\bauthor{\bsnm{{Cagnoli}}, \binits{G.}},
\bauthor{\bsnm{{Cahillane}}, \binits{C.}},
\bauthor{\bsnm{{Calder{\'o}n Bustillo}}, \binits{J.}},
\bauthor{\bsnm{{Callister}}, \binits{T.A.}},
\bauthor{\bsnm{{Calloni}}, \binits{E.}},
\bauthor{\bsnm{{Camp}}, \binits{J.B.}},
\bauthor{\bsnm{{Cannon}}, \binits{K.C.}},
\bauthor{\bsnm{{Cao}}, \binits{H.}},
\bauthor{\bsnm{{Cao}}, \binits{J.}},
\bauthor{\bsnm{{Capano}}, \binits{C.D.}},
\bauthor{\bsnm{{Capocasa}}, \binits{E.}},
\bauthor{\bsnm{{Carbognani}}, \binits{F.}},
\bauthor{\bsnm{{Caride}}, \binits{S.}},
\bauthor{\bsnm{{Casanueva Diaz}}, \binits{J.}},
\bauthor{\bsnm{{Casentini}}, \binits{C.}},
\bauthor{\bsnm{{Caudill}}, \binits{S.}},
\bauthor{\bsnm{{Cavagli{\`a}}}, \binits{M.}},
\bauthor{\bsnm{{Cavalier}}, \binits{F.}},
\bauthor{\bsnm{{Cavalieri}}, \binits{R.}},
\bauthor{\bsnm{{Cella}}, \binits{G.}},
\bauthor{\bsnm{{Cepeda}}, \binits{C.B.}},
\bauthor{\bsnm{{Cerboni Baiardi}}, \binits{L.}},
\bauthor{\bsnm{{Cerretani}}, \binits{G.}},
\bauthor{\bsnm{{Cesarini}}, \binits{E.}},
\bauthor{\bsnm{{Chamberlin}}, \binits{S.J.}},
\bauthor{\bsnm{{Chan}}, \binits{M.}},
\bauthor{\bsnm{{Chao}}, \binits{S.}},
\bauthor{\bsnm{{Charlton}}, \binits{P.}},
\bauthor{\bsnm{{Chassande-Mottin}}, \binits{E.}},
\bauthor{\bsnm{{Cheeseboro}}, \binits{B.D.}},
\bauthor{\bsnm{{Chen}}, \binits{H.-Y.}},
\bauthor{\bsnm{{Chen}}, \binits{Y.}},
\bauthor{\bsnm{{Cheng}}, \binits{H.-P.}},
\bauthor{\bsnm{{Chincarini}}, \binits{A.}},
\bauthor{\bsnm{{Chiummo}}, \binits{A.}},
\bauthor{\bsnm{{Chmiel}}, \binits{T.}},
\bauthor{\bsnm{{Cho}}, \binits{H.S.}},
\bauthor{\bsnm{{Cho}}, \binits{M.}},
\bauthor{\bsnm{{Chow}}, \binits{J.H.}},
\bauthor{\bsnm{{Christensen}}, \binits{N.}},
\bauthor{\bsnm{{Chu}}, \binits{Q.}},
\bauthor{\bsnm{{Chua}}, \binits{A.J.K.}},
\bauthor{\bsnm{{Chua}}, \binits{S.}},
\bauthor{\bsnm{{Chung}}, \binits{S.}},
\bauthor{\bsnm{{Ciani}}, \binits{G.}},
\bauthor{\bsnm{{Clara}}, \binits{F.}},
\bauthor{\bsnm{{Clark}}, \binits{J.A.}},
\bauthor{\bsnm{{Cleva}}, \binits{F.}},
\bauthor{\bsnm{{Cocchieri}}, \binits{C.}},
\bauthor{\bsnm{{Coccia}}, \binits{E.}},
\bauthor{\bsnm{{Cohadon}}, \binits{P.-F.}},
\bauthor{\bsnm{{Colla}}, \binits{A.}},
\bauthor{\bsnm{{Collette}}, \binits{C.G.}},
\bauthor{\bsnm{{Cominsky}}, \binits{L.}},
\bauthor{\bsnm{{Constancio}}, \binits{M.}},
\bauthor{\bsnm{{Conti}}, \binits{L.}},
\bauthor{\bsnm{{Cooper}}, \binits{S.J.}},
\bauthor{\bsnm{{Corbitt}}, \binits{T.R.}},
\bauthor{\bsnm{{Cornish}}, \binits{N.}},
\bauthor{\bsnm{{Corsi}}, \binits{A.}},
\bauthor{\bsnm{{Cortese}}, \binits{S.}},
\bauthor{\bsnm{{Costa}}, \binits{C.A.}},
\bauthor{\bsnm{{Coughlin}}, \binits{M.W.}},
\bauthor{\bsnm{{Coughlin}}, \binits{S.B.}},
\bauthor{\bsnm{{Coulon}}, \binits{J.-P.}},
\bauthor{\bsnm{{Countryman}}, \binits{S.T.}},
\bauthor{\bsnm{{Couvares}}, \binits{P.}},
\bauthor{\bsnm{{Covas}}, \binits{P.B.}},
\bauthor{\bsnm{{Cowan}}, \binits{E.E.}},
\bauthor{\bsnm{{Coward}}, \binits{D.M.}},
\bauthor{\bsnm{{Cowart}}, \binits{M.J.}},
\bauthor{\bsnm{{Coyne}}, \binits{D.C.}},
\bauthor{\bsnm{{Coyne}}, \binits{R.}},
\bauthor{\bsnm{{Creighton}}, \binits{J.D.E.}},
\bauthor{\bsnm{{Creighton}}, \binits{T.D.}},
\bauthor{\bsnm{{Cripe}}, \binits{J.}},
\bauthor{\bsnm{{Crowder}}, \binits{S.G.}},
\bauthor{\bsnm{{Cullen}}, \binits{T.J.}},
\bauthor{\bsnm{{Cumming}}, \binits{A.}},
\bauthor{\bsnm{{Cunningham}}, \binits{L.}},
\bauthor{\bsnm{{Cuoco}}, \binits{E.}},
\bauthor{\bsnm{{Dal Canton}}, \binits{T.}},
\bauthor{\bsnm{{Danilishin}}, \binits{S.L.}},
\bauthor{\bsnm{{D'Antonio}}, \binits{S.}},
\bauthor{\bsnm{{Danzmann}}, \binits{K.}},
\bauthor{\bsnm{{Dasgupta}}, \binits{A.}},
\bauthor{\bsnm{{Da Silva Costa}}, \binits{C.F.}},
\bauthor{\bsnm{{Dattilo}}, \binits{V.}},
\bauthor{\bsnm{{Dave}}, \binits{I.}},
\bauthor{\bsnm{{Davier}}, \binits{M.}},
\bauthor{\bsnm{{Davies}}, \binits{G.S.}},
\bauthor{\bsnm{{Davis}}, \binits{D.}},
\bauthor{\bsnm{{Daw}}, \binits{E.J.}},
\bauthor{\bsnm{{Day}}, \binits{B.}},
\bauthor{\bsnm{{Day}}, \binits{R.}},
\bauthor{\bsnm{{De}}, \binits{S.}},
\bauthor{\bsnm{{DeBra}}, \binits{D.}},
\bauthor{\bsnm{{Debreczeni}}, \binits{G.}},
\bauthor{\bsnm{{Degallaix}}, \binits{J.}},
\bauthor{\bsnm{{De Laurentis}}, \binits{M.}},
\bauthor{\bsnm{{Del{\'e}glise}}, \binits{S.}},
\bauthor{\bsnm{{Del Pozzo}}, \binits{W.}},
\bauthor{\bsnm{{Denker}}, \binits{T.}},
\bauthor{\bsnm{{Dent}}, \binits{T.}},
\bauthor{\bsnm{{Dergachev}}, \binits{V.}},
\bauthor{\bsnm{{De Rosa}}, \binits{R.}},
\bauthor{\bsnm{{DeRosa}}, \binits{R.T.}},
\bauthor{\bsnm{{DeSalvo}}, \binits{R.}},
\bauthor{\bsnm{{Devine}}, \binits{R.C.}},
\bauthor{\bsnm{{Dhurandhar}}, \binits{S.}},
\bauthor{\bsnm{{D{\'\i}az}}, \binits{M.C.}},
\bauthor{\bsnm{{Fiore}}, \binits{L.D.}},
\bauthor{\bsnm{{Giovanni}}, \binits{M.D.}},
\bauthor{\bsnm{{Girolamo}}, \binits{T.D.}},
\bauthor{\bsnm{{Lieto}}, \binits{A.D.}},
\bauthor{\bsnm{{Pace}}, \binits{S.D.}},
\bauthor{\bsnm{{Palma}}, \binits{I.D.}},
\bauthor{\bsnm{{Virgilio}}, \binits{A.D.}},
\bauthor{\bsnm{{Doctor}}, \binits{Z.}},
\bauthor{\bsnm{{Doi}}, \binits{K.}},
\bauthor{\bsnm{{Dolique}}, \binits{V.}},
\bauthor{\bsnm{{Donovan}}, \binits{F.}},
\bauthor{\bsnm{{Dooley}}, \binits{K.L.}},
\bauthor{\bsnm{{Doravari}}, \binits{S.}},
\bauthor{\bsnm{{Dorrington}}, \binits{I.}},
\bauthor{\bsnm{{Douglas}}, \binits{R.}},
\bauthor{\bsnm{{Dovale {\'A}lvarez}}, \binits{M.}},
\bauthor{\bsnm{{Downes}}, \binits{T.P.}},
\bauthor{\bsnm{{Drago}}, \binits{M.}},
\bauthor{\bsnm{{Drever}}, \binits{R.W.P.}},
\bauthor{\bsnm{{Driggers}}, \binits{J.C.}},
\bauthor{\bsnm{{Du}}, \binits{Z.}},
\bauthor{\bsnm{{Ducrot}}, \binits{M.}},
\bauthor{\bsnm{{Dwyer}}, \binits{S.E.}},
\bauthor{\bsnm{{Eda}}, \binits{K.}},
\bauthor{\bsnm{{Edo}}, \binits{T.B.}},
\bauthor{\bsnm{{Edwards}}, \binits{M.C.}},
\bauthor{\bsnm{{Effler}}, \binits{A.}},
\bauthor{\bsnm{{Eggenstein}}, \binits{H.-B.}},
\bauthor{\bsnm{{Ehrens}}, \binits{P.}},
\bauthor{\bsnm{{Eichholz}}, \binits{J.}},
\bauthor{\bsnm{{Eikenberry}}, \binits{S.S.}},
\bauthor{\bsnm{{Eisenstein}}, \binits{R.A.}},
\bauthor{\bsnm{{Essick}}, \binits{R.C.}},
\bauthor{\bsnm{{Etienne}}, \binits{Z.}},
\bauthor{\bsnm{{Etzel}}, \binits{T.}},
\bauthor{\bsnm{{Evans}}, \binits{M.}},
\bauthor{\bsnm{{Evans}}, \binits{T.M.}},
\bauthor{\bsnm{{Everett}}, \binits{R.}},
\bauthor{\bsnm{{Factourovich}}, \binits{M.}},
\bauthor{\bsnm{{Fafone}}, \binits{V.}},
\bauthor{\bsnm{{Fair}}, \binits{H.}},
\bauthor{\bsnm{{Fairhurst}}, \binits{S.}},
\bauthor{\bsnm{{Fan}}, \binits{X.}},
\bauthor{\bsnm{{Farinon}}, \binits{S.}},
\bauthor{\bsnm{{Farr}}, \binits{B.}},
\bauthor{\bsnm{{Farr}}, \binits{W.M.}},
\bauthor{\bsnm{{Fauchon-Jones}}, \binits{E.J.}},
\bauthor{\bsnm{{Favata}}, \binits{M.}},
\bauthor{\bsnm{{Fays}}, \binits{M.}},
\bauthor{\bsnm{{Fehrmann}}, \binits{H.}},
\bauthor{\bsnm{{Fejer}}, \binits{M.M.}},
\bauthor{\bsnm{{Fern{\'a}ndez Galiana}}, \binits{A.}},
\bauthor{\bsnm{{Ferrante}}, \binits{I.}},
\bauthor{\bsnm{{Ferreira}}, \binits{E.C.}},
\bauthor{\bsnm{{Ferrini}}, \binits{F.}},
\bauthor{\bsnm{{Fidecaro}}, \binits{F.}},
\bauthor{\bsnm{{Fiori}}, \binits{I.}},
\bauthor{\bsnm{{Fiorucci}}, \binits{D.}},
\bauthor{\bsnm{{Fisher}}, \binits{R.P.}},
\bauthor{\bsnm{{Flaminio}}, \binits{R.}},
\bauthor{\bsnm{{Fletcher}}, \binits{M.}},
\bauthor{\bsnm{{Fong}}, \binits{H.}},
\bauthor{\bsnm{{Forsyth}}, \binits{S.S.}},
\bauthor{\bsnm{{Fournier}}, \binits{J.-D.}},
\bauthor{\bsnm{{Frasca}}, \binits{S.}},
\bauthor{\bsnm{{Frasconi}}, \binits{F.}},
\bauthor{\bsnm{{Frei}}, \binits{Z.}},
\bauthor{\bsnm{{Freise}}, \binits{A.}},
\bauthor{\bsnm{{Frey}}, \binits{R.}},
\bauthor{\bsnm{{Frey}}, \binits{V.}},
\bauthor{\bsnm{{Fries}}, \binits{E.M.}},
\bauthor{\bsnm{{Fritschel}}, \binits{P.}},
\bauthor{\bsnm{{Frolov}}, \binits{V.V.}},
\bauthor{\bsnm{{Fujii}}, \binits{Y.}},
\bauthor{\bsnm{{Fujimoto}}, \binits{M.-K.}},
\bauthor{\bsnm{{Fulda}}, \binits{P.}},
\bauthor{\bsnm{{Fyffe}}, \binits{M.}},
\bauthor{\bsnm{{Gabbard}}, \binits{H.}},
\bauthor{\bsnm{{Gadre}}, \binits{B.U.}},
\bauthor{\bsnm{{Gaebel}}, \binits{S.M.}},
\bauthor{\bsnm{{Gair}}, \binits{J.R.}},
\bauthor{\bsnm{{Gammaitoni}}, \binits{L.}},
\bauthor{\bsnm{{Gaonkar}}, \binits{S.G.}},
\bauthor{\bsnm{{Garufi}}, \binits{F.}},
\bauthor{\bsnm{{Gaur}}, \binits{G.}},
\bauthor{\bsnm{{Gayathri}}, \binits{V.}},
\bauthor{\bsnm{{Gehrels}}, \binits{N.}},
\bauthor{\bsnm{{Gemme}}, \binits{G.}},
\bauthor{\bsnm{{Genin}}, \binits{E.}},
\bauthor{\bsnm{{Gennai}}, \binits{A.}},
\bauthor{\bsnm{{George}}, \binits{J.}},
\bauthor{\bsnm{{Gergely}}, \binits{L.}},
\bauthor{\bsnm{{Germain}}, \binits{V.}},
\bauthor{\bsnm{{Ghonge}}, \binits{S.}},
\bauthor{\bsnm{{Ghosh}}, \binits{A.}},
\bauthor{\bsnm{{Ghosh}}, \binits{A.}},
\bauthor{\bsnm{{Ghosh}}, \binits{S.}},
\bauthor{\bsnm{{Giaime}}, \binits{J.A.}},
\bauthor{\bsnm{{Giardina}}, \binits{K.D.}},
\bauthor{\bsnm{{Giazotto}}, \binits{A.}},
\bauthor{\bsnm{{Gill}}, \binits{K.}},
\bauthor{\bsnm{{Glaefke}}, \binits{A.}},
\bauthor{\bsnm{{Goetz}}, \binits{E.}},
\bauthor{\bsnm{{Goetz}}, \binits{R.}},
\bauthor{\bsnm{{Gondan}}, \binits{L.}},
\bauthor{\bsnm{{Gonz{\'a}lez}}, \binits{G.}},
\bauthor{\bsnm{{Gonzalez Castro}}, \binits{J.M.}},
\bauthor{\bsnm{{Gopakumar}}, \binits{A.}},
\bauthor{\bsnm{{Gorodetsky}}, \binits{M.L.}},
\bauthor{\bsnm{{Gossan}}, \binits{S.E.}},
\bauthor{\bsnm{{Gosselin}}, \binits{M.}},
\bauthor{\bsnm{{Gouaty}}, \binits{R.}},
\bauthor{\bsnm{{Grado}}, \binits{A.}},
\bauthor{\bsnm{{Graef}}, \binits{C.}},
\bauthor{\bsnm{{Granata}}, \binits{M.}},
\bauthor{\bsnm{{Grant}}, \binits{A.}},
\bauthor{\bsnm{{Gras}}, \binits{S.}},
\bauthor{\bsnm{{Gray}}, \binits{C.}},
\bauthor{\bsnm{{Greco}}, \binits{G.}},
\bauthor{\bsnm{{Green}}, \binits{A.C.}},
\bauthor{\bsnm{{Groot}}, \binits{P.}},
\bauthor{\bsnm{{Grote}}, \binits{H.}},
\bauthor{\bsnm{{Grunewald}}, \binits{S.}},
\bauthor{\bsnm{{Guidi}}, \binits{G.M.}},
\bauthor{\bsnm{{Guo}}, \binits{X.}},
\bauthor{\bsnm{{Gupta}}, \binits{A.}},
\bauthor{\bsnm{{Gupta}}, \binits{M.K.}},
\bauthor{\bsnm{{Gushwa}}, \binits{K.E.}},
\bauthor{\bsnm{{Gustafson}}, \binits{E.K.}},
\bauthor{\bsnm{{Gustafson}}, \binits{R.}},
\bauthor{\bsnm{{Hacker}}, \binits{J.J.}},
\bauthor{\bsnm{{Hagiwara}}, \binits{A.}},
\bauthor{\bsnm{{Hall}}, \binits{B.R.}},
\bauthor{\bsnm{{Hall}}, \binits{E.D.}},
\bauthor{\bsnm{{Hammond}}, \binits{G.}},
\bauthor{\bsnm{{Haney}}, \binits{M.}},
\bauthor{\bsnm{{Hanke}}, \binits{M.M.}},
\bauthor{\bsnm{{Hanks}}, \binits{J.}},
\bauthor{\bsnm{{Hanna}}, \binits{C.}},
\bauthor{\bsnm{{Hannam}}, \binits{M.D.}},
\bauthor{\bsnm{{Hanson}}, \binits{J.}},
\bauthor{\bsnm{{Hardwick}}, \binits{T.}},
\bauthor{\bsnm{{Harms}}, \binits{J.}},
\bauthor{\bsnm{{Harry}}, \binits{G.M.}},
\bauthor{\bsnm{{Harry}}, \binits{I.W.}},
\bauthor{\bsnm{{Hart}}, \binits{M.J.}},
\bauthor{\bsnm{{Hartman}}, \binits{M.T.}},
\bauthor{\bsnm{{Haster}}, \binits{C.-J.}},
\bauthor{\bsnm{{Haughian}}, \binits{K.}},
\bauthor{\bsnm{{Hayama}}, \binits{K.}},
\bauthor{\bsnm{{Healy}}, \binits{J.}},
\bauthor{\bsnm{{Heidmann}}, \binits{A.}},
\bauthor{\bsnm{{Heintze}}, \binits{M.C.}},
\bauthor{\bsnm{{Heitmann}}, \binits{H.}},
\bauthor{\bsnm{{Hello}}, \binits{P.}},
\bauthor{\bsnm{{Hemming}}, \binits{G.}},
\bauthor{\bsnm{{Hendry}}, \binits{M.}},
\bauthor{\bsnm{{Heng}}, \binits{I.S.}},
\bauthor{\bsnm{{Hennig}}, \binits{J.}},
\bauthor{\bsnm{{Henry}}, \binits{J.}},
\bauthor{\bsnm{{Heptonstall}}, \binits{A.W.}},
\bauthor{\bsnm{{Heurs}}, \binits{M.}},
\bauthor{\bsnm{{Hild}}, \binits{S.}},
\bauthor{\bsnm{{Hirose}}, \binits{E.}},
\bauthor{\bsnm{{Hoak}}, \binits{D.}},
\bauthor{\bsnm{{Hofman}}, \binits{D.}},
\bauthor{\bsnm{{Holt}}, \binits{K.}},
\bauthor{\bsnm{{Holz}}, \binits{D.E.}},
\bauthor{\bsnm{{Hopkins}}, \binits{P.}},
\bauthor{\bsnm{{Hough}}, \binits{J.}},
\bauthor{\bsnm{{Houston}}, \binits{E.A.}},
\bauthor{\bsnm{{Howell}}, \binits{E.J.}},
\bauthor{\bsnm{{Hu}}, \binits{Y.M.}},
\bauthor{\bsnm{{Huerta}}, \binits{E.A.}},
\bauthor{\bsnm{{Huet}}, \binits{D.}},
\bauthor{\bsnm{{Hughey}}, \binits{B.}},
\bauthor{\bsnm{{Husa}}, \binits{S.}},
\bauthor{\bsnm{{Huttner}}, \binits{S.H.}},
\bauthor{\bsnm{{Huynh-Dinh}}, \binits{T.}},
\bauthor{\bsnm{{Indik}}, \binits{N.}},
\bauthor{\bsnm{{Ingram}}, \binits{D.R.}},
\bauthor{\bsnm{{Inta}}, \binits{R.}},
\bauthor{\bsnm{{Ioka}}, \binits{K.}},
\bauthor{\bsnm{{Isa}}, \binits{H.N.}},
\bauthor{\bsnm{{Isac}}, \binits{J.-M.}},
\bauthor{\bsnm{{Isi}}, \binits{M.}},
\bauthor{\bsnm{{Isogai}}, \binits{T.}},
\bauthor{\bsnm{{Itoh}}, \binits{Y.}},
\bauthor{\bsnm{{Iyer}}, \binits{B.R.}},
\bauthor{\bsnm{{Izumi}}, \binits{K.}},
\bauthor{\bsnm{{Jacqmin}}, \binits{T.}},
\bauthor{\bsnm{{Jani}}, \binits{K.}},
\bauthor{\bsnm{{Jaranowski}}, \binits{P.}},
\bauthor{\bsnm{{Jawahar}}, \binits{S.}},
\bauthor{\bsnm{{Jim{\'e}nez-Forteza}}, \binits{F.}},
\bauthor{\bsnm{{Johnson}}, \binits{W.W.}},
\bauthor{\bsnm{{Jones}}, \binits{D.I.}},
\bauthor{\bsnm{{Jones}}, \binits{R.}},
\bauthor{\bsnm{{Jonker}}, \binits{R.J.G.}},
\bauthor{\bsnm{{Ju}}, \binits{L.}},
\bauthor{\bsnm{{Junker}}, \binits{J.}},
\bauthor{\bsnm{{Kagawa}}, \binits{T.}},
\bauthor{\bsnm{{Kajita}}, \binits{T.}},
\bauthor{\bsnm{{Kakizaki}}, \binits{M.}},
\bauthor{\bsnm{{Kalaghatgi}}, \binits{C.V.}},
\bauthor{\bsnm{{Kalogera}}, \binits{V.}},
\bauthor{\bsnm{{Kamiizumi}}, \binits{M.}},
\bauthor{\bsnm{{Kanda}}, \binits{N.}},
\bauthor{\bsnm{{Kand hasamy}}, \binits{S.}},
\bauthor{\bsnm{{Kanemura}}, \binits{S.}},
\bauthor{\bsnm{{Kaneyama}}, \binits{M.}},
\bauthor{\bsnm{{Kang}}, \binits{G.}},
\bauthor{\bsnm{{Kanner}}, \binits{J.B.}},
\bauthor{\bsnm{{Karki}}, \binits{S.}},
\bauthor{\bsnm{{Karvinen}}, \binits{K.S.}},
\bauthor{\bsnm{{Kasprzack}}, \binits{M.}},
\bauthor{\bsnm{{Kataoka}}, \binits{Y.}},
\bauthor{\bsnm{{Katsavounidis}}, \binits{E.}},
\bauthor{\bsnm{{Katzman}}, \binits{W.}},
\bauthor{\bsnm{{Kaufer}}, \binits{S.}},
\bauthor{\bsnm{{Kaur}}, \binits{T.}},
\bauthor{\bsnm{{Kawabe}}, \binits{K.}},
\bauthor{\bsnm{{Kawai}}, \binits{N.}},
\bauthor{\bsnm{{Kawamura}}, \binits{S.}},
\bauthor{\bsnm{{K{\'e}f{\'e}lian}}, \binits{F.}},
\bauthor{\bsnm{{Keitel}}, \binits{D.}},
\bauthor{\bsnm{{Kelley}}, \binits{D.B.}},
\bauthor{\bsnm{{Kennedy}}, \binits{R.}},
\bauthor{\bsnm{{Key}}, \binits{J.S.}},
\bauthor{\bsnm{{Khalili}}, \binits{F.Y.}},
\bauthor{\bsnm{{Khan}}, \binits{I.}},
\bauthor{\bsnm{{Khan}}, \binits{S.}},
\bauthor{\bsnm{{Khan}}, \binits{Z.}},
\bauthor{\bsnm{{Khazanov}}, \binits{E.A.}},
\bauthor{\bsnm{{Kijbunchoo}}, \binits{N.}},
\bauthor{\bsnm{{Kim}}, \binits{C.}},
\bauthor{\bsnm{{Kim}}, \binits{H.}},
\bauthor{\bsnm{{Kim}}, \binits{J.C.}},
\bauthor{\bsnm{{Kim}}, \binits{J.}},
\bauthor{\bsnm{{Kim}}, \binits{W.}},
\bauthor{\bsnm{{Kim}}, \binits{Y.-M.}},
\bauthor{\bsnm{{Kimbrell}}, \binits{S.J.}},
\bauthor{\bsnm{{Kimura}}, \binits{N.}},
\bauthor{\bsnm{{King}}, \binits{E.J.}},
\bauthor{\bsnm{{King}}, \binits{P.J.}},
\bauthor{\bsnm{{Kirchhoff}}, \binits{R.}},
\bauthor{\bsnm{{Kissel}}, \binits{J.S.}},
\bauthor{\bsnm{{Klein}}, \binits{B.}},
\bauthor{\bsnm{{Kleybolte}}, \binits{L.}},
\bauthor{\bsnm{{Klimenko}}, \binits{S.}},
\bauthor{\bsnm{{Koch}}, \binits{P.}},
\bauthor{\bsnm{{Koehlenbeck}}, \binits{S.M.}},
\bauthor{\bsnm{{Kojima}}, \binits{Y.}},
\bauthor{\bsnm{{Kokeyama}}, \binits{K.}},
\bauthor{\bsnm{{Koley}}, \binits{S.}},
\bauthor{\bsnm{{Komori}}, \binits{K.}},
\bauthor{\bsnm{{Kondrashov}}, \binits{V.}},
\bauthor{\bsnm{{Kontos}}, \binits{A.}},
\bauthor{\bsnm{{Korobko}}, \binits{M.}},
\bauthor{\bsnm{{Korth}}, \binits{W.Z.}},
\bauthor{\bsnm{{Kotake}}, \binits{K.}},
\bauthor{\bsnm{{Kowalska}}, \binits{I.}},
\bauthor{\bsnm{{Kozak}}, \binits{D.B.}},
\bauthor{\bsnm{{Kr{\"a}mer}}, \binits{C.}},
\bauthor{\bsnm{{Kringel}}, \binits{V.}},
\bauthor{\bsnm{{Krishnan}}, \binits{B.}},
\bauthor{\bsnm{{Kr{\'o}lak}}, \binits{A.}},
\bauthor{\bsnm{{Kuehn}}, \binits{G.}},
\bauthor{\bsnm{{Kumar}}, \binits{P.}},
\bauthor{\bsnm{{Kumar}}, \binits{R.}},
\bauthor{\bsnm{{Kumar}}, \binits{R.}},
\bauthor{\bsnm{{Kuo}}, \binits{L.}},
\bauthor{\bsnm{{Kuroda}}, \binits{K.}},
\bauthor{\bsnm{{Kutynia}}, \binits{A.}},
\bauthor{\bsnm{{Kuwahara}}, \binits{Y.}},
\bauthor{\bsnm{{Lackey}}, \binits{B.D.}},
\bauthor{\bsnm{{Landry}}, \binits{M.}},
\bauthor{\bsnm{{Lang}}, \binits{R.N.}},
\bauthor{\bsnm{{Lange}}, \binits{J.}},
\bauthor{\bsnm{{Lantz}}, \binits{B.}},
\bauthor{\bsnm{{Lanza}}, \binits{R.K.}},
\bauthor{\bsnm{{Lartaux-Vollard}}, \binits{A.}},
\bauthor{\bsnm{{Lasky}}, \binits{P.D.}},
\bauthor{\bsnm{{Laxen}}, \binits{M.}},
\bauthor{\bsnm{{Lazzarini}}, \binits{A.}},
\bauthor{\bsnm{{Lazzaro}}, \binits{C.}},
\bauthor{\bsnm{{Leaci}}, \binits{P.}},
\bauthor{\bsnm{{Leavey}}, \binits{S.}},
\bauthor{\bsnm{{Lebigot}}, \binits{E.O.}},
\bauthor{\bsnm{{Lee}}, \binits{C.H.}},
\bauthor{\bsnm{{Lee}}, \binits{H.K.}},
\bauthor{\bsnm{{Lee}}, \binits{H.M.}},
\bauthor{\bsnm{{Lee}}, \binits{H.W.}},
\bauthor{\bsnm{{Lee}}, \binits{K.}},
\bauthor{\bsnm{{Lehmann}}, \binits{J.}},
\bauthor{\bsnm{{Lenon}}, \binits{A.}},
\bauthor{\bsnm{{Leonardi}}, \binits{M.}},
\bauthor{\bsnm{{Leong}}, \binits{J.R.}},
\bauthor{\bsnm{{Leroy}}, \binits{N.}},
\bauthor{\bsnm{{Letendre}}, \binits{N.}},
\bauthor{\bsnm{{Levin}}, \binits{Y.}},
\bauthor{\bsnm{{Li}}, \binits{T.G.F.}},
\bauthor{\bsnm{{Libson}}, \binits{A.}},
\bauthor{\bsnm{{Littenberg}}, \binits{T.B.}},
\bauthor{\bsnm{{Liu}}, \binits{J.}},
\bauthor{\bsnm{{Lockerbie}}, \binits{N.A.}},
\bauthor{\bsnm{{Lombardi}}, \binits{A.L.}},
\bauthor{\bsnm{{London}}, \binits{L.T.}},
\bauthor{\bsnm{{Lord}}, \binits{J.E.}},
\bauthor{\bsnm{{Lorenzini}}, \binits{M.}},
\bauthor{\bsnm{{Loriette}}, \binits{V.}},
\bauthor{\bsnm{{Lormand}}, \binits{M.}},
\bauthor{\bsnm{{Losurdo}}, \binits{G.}},
\bauthor{\bsnm{{Lough}}, \binits{J.D.}},
\bauthor{\bsnm{{Lousto}}, \binits{C.O.}},
\bauthor{\bsnm{{Lovelace}}, \binits{G.}},
\bauthor{\bsnm{{L{\"u}ck}}, \binits{H.}},
\bauthor{\bsnm{{Lundgren}}, \binits{A.P.}},
\bauthor{\bsnm{{Lynch}}, \binits{R.}},
\bauthor{\bsnm{{Ma}}, \binits{Y.}},
\bauthor{\bsnm{{Macfoy}}, \binits{S.}},
\bauthor{\bsnm{{Machenschalk}}, \binits{B.}},
\bauthor{\bsnm{{MacInnis}}, \binits{M.}},
\bauthor{\bsnm{{Macleod}}, \binits{D.M.}},
\bauthor{\bsnm{{Maga{\~n}a-Sandoval}}, \binits{F.}},
\bauthor{\bsnm{{Majorana}}, \binits{E.}},
\bauthor{\bsnm{{Maksimovic}}, \binits{I.}},
\bauthor{\bsnm{{Malvezzi}}, \binits{V.}},
\bauthor{\bsnm{{Man}}, \binits{N.}},
\bauthor{\bsnm{{Mandic}}, \binits{V.}},
\bauthor{\bsnm{{Mangano}}, \binits{V.}},
\bauthor{\bsnm{{Mano}}, \binits{S.}},
\bauthor{\bsnm{{Mansell}}, \binits{G.L.}},
\bauthor{\bsnm{{Manske}}, \binits{M.}},
\bauthor{\bsnm{{Mantovani}}, \binits{M.}},
\bauthor{\bsnm{{Marchesoni}}, \binits{F.}},
\bauthor{\bsnm{{Marchio}}, \binits{M.}},
\bauthor{\bsnm{{Marion}}, \binits{F.}},
\bauthor{\bsnm{{M{\'a}rka}}, \binits{S.}},
\bauthor{\bsnm{{M{\'a}rka}}, \binits{Z.}},
\bauthor{\bsnm{{Markosyan}}, \binits{A.S.}},
\bauthor{\bsnm{{Maros}}, \binits{E.}},
\bauthor{\bsnm{{Martelli}}, \binits{F.}},
\bauthor{\bsnm{{Martellini}}, \binits{L.}},
\bauthor{\bsnm{{Martin}}, \binits{I.W.}},
\bauthor{\bsnm{{Martynov}}, \binits{D.V.}},
\bauthor{\bsnm{{Mason}}, \binits{K.}},
\bauthor{\bsnm{{Masserot}}, \binits{A.}},
\bauthor{\bsnm{{Massinger}}, \binits{T.J.}},
\bauthor{\bsnm{{Masso-Reid}}, \binits{M.}},
\bauthor{\bsnm{{Mastrogiovanni}}, \binits{S.}},
\bauthor{\bsnm{{Matichard}}, \binits{F.}},
\bauthor{\bsnm{{Matone}}, \binits{L.}},
\bauthor{\bsnm{{Matsumoto}}, \binits{N.}},
\bauthor{\bsnm{{Matsushima}}, \binits{F.}},
\bauthor{\bsnm{{Mavalvala}}, \binits{N.}},
\bauthor{\bsnm{{Mazumder}}, \binits{N.}},
\bauthor{\bsnm{{McCarthy}}, \binits{R.}},
\bauthor{\bsnm{{McClelland }}, \binits{D.E.}},
\bauthor{\bsnm{{McCormick}}, \binits{S.}},
\bauthor{\bsnm{{McGrath}}, \binits{C.}},
\bauthor{\bsnm{{McGuire}}, \binits{S.C.}},
\bauthor{\bsnm{{McIntyre}}, \binits{G.}},
\bauthor{\bsnm{{McIver}}, \binits{J.}},
\bauthor{\bsnm{{McManus}}, \binits{D.J.}},
\bauthor{\bsnm{{McRae}}, \binits{T.}},
\bauthor{\bsnm{{McWilliams}}, \binits{S.T.}},
\bauthor{\bsnm{{Meacher}}, \binits{D.}},
\bauthor{\bsnm{{Meadors}}, \binits{G.D.}},
\bauthor{\bsnm{{Meidam}}, \binits{J.}},
\bauthor{\bsnm{{Melatos}}, \binits{A.}},
\bauthor{\bsnm{{Mendell}}, \binits{G.}},
\bauthor{\bsnm{{Mendoza-Gand ara}}, \binits{D.}},
\bauthor{\bsnm{{Mercer}}, \binits{R.A.}},
\bauthor{\bsnm{{Merilh}}, \binits{E.L.}},
\bauthor{\bsnm{{Merzougui}}, \binits{M.}},
\bauthor{\bsnm{{Meshkov}}, \binits{S.}},
\bauthor{\bsnm{{Messenger}}, \binits{C.}},
\bauthor{\bsnm{{Messick}}, \binits{C.}},
\bauthor{\bsnm{{Metzdorff}}, \binits{R.}},
\bauthor{\bsnm{{Meyers}}, \binits{P.M.}},
\bauthor{\bsnm{{Mezzani}}, \binits{F.}},
\bauthor{\bsnm{{Miao}}, \binits{H.}},
\bauthor{\bsnm{{Michel}}, \binits{C.}},
\bauthor{\bsnm{{Michimura}}, \binits{Y.}},
\bauthor{\bsnm{{Middleton}}, \binits{H.}},
\bauthor{\bsnm{{Mikhailov}}, \binits{E.E.}},
\bauthor{\bsnm{{Milano}}, \binits{L.}},
\bauthor{\bsnm{{Miller}}, \binits{A.L.}},
\bauthor{\bsnm{{Miller}}, \binits{A.}},
\bauthor{\bsnm{{Miller}}, \binits{B.B.}},
\bauthor{\bsnm{{Miller}}, \binits{J.}},
\bauthor{\bsnm{{Millhouse}}, \binits{M.}},
\bauthor{\bsnm{{Minenkov}}, \binits{Y.}},
\bauthor{\bsnm{{Ming}}, \binits{J.}},
\bauthor{\bsnm{{Mirshekari}}, \binits{S.}},
\bauthor{\bsnm{{Mishra}}, \binits{C.}},
\bauthor{\bsnm{{Mitrofanov}}, \binits{V.P.}},
\bauthor{\bsnm{{Mitselmakher}}, \binits{G.}},
\bauthor{\bsnm{{Mittleman}}, \binits{R.}},
\bauthor{\bsnm{{Miyakawa}}, \binits{O.}},
\bauthor{\bsnm{{Miyamoto}}, \binits{A.}},
\bauthor{\bsnm{{Miyamoto}}, \binits{T.}},
\bauthor{\bsnm{{Miyoki}}, \binits{S.}},
\bauthor{\bsnm{{Moggi}}, \binits{A.}},
\bauthor{\bsnm{{Mohan}}, \binits{M.}},
\bauthor{\bsnm{{Mohapatra}}, \binits{S.R.P.}},
\bauthor{\bsnm{{Montani}}, \binits{M.}},
\bauthor{\bsnm{{Moore}}, \binits{B.C.}},
\bauthor{\bsnm{{Moore}}, \binits{C.J.}},
\bauthor{\bsnm{{Moraru}}, \binits{D.}},
\bauthor{\bsnm{{Moreno}}, \binits{G.}},
\bauthor{\bsnm{{Morii}}, \binits{W.}},
\bauthor{\bsnm{{Morisaki}}, \binits{S.}},
\bauthor{\bsnm{{Moriwaki}}, \binits{Y.}},
\bauthor{\bsnm{{Morriss}}, \binits{S.R.}},
\bauthor{\bsnm{{Mours}}, \binits{B.}},
\bauthor{\bsnm{{Mow-Lowry}}, \binits{C.M.}},
\bauthor{\bsnm{{Mueller}}, \binits{G.}},
\bauthor{\bsnm{{Muir}}, \binits{A.W.}},
\bauthor{\bsnm{{Mukherjee}}, \binits{A.}},
\bauthor{\bsnm{{Mukherjee}}, \binits{D.}},
\bauthor{\bsnm{{Mukherjee}}, \binits{S.}},
\bauthor{\bsnm{{Mukund}}, \binits{N.}},
\bauthor{\bsnm{{Mullavey}}, \binits{A.}},
\bauthor{\bsnm{{Munch}}, \binits{J.}},
\bauthor{\bsnm{{Muniz}}, \binits{E.A.M.}},
\bauthor{\bsnm{{Murray}}, \binits{P.G.}},
\bauthor{\bsnm{{Mytidis}}, \binits{A.}},
\bauthor{\bsnm{{Nagano}}, \binits{S.}},
\bauthor{\bsnm{{Nakamura}}, \binits{K.}},
\bauthor{\bsnm{{Nakamura}}, \binits{T.}},
\bauthor{\bsnm{{Nakano}}, \binits{H.}},
\bauthor{\bsnm{{Nakano}}, \binits{M.}},
\bauthor{\bsnm{{Nakano}}, \binits{M.}},
\bauthor{\bsnm{{Nakao}}, \binits{K.}},
\bauthor{\bsnm{{Napier}}, \binits{K.}},
\bauthor{\bsnm{{Nardecchia}}, \binits{I.}},
\bauthor{\bsnm{{Narikawa}}, \binits{T.}},
\bauthor{\bsnm{{Naticchioni}}, \binits{L.}},
\bauthor{\bsnm{{Nelemans}}, \binits{G.}},
\bauthor{\bsnm{{Nelson}}, \binits{T.J.N.}},
\bauthor{\bsnm{{Neri}}, \binits{M.}},
\bauthor{\bsnm{{Nery}}, \binits{M.}},
\bauthor{\bsnm{{Neunzert}}, \binits{A.}},
\bauthor{\bsnm{{Newport}}, \binits{J.M.}},
\bauthor{\bsnm{{Newton}}, \binits{G.}},
\bauthor{\bsnm{{Nguyen}}, \binits{T.T.}},
\bauthor{\bsnm{{Ni}}, \binits{W.-T.}},
\bauthor{\bsnm{{Nielsen}}, \binits{A.B.}},
\bauthor{\bsnm{{Nissanke}}, \binits{S.}},
\bauthor{\bsnm{{Nitz}}, \binits{A.}},
\bauthor{\bsnm{{Noack}}, \binits{A.}},
\bauthor{\bsnm{{Nocera}}, \binits{F.}},
\bauthor{\bsnm{{Nolting}}, \binits{D.}},
\bauthor{\bsnm{{Normandin}}, \binits{M.E.N.}},
\bauthor{\bsnm{{Nuttall}}, \binits{L.K.}},
\bauthor{\bsnm{{Oberling}}, \binits{J.}},
\bauthor{\bsnm{{Ochsner}}, \binits{E.}},
\bauthor{\bsnm{{Oelker}}, \binits{E.}},
\bauthor{\bsnm{{Ogin}}, \binits{G.H.}},
\bauthor{\bsnm{{Oh}}, \binits{J.J.}},
\bauthor{\bsnm{{Oh}}, \binits{S.H.}},
\bauthor{\bsnm{{Ohashi}}, \binits{M.}},
\bauthor{\bsnm{{Ohishi}}, \binits{N.}},
\bauthor{\bsnm{{Ohkawa}}, \binits{M.}},
\bauthor{\bsnm{{Ohme}}, \binits{F.}},
\bauthor{\bsnm{{Okutomi}}, \binits{K.}},
\bauthor{\bsnm{{Oliver}}, \binits{M.}},
\bauthor{\bsnm{{Ono}}, \binits{K.}},
\bauthor{\bsnm{{Ono}}, \binits{Y.}},
\bauthor{\bsnm{{Oohara}}, \binits{K.}},
\bauthor{\bsnm{{Oppermann}}, \binits{P.}},
\bauthor{\bsnm{{Oram}}, \binits{R.J.}},
\bauthor{\bsnm{{O'Reilly}}, \binits{B.}},
\bauthor{\bsnm{{O'Shaughnessy}}, \binits{R.}},
\bauthor{\bsnm{{Ottaway}}, \binits{D.J.}},
\bauthor{\bsnm{{Overmier}}, \binits{H.}},
\bauthor{\bsnm{{Owen}}, \binits{B.J.}},
\bauthor{\bsnm{{Pace}}, \binits{A.E.}},
\bauthor{\bsnm{{Page}}, \binits{J.}},
\bauthor{\bsnm{{Pai}}, \binits{A.}},
\bauthor{\bsnm{{Pai}}, \binits{S.A.}},
\bauthor{\bsnm{{Palamos}}, \binits{J.R.}},
\bauthor{\bsnm{{Palashov}}, \binits{O.}},
\bauthor{\bsnm{{Palomba}}, \binits{C.}},
\bauthor{\bsnm{{Pal-Singh}}, \binits{A.}},
\bauthor{\bsnm{{Pan}}, \binits{H.}},
\bauthor{\bsnm{{Pankow}}, \binits{C.}},
\bauthor{\bsnm{{Pannarale}}, \binits{F.}},
\bauthor{\bsnm{{Pant}}, \binits{B.C.}},
\bauthor{\bsnm{{Paoletti}}, \binits{F.}},
\bauthor{\bsnm{{Paoli}}, \binits{A.}},
\bauthor{\bsnm{{Papa}}, \binits{M.A.}},
\bauthor{\bsnm{{Paris}}, \binits{H.R.}},
\bauthor{\bsnm{{Parker}}, \binits{W.}},
\bauthor{\bsnm{{Pascucci}}, \binits{D.}},
\bauthor{\bsnm{{Pasqualetti}}, \binits{A.}},
\bauthor{\bsnm{{Passaquieti}}, \binits{R.}},
\bauthor{\bsnm{{Passuello}}, \binits{D.}},
\bauthor{\bsnm{{Patricelli}}, \binits{B.}},
\bauthor{\bsnm{{Pearlstone}}, \binits{B.L.}},
\bauthor{\bsnm{{Pedraza}}, \binits{M.}},
\bauthor{\bsnm{{Pedurand }}, \binits{R.}},
\bauthor{\bsnm{{Pekowsky}}, \binits{L.}},
\bauthor{\bsnm{{Pele}}, \binits{A.}},
\bauthor{\bsnm{{Pe{\~n}a Arellano}}, \binits{F.E.}},
\bauthor{\bsnm{{Penn}}, \binits{S.}},
\bauthor{\bsnm{{Perez}}, \binits{C.J.}},
\bauthor{\bsnm{{Perreca}}, \binits{A.}},
\bauthor{\bsnm{{Perri}}, \binits{L.M.}},
\bauthor{\bsnm{{Pfeiffer}}, \binits{H.P.}},
\bauthor{\bsnm{{Phelps}}, \binits{M.}},
\bauthor{\bsnm{{Piccinni}}, \binits{O.J.}},
\bauthor{\bsnm{{Pichot}}, \binits{M.}},
\bauthor{\bsnm{{Piergiovanni}}, \binits{F.}},
\bauthor{\bsnm{{Pierro}}, \binits{V.}},
\bauthor{\bsnm{{Pillant}}, \binits{G.}},
\bauthor{\bsnm{{Pinard}}, \binits{L.}},
\bauthor{\bsnm{{Pinto}}, \binits{I.M.}},
\bauthor{\bsnm{{Pitkin}}, \binits{M.}},
\bauthor{\bsnm{{Poe}}, \binits{M.}},
\bauthor{\bsnm{{Poggiani}}, \binits{R.}},
\bauthor{\bsnm{{Popolizio}}, \binits{P.}},
\bauthor{\bsnm{{Post}}, \binits{A.}},
\bauthor{\bsnm{{Powell}}, \binits{J.}},
\bauthor{\bsnm{{Prasad}}, \binits{J.}},
\bauthor{\bsnm{{Pratt}}, \binits{J.W.W.}},
\bauthor{\bsnm{{Predoi}}, \binits{V.}},
\bauthor{\bsnm{{Prestegard}}, \binits{T.}},
\bauthor{\bsnm{{Prijatelj}}, \binits{M.}},
\bauthor{\bsnm{{Principe}}, \binits{M.}},
\bauthor{\bsnm{{Privitera}}, \binits{S.}},
\bauthor{\bsnm{{Prodi}}, \binits{G.A.}},
\bauthor{\bsnm{{Prokhorov}}, \binits{L.G.}},
\bauthor{\bsnm{{Puncken}}, \binits{O.}},
\bauthor{\bsnm{{Punturo}}, \binits{M.}},
\bauthor{\bsnm{{Puppo}}, \binits{P.}},
\bauthor{\bsnm{{P{\"u}rrer}}, \binits{M.}},
\bauthor{\bsnm{{Qi}}, \binits{H.}},
\bauthor{\bsnm{{Qin}}, \binits{J.}},
\bauthor{\bsnm{{Qiu}}, \binits{S.}},
\bauthor{\bsnm{{Quetschke}}, \binits{V.}},
\bauthor{\bsnm{{Quintero}}, \binits{E.A.}},
\bauthor{\bsnm{{Quitzow-James}}, \binits{R.}},
\bauthor{\bsnm{{Raab}}, \binits{F.J.}},
\bauthor{\bsnm{{Rabeling}}, \binits{D.S.}},
\bauthor{\bsnm{{Radkins}}, \binits{H.}},
\bauthor{\bsnm{{Raffai}}, \binits{P.}},
\bauthor{\bsnm{{Raja}}, \binits{S.}},
\bauthor{\bsnm{{Rajan}}, \binits{C.}},
\bauthor{\bsnm{{Rakhmanov}}, \binits{M.}},
\bauthor{\bsnm{{Rapagnani}}, \binits{P.}},
\bauthor{\bsnm{{Raymond}}, \binits{V.}},
\bauthor{\bsnm{{Razzano}}, \binits{M.}},
\bauthor{\bsnm{{Re}}, \binits{V.}},
\bauthor{\bsnm{{Read}}, \binits{J.}},
\bauthor{\bsnm{{Regimbau}}, \binits{T.}},
\bauthor{\bsnm{{Rei}}, \binits{L.}},
\bauthor{\bsnm{{Reid}}, \binits{S.}},
\bauthor{\bsnm{{Reitze}}, \binits{D.H.}},
\bauthor{\bsnm{{Rew}}, \binits{H.}},
\bauthor{\bsnm{{Reyes}}, \binits{S.D.}},
\bauthor{\bsnm{{Rhoades}}, \binits{E.}},
\bauthor{\bsnm{{Ricci}}, \binits{F.}},
\bauthor{\bsnm{{Riles}}, \binits{K.}},
\bauthor{\bsnm{{Rizzo}}, \binits{M.}},
\bauthor{\bsnm{{Robertson}}, \binits{N.A.}},
\bauthor{\bsnm{{Robie}}, \binits{R.}},
\bauthor{\bsnm{{Robinet}}, \binits{F.}},
\bauthor{\bsnm{{Rocchi}}, \binits{A.}},
\bauthor{\bsnm{{Rolland}}, \binits{L.}},
\bauthor{\bsnm{{Rollins}}, \binits{J.G.}},
\bauthor{\bsnm{{Roma}}, \binits{V.J.}},
\bauthor{\bsnm{{Romano}}, \binits{R.}},
\bauthor{\bsnm{{Romie}}, \binits{J.H.}},
\bauthor{\bsnm{{Rosi{\'n}ska}}, \binits{D.}},
\bauthor{\bsnm{{Rowan}}, \binits{S.}},
\bauthor{\bsnm{{R{\"u}diger}}, \binits{A.}},
\bauthor{\bsnm{{Ruggi}}, \binits{P.}},
\bauthor{\bsnm{{Ryan}}, \binits{K.}},
\bauthor{\bsnm{{Sachdev}}, \binits{S.}},
\bauthor{\bsnm{{Sadecki}}, \binits{T.}},
\bauthor{\bsnm{{Sadeghian}}, \binits{L.}},
\bauthor{\bsnm{{Sago}}, \binits{N.}},
\bauthor{\bsnm{{Saijo}}, \binits{M.}},
\bauthor{\bsnm{{Saito}}, \binits{Y.}},
\bauthor{\bsnm{{Sakai}}, \binits{K.}},
\bauthor{\bsnm{{Sakellariadou}}, \binits{M.}},
\bauthor{\bsnm{{Salconi}}, \binits{L.}},
\bauthor{\bsnm{{Saleem}}, \binits{M.}},
\bauthor{\bsnm{{Salemi}}, \binits{F.}},
\bauthor{\bsnm{{Samajdar}}, \binits{A.}},
\bauthor{\bsnm{{Sammut}}, \binits{L.}},
\bauthor{\bsnm{{Sampson}}, \binits{L.M.}},
\bauthor{\bsnm{{Sanchez}}, \binits{E.J.}},
\bauthor{\bsnm{{Sandberg}}, \binits{V.}},
\bauthor{\bsnm{{Sand ers}}, \binits{J.R.}},
\bauthor{\bsnm{{Sasaki}}, \binits{Y.}},
\bauthor{\bsnm{{Sassolas}}, \binits{B.}},
\bauthor{\bsnm{{Sathyaprakash}}, \binits{B.S.}},
\bauthor{\bsnm{{Sato}}, \binits{S.}},
\bauthor{\bsnm{{Sato}}, \binits{T.}},
\bauthor{\bsnm{{Saulson}}, \binits{P.R.}},
\bauthor{\bsnm{{Sauter}}, \binits{O.}},
\bauthor{\bsnm{{Savage}}, \binits{R.L.}},
\bauthor{\bsnm{{Sawadsky}}, \binits{A.}},
\bauthor{\bsnm{{Schale}}, \binits{P.}},
\bauthor{\bsnm{{Scheuer}}, \binits{J.}},
\bauthor{\bsnm{{Schmidt}}, \binits{E.}},
\bauthor{\bsnm{{Schmidt}}, \binits{J.}},
\bauthor{\bsnm{{Schmidt}}, \binits{P.}},
\bauthor{\bsnm{{Schnabel}}, \binits{R.}},
\bauthor{\bsnm{{Schofield}}, \binits{R.M.S.}},
\bauthor{\bsnm{{Sch{\"o}nbeck}}, \binits{A.}},
\bauthor{\bsnm{{Schreiber}}, \binits{E.}},
\bauthor{\bsnm{{Schuette}}, \binits{D.}},
\bauthor{\bsnm{{Schutz}}, \binits{B.F.}},
\bauthor{\bsnm{{Schwalbe}}, \binits{S.G.}},
\bauthor{\bsnm{{Scott}}, \binits{J.}},
\bauthor{\bsnm{{Scott}}, \binits{S.M.}},
\bauthor{\bsnm{{Sekiguchi}}, \binits{T.}},
\bauthor{\bsnm{{Sekiguchi}}, \binits{Y.}},
\bauthor{\bsnm{{Sellers}}, \binits{D.}},
\bauthor{\bsnm{{Sengupta}}, \binits{A.S.}},
\bauthor{\bsnm{{Sentenac}}, \binits{D.}},
\bauthor{\bsnm{{Sequino}}, \binits{V.}},
\bauthor{\bsnm{{Sergeev}}, \binits{A.}},
\bauthor{\bsnm{{Setyawati}}, \binits{Y.}},
\bauthor{\bsnm{{Shaddock}}, \binits{D.A.}},
\bauthor{\bsnm{{Shaffer}}, \binits{T.J.}},
\bauthor{\bsnm{{Shahriar}}, \binits{M.S.}},
\bauthor{\bsnm{{Shapiro}}, \binits{B.}},
\bauthor{\bsnm{{Shawhan}}, \binits{P.}},
\bauthor{\bsnm{{Sheperd}}, \binits{A.}},
\bauthor{\bsnm{{Shibata}}, \binits{M.}},
\bauthor{\bsnm{{Shikano}}, \binits{Y.}},
\bauthor{\bsnm{{Shimoda}}, \binits{T.}},
\bauthor{\bsnm{{Shoda}}, \binits{A.}},
\bauthor{\bsnm{{Shoemaker}}, \binits{D.H.}},
\bauthor{\bsnm{{Shoemaker}}, \binits{D.M.}},
\bauthor{\bsnm{{Siellez}}, \binits{K.}},
\bauthor{\bsnm{{Siemens}}, \binits{X.}},
\bauthor{\bsnm{{Sieniawska}}, \binits{M.}},
\bauthor{\bsnm{{Sigg}}, \binits{D.}},
\bauthor{\bsnm{{Silva}}, \binits{A.D.}},
\bauthor{\bsnm{{Singer}}, \binits{A.}},
\bauthor{\bsnm{{Singer}}, \binits{L.P.}},
\bauthor{\bsnm{{Singh}}, \binits{A.}},
\bauthor{\bsnm{{Singh}}, \binits{R.}},
\bauthor{\bsnm{{Singhal}}, \binits{A.}},
\bauthor{\bsnm{{Sintes}}, \binits{A.M.}},
\bauthor{\bsnm{{Slagmolen}}, \binits{B.J.J.}},
\bauthor{\bsnm{{Smith}}, \binits{B.}},
\bauthor{\bsnm{{Smith}}, \binits{J.R.}},
\bauthor{\bsnm{{Smith}}, \binits{R.J.E.}},
\bauthor{\bsnm{{Somiya}}, \binits{K.}},
\bauthor{\bsnm{{Son}}, \binits{E.J.}},
\bauthor{\bsnm{{Sorazu}}, \binits{B.}},
\bauthor{\bsnm{{Sorrentino}}, \binits{F.}},
\bauthor{\bsnm{{Souradeep}}, \binits{T.}},
\bauthor{\bsnm{{Spencer}}, \binits{A.P.}},
\bauthor{\bsnm{{Srivastava}}, \binits{A.K.}},
\bauthor{\bsnm{{Staley}}, \binits{A.}},
\bauthor{\bsnm{{Steinke}}, \binits{M.}},
\bauthor{\bsnm{{Steinlechner}}, \binits{J.}},
\bauthor{\bsnm{{Steinlechner}}, \binits{S.}},
\bauthor{\bsnm{{Steinmeyer}}, \binits{D.}},
\bauthor{\bsnm{{Stephens}}, \binits{B.C.}},
\bauthor{\bsnm{{Stevenson}}, \binits{S.P.}},
\bauthor{\bsnm{{Stone}}, \binits{R.}},
\bauthor{\bsnm{{Strain}}, \binits{K.A.}},
\bauthor{\bsnm{{Straniero}}, \binits{N.}},
\bauthor{\bsnm{{Stratta}}, \binits{G.}},
\bauthor{\bsnm{{Strigin}}, \binits{S.E.}},
\bauthor{\bsnm{{Sturani}}, \binits{R.}},
\bauthor{\bsnm{{Stuver}}, \binits{A.L.}},
\bauthor{\bsnm{{Sugimoto}}, \binits{Y.}},
\bauthor{\bsnm{{Summerscales}}, \binits{T.Z.}},
\bauthor{\bsnm{{Sun}}, \binits{L.}},
\bauthor{\bsnm{{Sunil}}, \binits{S.}},
\bauthor{\bsnm{{Sutton}}, \binits{P.J.}},
\bauthor{\bsnm{{Suzuki}}, \binits{T.}},
\bauthor{\bsnm{{Swinkels}}, \binits{B.L.}},
\bauthor{\bsnm{{Szczepa{\'n}czyk}}, \binits{M.J.}},
\bauthor{\bsnm{{Tacca}}, \binits{M.}},
\bauthor{\bsnm{{Tagoshi}}, \binits{H.}},
\bauthor{\bsnm{{Takada}}, \binits{S.}},
\bauthor{\bsnm{{Takahashi}}, \binits{H.}},
\bauthor{\bsnm{{Takahashi}}, \binits{R.}},
\bauthor{\bsnm{{Takamori}}, \binits{A.}},
\bauthor{\bsnm{{Talukder}}, \binits{D.}},
\bauthor{\bsnm{{Tanaka}}, \binits{H.}},
\bauthor{\bsnm{{Tanaka}}, \binits{K.}},
\bauthor{\bsnm{{Tanaka}}, \binits{T.}},
\bauthor{\bsnm{{Tanner}}, \binits{D.B.}},
\bauthor{\bsnm{{T{\'a}pai}}, \binits{M.}},
\bauthor{\bsnm{{Taracchini}}, \binits{A.}},
\bauthor{\bsnm{{Tatsumi}}, \binits{D.}},
\bauthor{\bsnm{{Taylor}}, \binits{R.}},
\bauthor{\bsnm{{Telada}}, \binits{S.}},
\bauthor{\bsnm{{Theeg}}, \binits{T.}},
\bauthor{\bsnm{{Thomas}}, \binits{E.G.}},
\bauthor{\bsnm{{Thomas}}, \binits{M.}},
\bauthor{\bsnm{{Thomas}}, \binits{P.}},
\bauthor{\bsnm{{Thorne}}, \binits{K.A.}},
\bauthor{\bsnm{{Thrane}}, \binits{E.}},
\bauthor{\bsnm{{Tippens}}, \binits{T.}},
\bauthor{\bsnm{{Tiwari}}, \binits{S.}},
\bauthor{\bsnm{{Tiwari}}, \binits{V.}},
\bauthor{\bsnm{{Tokmakov}}, \binits{K.V.}},
\bauthor{\bsnm{{Toland}}, \binits{K.}},
\bauthor{\bsnm{{Tomaru}}, \binits{T.}},
\bauthor{\bsnm{{Tomlinson}}, \binits{C.}},
\bauthor{\bsnm{{Tonelli}}, \binits{M.}},
\bauthor{\bsnm{{Tornasi}}, \binits{Z.}},
\bauthor{\bsnm{{Torrie}}, \binits{C.I.}},
\bauthor{\bsnm{{T{\"o}yr{\"a}}}, \binits{D.}},
\bauthor{\bsnm{{Travasso}}, \binits{F.}},
\bauthor{\bsnm{{Traylor}}, \binits{G.}},
\bauthor{\bsnm{{Trifir{\`o}}}, \binits{D.}},
\bauthor{\bsnm{{Trinastic}}, \binits{J.}},
\bauthor{\bsnm{{Tringali}}, \binits{M.C.}},
\bauthor{\bsnm{{Trozzo}}, \binits{L.}},
\bauthor{\bsnm{{Tse}}, \binits{M.}},
\bauthor{\bsnm{{Tso}}, \binits{R.}},
\bauthor{\bsnm{{Tsubono}}, \binits{K.}},
\bauthor{\bsnm{{Tsuzuki}}, \binits{T.}},
\bauthor{\bsnm{{Turconi}}, \binits{M.}},
\bauthor{\bsnm{{Tuyenbayev}}, \binits{D.}},
\bauthor{\bsnm{{Uchiyama}}, \binits{T.}},
\bauthor{\bsnm{{Uehara}}, \binits{T.}},
\bauthor{\bsnm{{Ueki}}, \binits{S.}},
\bauthor{\bsnm{{Ueno}}, \binits{K.}},
\bauthor{\bsnm{{Ugolini}}, \binits{D.}},
\bauthor{\bsnm{{Unnikrishnan}}, \binits{C.S.}},
\bauthor{\bsnm{{Urban}}, \binits{A.L.}},
\bauthor{\bsnm{{Ushiba}}, \binits{T.}},
\bauthor{\bsnm{{Usman}}, \binits{S.A.}},
\bauthor{\bsnm{{Vahlbruch}}, \binits{H.}},
\bauthor{\bsnm{{Vajente}}, \binits{G.}},
\bauthor{\bsnm{{Valdes}}, \binits{G.}},
\bauthor{\bsnm{{van Bakel}}, \binits{N.}},
\bauthor{\bsnm{{van Beuzekom}}, \binits{M.}},
\bauthor{\bsnm{{van den Brand}}}:
\bjtitle{Living Reviews in Relativity}
\bvolume{21}(\bissue{1}),
\bfpage{3}
(\byear{2018}a).
\arxivurl{1304.0670}.
doi:\doiurl{10.1007/s41114-018-0012-9}
\end{barticle}
\endbibitem

\bibitem[\protect\citeauthoryear{{Abbott} et~al.}{2018b}]{Abbott2018a}
\begin{barticle}
\bauthor{\bsnm{{Abbott}}, \binits{B.P.}},
\bauthor{\bsnm{{Abbott}}, \binits{R.}},
\bauthor{\bsnm{{Abbott}}, \binits{T.D.}},
\bauthor{\bsnm{{Abernathy}}, \binits{M.R.}},
\bauthor{\bsnm{{Acernese}}, \binits{F.}},
\bauthor{\bsnm{{Ackley}}, \binits{K.}},
\bauthor{\bsnm{{Adams}}, \binits{C.}},
\bauthor{\bsnm{{Adams}}, \binits{T.}},
\bauthor{\bsnm{{Addesso}}, \binits{P.}},
\bauthor{\bsnm{{Adhikari}}, \binits{R.X.}},
\bauthor{\bsnm{{Adya}}, \binits{V.B.}},
\bauthor{\bsnm{{Affeldt}}, \binits{C.}},
\bauthor{\bsnm{{Agathos}}, \binits{M.}},
\bauthor{\bsnm{{Agatsuma}}, \binits{K.}},
\bauthor{\bsnm{{Aggarwal}}, \binits{N.}},
\bauthor{\bsnm{{Aguiar}}, \binits{O.D.}},
\bauthor{\bsnm{{Aiello}}, \binits{L.}},
\bauthor{\bsnm{{Ain}}, \binits{A.}},
\bauthor{\bsnm{{Ajith}}, \binits{P.}},
\bauthor{\bsnm{{Akutsu}}, \binits{T.}},
\bauthor{\bsnm{{Allen}}, \binits{B.}},
\bauthor{\bsnm{{Allocca}}, \binits{A.}},
\bauthor{\bsnm{{Altin}}, \binits{P.A.}},
\bauthor{\bsnm{{Ananyeva}}, \binits{A.}},
\bauthor{\bsnm{{Anderson}}, \binits{S.B.}},
\bauthor{\bsnm{{Anderson}}, \binits{W.G.}},
\bauthor{\bsnm{{Ando}}, \binits{M.}},
\bauthor{\bsnm{{Appert}}, \binits{S.}},
\bauthor{\bsnm{{Arai}}, \binits{K.}},
\bauthor{\bsnm{{Araya}}, \binits{A.}},
\bauthor{\bsnm{{Araya}}, \binits{M.C.}},
\bauthor{\bsnm{{Areeda}}, \binits{J.S.}},
\bauthor{\bsnm{{Arnaud}}, \binits{N.}},
\bauthor{\bsnm{{Arun}}, \binits{K.G.}},
\bauthor{\bsnm{{Asada}}, \binits{H.}},
\bauthor{\bsnm{{Ascenzi}}, \binits{S.}},
\bauthor{\bsnm{{Ashton}}, \binits{G.}},
\bauthor{\bsnm{{Aso}}, \binits{Y.}},
\bauthor{\bsnm{{Ast}}, \binits{M.}},
\bauthor{\bsnm{{Aston}}, \binits{S.M.}},
\bauthor{\bsnm{{Astone}}, \binits{P.}},
\bauthor{\bsnm{{Atsuta}}, \binits{S.}},
\bauthor{\bsnm{{Aufmuth}}, \binits{P.}},
\bauthor{\bsnm{{Aulbert}}, \binits{C.}},
\bauthor{\bsnm{{Avila-Alvarez}}, \binits{A.}},
\bauthor{\bsnm{{Awai}}, \binits{K.}},
\bauthor{\bsnm{{Babak}}, \binits{S.}},
\bauthor{\bsnm{{Bacon}}, \binits{P.}},
\bauthor{\bsnm{{Bader}}, \binits{M.K.M.}},
\bauthor{\bsnm{{Baiotti}}, \binits{L.}},
\bauthor{\bsnm{{Baker}}, \binits{P.T.}},
\bauthor{\bsnm{{Baldaccini}}, \binits{F.}},
\bauthor{\bsnm{{Ballardin}}, \binits{G.}},
\bauthor{\bsnm{{Ballmer}}, \binits{S.W.}},
\bauthor{\bsnm{{Barayoga}}, \binits{J.C.}},
\bauthor{\bsnm{{Barclay}}, \binits{S.E.}},
\bauthor{\bsnm{{Barish}}, \binits{B.C.}},
\bauthor{\bsnm{{Barker}}, \binits{D.}},
\bauthor{\bsnm{{Barone}}, \binits{F.}},
\bauthor{\bsnm{{Barr}}, \binits{B.}},
\bauthor{\bsnm{{Barsotti}}, \binits{L.}},
\bauthor{\bsnm{{Barsuglia}}, \binits{M.}},
\bauthor{\bsnm{{Barta}}, \binits{D.}},
\bauthor{\bsnm{{Bartlett}}, \binits{J.}},
\bauthor{\bsnm{{Barton}}, \binits{M.A.}},
\bauthor{\bsnm{{Bartos}}, \binits{I.}},
\bauthor{\bsnm{{Bassiri}}, \binits{R.}},
\bauthor{\bsnm{{Basti}}, \binits{A.}},
\bauthor{\bsnm{{Batch}}, \binits{J.C.}},
\bauthor{\bsnm{{Baune}}, \binits{C.}},
\bauthor{\bsnm{{Bavigadda}}, \binits{V.}},
\bauthor{\bsnm{{Bazzan}}, \binits{M.}},
\bauthor{\bsnm{{B{\'e}csy}}, \binits{B.}},
\bauthor{\bsnm{{Beer}}, \binits{C.}},
\bauthor{\bsnm{{Bejger}}, \binits{M.}},
\bauthor{\bsnm{{Belahcene}}, \binits{I.}},
\bauthor{\bsnm{{Belgin}}, \binits{M.}},
\bauthor{\bsnm{{Bell}}, \binits{A.S.}},
\bauthor{\bsnm{{Berger}}, \binits{B.K.}},
\bauthor{\bsnm{{Bergmann}}, \binits{G.}},
\bauthor{\bsnm{{Berry}}, \binits{C.P.L.}},
\bauthor{\bsnm{{Bersanetti}}, \binits{D.}},
\bauthor{\bsnm{{Bertolini}}, \binits{A.}},
\bauthor{\bsnm{{Betzwieser}}, \binits{J.}},
\bauthor{\bsnm{{Bhagwat}}, \binits{S.}},
\bauthor{\bsnm{{Bhandare}}, \binits{R.}},
\bauthor{\bsnm{{Bilenko}}, \binits{I.A.}},
\bauthor{\bsnm{{Billingsley}}, \binits{G.}},
\bauthor{\bsnm{{Billman}}, \binits{C.R.}},
\bauthor{\bsnm{{Birch}}, \binits{J.}},
\bauthor{\bsnm{{Birney}}, \binits{R.}},
\bauthor{\bsnm{{Birnholtz}}, \binits{O.}},
\bauthor{\bsnm{{Biscans}}, \binits{S.}},
\bauthor{\bsnm{{Bisht}}, \binits{A.}},
\bauthor{\bsnm{{Bitossi}}, \binits{M.}},
\bauthor{\bsnm{{Biwer}}, \binits{C.}},
\bauthor{\bsnm{{Bizouard}}, \binits{M.A.}},
\bauthor{\bsnm{{Blackburn}}, \binits{J.K.}},
\bauthor{\bsnm{{Blackman}}, \binits{J.}},
\bauthor{\bsnm{{Blair}}, \binits{C.D.}},
\bauthor{\bsnm{{Blair}}, \binits{D.G.}},
\bauthor{\bsnm{{Blair}}, \binits{R.M.}},
\bauthor{\bsnm{{Bloemen}}, \binits{S.}},
\bauthor{\bsnm{{Bock}}, \binits{O.}},
\bauthor{\bsnm{{Boer}}, \binits{M.}},
\bauthor{\bsnm{{Bogaert}}, \binits{G.}},
\bauthor{\bsnm{{Bohe}}, \binits{A.}},
\bauthor{\bsnm{{Bondu}}, \binits{F.}},
\bauthor{\bsnm{{Bonnand }}, \binits{R.}},
\bauthor{\bsnm{{Boom}}, \binits{B.A.}},
\bauthor{\bsnm{{Bork}}, \binits{R.}},
\bauthor{\bsnm{{Boschi}}, \binits{V.}},
\bauthor{\bsnm{{Bose}}, \binits{S.}},
\bauthor{\bsnm{{Bouffanais}}, \binits{Y.}},
\bauthor{\bsnm{{Bozzi}}, \binits{A.}},
\bauthor{\bsnm{{Bradaschia}}, \binits{C.}},
\bauthor{\bsnm{{Brady}}, \binits{P.R.}},
\bauthor{\bsnm{{Braginsky}}, \binits{V.B.}},
\bauthor{\bsnm{{Branchesi}}, \binits{M.}},
\bauthor{\bsnm{{Brau}}, \binits{J.E.}},
\bauthor{\bsnm{{Briant}}, \binits{T.}},
\bauthor{\bsnm{{Brillet}}, \binits{A.}},
\bauthor{\bsnm{{Brinkmann}}, \binits{M.}},
\bauthor{\bsnm{{Brisson}}, \binits{V.}},
\bauthor{\bsnm{{Brockill}}, \binits{P.}},
\bauthor{\bsnm{{Broida}}, \binits{J.E.}},
\bauthor{\bsnm{{Brooks}}, \binits{A.F.}},
\bauthor{\bsnm{{Brown}}, \binits{D.A.}},
\bauthor{\bsnm{{Brown}}, \binits{D.D.}},
\bauthor{\bsnm{{Brown}}, \binits{N.M.}},
\bauthor{\bsnm{{Brunett}}, \binits{S.}},
\bauthor{\bsnm{{Buchanan}}, \binits{C.C.}},
\bauthor{\bsnm{{Buikema}}, \binits{A.}},
\bauthor{\bsnm{{Bulik}}, \binits{T.}},
\bauthor{\bsnm{{Bulten}}, \binits{H.J.}},
\bauthor{\bsnm{{Buonanno}}, \binits{A.}},
\bauthor{\bsnm{{Buskulic}}, \binits{D.}},
\bauthor{\bsnm{{Buy}}, \binits{C.}},
\bauthor{\bsnm{{Byer}}, \binits{R.L.}}:
\bjtitle{Living Reviews in Relativity}
\bvolume{21}(\bissue{1}),
\bfpage{3}
(\byear{2018}b).
\arxivurl{1304.0670}.
doi:\doiurl{10.1007/s41114-018-0012-9}
\end{barticle}
\endbibitem

\bibitem[\protect\citeauthoryear{{Abbott} et~al.}{2008}]{Abbott2008_070201}
\begin{barticle}
\bauthor{\bsnm{{Abbott}}, \binits{B.}},
\bauthor{\bsnm{{Abbott}}, \binits{R.}},
\bauthor{\bsnm{{Adhikari}}, \binits{R.}},
\bauthor{\bsnm{{Agresti}}, \binits{J.}},
\bauthor{\bsnm{{Ajith}}, \binits{P.}},
\bauthor{\bsnm{{Allen}}, \binits{B.}},
\bauthor{\bsnm{{Amin}}, \binits{R.}},
\bauthor{\bsnm{{Anderson}}, \binits{S.B.}},
\bauthor{\bsnm{{Anderson}}, \binits{W.G.}},
\bauthor{\bsnm{{Arain}}, \binits{M.}},
\bauthor{\bsnm{{Araya}}, \binits{M.}},
\bauthor{\bsnm{{Armandula}}, \binits{H.}},
\bauthor{\bsnm{{Ashley}}, \binits{M.}},
\bauthor{\bsnm{{Aston}}, \binits{S.}},
\bauthor{\bsnm{{Aufmuth}}, \binits{P.}},
\bauthor{\bsnm{{Aulbert}}, \binits{C.}},
\bauthor{\bsnm{{Babak}}, \binits{S.}},
\bauthor{\bsnm{{Ballmer}}, \binits{S.}},
\bauthor{\bsnm{{Bantilan}}, \binits{H.}},
\bauthor{\bsnm{{Barish}}, \binits{B.C.}},
\bauthor{\bsnm{{Barker}}, \binits{C.}},
\bauthor{\bsnm{{Barker}}, \binits{D.}},
\bauthor{\bsnm{{Barr}}, \binits{B.}},
\bauthor{\bsnm{{Barriga}}, \binits{P.}},
\bauthor{\bsnm{{Barton}}, \binits{M.A.}},
\bauthor{\bsnm{{Bayer}}, \binits{K.}},
\bauthor{\bsnm{{Betzwieser}}, \binits{J.}},
\bauthor{\bsnm{{Beyersdorf}}, \binits{P.T.}},
\bauthor{\bsnm{{Bhawal}}, \binits{B.}},
\bauthor{\bsnm{{Bilenko}}, \binits{I.A.}},
\bauthor{\bsnm{{Billingsley}}, \binits{G.}},
\bauthor{\bsnm{{Biswas}}, \binits{R.}},
\bauthor{\bsnm{{Black}}, \binits{E.}},
\bauthor{\bsnm{{Blackburn}}, \binits{K.}},
\bauthor{\bsnm{{Blackburn}}, \binits{L.}},
\bauthor{\bsnm{{Blair}}, \binits{D.}},
\bauthor{\bsnm{{Bland}}, \binits{B.}},
\bauthor{\bsnm{{Bogenstahl}}, \binits{J.}},
\bauthor{\bsnm{{Bogue}}, \binits{L.}},
\bauthor{\bsnm{{Bork}}, \binits{R.}},
\bauthor{\bsnm{{Boschi}}, \binits{V.}},
\bauthor{\bsnm{{Bose}}, \binits{S.}},
\bauthor{\bsnm{{Brady}}, \binits{P.R.}},
\bauthor{\bsnm{{Braginsky}}, \binits{V.B.}},
\bauthor{\bsnm{{Brau}}, \binits{J.E.}},
\bauthor{\bsnm{{Brinkmann}}, \binits{M.}},
\bauthor{\bsnm{{Brooks}}, \binits{A.}},
\bauthor{\bsnm{{Brown}}, \binits{D.A.}},
\bauthor{\bsnm{{Bullington}}, \binits{A.}},
\bauthor{\bsnm{{Bunkowski}}, \binits{A.}},
\bauthor{\bsnm{{Buonanno}}, \binits{A.}},
\bauthor{\bsnm{{Burmeister}}, \binits{O.}},
\bauthor{\bsnm{{Busby}}, \binits{D.}},
\bauthor{\bsnm{{Byer}}, \binits{R.L.}},
\bauthor{\bsnm{{Cadonati}}, \binits{L.}},
\bauthor{\bsnm{{Cagnoli}}, \binits{G.}},
\bauthor{\bsnm{{Camp}}, \binits{J.B.}},
\bauthor{\bsnm{{Cannizzo}}, \binits{J.}},
\bauthor{\bsnm{{Cannon}}, \binits{K.}},
\bauthor{\bsnm{{Cantley}}, \binits{C.A.}},
\bauthor{\bsnm{{Cao}}, \binits{J.}},
\bauthor{\bsnm{{Cardenas}}, \binits{L.}},
\bauthor{\bsnm{{Castaldi}}, \binits{G.}},
\bauthor{\bsnm{{Cepeda}}, \binits{C.}},
\bauthor{\bsnm{{Chalkley}}, \binits{E.}},
\bauthor{\bsnm{{Charlton}}, \binits{P.}},
\bauthor{\bsnm{{Chatterji}}, \binits{S.}},
\bauthor{\bsnm{{Chelkowski}}, \binits{S.}},
\bauthor{\bsnm{{Chen}}, \binits{Y.}},
\bauthor{\bsnm{{Chiadini}}, \binits{F.}},
\bauthor{\bsnm{{Christensen}}, \binits{N.}},
\bauthor{\bsnm{{Clark}}, \binits{J.}},
\bauthor{\bsnm{{Cochrane}}, \binits{P.}},
\bauthor{\bsnm{{Cokelaer}}, \binits{T.}},
\bauthor{\bsnm{{Coldwell}}, \binits{R.}},
\bauthor{\bsnm{{Conte}}, \binits{R.}},
\bauthor{\bsnm{{Cook}}, \binits{D.}},
\bauthor{\bsnm{{Corbitt}}, \binits{T.}},
\bauthor{\bsnm{{Coyne}}, \binits{D.}},
\bauthor{\bsnm{{Creighton}}, \binits{J.D.E.}},
\bauthor{\bsnm{{Croce}}, \binits{R.P.}},
\bauthor{\bsnm{{Crooks}}, \binits{D.R.M.}},
\bauthor{\bsnm{{Cruise}}, \binits{A.M.}},
\bauthor{\bsnm{{Cumming}}, \binits{A.}},
\bauthor{\bsnm{{Dalrymple}}, \binits{J.}},
\bauthor{\bsnm{{D'Ambrosio}}, \binits{E.}},
\bauthor{\bsnm{{Danzmann}}, \binits{K.}},
\bauthor{\bsnm{{Davies}}, \binits{G.}},
\bauthor{\bsnm{{DeBra}}, \binits{D.}},
\bauthor{\bsnm{{Degallaix}}, \binits{J.}},
\bauthor{\bsnm{{Degree}}, \binits{M.}},
\bauthor{\bsnm{{Demma}}, \binits{T.}},
\bauthor{\bsnm{{Dergachev}}, \binits{V.}},
\bauthor{\bsnm{{Desai}}, \binits{S.}},
\bauthor{\bsnm{{DeSalvo}}, \binits{R.}},
\bauthor{\bsnm{{Dhurandhar}}, \binits{S.}},
\bauthor{\bsnm{{D{\'\i}az}}, \binits{M.}},
\bauthor{\bsnm{{Dickson}}, \binits{J.}},
\bauthor{\bsnm{{Di Credico}}, \binits{A.}},
\bauthor{\bsnm{{Diederichs}}, \binits{G.}},
\bauthor{\bsnm{{Dietz}}, \binits{A.}},
\bauthor{\bsnm{{Doomes}}, \binits{E.E.}},
\bauthor{\bsnm{{Drever}}, \binits{R.W.P.}},
\bauthor{\bsnm{{Dumas}}, \binits{J.-C.}},
\bauthor{\bsnm{{Dupuis}}, \binits{R.J.}},
\bauthor{\bsnm{{Dwyer}}, \binits{J.G.}},
\bauthor{\bsnm{{Ehrens}}, \binits{P.}},
\bauthor{\bsnm{{Espinoza}}, \binits{E.}},
\bauthor{\bsnm{{Etzel}}, \binits{T.}},
\bauthor{\bsnm{{Evans}}, \binits{M.}},
\bauthor{\bsnm{{Evans}}, \binits{T.}},
\bauthor{\bsnm{{Fairhurst}}, \binits{S.}},
\bauthor{\bsnm{{Fan}}, \binits{Y.}},
\bauthor{\bsnm{{Fazi}}, \binits{D.}},
\bauthor{\bsnm{{Fejer}}, \binits{M.M.}},
\bauthor{\bsnm{{Finn}}, \binits{L.S.}},
\bauthor{\bsnm{{Fiumara}}, \binits{V.}},
\bauthor{\bsnm{{Fotopoulos}}, \binits{N.}},
\bauthor{\bsnm{{Franzen}}, \binits{A.}},
\bauthor{\bsnm{{Franzen}}, \binits{K.Y.}},
\bauthor{\bsnm{{Freise}}, \binits{A.}},
\bauthor{\bsnm{{Frey}}, \binits{R.}},
\bauthor{\bsnm{{Fricke}}, \binits{T.}},
\bauthor{\bsnm{{Fritschel}}, \binits{P.}},
\bauthor{\bsnm{{Frolov}}, \binits{V.V.}},
\bauthor{\bsnm{{Fyffe}}, \binits{M.}},
\bauthor{\bsnm{{Galdi}}, \binits{V.}},
\bauthor{\bsnm{{Garofoli}}, \binits{J.}},
\bauthor{\bsnm{{Gholami}}, \binits{I.}},
\bauthor{\bsnm{{Giaime}}, \binits{J.A.}},
\bauthor{\bsnm{{Giampanis}}, \binits{S.}},
\bauthor{\bsnm{{Giardina}}, \binits{K.D.}},
\bauthor{\bsnm{{Goda}}, \binits{K.}},
\bauthor{\bsnm{{Goetz}}, \binits{E.}},
\bauthor{\bsnm{{Goggin}}, \binits{L.M.}},
\bauthor{\bsnm{{Gonz{\'a}lez}}, \binits{G.}},
\bauthor{\bsnm{{Gossler}}, \binits{S.}},
\bauthor{\bsnm{{Grant}}, \binits{A.}},
\bauthor{\bsnm{{Gras}}, \binits{S.}},
\bauthor{\bsnm{{Gray}}, \binits{C.}},
\bauthor{\bsnm{{Gray}}, \binits{M.}},
\bauthor{\bsnm{{Greenhalgh}}, \binits{J.}},
\bauthor{\bsnm{{Gretarsson}}, \binits{A.M.}},
\bauthor{\bsnm{{Grosso}}, \binits{R.}},
\bauthor{\bsnm{{Grote}}, \binits{H.}},
\bauthor{\bsnm{{Grunewald}}, \binits{S.}},
\bauthor{\bsnm{{Guenther}}, \binits{M.}},
\bauthor{\bsnm{{Gustafson}}, \binits{R.}},
\bauthor{\bsnm{{Hage}}, \binits{B.}},
\bauthor{\bsnm{{Hammer}}, \binits{D.}},
\bauthor{\bsnm{{Hanna}}, \binits{C.}},
\bauthor{\bsnm{{Hanson}}, \binits{J.}},
\bauthor{\bsnm{{Harms}}, \binits{J.}},
\bauthor{\bsnm{{Harry}}, \binits{G.}},
\bauthor{\bsnm{{Harstad}}, \binits{E.}},
\bauthor{\bsnm{{Hayler}}, \binits{T.}},
\bauthor{\bsnm{{Heefner}}, \binits{J.}},
\bauthor{\bsnm{{Heng}}, \binits{I.S.}},
\bauthor{\bsnm{{Heptonstall}}, \binits{A.}},
\bauthor{\bsnm{{Heurs}}, \binits{M.}},
\bauthor{\bsnm{{Hewitson}}, \binits{M.}},
\bauthor{\bsnm{{Hild}}, \binits{S.}},
\bauthor{\bsnm{{Hirose}}, \binits{E.}},
\bauthor{\bsnm{{Hoak}}, \binits{D.}},
\bauthor{\bsnm{{Hosken}}, \binits{D.}},
\bauthor{\bsnm{{Hough}}, \binits{J.}},
\bauthor{\bsnm{{Hoyland}}, \binits{D.}},
\bauthor{\bsnm{{Huttner}}, \binits{S.H.}},
\bauthor{\bsnm{{Ingram}}, \binits{D.}},
\bauthor{\bsnm{{Innerhofer}}, \binits{E.}},
\bauthor{\bsnm{{Ito}}, \binits{M.}},
\bauthor{\bsnm{{Itoh}}, \binits{Y.}},
\bauthor{\bsnm{{Ivanov}}, \binits{A.}},
\bauthor{\bsnm{{Johnson}}, \binits{B.}},
\bauthor{\bsnm{{Johnson}}, \binits{W.W.}},
\bauthor{\bsnm{{Jones}}, \binits{D.I.}},
\bauthor{\bsnm{{Jones}}, \binits{G.}},
\bauthor{\bsnm{{Jones}}, \binits{R.}},
\bauthor{\bsnm{{Ju}}, \binits{L.}},
\bauthor{\bsnm{{Kalmus}}, \binits{P.}},
\bauthor{\bsnm{{Kalogera}}, \binits{V.}},
\bauthor{\bsnm{{Kasprzyk}}, \binits{D.}},
\bauthor{\bsnm{{Katsavounidis}}, \binits{E.}},
\bauthor{\bsnm{{Kawabe}}, \binits{K.}},
\bauthor{\bsnm{{Kawamura}}, \binits{S.}},
\bauthor{\bsnm{{Kawazoe}}, \binits{F.}},
\bauthor{\bsnm{{Kells}}, \binits{W.}},
\bauthor{\bsnm{{Keppel}}, \binits{D.G.}},
\bauthor{\bsnm{{Khalili}}, \binits{F.Y.}},
\bauthor{\bsnm{{Kim}}, \binits{C.}},
\bauthor{\bsnm{{King}}, \binits{P.}},
\bauthor{\bsnm{{Kissel}}, \binits{J.S.}},
\bauthor{\bsnm{{Klimenko}}, \binits{S.}},
\bauthor{\bsnm{{Kokeyama}}, \binits{K.}},
\bauthor{\bsnm{{Kondrashov}}, \binits{V.}},
\bauthor{\bsnm{{Kopparapu}}, \binits{R.K.}},
\bauthor{\bsnm{{Kozak}}, \binits{D.}},
\bauthor{\bsnm{{Krishnan}}, \binits{B.}},
\bauthor{\bsnm{{Kwee}}, \binits{P.}},
\bauthor{\bsnm{{Lam}}, \binits{P.K.}},
\bauthor{\bsnm{{Landry}}, \binits{M.}},
\bauthor{\bsnm{{Lantz}}, \binits{B.}},
\bauthor{\bsnm{{Lazzarini}}, \binits{A.}},
\bauthor{\bsnm{{Lei}}, \binits{M.}},
\bauthor{\bsnm{{Leiner}}, \binits{J.}},
\bauthor{\bsnm{{Leonhardt}}, \binits{V.}},
\bauthor{\bsnm{{Leonor}}, \binits{I.}},
\bauthor{\bsnm{{Libbrecht}}, \binits{K.}},
\bauthor{\bsnm{{Lindquist}}, \binits{P.}},
\bauthor{\bsnm{{Lockerbie}}, \binits{N.A.}},
\bauthor{\bsnm{{Longo}}, \binits{M.}},
\bauthor{\bsnm{{Lormand}}, \binits{M.}},
\bauthor{\bsnm{{Lubinski}}, \binits{M.}},
\bauthor{\bsnm{{L{\"u}ck}}, \binits{H.}},
\bauthor{\bsnm{{Machenschalk}}, \binits{B.}},
\bauthor{\bsnm{{MacInnis}}, \binits{M.}},
\bauthor{\bsnm{{Mageswaran}}, \binits{M.}},
\bauthor{\bsnm{{Mailand }}, \binits{K.}},
\bauthor{\bsnm{{Malec}}, \binits{M.}},
\bauthor{\bsnm{{Mandic}}, \binits{V.}},
\bauthor{\bsnm{{Marano}}, \binits{S.}},
\bauthor{\bsnm{{M{\'a}rka}}, \binits{S.}},
\bauthor{\bsnm{{Markowitz}}, \binits{J.}},
\bauthor{\bsnm{{Maros}}, \binits{E.}},
\bauthor{\bsnm{{Martin}}, \binits{I.}},
\bauthor{\bsnm{{Marx}}, \binits{J.N.}},
\bauthor{\bsnm{{Mason}}, \binits{K.}},
\bauthor{\bsnm{{Matone}}, \binits{L.}},
\bauthor{\bsnm{{Matta}}, \binits{V.}},
\bauthor{\bsnm{{Mavalvala}}, \binits{N.}},
\bauthor{\bsnm{{McCarthy}}, \binits{R.}},
\bauthor{\bsnm{{McClelland}}, \binits{D.E.}},
\bauthor{\bsnm{{McGuire}}, \binits{S.C.}},
\bauthor{\bsnm{{McHugh}}, \binits{M.}},
\bauthor{\bsnm{{McKenzie}}, \binits{K.}},
\bauthor{\bsnm{{McWilliams}}, \binits{S.}},
\bauthor{\bsnm{{Meier}}, \binits{T.}},
\bauthor{\bsnm{{Melissinos}}, \binits{A.}},
\bauthor{\bsnm{{Mendell}}, \binits{G.}},
\bauthor{\bsnm{{Mercer}}, \binits{R.A.}},
\bauthor{\bsnm{{Meshkov}}, \binits{S.}},
\bauthor{\bsnm{{Messenger}}, \binits{C.J.}},
\bauthor{\bsnm{{Meyers}}, \binits{D.}},
\bauthor{\bsnm{{Mikhailov}}, \binits{E.}},
\bauthor{\bsnm{{Mitra}}, \binits{S.}},
\bauthor{\bsnm{{Mitrofanov}}, \binits{V.P.}},
\bauthor{\bsnm{{Mitselmakher}}, \binits{G.}},
\bauthor{\bsnm{{Mittleman}}, \binits{R.}},
\bauthor{\bsnm{{Miyakawa}}, \binits{O.}},
\bauthor{\bsnm{{Mohanty}}, \binits{S.}},
\bauthor{\bsnm{{Moreno}}, \binits{G.}},
\bauthor{\bsnm{{Mossavi}}, \binits{K.}},
\bauthor{\bsnm{{MowLowry}}, \binits{C.}},
\bauthor{\bsnm{{Moylan}}, \binits{A.}},
\bauthor{\bsnm{{Mudge}}, \binits{D.}},
\bauthor{\bsnm{{Mueller}}, \binits{G.}},
\bauthor{\bsnm{{Mukherjee}}, \binits{S.}},
\bauthor{\bsnm{{Muller-}}, \binits{H.}},
\bauthor{\bsnm{{Munch}}, \binits{J.}},
\bauthor{\bsnm{{Murray}}, \binits{P.}},
\bauthor{\bsnm{{Myers}}, \binits{E.}},
\bauthor{\bsnm{{Myers}}, \binits{J.}},
\bauthor{\bsnm{{Nash}}, \binits{T.}},
\bauthor{\bsnm{{Newton}}, \binits{G.}},
\bauthor{\bsnm{{Nishizawa}}, \binits{A.}},
\bauthor{\bsnm{{Numata}}, \binits{K.}},
\bauthor{\bsnm{{O'Reilly}}, \binits{B.}},
\bauthor{\bsnm{{O'Shaughnessy}}, \binits{R.}},
\bauthor{\bsnm{{Ottaway}}, \binits{D.J.}},
\bauthor{\bsnm{{Overmier}}, \binits{H.}},
\bauthor{\bsnm{{Owen}}, \binits{B.J.}},
\bauthor{\bsnm{{Pan}}, \binits{Y.}},
\bauthor{\bsnm{{Papa}}, \binits{M.A.}},
\bauthor{\bsnm{{Parameshwaraiah}}, \binits{V.}},
\bauthor{\bsnm{{Patel}}, \binits{P.}},
\bauthor{\bsnm{{Pedraza}}, \binits{M.}},
\bauthor{\bsnm{{Penn}}, \binits{S.}},
\bauthor{\bsnm{{Pierro}}, \binits{V.}},
\bauthor{\bsnm{{Pinto}}, \binits{I.M.}},
\bauthor{\bsnm{{Pitkin}}, \binits{M.}},
\bauthor{\bsnm{{Pletsch}}, \binits{H.}},
\bauthor{\bsnm{{Plissi}}, \binits{M.V.}},
\bauthor{\bsnm{{Postiglione}}, \binits{F.}},
\bauthor{\bsnm{{Prix}}, \binits{R.}},
\bauthor{\bsnm{{Quetschke}}, \binits{V.}},
\bauthor{\bsnm{{Raab}}, \binits{F.}},
\bauthor{\bsnm{{Rabeling}}, \binits{D.}},
\bauthor{\bsnm{{Radkins}}, \binits{H.}},
\bauthor{\bsnm{{Rahkola}}, \binits{R.}},
\bauthor{\bsnm{{Rainer}}, \binits{N.}},
\bauthor{\bsnm{{Rakhmanov}}, \binits{M.}},
\bauthor{\bsnm{{Ramsunder}}, \binits{M.}},
\bauthor{\bsnm{{Ray-Majumder}}, \binits{S.}},
\bauthor{\bsnm{{Re}}, \binits{V.}},
\bauthor{\bsnm{{Rehbein}}, \binits{H.}},
\bauthor{\bsnm{{Reid}}, \binits{S.}},
\bauthor{\bsnm{{Reitze}}, \binits{D.H.}},
\bauthor{\bsnm{{Ribichini}}, \binits{L.}},
\bauthor{\bsnm{{Riesen}}, \binits{R.}},
\bauthor{\bsnm{{Riles}}, \binits{K.}},
\bauthor{\bsnm{{Rivera}}, \binits{B.}},
\bauthor{\bsnm{{Robertson}}, \binits{N.A.}},
\bauthor{\bsnm{{Robinson}}, \binits{C.}},
\bauthor{\bsnm{{Robinson}}, \binits{E.L.}},
\bauthor{\bsnm{{Roddy}}, \binits{S.}},
\bauthor{\bsnm{{Rodriguez}}, \binits{A.}},
\bauthor{\bsnm{{Rogan}}, \binits{A.M.}},
\bauthor{\bsnm{{Rollins}}, \binits{J.}},
\bauthor{\bsnm{{Romano}}, \binits{J.D.}},
\bauthor{\bsnm{{Romie}}, \binits{J.}},
\bauthor{\bsnm{{Route}}, \binits{R.}},
\bauthor{\bsnm{{Rowan}}, \binits{S.}},
\bauthor{\bsnm{{R{\"u}diger}}, \binits{A.}},
\bauthor{\bsnm{{Ruet}}, \binits{L.}},
\bauthor{\bsnm{{Russell}}, \binits{P.}},
\bauthor{\bsnm{{Ryan}}, \binits{K.}},
\bauthor{\bsnm{{Sakata}}, \binits{S.}},
\bauthor{\bsnm{{Samidi}}, \binits{M.}},
\bauthor{\bsnm{{Sancho de la Jordana}}, \binits{L.}},
\bauthor{\bsnm{{Sandberg}}, \binits{V.}},
\bauthor{\bsnm{{Sannibale}}, \binits{V.}},
\bauthor{\bsnm{{Saraf}}, \binits{S.}},
\bauthor{\bsnm{{Sarin}}, \binits{P.}},
\bauthor{\bsnm{{Sathyaprakash}}, \binits{B.S.}},
\bauthor{\bsnm{{Sato}}, \binits{S.}},
\bauthor{\bsnm{{Saulson}}, \binits{P.R.}},
\bauthor{\bsnm{{Savage}}, \binits{R.}},
\bauthor{\bsnm{{Savov}}, \binits{P.}},
\bauthor{\bsnm{{Schediwy}}, \binits{S.}},
\bauthor{\bsnm{{Schilling}}, \binits{R.}},
\bauthor{\bsnm{{Schnabel}}, \binits{R.}},
\bauthor{\bsnm{{Schofield}}, \binits{R.}},
\bauthor{\bsnm{{Schutz}}, \binits{B.F.}},
\bauthor{\bsnm{{Schwinberg}}, \binits{P.}},
\bauthor{\bsnm{{Scott}}, \binits{S.M.}},
\bauthor{\bsnm{{Searle}}, \binits{A.C.}},
\bauthor{\bsnm{{Sears}}, \binits{B.}},
\bauthor{\bsnm{{Seifert}}, \binits{F.}},
\bauthor{\bsnm{{Sellers}}, \binits{D.}},
\bauthor{\bsnm{{Sengupta}}, \binits{A.S.}},
\bauthor{\bsnm{{Shawhan}}, \binits{P.}},
\bauthor{\bsnm{{Shoemaker}}, \binits{D.H.}},
\bauthor{\bsnm{{Sibley}}, \binits{A.}},
\bauthor{\bsnm{{Siemens}}, \binits{X.}},
\bauthor{\bsnm{{Sigg}}, \binits{D.}},
\bauthor{\bsnm{{Sinha}}, \binits{S.}},
\bauthor{\bsnm{{Sintes}}, \binits{A.M.}},
\bauthor{\bsnm{{Slagmolen}}, \binits{B.J.J.}},
\bauthor{\bsnm{{Slutsky}}, \binits{J.}},
\bauthor{\bsnm{{Smith}}, \binits{J.R.}},
\bauthor{\bsnm{{Smith}}, \binits{M.R.}},
\bauthor{\bsnm{{Somiya}}, \binits{K.}},
\bauthor{\bsnm{{Strain}}, \binits{K.A.}},
\bauthor{\bsnm{{Strom}}, \binits{D.M.}},
\bauthor{\bsnm{{Stuver}}, \binits{A.}},
\bauthor{\bsnm{{Summerscales}}, \binits{T.Z.}},
\bauthor{\bsnm{{Sun}}, \binits{K.-X.}},
\bauthor{\bsnm{{Sung}}, \binits{M.}},
\bauthor{\bsnm{{Sutton}}, \binits{P.J.}},
\bauthor{\bsnm{{Takahashi}}, \binits{H.}},
\bauthor{\bsnm{{Tanner}}, \binits{D.B.}},
\bauthor{\bsnm{{Taylor}}, \binits{R.}},
\bauthor{\bsnm{{Taylor}}, \binits{R.}},
\bauthor{\bsnm{{Thacker}}, \binits{J.}},
\bauthor{\bsnm{{Thorne}}, \binits{K.A.}},
\bauthor{\bsnm{{Thorne}}, \binits{K.S.}},
\bauthor{\bsnm{{Th{\"u}ring}}, \binits{A.}},
\bauthor{\bsnm{{Tokmakov}}, \binits{K.V.}},
\bauthor{\bsnm{{Torres}}, \binits{C.}},
\bauthor{\bsnm{{Torrie}}, \binits{C.}},
\bauthor{\bsnm{{Traylor}}, \binits{G.}},
\bauthor{\bsnm{{Trias}}, \binits{M.}},
\bauthor{\bsnm{{Tyler}}, \binits{W.}},
\bauthor{\bsnm{{Ugolini}}, \binits{D.}},
\bauthor{\bsnm{{Urbanek}}, \binits{K.}},
\bauthor{\bsnm{{Vahlbruch}}, \binits{H.}},
\bauthor{\bsnm{{Vallisneri}}, \binits{M.}},
\bauthor{\bsnm{{Van Den Broeck}}, \binits{C.}},
\bauthor{\bsnm{{Varvella}}, \binits{M.}},
\bauthor{\bsnm{{Vass}}, \binits{S.}},
\bauthor{\bsnm{{Vecchio}}, \binits{A.}},
\bauthor{\bsnm{{Veitch}}, \binits{J.}},
\bauthor{\bsnm{{Veitch}}, \binits{P.}},
\bauthor{\bsnm{{Villar}}, \binits{A.}},
\bauthor{\bsnm{{Vorvick}}, \binits{C.}},
\bauthor{\bsnm{{Vyachanin}}, \binits{S.P.}},
\bauthor{\bsnm{{Waldman}}, \binits{S.J.}},
\bauthor{\bsnm{{Wallace}}, \binits{L.}},
\bauthor{\bsnm{{Ward}}, \binits{H.}},
\bauthor{\bsnm{{Ward}}, \binits{R.}},
\bauthor{\bsnm{{Watts}}, \binits{K.}},
\bauthor{\bsnm{{Weidner}}, \binits{A.}},
\bauthor{\bsnm{{Weinert}}, \binits{M.}},
\bauthor{\bsnm{{Weinstein}}, \binits{A.}},
\bauthor{\bsnm{{Weiss}}, \binits{R.}},
\bauthor{\bsnm{{Wen}}, \binits{S.}},
\bauthor{\bsnm{{Wette}}, \binits{K.}},
\bauthor{\bsnm{{Whelan}}, \binits{J.T.}},
\bauthor{\bsnm{{Whitcomb}}, \binits{S.E.}},
\bauthor{\bsnm{{Whiting}}, \binits{B.F.}},
\bauthor{\bsnm{{Wilkinson}}, \binits{C.}},
\bauthor{\bsnm{{Willems}}, \binits{P.A.}},
\bauthor{\bsnm{{Williams}}, \binits{L.}},
\bauthor{\bsnm{{Willke}}, \binits{B.}},
\bauthor{\bsnm{{Wilmut}}, \binits{I.}},
\bauthor{\bsnm{{Winkler}}, \binits{W.}},
\bauthor{\bsnm{{Wipf}}, \binits{C.C.}},
\bauthor{\bsnm{{Wise}}, \binits{S.}},
\bauthor{\bsnm{{Wiseman}}, \binits{A.G.}},
\bauthor{\bsnm{{Woan}}, \binits{G.}},
\bauthor{\bsnm{{Woods}}, \binits{D.}},
\bauthor{\bsnm{{Wooley}}, \binits{R.}},
\bauthor{\bsnm{{Worden}}, \binits{J.}},
\bauthor{\bsnm{{Wu}}, \binits{W.}},
\bauthor{\bsnm{{Yakushin}}, \binits{I.}},
\bauthor{\bsnm{{Yamamoto}}, \binits{H.}},
\bauthor{\bsnm{{Yan}}, \binits{Z.}},
\bauthor{\bsnm{{Yoshida}}, \binits{S.}},
\bauthor{\bsnm{{Yunes}}, \binits{N.}},
\bauthor{\bsnm{{Zanolin}}, \binits{M.}},
\bauthor{\bsnm{{Zhang}}, \binits{J.}},
\bauthor{\bsnm{{Zhang}}, \binits{L.}},
\bauthor{\bsnm{{Zhao}}, \binits{C.}},
\bauthor{\bsnm{{Zotov}}, \binits{N.}},
\bauthor{\bsnm{{Zucker}}, \binits{M.}},
\bauthor{\bsnm{{zur M{\"u}hlen}}, \binits{H.}},
\bauthor{\bsnm{{Zweizig}}, \binits{J.}},
\bauthor{\bsnm{{LIGO Scientific Collaboration}}},
\bauthor{\bsnm{{Hurley}}, \binits{K.C.}}:
\bjtitle{\apj}
\bvolume{681}(\bissue{2}),
\bfpage{1419}
(\byear{2008}).
\arxivurl{0711.1163}.
doi:\doiurl{10.1086/587954}
\end{barticle}
\endbibitem

\bibitem[\protect\citeauthoryear{{Ajello} et~al.}{2008}]{Ajello2008}
\begin{barticle}
\bauthor{\bsnm{{Ajello}}, \binits{M.}},
\bauthor{\bsnm{{Greiner}}, \binits{J.}},
\bauthor{\bsnm{{Sato}}, \binits{G.}},
\bauthor{\bsnm{{Willis}}, \binits{D.R.}},
\bauthor{\bsnm{{Kanbach}}, \binits{G.}},
\bauthor{\bsnm{{Strong}}, \binits{A.W.}},
\bauthor{\bsnm{{Diehl}}, \binits{R.}},
\bauthor{\bsnm{{Hasinger}}, \binits{G.}},
\bauthor{\bsnm{{Gehrels}}, \binits{N.}},
\bauthor{\bsnm{{Markwardt}}, \binits{C.B.}},
\bauthor{\bsnm{{Tueller}}, \binits{J.}}:
\bjtitle{\apj}
\bvolume{689}(\bissue{2}),
\bfpage{666}
(\byear{2008}).
\arxivurl{0808.3377}.
doi:\doiurl{10.1086/592595}
\end{barticle}
\endbibitem

\bibitem[\protect\citeauthoryear{Alexander
  et~al.}{2016}]{Alexander2016_161219B}
\begin{barticle}
\bauthor{\bsnm{Alexander}, \binits{K.D.}},
\bauthor{\bsnm{Laskar}, \binits{T.}},
\bauthor{\bsnm{Berger}, \binits{E.}}:
\bjtitle{GRB Coordinates Network}
\bvolume{20313},
\bfpage{1}
(\byear{2016})
\end{barticle}
\endbibitem

\bibitem[\protect\citeauthoryear{{Allison} et~al.}{2016}]{Allison2016_GEANT4}
\begin{barticle}
\bauthor{\bsnm{{Allison}}, \binits{J.}},
\bauthor{\bsnm{{Amako}}, \binits{K.}},
\bauthor{\bsnm{{Apostolakis}}, \binits{J.}},
\bauthor{\bsnm{{Arce}}, \binits{P.}},
\bauthor{\bsnm{{Asai}}, \binits{M.}},
\bauthor{\bsnm{{Aso}}, \binits{T.}},
\bauthor{\bsnm{{Bagli}}, \binits{E.}},
\bauthor{\bsnm{{Bagulya}}, \binits{A.}},
\bauthor{\bsnm{{Banerjee}}, \binits{S.}},
\bauthor{\bsnm{{Barrand}}, \binits{G.}},
\bauthor{\bsnm{{Beck}}, \binits{B.R.}},
\bauthor{\bsnm{{Bogdanov}}, \binits{A.G.}},
\bauthor{\bsnm{{Brandt}}, \binits{D.}},
\bauthor{\bsnm{{Brown}}, \binits{J.M.C.}},
\bauthor{\bsnm{{Burkhardt}}, \binits{H.}},
\bauthor{\bsnm{{Canal}}, \binits{P.}},
\bauthor{\bsnm{{Cano-Ott}}, \binits{D.}},
\bauthor{\bsnm{{Chauvie}}, \binits{S.}},
\bauthor{\bsnm{{Cho}}, \binits{K.}},
\bauthor{\bsnm{{Cirrone}}, \binits{G.A.P.}},
\bauthor{\bsnm{{Cooperman}}, \binits{G.}},
\bauthor{\bsnm{{Cort{\'e}s-Giraldo}}, \binits{M.A.}},
\bauthor{\bsnm{{Cosmo}}, \binits{G.}},
\bauthor{\bsnm{{Cuttone}}, \binits{G.}},
\bauthor{\bsnm{{Depaola}}, \binits{G.}},
\bauthor{\bsnm{{Desorgher}}, \binits{L.}},
\bauthor{\bsnm{{Dong}}, \binits{X.}},
\bauthor{\bsnm{{Dotti}}, \binits{A.}},
\bauthor{\bsnm{{Elvira}}, \binits{V.D.}},
\bauthor{\bsnm{{Folger}}, \binits{G.}},
\bauthor{\bsnm{{Francis}}, \binits{Z.}},
\bauthor{\bsnm{{Galoyan}}, \binits{A.}},
\bauthor{\bsnm{{Garnier}}, \binits{L.}},
\bauthor{\bsnm{{Gayer}}, \binits{M.}},
\bauthor{\bsnm{{Genser}}, \binits{K.L.}},
\bauthor{\bsnm{{Grichine}}, \binits{V.M.}},
\bauthor{\bsnm{{Guatelli}}, \binits{S.}},
\bauthor{\bsnm{{Gu{\`e}ye}}, \binits{P.}},
\bauthor{\bsnm{{Gumplinger}}, \binits{P.}},
\bauthor{\bsnm{{Howard}}, \binits{A.S.}},
\bauthor{\bsnm{{H{\v{r}}ivn{\'a}{\v{c}}ov{\'a}}}, \binits{I.}},
\bauthor{\bsnm{{Hwang}}, \binits{S.}},
\bauthor{\bsnm{{Incerti}}, \binits{S.}},
\bauthor{\bsnm{{Ivanchenko}}, \binits{A.}},
\bauthor{\bsnm{{Ivanchenko}}, \binits{V.N.}},
\bauthor{\bsnm{{Jones}}, \binits{F.W.}},
\bauthor{\bsnm{{Jun}}, \binits{S.Y.}},
\bauthor{\bsnm{{Kaitaniemi}}, \binits{P.}},
\bauthor{\bsnm{{Karakatsanis}}, \binits{N.}},
\bauthor{\bsnm{{Karamitrosi}}, \binits{M.}},
\bauthor{\bsnm{{Kelsey}}, \binits{M.}},
\bauthor{\bsnm{{Kimura}}, \binits{A.}},
\bauthor{\bsnm{{Koi}}, \binits{T.}},
\bauthor{\bsnm{{Kurashige}}, \binits{H.}},
\bauthor{\bsnm{{Lechner}}, \binits{A.}},
\bauthor{\bsnm{{Lee}}, \binits{S.B.}},
\bauthor{\bsnm{{Longo}}, \binits{F.}},
\bauthor{\bsnm{{Maire}}, \binits{M.}},
\bauthor{\bsnm{{Mancusi}}, \binits{D.}},
\bauthor{\bsnm{{Mantero}}, \binits{A.}},
\bauthor{\bsnm{{Mendoza}}, \binits{E.}},
\bauthor{\bsnm{{Morgan}}, \binits{B.}},
\bauthor{\bsnm{{Murakami}}, \binits{K.}},
\bauthor{\bsnm{{Nikitina}}, \binits{T.}},
\bauthor{\bsnm{{Pandola}}, \binits{L.}},
\bauthor{\bsnm{{Paprocki}}, \binits{P.}},
\bauthor{\bsnm{{Perl}}, \binits{J.}},
\bauthor{\bsnm{{Petrovi{\'c}}}, \binits{I.}},
\bauthor{\bsnm{{Pia}}, \binits{M.G.}},
\bauthor{\bsnm{{Pokorski}}, \binits{W.}},
\bauthor{\bsnm{{Quesada}}, \binits{J.M.}},
\bauthor{\bsnm{{Raine}}, \binits{M.}},
\bauthor{\bsnm{{Reis}}, \binits{M.A.}},
\bauthor{\bsnm{{Ribon}}, \binits{A.}},
\bauthor{\bsnm{{Risti{\'c} Fira}}, \binits{A.}},
\bauthor{\bsnm{{Romano}}, \binits{F.}},
\bauthor{\bsnm{{Russo}}, \binits{G.}},
\bauthor{\bsnm{{Santin}}, \binits{G.}},
\bauthor{\bsnm{{Sasaki}}, \binits{T.}},
\bauthor{\bsnm{{Sawkey}}, \binits{D.}},
\bauthor{\bsnm{{Shin}}, \binits{J.I.}},
\bauthor{\bsnm{{Strakovsky}}, \binits{I.I.}},
\bauthor{\bsnm{{Taborda}}, \binits{A.}},
\bauthor{\bsnm{{Tanaka}}, \binits{S.}},
\bauthor{\bsnm{{Tom{\'e}}}, \binits{B.}},
\bauthor{\bsnm{{Toshito}}, \binits{T.}},
\bauthor{\bsnm{{Tran}}, \binits{H.N.}},
\bauthor{\bsnm{{Truscott}}, \binits{P.R.}},
\bauthor{\bsnm{{Urban}}, \binits{L.}},
\bauthor{\bsnm{{Uzhinsky}}, \binits{V.}},
\bauthor{\bsnm{{Verbeke}}, \binits{J.M.}},
\bauthor{\bsnm{{Verderi}}, \binits{M.}},
\bauthor{\bsnm{{Wendt}}, \binits{B.L.}},
\bauthor{\bsnm{{Wenzel}}, \binits{H.}},
\bauthor{\bsnm{{Wright}}, \binits{D.H.}},
\bauthor{\bsnm{{Wright}}, \binits{D.M.}},
\bauthor{\bsnm{{Yamashita}}, \binits{T.}},
\bauthor{\bsnm{{Yarba}}, \binits{J.}},
\bauthor{\bsnm{{Yoshida}}, \binits{H.}}:
\bjtitle{Nuclear Instruments and Methods in Physics Research A}
\bvolume{835},
\bfpage{186}
(\byear{2016}).
doi:\doiurl{10.1016/j.nima.2016.06.125}
\end{barticle}
\endbibitem

\bibitem[\protect\citeauthoryear{{Amati} et~al.}{2007}]{Amati2007_040701}
\begin{barticle}
\bauthor{\bsnm{{Amati}}, \binits{L.}},
\bauthor{\bsnm{{Della Valle}}, \binits{M.}},
\bauthor{\bsnm{{Frontera}}, \binits{F.}},
\bauthor{\bsnm{{Malesani}}, \binits{D.}},
\bauthor{\bsnm{{Guidorzi}}, \binits{C.}},
\bauthor{\bsnm{{Montanari}}, \binits{E.}},
\bauthor{\bsnm{{Pian}}, \binits{E.}}:
\bjtitle{\aap}
\bvolume{463}(\bissue{3}),
\bfpage{913}
(\byear{2007}).
\arxivurl{astro-ph/0607148}.
doi:\doiurl{10.1051/0004-6361:20065994}
\end{barticle}
\endbibitem

\bibitem[\protect\citeauthoryear{{Amati}}{2006}]{Amati2006}
\begin{barticle}
\bauthor{\bsnm{{Amati}}, \binits{L.}}:
\bjtitle{\mnras}
\bvolume{372}(\bissue{1}),
\bfpage{233}
(\byear{2006}).
\arxivurl{astro-ph/0601553}.
doi:\doiurl{10.1111/j.1365-2966.2006.10840.x}
\end{barticle}
\endbibitem

\bibitem[\protect\citeauthoryear{Andersen et~al.}{2004}]{Andersen2004}
\begin{barticle}
\bauthor{\bsnm{Andersen}, \binits{M.I.}},
\bauthor{\bsnm{Hjorth}, \binits{J.}},
\bauthor{\bsnm{Sollerman}, \binits{J.}},
\bauthor{\bsnm{Møller}, \binits{P.}},
\bauthor{\bsnm{Fynbo}, \binits{J.U.P.}}:
\bjtitle{Baltic Astronomy}
\bvolume{13},
\bfpage{247}
(\byear{2004})
\end{barticle}
\endbibitem

\bibitem[\protect\citeauthoryear{Antier-Farfar}{2016}]{Antier2016}
\begin{botherref}
\oauthor{\bsnm{Antier-Farfar}, \binits{S.}}:
La d\'etection des sursauts gamma par le t\'elescope eclairs pour la mission
  spatiale svom.
PhD thesis,
CEA-Saclay
(2016).
\url{http://www.theses.fr/2016SACLS467}
\end{botherref}
\endbibitem

\bibitem[\protect\citeauthoryear{{Aptekar} et~al.}{1995}]{Aptekar1995_konus}
\begin{barticle}
\bauthor{\bsnm{{Aptekar}}, \binits{R.L.}},
\bauthor{\bsnm{{Frederiks}}, \binits{D.D.}},
\bauthor{\bsnm{{Golenetskii}}, \binits{S.V.}},
\bauthor{\bsnm{{Ilynskii}}, \binits{V.N.}},
\bauthor{\bsnm{{Mazets}}, \binits{E.P.}},
\bauthor{\bsnm{{Panov}}, \binits{V.N.}},
\bauthor{\bsnm{{Sokolova}}, \binits{Z.J.}},
\bauthor{\bsnm{{Terekhov}}, \binits{M.M.}},
\bauthor{\bsnm{{Sheshin}}, \binits{L.O.}},
\bauthor{\bsnm{{Cline}}, \binits{T.L.}},
\bauthor{\bsnm{{Stilwell}}, \binits{D.E.}}:
\bjtitle{\ssr}
\bvolume{71}(\bissue{1-4}),
\bfpage{265}
(\byear{1995}).
doi:\doiurl{10.1007/BF00751332}
\end{barticle}
\endbibitem

\bibitem[\protect\citeauthoryear{{Aso} et~al.}{2013}]{Aso2013}
\begin{barticle}
\bauthor{\bsnm{{Aso}}, \binits{Y.}},
\bauthor{\bsnm{{Michimura}}, \binits{Y.}},
\bauthor{\bsnm{{Somiya}}, \binits{K.}},
\bauthor{\bsnm{{Ando}}, \binits{M.}},
\bauthor{\bsnm{{Miyakawa}}, \binits{O.}},
\bauthor{\bsnm{{Sekiguchi}}, \binits{T.}},
\bauthor{\bsnm{{Tatsumi}}, \binits{D.}},
\bauthor{\bsnm{{Yamamoto}}, \binits{H.}}:
\bjtitle{\prd}
\bvolume{88}(\bissue{4}),
\bfpage{043007}
(\byear{2013}).
\arxivurl{1306.6747}.
doi:\doiurl{10.1103/PhysRevD.88.043007}
\end{barticle}
\endbibitem

\bibitem[\protect\citeauthoryear{{Astropy Collaboration}
  et~al.}{2018}]{Astropy}
\begin{barticle}
\bauthor{\bsnm{{Astropy Collaboration}}},
\bauthor{\bsnm{{Price-Whelan}}, \binits{A.M.}},
\bauthor{\bsnm{{Sip{\H{o}}cz}}, \binits{B.M.}},
\bauthor{\bsnm{{G{\"u}nther}}, \binits{H.M.}},
\bauthor{\bsnm{{Lim}}, \binits{P.L.}},
\bauthor{\bsnm{{Crawford}}, \binits{S.M.}},
\bauthor{\bsnm{{Conseil}}, \binits{S.}},
\bauthor{\bsnm{{Shupe}}, \binits{D.L.}},
\bauthor{\bsnm{{Craig}}, \binits{M.W.}},
\bauthor{\bsnm{{Dencheva}}, \binits{N.}},
\bauthor{\bsnm{{Ginsburg}}, \binits{A.}},
\bauthor{\bsnm{{Vand erPlas}}, \binits{J.T.}},
\bauthor{\bsnm{{Bradley}}, \binits{L.D.}},
\bauthor{\bsnm{{P{\'e}rez-Su{\'a}rez}}, \binits{D.}},
\bauthor{\bsnm{{de Val-Borro}}, \binits{M.}},
\bauthor{\bsnm{{Aldcroft}}, \binits{T.L.}},
\bauthor{\bsnm{{Cruz}}, \binits{K.L.}},
\bauthor{\bsnm{{Robitaille}}, \binits{T.P.}},
\bauthor{\bsnm{{Tollerud}}, \binits{E.J.}},
\bauthor{\bsnm{{Ardelean}}, \binits{C.}},
\bauthor{\bsnm{{Babej}}, \binits{T.}},
\bauthor{\bsnm{{Bach}}, \binits{Y.P.}},
\bauthor{\bsnm{{Bachetti}}, \binits{M.}},
\bauthor{\bsnm{{Bakanov}}, \binits{A.V.}},
\bauthor{\bsnm{{Bamford}}, \binits{S.P.}},
\bauthor{\bsnm{{Barentsen}}, \binits{G.}},
\bauthor{\bsnm{{Barmby}}, \binits{P.}},
\bauthor{\bsnm{{Baumbach}}, \binits{A.}},
\bauthor{\bsnm{{Berry}}, \binits{K.L.}},
\bauthor{\bsnm{{Biscani}}, \binits{F.}},
\bauthor{\bsnm{{Boquien}}, \binits{M.}},
\bauthor{\bsnm{{Bostroem}}, \binits{K.A.}},
\bauthor{\bsnm{{Bouma}}, \binits{L.G.}},
\bauthor{\bsnm{{Brammer}}, \binits{G.B.}},
\bauthor{\bsnm{{Bray}}, \binits{E.M.}},
\bauthor{\bsnm{{Breytenbach}}, \binits{H.}},
\bauthor{\bsnm{{Buddelmeijer}}, \binits{H.}},
\bauthor{\bsnm{{Burke}}, \binits{D.J.}},
\bauthor{\bsnm{{Calderone}}, \binits{G.}},
\bauthor{\bsnm{{Cano Rodr{\'\i}guez}}, \binits{J.L.}},
\bauthor{\bsnm{{Cara}}, \binits{M.}},
\bauthor{\bsnm{{Cardoso}}, \binits{J.V.M.}},
\bauthor{\bsnm{{Cheedella}}, \binits{S.}},
\bauthor{\bsnm{{Copin}}, \binits{Y.}},
\bauthor{\bsnm{{Corrales}}, \binits{L.}},
\bauthor{\bsnm{{Crichton}}, \binits{D.}},
\bauthor{\bsnm{{D'Avella}}, \binits{D.}},
\bauthor{\bsnm{{Deil}}, \binits{C.}},
\bauthor{\bsnm{{Depagne}}, \binits{{\'E}.}},
\bauthor{\bsnm{{Dietrich}}, \binits{J.P.}},
\bauthor{\bsnm{{Donath}}, \binits{A.}},
\bauthor{\bsnm{{Droettboom}}, \binits{M.}},
\bauthor{\bsnm{{Earl}}, \binits{N.}},
\bauthor{\bsnm{{Erben}}, \binits{T.}},
\bauthor{\bsnm{{Fabbro}}, \binits{S.}},
\bauthor{\bsnm{{Ferreira}}, \binits{L.A.}},
\bauthor{\bsnm{{Finethy}}, \binits{T.}},
\bauthor{\bsnm{{Fox}}, \binits{R.T.}},
\bauthor{\bsnm{{Garrison}}, \binits{L.H.}},
\bauthor{\bsnm{{Gibbons}}, \binits{S.L.J.}},
\bauthor{\bsnm{{Goldstein}}, \binits{D.A.}},
\bauthor{\bsnm{{Gommers}}, \binits{R.}},
\bauthor{\bsnm{{Greco}}, \binits{J.P.}},
\bauthor{\bsnm{{Greenfield}}, \binits{P.}},
\bauthor{\bsnm{{Groener}}, \binits{A.M.}},
\bauthor{\bsnm{{Grollier}}, \binits{F.}},
\bauthor{\bsnm{{Hagen}}, \binits{A.}},
\bauthor{\bsnm{{Hirst}}, \binits{P.}},
\bauthor{\bsnm{{Homeier}}, \binits{D.}},
\bauthor{\bsnm{{Horton}}, \binits{A.J.}},
\bauthor{\bsnm{{Hosseinzadeh}}, \binits{G.}},
\bauthor{\bsnm{{Hu}}, \binits{L.}},
\bauthor{\bsnm{{Hunkeler}}, \binits{J.S.}},
\bauthor{\bsnm{{Ivezi{\'c}}}, \binits{{\v{Z}}.}},
\bauthor{\bsnm{{Jain}}, \binits{A.}},
\bauthor{\bsnm{{Jenness}}, \binits{T.}},
\bauthor{\bsnm{{Kanarek}}, \binits{G.}},
\bauthor{\bsnm{{Kendrew}}, \binits{S.}},
\bauthor{\bsnm{{Kern}}, \binits{N.S.}},
\bauthor{\bsnm{{Kerzendorf}}, \binits{W.E.}},
\bauthor{\bsnm{{Khvalko}}, \binits{A.}},
\bauthor{\bsnm{{King}}, \binits{J.}},
\bauthor{\bsnm{{Kirkby}}, \binits{D.}},
\bauthor{\bsnm{{Kulkarni}}, \binits{A.M.}},
\bauthor{\bsnm{{Kumar}}, \binits{A.}},
\bauthor{\bsnm{{Lee}}, \binits{A.}},
\bauthor{\bsnm{{Lenz}}, \binits{D.}},
\bauthor{\bsnm{{Littlefair}}, \binits{S.P.}},
\bauthor{\bsnm{{Ma}}, \binits{Z.}},
\bauthor{\bsnm{{Macleod}}, \binits{D.M.}},
\bauthor{\bsnm{{Mastropietro}}, \binits{M.}},
\bauthor{\bsnm{{McCully}}, \binits{C.}},
\bauthor{\bsnm{{Montagnac}}, \binits{S.}},
\bauthor{\bsnm{{Morris}}, \binits{B.M.}},
\bauthor{\bsnm{{Mueller}}, \binits{M.}},
\bauthor{\bsnm{{Mumford}}, \binits{S.J.}},
\bauthor{\bsnm{{Muna}}, \binits{D.}},
\bauthor{\bsnm{{Murphy}}, \binits{N.A.}},
\bauthor{\bsnm{{Nelson}}, \binits{S.}},
\bauthor{\bsnm{{Nguyen}}, \binits{G.H.}},
\bauthor{\bsnm{{Ninan}}, \binits{J.P.}},
\bauthor{\bsnm{{N{\"o}the}}, \binits{M.}},
\bauthor{\bsnm{{Ogaz}}, \binits{S.}},
\bauthor{\bsnm{{Oh}}, \binits{S.}},
\bauthor{\bsnm{{Parejko}}, \binits{J.K.}},
\bauthor{\bsnm{{Parley}}, \binits{N.}},
\bauthor{\bsnm{{Pascual}}, \binits{S.}},
\bauthor{\bsnm{{Patil}}, \binits{R.}},
\bauthor{\bsnm{{Patil}}, \binits{A.A.}},
\bauthor{\bsnm{{Plunkett}}, \binits{A.L.}},
\bauthor{\bsnm{{Prochaska}}, \binits{J.X.}},
\bauthor{\bsnm{{Rastogi}}, \binits{T.}},
\bauthor{\bsnm{{Reddy Janga}}, \binits{V.}},
\bauthor{\bsnm{{Sabater}}, \binits{J.}},
\bauthor{\bsnm{{Sakurikar}}, \binits{P.}},
\bauthor{\bsnm{{Seifert}}, \binits{M.}},
\bauthor{\bsnm{{Sherbert}}, \binits{L.E.}},
\bauthor{\bsnm{{Sherwood-Taylor}}, \binits{H.}},
\bauthor{\bsnm{{Shih}}, \binits{A.Y.}},
\bauthor{\bsnm{{Sick}}, \binits{J.}},
\bauthor{\bsnm{{Silbiger}}, \binits{M.T.}},
\bauthor{\bsnm{{Singanamalla}}, \binits{S.}},
\bauthor{\bsnm{{Singer}}, \binits{L.P.}},
\bauthor{\bsnm{{Sladen}}, \binits{P.H.}},
\bauthor{\bsnm{{Sooley}}, \binits{K.A.}},
\bauthor{\bsnm{{Sornarajah}}, \binits{S.}},
\bauthor{\bsnm{{Streicher}}, \binits{O.}},
\bauthor{\bsnm{{Teuben}}, \binits{P.}},
\bauthor{\bsnm{{Thomas}}, \binits{S.W.}},
\bauthor{\bsnm{{Tremblay}}, \binits{G.R.}},
\bauthor{\bsnm{{Turner}}, \binits{J.E.H.}},
\bauthor{\bsnm{{Terr{\'o}n}}, \binits{V.}},
\bauthor{\bsnm{{van Kerkwijk}}, \binits{M.H.}},
\bauthor{\bsnm{{de la Vega}}, \binits{A.}},
\bauthor{\bsnm{{Watkins}}, \binits{L.L.}},
\bauthor{\bsnm{{Weaver}}, \binits{B.A.}},
\bauthor{\bsnm{{Whitmore}}, \binits{J.B.}},
\bauthor{\bsnm{{Woillez}}, \binits{J.}},
\bauthor{\bsnm{{Zabalza}}, \binits{V.}},
\bauthor{\bsnm{{Astropy Contributors}}}:
\bjtitle{\aj}
\bvolume{156}(\bissue{3}),
\bfpage{123}
(\byear{2018}).
\arxivurl{1801.02634}.
doi:\doiurl{10.3847/1538-3881/aabc4f}
\end{barticle}
\endbibitem

\bibitem[\protect\citeauthoryear{{Atteia} et~al.}{2003}]{Atteia2003_fregate}
\begin{bchapter}
\bauthor{\bsnm{{Atteia}}, \binits{J.-L.}},
\bauthor{\bsnm{{Boer}}, \binits{M.}},
\bauthor{\bsnm{{Cotin}}, \binits{F.}},
\bauthor{\bsnm{{Couteret}}, \binits{J.}},
\bauthor{\bsnm{{Dezalay}}, \binits{J.-P.}},
\bauthor{\bsnm{{Ehanno}}, \binits{M.}},
\bauthor{\bsnm{{Evrard}}, \binits{J.}},
\bauthor{\bsnm{{Lagrange}}, \binits{D.}},
\bauthor{\bsnm{{Niel}}, \binits{M.}},
\bauthor{\bsnm{{Olive}}, \binits{J.-F.}},
\bauthor{\bsnm{{Rouaix}}, \binits{G.}},
\bauthor{\bsnm{{Souleille}}, \binits{P.}},
\bauthor{\bsnm{{Vedrenne}}, \binits{G.}},
\bauthor{\bsnm{{Hurley}}, \binits{K.}},
\bauthor{\bsnm{{Ricker}}, \binits{G.}},
\bauthor{\bsnm{{Vanderspek}}, \binits{R.}},
\bauthor{\bsnm{{Crew}}, \binits{G.}},
\bauthor{\bsnm{{Doty}}, \binits{J.}},
\bauthor{\bsnm{{Butler}}, \binits{N.}}:
In: \beditor{\bsnm{{Ricker}}, \binits{G.R.}},
\beditor{\bsnm{{Vanderspek}}, \binits{R.K.}} (eds.)
\bbtitle{Gamma-Ray Burst and Afterglow Astronomy 2001: A Workshop Celebrating
  the First Year of the HETE Mission}.
\bsertitle{American Institute of Physics Conference Series},
vol. \bseriesno{662},
p. \bfpage{17}
(\byear{2003}).
\arxivurl{astro-ph/0202515}.
doi:\doiurl{10.1063/1.1579292}
\end{bchapter}
\endbibitem

\bibitem[\protect\citeauthoryear{{Atteia} et~al.}{2017}]{Atteia2017}
\begin{barticle}
\bauthor{\bsnm{{Atteia}}, \binits{J.-L.}},
\bauthor{\bsnm{{Heussaff}}, \binits{V.}},
\bauthor{\bsnm{{Dezalay}}, \binits{J.-P.}},
\bauthor{\bsnm{{Klotz}}, \binits{A.}},
\bauthor{\bsnm{{Turpin}}, \binits{D.}},
\bauthor{\bsnm{{Tsvetkova}}, \binits{A.E.}},
\bauthor{\bsnm{{Frederiks}}, \binits{D.D.}},
\bauthor{\bsnm{{Zolnierowski}}, \binits{Y.}},
\bauthor{\bsnm{{Daigne}}, \binits{F.}},
\bauthor{\bsnm{{Mochkovitch}}, \binits{R.}}:
\bjtitle{\apj}
\bvolume{837}(\bissue{2}),
\bfpage{119}
(\byear{2017}).
\arxivurl{1702.02961}.
doi:\doiurl{10.3847/1538-4357/aa5ffa}
\end{barticle}
\endbibitem

\bibitem[\protect\citeauthoryear{{Band} et~al.}{1993}]{Band1993}
\begin{barticle}
\bauthor{\bsnm{{Band}}, \binits{D.}},
\bauthor{\bsnm{{Matteson}}, \binits{J.}},
\bauthor{\bsnm{{Ford}}, \binits{L.}},
\bauthor{\bsnm{{Schaefer}}, \binits{B.}},
\bauthor{\bsnm{{Palmer}}, \binits{D.}},
\bauthor{\bsnm{{Teegarden}}, \binits{B.}},
\bauthor{\bsnm{{Cline}}, \binits{T.}},
\bauthor{\bsnm{{Briggs}}, \binits{M.}},
\bauthor{\bsnm{{Paciesas}}, \binits{W.}},
\bauthor{\bsnm{{Pendleton}}, \binits{G.}},
\bauthor{\bsnm{{Fishman}}, \binits{G.}},
\bauthor{\bsnm{{Kouveliotou}}, \binits{C.}},
\bauthor{\bsnm{{Meegan}}, \binits{C.}},
\bauthor{\bsnm{{Wilson}}, \binits{R.}},
\bauthor{\bsnm{{Lestrade}}, \binits{P.}}:
\bjtitle{\apj}
\bvolume{413},
\bfpage{281}
(\byear{1993}).
doi:\doiurl{10.1086/172995}
\end{barticle}
\endbibitem

\bibitem[\protect\citeauthoryear{Barkov and Baushev}{2011}]{Barkov2011}
\begin{botherref}
\oauthor{\bsnm{Barkov}, \binits{M.V.}},
\oauthor{\bsnm{Baushev}, \binits{A.N.}}:
Accretion of a massive magnetized torus on a rotating black hole
\textbf{16},
46
(2011).
doi:\doiurl{10.1016/j.newast.2010.07.001}
\end{botherref}
\endbibitem

\bibitem[\protect\citeauthoryear{{Barraud} et~al.}{2003}]{Barraud2003}
\begin{barticle}
\bauthor{\bsnm{{Barraud}}, \binits{C.}},
\bauthor{\bsnm{{Olive}}, \binits{J.-F.}},
\bauthor{\bsnm{{Lestrade}}, \binits{J.P.}},
\bauthor{\bsnm{{Atteia}}, \binits{J.-L.}},
\bauthor{\bsnm{{Hurley}}, \binits{K.}},
\bauthor{\bsnm{{Ricker}}, \binits{G.}},
\bauthor{\bsnm{{Lamb}}, \binits{D.Q.}},
\bauthor{\bsnm{{Kawai}}, \binits{N.}},
\bauthor{\bsnm{{Boer}}, \binits{M.}},
\bauthor{\bsnm{{Dezalay}}, \binits{J.-P.}},
\bauthor{\bsnm{{Pizzichini}}, \binits{G.}},
\bauthor{\bsnm{{Vanderspek}}, \binits{R.}},
\bauthor{\bsnm{{Crew}}, \binits{G.}},
\bauthor{\bsnm{{Doty}}, \binits{J.}},
\bauthor{\bsnm{{Monnelly}}, \binits{G.}},
\bauthor{\bsnm{{Villasenor}}, \binits{J.}},
\bauthor{\bsnm{{Butler}}, \binits{N.}},
\bauthor{\bsnm{{Levine}}, \binits{A.}},
\bauthor{\bsnm{{Yoshida}}, \binits{A.}},
\bauthor{\bsnm{{Shirasaki}}, \binits{Y.}},
\bauthor{\bsnm{{Sakamoto}}, \binits{T.}},
\bauthor{\bsnm{{Tamagawa}}, \binits{T.}},
\bauthor{\bsnm{{Torii}}, \binits{K.}},
\bauthor{\bsnm{{Matsuoka}}, \binits{M.}},
\bauthor{\bsnm{{Fenimore}}, \binits{E.E.}},
\bauthor{\bsnm{{Galassi}}, \binits{M.}},
\bauthor{\bsnm{{Tavenner}}, \binits{T.}},
\bauthor{\bsnm{{Donaghy}}, \binits{T.Q.}},
\bauthor{\bsnm{{Graziani}}, \binits{C.}},
\bauthor{\bsnm{{Jernigan}}, \binits{J.G.}}:
\bjtitle{\aap}
\bvolume{400},
\bfpage{1021}
(\byear{2003}).
\arxivurl{astro-ph/0206380}.
doi:\doiurl{10.1051/0004-6361:20030074}
\end{barticle}
\endbibitem

\bibitem[\protect\citeauthoryear{{Barraud} et~al.}{2004}]{Barraud2004_040701}
\begin{barticle}
\bauthor{\bsnm{{Barraud}}, \binits{C.}},
\bauthor{\bsnm{{Ricker}}, \binits{G.}},
\bauthor{\bsnm{{Atteia}}, \binits{J.-L.}},
\bauthor{\bsnm{{Kawai}}, \binits{N.}},
\bauthor{\bsnm{{Lamb}}, \binits{D.}},
\bauthor{\bsnm{{Woosley}}, \binits{S.}},
\bauthor{\bsnm{{Donaghy}}, \binits{T.}},
\bauthor{\bsnm{{Fenimore}}, \binits{E.}},
\bauthor{\bsnm{{Galassi}}, \binits{M.}},
\bauthor{\bsnm{{Graziani}}, \binits{C.}},
\bauthor{\bsnm{{Matsuoka}}, \binits{M.}},
\bauthor{\bsnm{{Nakagawa}}, \binits{Y.}},
\bauthor{\bsnm{{Sakamoto}}, \binits{T.}},
\bauthor{\bsnm{{Sato}}, \binits{R.}},
\bauthor{\bsnm{{Shirasaki}}, \binits{Y.}},
\bauthor{\bsnm{{Suzuki}}, \binits{M.}},
\bauthor{\bsnm{{Tamagawa}}, \binits{T.}},
\bauthor{\bsnm{{Torii}}, \binits{K.}},
\bauthor{\bsnm{{Urata}}, \binits{Y.}},
\bauthor{\bsnm{{Yamazaki}}, \binits{T.}},
\bauthor{\bsnm{{Yamamoto}}, \binits{Y.}},
\bauthor{\bsnm{{Yoshida}}, \binits{A.}},
\bauthor{\bsnm{{Butler}}, \binits{N.}},
\bauthor{\bsnm{{Crew}}, \binits{G.}},
\bauthor{\bsnm{{Doty}}, \binits{J.}},
\bauthor{\bsnm{{Dullighan}}, \binits{A.}},
\bauthor{\bsnm{{Prigozhin}}, \binits{G.}},
\bauthor{\bsnm{{Vand erspek}}, \binits{R.}},
\bauthor{\bsnm{{Villasenor}}, \binits{J.}},
\bauthor{\bsnm{{Jernigan}}, \binits{J.G.}},
\bauthor{\bsnm{{Levine}}, \binits{A.}},
\bauthor{\bsnm{{Azzibrouck}}, \binits{G.}},
\bauthor{\bsnm{{Braga}}, \binits{J.}},
\bauthor{\bsnm{{Manchanda}}, \binits{R.}},
\bauthor{\bsnm{{Pizzichini}}, \binits{G.}},
\bauthor{\bsnm{{Boer}}, \binits{M.}},
\bauthor{\bsnm{{Olive}}, \binits{J.-F.}},
\bauthor{\bsnm{{Dezalay}}, \binits{J.-P.}},
\bauthor{\bsnm{{Hurley}}, \binits{K.}}:
\bjtitle{GRB Coordinates Network}
\bvolume{2620},
\bfpage{1}
(\byear{2004})
\end{barticle}
\endbibitem

\bibitem[\protect\citeauthoryear{{Barraud} et~al.}{2005}]{Barraud2005}
\begin{barticle}
\bauthor{\bsnm{{Barraud}}, \binits{C.}},
\bauthor{\bsnm{{Daigne}}, \binits{F.}},
\bauthor{\bsnm{{Mochkovitch}}, \binits{R.}},
\bauthor{\bsnm{{Atteia}}, \binits{J.-L.}}:
\bjtitle{\aap}
\bvolume{440}(\bissue{3}),
\bfpage{809}
(\byear{2005}).
\arxivurl{astro-ph/0507173}.
doi:\doiurl{10.1051/0004-6361:20041572}
\end{barticle}
\endbibitem

\bibitem[\protect\citeauthoryear{{Barthelmy}
  et~al.}{2005a}]{Barthelmy2005_050724}
\begin{barticle}
\bauthor{\bsnm{{Barthelmy}}, \binits{S.D.}},
\bauthor{\bsnm{{Chincarini}}, \binits{G.}},
\bauthor{\bsnm{{Burrows}}, \binits{D.N.}},
\bauthor{\bsnm{{Gehrels}}, \binits{N.}},
\bauthor{\bsnm{{Covino}}, \binits{S.}},
\bauthor{\bsnm{{Moretti}}, \binits{A.}},
\bauthor{\bsnm{{Romano}}, \binits{P.}},
\bauthor{\bsnm{{O'Brien}}, \binits{P.T.}},
\bauthor{\bsnm{{Sarazin}}, \binits{C.L.}},
\bauthor{\bsnm{{Kouveliotou}}, \binits{C.}},
\bauthor{\bsnm{{Goad}}, \binits{M.}},
\bauthor{\bsnm{{Vaughan}}, \binits{S.}},
\bauthor{\bsnm{{Tagliaferri}}, \binits{G.}},
\bauthor{\bsnm{{Zhang}}, \binits{B.}},
\bauthor{\bsnm{{Antonelli}}, \binits{L.A.}},
\bauthor{\bsnm{{Campana}}, \binits{S.}},
\bauthor{\bsnm{{Cummings}}, \binits{J.R.}},
\bauthor{\bsnm{{D'Avanzo}}, \binits{P.}},
\bauthor{\bsnm{{Davies}}, \binits{M.B.}},
\bauthor{\bsnm{{Giommi}}, \binits{P.}},
\bauthor{\bsnm{{Grupe}}, \binits{D.}},
\bauthor{\bsnm{{Kaneko}}, \binits{Y.}},
\bauthor{\bsnm{{Kennea}}, \binits{J.A.}},
\bauthor{\bsnm{{King}}, \binits{A.}},
\bauthor{\bsnm{{Kobayashi}}, \binits{S.}},
\bauthor{\bsnm{{Melandri}}, \binits{A.}},
\bauthor{\bsnm{{Meszaros}}, \binits{P.}},
\bauthor{\bsnm{{Nousek}}, \binits{J.A.}},
\bauthor{\bsnm{{Patel}}, \binits{S.}},
\bauthor{\bsnm{{Sakamoto}}, \binits{T.}},
\bauthor{\bsnm{{Wijers}}, \binits{R.A.M.J.}}:
\bjtitle{\nat}
\bvolume{438}(\bissue{7070}),
\bfpage{994}
(\byear{2005}a).
\arxivurl{astro-ph/0511579}.
doi:\doiurl{10.1038/nature04392}
\end{barticle}
\endbibitem

\bibitem[\protect\citeauthoryear{{Barthelmy} et~al.}{2005b}]{Barthelmy2005}
\begin{barticle}
\bauthor{\bsnm{{Barthelmy}}, \binits{S.D.}},
\bauthor{\bsnm{{Barbier}}, \binits{L.M.}},
\bauthor{\bsnm{{Cummings}}, \binits{J.R.}},
\bauthor{\bsnm{{Fenimore}}, \binits{E.E.}},
\bauthor{\bsnm{{Gehrels}}, \binits{N.}},
\bauthor{\bsnm{{Hullinger}}, \binits{D.}},
\bauthor{\bsnm{{Krimm}}, \binits{H.A.}},
\bauthor{\bsnm{{Markwardt}}, \binits{C.B.}},
\bauthor{\bsnm{{Palmer}}, \binits{D.M.}},
\bauthor{\bsnm{{Parsons}}, \binits{A.}},
\bauthor{\bsnm{{Sato}}, \binits{G.}},
\bauthor{\bsnm{{Suzuki}}, \binits{M.}},
\bauthor{\bsnm{{Takahashi}}, \binits{T.}},
\bauthor{\bsnm{{Tashiro}}, \binits{M.}},
\bauthor{\bsnm{{Tueller}}, \binits{J.}}:
\bjtitle{\ssr}
\bvolume{120}(\bissue{3-4}),
\bfpage{143}
(\byear{2005}b).
\arxivurl{astro-ph/0507410}.
doi:\doiurl{10.1007/s11214-005-5096-3}
\end{barticle}
\endbibitem

\bibitem[\protect\citeauthoryear{Beardmore
  et~al.}{2012}]{Beardmore2012_120422A}
\begin{barticle}
\bauthor{\bsnm{Beardmore}, \binits{A.P.}},
\bauthor{\bsnm{Evans}, \binits{P.A.}},
\bauthor{\bsnm{Goad}, \binits{M.R.}},
\bauthor{\bsnm{Osborne}, \binits{J.P.}}:
\bjtitle{GRB Coordinates Network}
\bvolume{13247},
\bfpage{1}
(\byear{2012})
\end{barticle}
\endbibitem

\bibitem[\protect\citeauthoryear{Beardmore
  et~al.}{2016}]{Beardmore2016_161219B}
\begin{barticle}
\bauthor{\bsnm{Beardmore}, \binits{A.P.}},
\bauthor{\bsnm{Evans}, \binits{P.A.}},
\bauthor{\bsnm{Goad}, \binits{M.R.}},
\bauthor{\bsnm{Osborne}, \binits{J.P.}}:
\bjtitle{GRB Coordinates Network}
\bvolume{20297},
\bfpage{1}
(\byear{2016})
\end{barticle}
\endbibitem

\bibitem[\protect\citeauthoryear{Berger and
  Gonzalez}{2005}]{Berger2005_050219A}
\begin{barticle}
\bauthor{\bsnm{Berger}, \binits{E.}},
\bauthor{\bsnm{Gonzalez}, \binits{S.}}:
\bjtitle{GRB Coordinates Network}
\bvolume{3048},
\bfpage{1}
(\byear{2005})
\end{barticle}
\endbibitem

\bibitem[\protect\citeauthoryear{Berger et~al.}{2006}]{Berger2006_061201}
\begin{barticle}
\bauthor{\bsnm{Berger}, \binits{E.}},
\bauthor{\bsnm{Berger}},
\bauthor{\bsnm{E.}}:
\bjtitle{GCN}
\bvolume{5952},
\bfpage{1}
(\byear{2006})
\end{barticle}
\endbibitem

\bibitem[\protect\citeauthoryear{Berger et~al.}{2003}]{Berger2003}
\begin{botherref}
\oauthor{\bsnm{Berger}, \binits{E.}},
\oauthor{\bsnm{Kulkarni}, \binits{S.R.}},
\oauthor{\bsnm{Frail}, \binits{D.A.}},
\oauthor{\bsnm{Soderberg}, \binits{A.M.}}:
A radio survey of type ib and ic supernovae: Searching for engine-driven
  supernovae
\textbf{599},
408
(2003).
doi:\doiurl{10.1086/379214}
\end{botherref}
\endbibitem

\bibitem[\protect\citeauthoryear{Berger et~al.}{2005}]{Berger2005_050724}
\begin{barticle}
\bauthor{\bsnm{Berger}, \binits{E.}},
\bauthor{\bsnm{Price}, \binits{P.A.}},
\bauthor{\bsnm{Cenko}, \binits{S.B.}},
\bauthor{\bsnm{Gal-Yam}, \binits{A.}},
\bauthor{\bsnm{Soderberg}, \binits{A.M.}},
\bauthor{\bsnm{Kasliwal}, \binits{M.}},
\bauthor{\bsnm{Leonard}, \binits{D.C.}},
\bauthor{\bsnm{Cameron}, \binits{P.B.}},
\bauthor{\bsnm{Frail}, \binits{D.A.}},
\bauthor{\bsnm{Kulkarni}, \binits{S.R.}},
\bauthor{\bsnm{Murphy}, \binits{D.C.}},
\bauthor{\bsnm{Krzeminski}, \binits{W.}},
\bauthor{\bsnm{Piran}, \binits{T.}},
\bauthor{\bsnm{Lee}, \binits{B.L.}},
\bauthor{\bsnm{Roth}, \binits{K.C.}},
\bauthor{\bsnm{Moon}, \binits{D.-S.}},
\bauthor{\bsnm{Fox}, \binits{D.B.}},
\bauthor{\bsnm{Harrison}, \binits{F.A.}},
\bauthor{\bsnm{Persson}, \binits{S.E.}},
\bauthor{\bsnm{Schmidt}, \binits{B.P.}},
\bauthor{\bsnm{Penprase}, \binits{B.E.}},
\bauthor{\bsnm{Rich}, \binits{J.}},
\bauthor{\bsnm{Peterson}, \binits{B.A.}},
\bauthor{\bsnm{Cowie}, \binits{L.L.}}:
\bjtitle{\nat}
\bvolume{438},
\bfpage{988}
(\byear{2005}).
doi:\doiurl{10.1038/nature04238}
\end{barticle}
\endbibitem

\bibitem[\protect\citeauthoryear{{Bernardini} et~al.}{2017}]{Bernardini2017}
\begin{barticle}
\bauthor{\bsnm{{Bernardini}}, \binits{M.G.}},
\bauthor{\bsnm{{Xie}}, \binits{F.}},
\bauthor{\bsnm{{Sizun}}, \binits{P.}},
\bauthor{\bsnm{{Piron}}, \binits{F.}},
\bauthor{\bsnm{{Dong}}, \binits{Y.}},
\bauthor{\bsnm{{Atteia}}, \binits{J.-L.}},
\bauthor{\bsnm{{Antier}}, \binits{S.}},
\bauthor{\bsnm{{Daigne}}, \binits{F.}},
\bauthor{\bsnm{{Godet}}, \binits{O.}},
\bauthor{\bsnm{{Cordier}}, \binits{B.}},
\bauthor{\bsnm{{Wei}}, \binits{J.}}:
\bjtitle{Experimental Astronomy}
\bvolume{44}(\bissue{1}),
\bfpage{113}
(\byear{2017}).
doi:\doiurl{10.1007/s10686-017-9551-4}
\end{barticle}
\endbibitem

\bibitem[\protect\citeauthoryear{Bersier et~al.}{2004}]{Bersier2004_031203}
\begin{barticle}
\bauthor{\bsnm{Bersier}, \binits{D.}},
\bauthor{\bsnm{Rhoads}, \binits{J.}},
\bauthor{\bsnm{Fruchter}, \binits{A.}},
\bauthor{\bsnm{Cerón}, \binits{J.M.C.}},
\bauthor{\bsnm{Strolger}, \binits{L.-G.}},
\bauthor{\bsnm{Malhotra}, \binits{S.}},
\bauthor{\bsnm{Gorosabel}, \binits{J.}},
\bauthor{\bsnm{Levan}, \binits{A.}},
\bauthor{\bsnm{Kouveliotou}, \binits{C.}},
\bauthor{\bsnm{Patel}, \binits{S.}},
\bauthor{\bsnm{Merrill}, \binits{M.}},
\bauthor{\bsnm{Gawiser}, \binits{E.}},
\bauthor{\bsnm{Duran}, \binits{M.F.}},
\bauthor{\bsnm{Gonzalez}, \binits{V.}}:
\bjtitle{GRB Coordinates Network}
\bvolume{2544},
\bfpage{1}
(\byear{2004})
\end{barticle}
\endbibitem

\bibitem[\protect\citeauthoryear{Bersier et~al.}{2006}]{Bersier2006_020903}
\begin{barticle}
\bauthor{\bsnm{Bersier}, \binits{D.}},
\bauthor{\bsnm{Fruchter}, \binits{A.S.}},
\bauthor{\bsnm{Strolger}, \binits{L.-G.}},
\bauthor{\bsnm{Gorosabel}, \binits{J.}},
\bauthor{\bsnm{Levan}, \binits{A.}},
\bauthor{\bsnm{Burud}, \binits{I.}},
\bauthor{\bsnm{Rhoads}, \binits{J.E.}},
\bauthor{\bsnm{Becker}, \binits{A.C.}},
\bauthor{\bsnm{Cassan}, \binits{A.}},
\bauthor{\bsnm{Chornock}, \binits{R.}},
\bauthor{\bsnm{Covino}, \binits{S.}},
\bauthor{\bparticle{de} \bsnm{Jong}, \binits{R.S.}},
\bauthor{\bsnm{Dominis}, \binits{D.}},
\bauthor{\bsnm{Filippenko}, \binits{A.V.}},
\bauthor{\bsnm{Hjorth}, \binits{J.}},
\bauthor{\bsnm{Holmberg}, \binits{J.}},
\bauthor{\bsnm{Malesani}, \binits{D.}},
\bauthor{\bsnm{Mobasher}, \binits{B.}},
\bauthor{\bsnm{Olsen}, \binits{K.A.G.}},
\bauthor{\bsnm{Stefanon}, \binits{M.}},
\bauthor{\bsnm{Cerón}, \binits{J.M.C.}},
\bauthor{\bsnm{Fynbo}, \binits{J.P.U.}},
\bauthor{\bsnm{Holland}, \binits{S.T.}},
\bauthor{\bsnm{Kouveliotou}, \binits{C.}},
\bauthor{\bsnm{Pedersen}, \binits{H.}},
\bauthor{\bsnm{Tanvir}, \binits{N.R.}},
\bauthor{\bsnm{Woosley}, \binits{S.E.}}:
\bjtitle{\apj}
\bvolume{643},
\bfpage{284}
(\byear{2006}).
doi:\doiurl{10.1086/502640}
\end{barticle}
\endbibitem

\bibitem[\protect\citeauthoryear{{Bhalerao} et~al.}{2017}]{Bhalerao2017}
\begin{barticle}
\bauthor{\bsnm{{Bhalerao}}, \binits{V.}},
\bauthor{\bsnm{{Bhattacharya}}, \binits{D.}},
\bauthor{\bsnm{{Vibhute}}, \binits{A.}},
\bauthor{\bsnm{{Pawar}}, \binits{P.}},
\bauthor{\bsnm{{Rao}}, \binits{A.R.}},
\bauthor{\bsnm{{Hingar}}, \binits{M.K.}},
\bauthor{\bsnm{{Khanna}}, \binits{R.}},
\bauthor{\bsnm{{Kutty}}, \binits{A.P.K.}},
\bauthor{\bsnm{{Malkar}}, \binits{J.P.}},
\bauthor{\bsnm{{Patil}}, \binits{M.H.}},
\bauthor{\bsnm{{Arora}}, \binits{Y.K.}},
\bauthor{\bsnm{{Sinha}}, \binits{S.}},
\bauthor{\bsnm{{Priya}}, \binits{P.}},
\bauthor{\bsnm{{Samuel}}, \binits{E.}},
\bauthor{\bsnm{{Sreekumar}}, \binits{S.}},
\bauthor{\bsnm{{Vinod}}, \binits{P.}},
\bauthor{\bsnm{{Mithun}}, \binits{N.P.S.}},
\bauthor{\bsnm{{Vadawale}}, \binits{S.V.}},
\bauthor{\bsnm{{Vagshette}}, \binits{N.}},
\bauthor{\bsnm{{Navalgund}}, \binits{K.H.}},
\bauthor{\bsnm{{Sarma}}, \binits{K.S.}},
\bauthor{\bsnm{{Pandiyan}}, \binits{R.}},
\bauthor{\bsnm{{Seetha}}, \binits{S.}},
\bauthor{\bsnm{{Subbarao}}, \binits{K.}}:
\bjtitle{Journal of Astrophysics and Astronomy}
\bvolume{38}(\bissue{2}),
\bfpage{31}
(\byear{2017}).
\arxivurl{1608.03408}.
doi:\doiurl{10.1007/s12036-017-9447-8}
\end{barticle}
\endbibitem

\bibitem[\protect\citeauthoryear{{Bissaldi}
  et~al.}{2020}]{Bissaldi2020_200415A}
\begin{barticle}
\bauthor{\bsnm{{Bissaldi}}, \binits{E.}},
\bauthor{\bsnm{{Briggs}}, \binits{M.}},
\bauthor{\bsnm{{Burns}}, \binits{E.}},
\bauthor{\bsnm{{Roberts}}, \binits{O.J.}},
\bauthor{\bsnm{{Veres}}, \binits{P.}},
\bauthor{\bsnm{{Fermi GBM Team}}}:
\bjtitle{GRB Coordinates Network}
\bvolume{27587},
\bfpage{1}
(\byear{2020})
\end{barticle}
\endbibitem

\bibitem[\protect\citeauthoryear{Blake and Bloom}{2004}]{Blake2004}
\begin{botherref}
\oauthor{\bsnm{Blake}, \binits{C.}},
\oauthor{\bsnm{Bloom}, \binits{J.S.}}:
Optical limits on precursor emission from gamma-ray bursts with known redshift
\textbf{606},
1019
(2004).
doi:\doiurl{10.1086/383134}
\end{botherref}
\endbibitem

\bibitem[\protect\citeauthoryear{Bloom et~al.}{2006a}]{Bloom2006_050509B}
\begin{barticle}
\bauthor{\bsnm{Bloom}, \binits{J.S.}},
\bauthor{\bsnm{Prochaska}, \binits{J.X.}},
\bauthor{\bsnm{Pooley}, \binits{D.}},
\bauthor{\bsnm{Blake}, \binits{C.H.}},
\bauthor{\bsnm{Foley}, \binits{R.J.}},
\bauthor{\bsnm{Jha}, \binits{S.}},
\bauthor{\bsnm{Ramirez‐Ruiz}, \binits{E.}},
\bauthor{\bsnm{Granot}, \binits{J.}},
\bauthor{\bsnm{Filippenko}, \binits{A.V.}},
\bauthor{\bsnm{Sigurdsson}, \binits{S.}},
\bauthor{\bsnm{Barth}, \binits{A.J.}},
\bauthor{\bsnm{Chen}, \binits{H.-W.}},
\bauthor{\bsnm{Cooper}, \binits{M.C.}},
\bauthor{\bsnm{Falco}, \binits{E.E.}},
\bauthor{\bsnm{Gal}, \binits{R.R.}},
\bauthor{\bsnm{Gerke}, \binits{B.F.}},
\bauthor{\bsnm{Gladders}, \binits{M.D.}},
\bauthor{\bsnm{Greene}, \binits{J.E.}},
\bauthor{\bsnm{Hennanwi}, \binits{J.}},
\bauthor{\bsnm{Ho}, \binits{L.C.}},
\bauthor{\bsnm{Hurley}, \binits{K.}},
\bauthor{\bsnm{Koester}, \binits{B.P.}},
\bauthor{\bsnm{Li}, \binits{W.}},
\bauthor{\bsnm{Lubin}, \binits{L.}},
\bauthor{\bsnm{Newman}, \binits{J.}},
\bauthor{\bsnm{Perley}, \binits{D.A.}},
\bauthor{\bsnm{Squires}, \binits{G.K.}},
\bauthor{\bsnm{Wood‐Vasey}, \binits{W.M.}}:
\bjtitle{Astrophys. J.}
\bvolume{638}(\bissue{1}),
\bfpage{354}
(\byear{2006}a).
\arxivurl{0505480}.
doi:\doiurl{10.1086/498107}
\end{barticle}
\endbibitem

\bibitem[\protect\citeauthoryear{Bloom et~al.}{2006b}]{Bloom2006_060502B}
\begin{barticle}
\bauthor{\bsnm{Bloom}, \binits{J.S.}},
\bauthor{\bsnm{Perley}, \binits{D.}},
\bauthor{\bsnm{Kocevski}, \binits{D.}},
\bauthor{\bsnm{Butler}, \binits{N.}},
\bauthor{\bsnm{Prochaska}, \binits{J.X.}},
\bauthor{\bsnm{Chen}, \binits{H.-W.}}:
\bjtitle{GRB Coordinates Network}
\bvolume{5238},
\bfpage{1}
(\byear{2006}b)
\end{barticle}
\endbibitem

\bibitem[\protect\citeauthoryear{Bloom et~al.}{2007}]{Bloom2007_060502B}
\begin{barticle}
\bauthor{\bsnm{Bloom}, \binits{J.S.}},
\bauthor{\bsnm{Perley}, \binits{D.A.}},
\bauthor{\bsnm{Chen}, \binits{H.-W.}},
\bauthor{\bsnm{Butler}, \binits{N.}},
\bauthor{\bsnm{Prochaska}, \binits{J.X.}},
\bauthor{\bsnm{Kocevski}, \binits{D.}},
\bauthor{\bsnm{Blake}, \binits{C.H.}},
\bauthor{\bsnm{Szentgyorgyi}, \binits{A.}},
\bauthor{\bsnm{Falco}, \binits{E.E.}},
\bauthor{\bsnm{Starr}, \binits{D.L.}}:
\bjtitle{\apj}
\bvolume{654},
\bfpage{878}
(\byear{2007}).
doi:\doiurl{10.1086/509114}
\end{barticle}
\endbibitem

\bibitem[\protect\citeauthoryear{{Boella} et~al.}{1997}]{Boella1997_beppo}
\begin{barticle}
\bauthor{\bsnm{{Boella}}, \binits{G.}},
\bauthor{\bsnm{{Butler}}, \binits{R.C.}},
\bauthor{\bsnm{{Perola}}, \binits{G.C.}},
\bauthor{\bsnm{{Piro}}, \binits{L.}},
\bauthor{\bsnm{{Scarsi}}, \binits{L.}},
\bauthor{\bsnm{{Bleeker}}, \binits{J.A.M.}}:
\bjtitle{\aaps}
\bvolume{122},
\bfpage{299}
(\byear{1997}).
doi:\doiurl{10.1051/aas:1997136}
\end{barticle}
\endbibitem

\bibitem[\protect\citeauthoryear{Boggs et~al.}{2007}]{Boggs2007}
\begin{botherref}
\oauthor{\bsnm{Boggs}, \binits{S.E.}},
\oauthor{\bsnm{Zoglauer}, \binits{A.}},
\oauthor{\bsnm{Bellm}, \binits{E.}},
\oauthor{\bsnm{Hurley}, \binits{K.}},
\oauthor{\bsnm{Lin}, \binits{R.P.}},
\oauthor{\bsnm{Smith}, \binits{D.M.}},
\oauthor{\bsnm{Wigger}, \binits{C.}},
\oauthor{\bsnm{Hajdas}, \binits{W.}}:
The giant flare of 2004 december 27 from sgr 1806-20
\textbf{661},
458
(2007).
doi:\doiurl{10.1086/516732}
\end{botherref}
\endbibitem

\bibitem[\protect\citeauthoryear{{Boselli} et~al.}{2016}]{Boselli2016}
\begin{barticle}
\bauthor{\bsnm{{Boselli}}, \binits{A.}},
\bauthor{\bsnm{{Boissier}}, \binits{S.}},
\bauthor{\bsnm{{Voyer}}, \binits{E.}},
\bauthor{\bsnm{{Ferrarese}}, \binits{L.}},
\bauthor{\bsnm{{Consolandi}}, \binits{G.}},
\bauthor{\bsnm{{Cortese}}, \binits{L.}},
\bauthor{\bsnm{{C{\^o}t{\'e}}}, \binits{P.}},
\bauthor{\bsnm{{Cuilland re}}, \binits{J.C.}},
\bauthor{\bsnm{{Gavazzi}}, \binits{G.}},
\bauthor{\bsnm{{Gwyn}}, \binits{S.}},
\bauthor{\bsnm{{Heinis}}, \binits{S.}},
\bauthor{\bsnm{{Ilbert}}, \binits{O.}},
\bauthor{\bsnm{{MacArthur}}, \binits{L.}},
\bauthor{\bsnm{{Roehlly}}, \binits{Y.}}:
\bjtitle{\aap}
\bvolume{585},
\bfpage{2}
(\byear{2016}).
\arxivurl{1509.05171}.
doi:\doiurl{10.1051/0004-6361/201526915}
\end{barticle}
\endbibitem

\bibitem[\protect\citeauthoryear{{Bromberg} et~al.}{2011}]{Bromberg2011}
\begin{barticle}
\bauthor{\bsnm{{Bromberg}}, \binits{O.}},
\bauthor{\bsnm{{Nakar}}, \binits{E.}},
\bauthor{\bsnm{{Piran}}, \binits{T.}}:
\bjtitle{\apjl}
\bvolume{739}(\bissue{2}),
\bfpage{55}
(\byear{2011}).
\arxivurl{1107.1346}.
doi:\doiurl{10.1088/2041-8205/739/2/L55}
\end{barticle}
\endbibitem

\bibitem[\protect\citeauthoryear{Burns et~al.}{2018}]{Burns2018_150101B}
\begin{barticle}
\bauthor{\bsnm{Burns}, \binits{E.}},
\bauthor{\bsnm{Veres}, \binits{P.}},
\bauthor{\bsnm{Connaughton}, \binits{V.}},
\bauthor{\bsnm{Racusin}, \binits{J.}},
\bauthor{\bsnm{Briggs}, \binits{M.S.}},
\bauthor{\bsnm{Christensen}, \binits{N.}},
\bauthor{\bsnm{Goldstein}, \binits{A.}},
\bauthor{\bsnm{Hamburg}, \binits{R.}},
\bauthor{\bsnm{Kocevski}, \binits{D.}},
\bauthor{\bsnm{McEnery}, \binits{J.}},
\bauthor{\bsnm{Bissaldi}, \binits{E.}},
\bauthor{\bsnm{Canton}, \binits{T.D.}},
\bauthor{\bsnm{Cleveland}, \binits{W.H.}},
\bauthor{\bsnm{Gibby}, \binits{M.H.}},
\bauthor{\bsnm{Hui}, \binits{C.M.}},
\bauthor{\bparticle{von} \bsnm{Kienlin}, \binits{A.}},
\bauthor{\bsnm{Mailyan}, \binits{B.}},
\bauthor{\bsnm{Paciesas}, \binits{W.S.}},
\bauthor{\bsnm{Roberts}, \binits{O.J.}},
\bauthor{\bsnm{Siellez}, \binits{K.}},
\bauthor{\bsnm{Stanbro}, \binits{M.}},
\bauthor{\bsnm{Wilson-Hodge}, \binits{C.A.}}:
\bjtitle{\apj}
\bvolume{863},
\bfpage{34}
(\byear{2018}).
doi:\doiurl{10.3847/2041-8213/aad813}
\end{barticle}
\endbibitem

\bibitem[\protect\citeauthoryear{Cameron and Frail}{2005}]{Cameron2005_051103}
\begin{barticle}
\bauthor{\bsnm{Cameron}, \binits{P.B.}},
\bauthor{\bsnm{Frail}, \binits{D.A.}}:
\bjtitle{GRB Coordinates Network}
\bvolume{4266},
\bfpage{1}
(\byear{2005})
\end{barticle}
\endbibitem

\bibitem[\protect\citeauthoryear{Campana et~al.}{2005}]{Campana2005_051109B}
\begin{barticle}
\bauthor{\bsnm{Campana}, \binits{S.}},
\bauthor{\bsnm{Mineo}, \binits{T.}},
\bauthor{\bsnm{Romano}, \binits{P.}},
\bauthor{\bsnm{Tagliaferri}, \binits{G.}},
\bauthor{\bsnm{Burrows}, \binits{D.N.}}:
\bjtitle{GRB Coordinates Network}
\bvolume{4226},
\bfpage{1}
(\byear{2005})
\end{barticle}
\endbibitem

\bibitem[\protect\citeauthoryear{Campana et~al.}{2006}]{Campana2006_060218}
\begin{barticle}
\bauthor{\bsnm{Campana}, \binits{S.}},
\bauthor{\bsnm{Mangano}, \binits{V.}},
\bauthor{\bsnm{Blustin}, \binits{A.J.}},
\bauthor{\bsnm{Brown}, \binits{P.}},
\bauthor{\bsnm{Burrows}, \binits{D.N.}},
\bauthor{\bsnm{Chincarini}, \binits{G.}},
\bauthor{\bsnm{Cummings}, \binits{J.R.}},
\bauthor{\bsnm{Cusumano}, \binits{G.}},
\bauthor{\bsnm{Valle}, \binits{M.D.}},
\bauthor{\bsnm{Malesani}, \binits{D.}},
\bauthor{\bsnm{M\'eszáros}, \binits{P.}},
\bauthor{\bsnm{Nousek}, \binits{J.A.}},
\bauthor{\bsnm{Page}, \binits{M.}},
\bauthor{\bsnm{Sakamoto}, \binits{T.}},
\bauthor{\bsnm{Waxman}, \binits{E.}},
\bauthor{\bsnm{Zhang}, \binits{B.}},
\bauthor{\bsnm{Dai}, \binits{Z.G.}},
\bauthor{\bsnm{Gehrels}, \binits{N.}},
\bauthor{\bsnm{Immler}, \binits{S.}},
\bauthor{\bsnm{Marshall}, \binits{F.E.}},
\bauthor{\bsnm{Mason}, \binits{K.O.}},
\bauthor{\bsnm{Moretti}, \binits{A.}},
\bauthor{\bsnm{O'Brien}, \binits{P.T.}},
\bauthor{\bsnm{Osborne}, \binits{J.P.}},
\bauthor{\bsnm{Page}, \binits{K.L.}},
\bauthor{\bsnm{Romano}, \binits{P.}},
\bauthor{\bsnm{Roming}, \binits{P.W.A.}},
\bauthor{\bsnm{Tagliaferri}, \binits{G.}},
\bauthor{\bsnm{Cominsky}, \binits{L.R.}},
\bauthor{\bsnm{Giommi}, \binits{P.}},
\bauthor{\bsnm{Godet}, \binits{O.}},
\bauthor{\bsnm{Kennea}, \binits{J.A.}},
\bauthor{\bsnm{Krimm}, \binits{H.}},
\bauthor{\bsnm{Angelini}, \binits{L.}},
\bauthor{\bsnm{Barthelmy}, \binits{S.D.}},
\bauthor{\bsnm{Boyd}, \binits{P.T.}},
\bauthor{\bsnm{Palmer}, \binits{D.M.}},
\bauthor{\bsnm{Wells}, \binits{A.A.}},
\bauthor{\bsnm{White}, \binits{N.E.}}:
\bjtitle{\nat}
\bvolume{442},
\bfpage{1008}
(\byear{2006}).
doi:\doiurl{10.1038/nature04892}
\end{barticle}
\endbibitem

\bibitem[\protect\citeauthoryear{{Cano} et~al.}{2011}]{Cano2011_100316D}
\begin{barticle}
\bauthor{\bsnm{{Cano}}, \binits{Z.}},
\bauthor{\bsnm{{Bersier}}, \binits{D.}},
\bauthor{\bsnm{{Guidorzi}}, \binits{C.}},
\bauthor{\bsnm{{Kobayashi}}, \binits{S.}},
\bauthor{\bsnm{{Levan}}, \binits{A.J.}},
\bauthor{\bsnm{{Tanvir}}, \binits{N.R.}},
\bauthor{\bsnm{{Wiersema}}, \binits{K.}},
\bauthor{\bsnm{{D'Avanzo}}, \binits{P.}},
\bauthor{\bsnm{{Fruchter}}, \binits{A.S.}},
\bauthor{\bsnm{{Garnavich}}, \binits{P.}},
\bauthor{\bsnm{{Gomboc}}, \binits{A.}},
\bauthor{\bsnm{{Gorosabel}}, \binits{J.}},
\bauthor{\bsnm{{Kasen}}, \binits{D.}},
\bauthor{\bsnm{{Kopa{\v{c}}}}, \binits{D.}},
\bauthor{\bsnm{{Margutti}}, \binits{R.}},
\bauthor{\bsnm{{Mazzali}}, \binits{P.A.}},
\bauthor{\bsnm{{Melandri}}, \binits{A.}},
\bauthor{\bsnm{{Mundell}}, \binits{C.G.}},
\bauthor{\bsnm{{Nugent}}, \binits{P.E.}},
\bauthor{\bsnm{{Pian}}, \binits{E.}},
\bauthor{\bsnm{{Smith}}, \binits{R.J.}},
\bauthor{\bsnm{{Steele}}, \binits{I.}},
\bauthor{\bsnm{{Wijers}}, \binits{R.A.M.J.}},
\bauthor{\bsnm{{Woosley}}, \binits{S.E.}}:
\bjtitle{\apj}
\bvolume{740}(\bissue{1}),
\bfpage{41}
(\byear{2011}).
\arxivurl{1104.5141}.
doi:\doiurl{10.1088/0004-637X/740/1/41}
\end{barticle}
\endbibitem

\bibitem[\protect\citeauthoryear{Cano et~al.}{2017}]{Cano2017_161219B}
\begin{barticle}
\bauthor{\bsnm{Cano}, \binits{Z.}},
\bauthor{\bsnm{Izzo}, \binits{L.}},
\bauthor{\bsnm{{de Ugarte Postigo}}, \binits{A.}},
\bauthor{\bsnm{Th{\"{o}}ne}, \binits{C.s.}},
\bauthor{\bsnm{Kr{\"{u}}hler}, \binits{T.}},
\bauthor{\bsnm{Heintz}, \binits{K.s.}},
\bauthor{\bsnm{Malesani}, \binits{D.}},
\bauthor{\bsnm{Geier}, \binits{S.}},
\bauthor{\bsnm{Fuentes}, \binits{C.}},
\bauthor{\bsnm{Chen}, \binits{T.-W.}},
\bauthor{\bsnm{Covino}, \binits{S.}},
\bauthor{\bsnm{D'Elia}, \binits{V.}},
\bauthor{\bsnm{Fynbo}, \binits{J.s.s.}},
\bauthor{\bsnm{Goldoni}, \binits{P.}},
\bauthor{\bsnm{Gomboc}, \binits{A.}},
\bauthor{\bsnm{Hjorth}, \binits{J.}},
\bauthor{\bsnm{Jakobsson}, \binits{P.}},
\bauthor{\bsnm{Kann}, \binits{D.s.}},
\bauthor{\bsnm{Milvang-Jensen}, \binits{B.}},
\bauthor{\bsnm{Pugliese}, \binits{G.}},
\bauthor{\bsnm{Sanchez-Ramirez}, \binits{R.}},
\bauthor{\bsnm{Schulze}, \binits{S.}},
\bauthor{\bsnm{Sollerman}, \binits{J.}},
\bauthor{\bsnm{Tanvir}, \binits{N.s.}},
\bauthor{\bsnm{Wiersema}, \binits{K.}}:
\bjtitle{\aap}
\bvolume{605},
\bfpage{107}
(\byear{2017}).
\arxivurl{1704.05401}.
doi:\doiurl{10.1051/0004-6361/201731005}
\end{barticle}
\endbibitem

\bibitem[\protect\citeauthoryear{Caroli et~al.}{1987}]{Caroli1987}
\begin{barticle}
\bauthor{\bsnm{Caroli}, \binits{E.}},
\bauthor{\bsnm{Stephen}, \binits{J.}},
\bauthor{\bsnm{Cocco}, \binits{G.}},
\bauthor{\bsnm{Natalucci}, \binits{L.}},
\bauthor{\bsnm{Spizzichino}, \binits{A.}}:
\bjtitle{Space Science Reviews}
\bvolume{45},
\bfpage{349}
(\byear{1987}).
doi:\doiurl{10.1007/BF00171998}
\end{barticle}
\endbibitem

\bibitem[\protect\citeauthoryear{Chand et~al.}{2020}]{Chand2020_190829A}
\begin{botherref}
\oauthor{\bsnm{Chand}, \binits{V.}},
\oauthor{\bsnm{Banerjee}, \binits{A.}},
\oauthor{\bsnm{Gupta}, \binits{R.}},
\oauthor{\bsnm{Dimple}},
\oauthor{\bsnm{Pal}, \binits{P.S.}},
\oauthor{\bsnm{Joshi}, \binits{J.C.}},
\oauthor{\bsnm{Zhang}, \binits{B.-B.}},
\oauthor{\bsnm{Basak}, \binits{R.}},
\oauthor{\bsnm{Tam}, \binits{P.H.T.}},
\oauthor{\bsnm{Sharma}, \binits{V.}},
\oauthor{\bsnm{ey}, \binits{S.B.P.}},
\oauthor{\bsnm{Kumar}, \binits{A.}},
\oauthor{\bsnm{Yang}, \binits{Y.-S.}}:
arXiv e-prints,
2001
(2020)
\end{botherref}
\endbibitem

\bibitem[\protect\citeauthoryear{Chandra et~al.}{2017}]{Chandra2017_171205A}
\begin{barticle}
\bauthor{\bsnm{Chandra}, \binits{P.}},
\bauthor{\bsnm{Nayana}, \binits{A.J.}},
\bauthor{\bsnm{Bhattacharya}, \binits{D.}},
\bauthor{\bsnm{Cenko}, \binits{S.B.}},
\bauthor{\bsnm{Corsi}, \binits{A.}}:
\bjtitle{GRB Coordinates Network}
\bvolume{22264},
\bfpage{1}
(\byear{2017})
\end{barticle}
\endbibitem

\bibitem[\protect\citeauthoryear{Chevalier et~al.}{2004}]{Chevalier2004}
\begin{botherref}
\oauthor{\bsnm{Chevalier}, \binits{R.A.}},
\oauthor{\bsnm{Li}, \binits{Z.-Y.}},
\oauthor{\bsnm{Fransson}, \binits{C.}}:
The diversity of gamma-ray burst afterglows and the surroundings of massive
  stars
\textbf{606},
369
(2004).
doi:\doiurl{10.1086/382867}
\end{botherref}
\endbibitem

\bibitem[\protect\citeauthoryear{{Churazov} et~al.}{2007}]{Churazov2007}
\begin{barticle}
\bauthor{\bsnm{{Churazov}}, \binits{E.}},
\bauthor{\bsnm{{Sunyaev}}, \binits{R.}},
\bauthor{\bsnm{{Revnivtsev}}, \binits{M.}},
\bauthor{\bsnm{{Sazonov}}, \binits{S.}},
\bauthor{\bsnm{{Molkov}}, \binits{S.}},
\bauthor{\bsnm{{Grebenev}}, \binits{S.}},
\bauthor{\bsnm{{Winkler}}, \binits{C.}},
\bauthor{\bsnm{{Parmar}}, \binits{A.}},
\bauthor{\bsnm{{Bazzano}}, \binits{A.}},
\bauthor{\bsnm{{Falanga}}, \binits{M.}},
\bauthor{\bsnm{{Gros}}, \binits{A.}},
\bauthor{\bsnm{{Lebrun}}, \binits{F.}},
\bauthor{\bsnm{{Natalucci}}, \binits{L.}},
\bauthor{\bsnm{{Ubertini}}, \binits{P.}},
\bauthor{\bsnm{{Roques}}, \binits{J.-P.}},
\bauthor{\bsnm{{Bouchet}}, \binits{L.}},
\bauthor{\bsnm{{Jourdain}}, \binits{E.}},
\bauthor{\bsnm{{Kn{\"o}dlseder}}, \binits{J.}},
\bauthor{\bsnm{{Diehl}}, \binits{R.}},
\bauthor{\bsnm{{Budtz-Jorgensen}}, \binits{C.}},
\bauthor{\bsnm{{Brandt}}, \binits{S.}},
\bauthor{\bsnm{{Lund}}, \binits{N.}},
\bauthor{\bsnm{{Westergaard}}, \binits{N.J.}},
\bauthor{\bsnm{{Neronov}}, \binits{A.}},
\bauthor{\bsnm{{T{\"u}rler}}, \binits{M.}},
\bauthor{\bsnm{{Chernyakova}}, \binits{M.}},
\bauthor{\bsnm{{Walter}}, \binits{R.}},
\bauthor{\bsnm{{Produit}}, \binits{N.}},
\bauthor{\bsnm{{Mowlavi}}, \binits{N.}},
\bauthor{\bsnm{{Mas-Hesse}}, \binits{J.M.}},
\bauthor{\bsnm{{Domingo}}, \binits{A.}},
\bauthor{\bsnm{{Gehrels}}, \binits{N.}},
\bauthor{\bsnm{{Kuulkers}}, \binits{E.}},
\bauthor{\bsnm{{Kretschmar}}, \binits{P.}},
\bauthor{\bsnm{{Schmidt}}, \binits{M.}}:
\bjtitle{\aap}
\bvolume{467}(\bissue{2}),
\bfpage{529}
(\byear{2007}).
\arxivurl{astro-ph/0608250}.
doi:\doiurl{10.1051/0004-6361:20066230}
\end{barticle}
\endbibitem

\bibitem[\protect\citeauthoryear{{Churazov} et~al.}{2008}]{Churazov2008}
\begin{barticle}
\bauthor{\bsnm{{Churazov}}, \binits{E.}},
\bauthor{\bsnm{{Sazonov}}, \binits{S.}},
\bauthor{\bsnm{{Sunyaev}}, \binits{R.}},
\bauthor{\bsnm{{Revnivtsev}}, \binits{M.}}:
\bjtitle{\mnras}
\bvolume{385}(\bissue{2}),
\bfpage{719}
(\byear{2008}).
doi:\doiurl{10.1111/j.1365-2966.2008.12918.x}
\end{barticle}
\endbibitem

\bibitem[\protect\citeauthoryear{{Ciolfi}}{2018}]{Ciolfi2018}
\begin{barticle}
\bauthor{\bsnm{{Ciolfi}}, \binits{R.}}:
\bjtitle{International Journal of Modern Physics D}
\bvolume{27}(\bissue{13}),
\bfpage{1842004}
(\byear{2018}).
\arxivurl{1804.03684}.
doi:\doiurl{10.1142/S021827181842004X}
\end{barticle}
\endbibitem

\bibitem[\protect\citeauthoryear{{Cline} et~al.}{1982}]{Cline1982_790305}
\begin{barticle}
\bauthor{\bsnm{{Cline}}, \binits{T.L.}},
\bauthor{\bsnm{{Desai}}, \binits{U.D.}},
\bauthor{\bsnm{{Teegarden}}, \binits{B.J.}},
\bauthor{\bsnm{{Evans}}, \binits{W.D.}},
\bauthor{\bsnm{{Klebesadel}}, \binits{R.W.}},
\bauthor{\bsnm{{Laros}}, \binits{J.G.}},
\bauthor{\bsnm{{Barat}}, \binits{C.}},
\bauthor{\bsnm{{Hurley}}, \binits{K.}},
\bauthor{\bsnm{{Niel}}, \binits{M.}},
\bauthor{\bsnm{{Bedrenne}}, \binits{G.}},
\bauthor{\bsnm{{Estulin}}, \binits{I.V.}},
\bauthor{\bsnm{{Kurt}}, \binits{V.G.}},
\bauthor{\bsnm{{Mersov}}, \binits{G.A.}},
\bauthor{\bsnm{{Zenchenko}}, \binits{V.M.}},
\bauthor{\bsnm{{Weisskopf}}, \binits{M.C.}},
\bauthor{\bsnm{{Grindlay}}, \binits{J.}}:
\bjtitle{\apjl}
\bvolume{255},
\bfpage{45}
(\byear{1982}).
doi:\doiurl{10.1086/183766}
\end{barticle}
\endbibitem

\bibitem[\protect\citeauthoryear{Conciatore
  et~al.}{2006}]{Conciatore2006_060505}
\begin{barticle}
\bauthor{\bsnm{Conciatore}, \binits{M.L.}},
\bauthor{\bsnm{Capalbi}, \binits{M.}},
\bauthor{\bsnm{Vetere}, \binits{L.}},
\bauthor{\bsnm{Palmer}, \binits{D.}},
\bauthor{\bsnm{Burrows}, \binits{D.}}:
\bjtitle{GRB Coordinates Network}
\bvolume{5078},
\bfpage{1}
(\byear{2006})
\end{barticle}
\endbibitem

\bibitem[\protect\citeauthoryear{Corsi et~al.}{2011}]{Corsi2011}
\begin{botherref}
\oauthor{\bsnm{Corsi}, \binits{A.}},
\oauthor{\bsnm{Ofek}, \binits{E.O.}},
\oauthor{\bsnm{Frail}, \binits{D.A.}},
\oauthor{\bsnm{Poznanski}, \binits{D.}},
\oauthor{\bsnm{Arcavi}, \binits{I.}},
\oauthor{\bsnm{Gal-Yam}, \binits{A.}},
\oauthor{\bsnm{Kulkarni}, \binits{S.R.}},
\oauthor{\bsnm{Hurley}, \binits{K.}},
\oauthor{\bsnm{Mazzali}, \binits{P.A.}},
\oauthor{\bsnm{Howell}, \binits{D.A.}},
\oauthor{\bsnm{Kasliwal}, \binits{M.M.}},
\oauthor{\bsnm{Green}, \binits{Y.}},
\oauthor{\bsnm{Murray}, \binits{D.}},
\oauthor{\bsnm{Sullivan}, \binits{M.}},
\oauthor{\bsnm{Xu}, \binits{D.}},
\oauthor{\bsnm{Ben-ami}, \binits{S.}},
\oauthor{\bsnm{Bloom}, \binits{J.S.}},
\oauthor{\bsnm{Cenko}, \binits{S.B.}},
\oauthor{\bsnm{Law}, \binits{N.M.}},
\oauthor{\bsnm{Nugent}, \binits{P.}},
\oauthor{\bsnm{Quimby}, \binits{R.M.}},
\oauthor{\bsnm{Pal'shin}, \binits{V.}},
\oauthor{\bsnm{Cummings}, \binits{J.}},
\oauthor{\bsnm{Connaughton}, \binits{V.}},
\oauthor{\bsnm{Yamaoka}, \binits{K.}},
\oauthor{\bsnm{Rau}, \binits{A.}},
\oauthor{\bsnm{Boynton}, \binits{W.}},
\oauthor{\bsnm{Mitrofanov}, \binits{I.}},
\oauthor{\bsnm{Goldsten}, \binits{J.}}:
Ptf 10bzf (sn 2010ah): A broad-line ic supernova discovered by the palomar
  transient factory
\textbf{741},
76
(2011).
doi:\doiurl{10.1088/0004-637X/741/2/76}
\end{botherref}
\endbibitem

\bibitem[\protect\citeauthoryear{Corsi et~al.}{2016}]{Corsi2016}
\begin{botherref}
\oauthor{\bsnm{Corsi}, \binits{A.}},
\oauthor{\bsnm{Gal-Yam}, \binits{A.}},
\oauthor{\bsnm{Kulkarni}, \binits{S.R.}},
\oauthor{\bsnm{Frail}, \binits{D.A.}},
\oauthor{\bsnm{Mazzali}, \binits{P.A.}},
\oauthor{\bsnm{Cenko}, \binits{S.B.}},
\oauthor{\bsnm{Kasliwal}, \binits{M.M.}},
\oauthor{\bsnm{Cao}, \binits{Y.}},
\oauthor{\bsnm{Horesh}, \binits{A.}},
\oauthor{\bsnm{Palliyaguru}, \binits{N.}},
\oauthor{\bsnm{Perley}, \binits{D.A.}},
\oauthor{\bsnm{Laher}, \binits{R.R.}},
\oauthor{\bsnm{Taddia}, \binits{F.}},
\oauthor{\bsnm{Leloudas}, \binits{G.}},
\oauthor{\bsnm{Maguire}, \binits{K.}},
\oauthor{\bsnm{Nugent}, \binits{P.E.}},
\oauthor{\bsnm{Sollerman}, \binits{J.}},
\oauthor{\bsnm{Sullivan}, \binits{M.}}:
Radio observations of a sample of broad-line type ic supernovae discovered by
  ptf/iptf: A search for relativistic explosions
\textbf{830},
42
(2016).
doi:\doiurl{10.3847/0004-637X/830/1/42}
\end{botherref}
\endbibitem

\bibitem[\protect\citeauthoryear{{Cowperthwaite}
  et~al.}{2017}]{Coperthwaite2017}
\begin{barticle}
\bauthor{\bsnm{{Cowperthwaite}}, \binits{P.S.}},
\bauthor{\bsnm{{Berger}}, \binits{E.}},
\bauthor{\bsnm{{Villar}}, \binits{V.A.}},
\bauthor{\bsnm{{Metzger}}, \binits{B.D.}},
\bauthor{\bsnm{{Nicholl}}, \binits{M.}},
\bauthor{\bsnm{{Chornock}}, \binits{R.}},
\bauthor{\bsnm{{Blanchard}}, \binits{P.K.}},
\bauthor{\bsnm{{Fong}}, \binits{W.}},
\bauthor{\bsnm{{Margutti}}, \binits{R.}},
\bauthor{\bsnm{{Soares-Santos}}, \binits{M.}},
\bauthor{\bsnm{{Alexander}}, \binits{K.D.}},
\bauthor{\bsnm{{Allam}}, \binits{S.}},
\bauthor{\bsnm{{Annis}}, \binits{J.}},
\bauthor{\bsnm{{Brout}}, \binits{D.}},
\bauthor{\bsnm{{Brown}}, \binits{D.A.}},
\bauthor{\bsnm{{Butler}}, \binits{R.E.}},
\bauthor{\bsnm{{Chen}}, \binits{H.-Y.}},
\bauthor{\bsnm{{Diehl}}, \binits{H.T.}},
\bauthor{\bsnm{{Doctor}}, \binits{Z.}},
\bauthor{\bsnm{{Drout}}, \binits{M.R.}},
\bauthor{\bsnm{{Eftekhari}}, \binits{T.}},
\bauthor{\bsnm{{Farr}}, \binits{B.}},
\bauthor{\bsnm{{Finley}}, \binits{D.A.}},
\bauthor{\bsnm{{Foley}}, \binits{R.J.}},
\bauthor{\bsnm{{Frieman}}, \binits{J.A.}},
\bauthor{\bsnm{{Fryer}}, \binits{C.L.}},
\bauthor{\bsnm{{Garc{\'\i}a-Bellido}}, \binits{J.}},
\bauthor{\bsnm{{Gill}}, \binits{M.S.S.}},
\bauthor{\bsnm{{Guillochon}}, \binits{J.}},
\bauthor{\bsnm{{Herner}}, \binits{K.}},
\bauthor{\bsnm{{Holz}}, \binits{D.E.}},
\bauthor{\bsnm{{Kasen}}, \binits{D.}},
\bauthor{\bsnm{{Kessler}}, \binits{R.}},
\bauthor{\bsnm{{Marriner}}, \binits{J.}},
\bauthor{\bsnm{{Matheson}}, \binits{T.}},
\bauthor{\bsnm{{Neilsen}}, \binits{J.} \bsuffix{E.~H.}},
\bauthor{\bsnm{{Quataert}}, \binits{E.}},
\bauthor{\bsnm{{Palmese}}, \binits{A.}},
\bauthor{\bsnm{{Rest}}, \binits{A.}},
\bauthor{\bsnm{{Sako}}, \binits{M.}},
\bauthor{\bsnm{{Scolnic}}, \binits{D.M.}},
\bauthor{\bsnm{{Smith}}, \binits{N.}},
\bauthor{\bsnm{{Tucker}}, \binits{D.L.}},
\bauthor{\bsnm{{Williams}}, \binits{P.K.G.}},
\bauthor{\bsnm{{Balbinot}}, \binits{E.}},
\bauthor{\bsnm{{Carlin}}, \binits{J.L.}},
\bauthor{\bsnm{{Cook}}, \binits{E.R.}},
\bauthor{\bsnm{{Durret}}, \binits{F.}},
\bauthor{\bsnm{{Li}}, \binits{T.S.}},
\bauthor{\bsnm{{Lopes}}, \binits{P.A.A.}},
\bauthor{\bsnm{{Louren{\c{c}}o}}, \binits{A.C.C.}},
\bauthor{\bsnm{{Marshall}}, \binits{J.L.}},
\bauthor{\bsnm{{Medina}}, \binits{G.E.}},
\bauthor{\bsnm{{Muir}}, \binits{J.}},
\bauthor{\bsnm{{Mu{\~n}oz}}, \binits{R.R.}},
\bauthor{\bsnm{{Sauseda}}, \binits{M.}},
\bauthor{\bsnm{{Schlegel}}, \binits{D.J.}},
\bauthor{\bsnm{{Secco}}, \binits{L.F.}},
\bauthor{\bsnm{{Vivas}}, \binits{A.K.}},
\bauthor{\bsnm{{Wester}}, \binits{W.}},
\bauthor{\bsnm{{Zenteno}}, \binits{A.}},
\bauthor{\bsnm{{Zhang}}, \binits{Y.}},
\bauthor{\bsnm{{Abbott}}, \binits{T.M.C.}},
\bauthor{\bsnm{{Banerji}}, \binits{M.}},
\bauthor{\bsnm{{Bechtol}}, \binits{K.}},
\bauthor{\bsnm{{Benoit-L{\'e}vy}}, \binits{A.}},
\bauthor{\bsnm{{Bertin}}, \binits{E.}},
\bauthor{\bsnm{{Buckley-Geer}}, \binits{E.}},
\bauthor{\bsnm{{Burke}}, \binits{D.L.}},
\bauthor{\bsnm{{Capozzi}}, \binits{D.}},
\bauthor{\bsnm{{Carnero Rosell}}, \binits{A.}},
\bauthor{\bsnm{{Carrasco Kind}}, \binits{M.}},
\bauthor{\bsnm{{Castander}}, \binits{F.J.}},
\bauthor{\bsnm{{Crocce}}, \binits{M.}},
\bauthor{\bsnm{{Cunha}}, \binits{C.E.}},
\bauthor{\bsnm{{D'Andrea}}, \binits{C.B.}},
\bauthor{\bsnm{{da Costa}}, \binits{L.N.}},
\bauthor{\bsnm{{Davis}}, \binits{C.}},
\bauthor{\bsnm{{DePoy}}, \binits{D.L.}},
\bauthor{\bsnm{{Desai}}, \binits{S.}},
\bauthor{\bsnm{{Dietrich}}, \binits{J.P.}},
\bauthor{\bsnm{{Drlica-Wagner}}, \binits{A.}},
\bauthor{\bsnm{{Eifler}}, \binits{T.F.}},
\bauthor{\bsnm{{Evrard}}, \binits{A.E.}},
\bauthor{\bsnm{{Fernand ez}}, \binits{E.}},
\bauthor{\bsnm{{Flaugher}}, \binits{B.}},
\bauthor{\bsnm{{Fosalba}}, \binits{P.}},
\bauthor{\bsnm{{Gaztanaga}}, \binits{E.}},
\bauthor{\bsnm{{Gerdes}}, \binits{D.W.}},
\bauthor{\bsnm{{Giannantonio}}, \binits{T.}},
\bauthor{\bsnm{{Goldstein}}, \binits{D.A.}},
\bauthor{\bsnm{{Gruen}}, \binits{D.}},
\bauthor{\bsnm{{Gruendl}}, \binits{R.A.}},
\bauthor{\bsnm{{Gutierrez}}, \binits{G.}},
\bauthor{\bsnm{{Honscheid}}, \binits{K.}},
\bauthor{\bsnm{{Jain}}, \binits{B.}},
\bauthor{\bsnm{{James}}, \binits{D.J.}},
\bauthor{\bsnm{{Jeltema}}, \binits{T.}},
\bauthor{\bsnm{{Johnson}}, \binits{M.W.G.}},
\bauthor{\bsnm{{Johnson}}, \binits{M.D.}},
\bauthor{\bsnm{{Kent}}, \binits{S.}},
\bauthor{\bsnm{{Krause}}, \binits{E.}},
\bauthor{\bsnm{{Kron}}, \binits{R.}},
\bauthor{\bsnm{{Kuehn}}, \binits{K.}},
\bauthor{\bsnm{{Nuropatkin}}, \binits{N.}},
\bauthor{\bsnm{{Lahav}}, \binits{O.}},
\bauthor{\bsnm{{Lima}}, \binits{M.}},
\bauthor{\bsnm{{Lin}}, \binits{H.}},
\bauthor{\bsnm{{Maia}}, \binits{M.A.G.}},
\bauthor{\bsnm{{March}}, \binits{M.}},
\bauthor{\bsnm{{Martini}}, \binits{P.}},
\bauthor{\bsnm{{McMahon}}, \binits{R.G.}},
\bauthor{\bsnm{{Menanteau}}, \binits{F.}},
\bauthor{\bsnm{{Miller}}, \binits{C.J.}},
\bauthor{\bsnm{{Miquel}}, \binits{R.}},
\bauthor{\bsnm{{Mohr}}, \binits{J.J.}},
\bauthor{\bsnm{{Neilsen}}, \binits{E.}},
\bauthor{\bsnm{{Nichol}}, \binits{R.C.}},
\bauthor{\bsnm{{Ogando}}, \binits{R.L.C.}},
\bauthor{\bsnm{{Plazas}}, \binits{A.A.}},
\bauthor{\bsnm{{Roe}}, \binits{N.}},
\bauthor{\bsnm{{Romer}}, \binits{A.K.}},
\bauthor{\bsnm{{Roodman}}, \binits{A.}},
\bauthor{\bsnm{{Rykoff}}, \binits{E.S.}},
\bauthor{\bsnm{{Sanchez}}, \binits{E.}},
\bauthor{\bsnm{{Scarpine}}, \binits{V.}},
\bauthor{\bsnm{{Schindler}}, \binits{R.}},
\bauthor{\bsnm{{Schubnell}}, \binits{M.}},
\bauthor{\bsnm{{Sevilla-Noarbe}}, \binits{I.}},
\bauthor{\bsnm{{Smith}}, \binits{M.}},
\bauthor{\bsnm{{Smith}}, \binits{R.C.}},
\bauthor{\bsnm{{Sobreira}}, \binits{F.}},
\bauthor{\bsnm{{Suchyta}}, \binits{E.}},
\bauthor{\bsnm{{Swanson}}, \binits{M.E.C.}},
\bauthor{\bsnm{{Tarle}}, \binits{G.}},
\bauthor{\bsnm{{Thomas}}, \binits{D.}},
\bauthor{\bsnm{{Thomas}}, \binits{R.C.}},
\bauthor{\bsnm{{Troxel}}, \binits{M.A.}},
\bauthor{\bsnm{{Vikram}}, \binits{V.}},
\bauthor{\bsnm{{Walker}}, \binits{A.R.}},
\bauthor{\bsnm{{Wechsler}}, \binits{R.H.}},
\bauthor{\bsnm{{Weller}}, \binits{J.}},
\bauthor{\bsnm{{Yanny}}, \binits{B.}},
\bauthor{\bsnm{{Zuntz}}, \binits{J.}}:
\bjtitle{\apjl}
\bvolume{848}(\bissue{2}),
\bfpage{17}
(\byear{2017}).
\arxivurl{1710.05840}.
doi:\doiurl{10.3847/2041-8213/aa8fc7}
\end{barticle}
\endbibitem

\bibitem[\protect\citeauthoryear{Crider}{2006}]{Crider2006_970110}
\begin{botherref}
\oauthor{\bsnm{Crider}, \binits{A.}}:
In: Holt, S.S., Gehrels, N., Nousek, J.A. (eds.)
A Magnetar Flare in the BATSE Catalog?,
vol. 836,
p. 64
(2006).
doi:\doiurl{10.1063/1.2207859}
\end{botherref}
\endbibitem

\bibitem[\protect\citeauthoryear{Cusumano et~al.}{2006}]{Cusumano2006_060218}
\begin{barticle}
\bauthor{\bsnm{Cusumano}, \binits{G.}},
\bauthor{\bsnm{Moretti}, \binits{A.}},
\bauthor{\bsnm{Tagliaferri}, \binits{G.}},
\bauthor{\bsnm{Kennea}, \binits{J.}},
\bauthor{\bsnm{Burrows}, \binits{D.}}:
\bjtitle{GRB Coordinates Network}
\bvolume{4781},
\bfpage{1}
(\byear{2006})
\end{barticle}
\endbibitem

\bibitem[\protect\citeauthoryear{{Dagoneau}}{2020}]{Dagoneau2020_xsources}
\begin{botherref}
\oauthor{\bsnm{{Dagoneau}}, \binits{N.}}:
\aap
(2020)
\end{botherref}
\endbibitem

\bibitem[\protect\citeauthoryear{{Dagoneau} et~al.}{2020}]{Dagoneau2020}
\begin{botherref}
\oauthor{\bsnm{{Dagoneau}}, \binits{N.}},
\oauthor{\bsnm{{Schanne}}, \binits{S.}},
\oauthor{\bsnm{{Atteia}}, \binits{J.-L.}},
\oauthor{\bsnm{{G{\"o}tz}}, \binits{D.}},
\oauthor{\bsnm{{Cordier}}, \binits{B.}}:
arXiv e-prints,
2005
(2020).
\arxivurl{2005.12560}
\end{botherref}
\endbibitem

\bibitem[\protect\citeauthoryear{D'Ai et~al.}{2016}]{DAi2016_161219B}
\begin{barticle}
\bauthor{\bsnm{D'Ai}, \binits{A.}},
\bauthor{\bsnm{Kennea}, \binits{J.A.}},
\bauthor{\bsnm{Krimm}, \binits{H.A.}},
\bauthor{\bsnm{Kuin}, \binits{N.P.M.}},
\bauthor{\bsnm{Marshall}, \binits{F.E.}},
\bauthor{\bsnm{Page}, \binits{K.L.}},
\bauthor{\bsnm{Palmer}, \binits{D.M.}},
\bauthor{\bsnm{Sbarufatti}, \binits{B.}},
\bauthor{\bsnm{Siegel}, \binits{M.H.}}:
\bjtitle{GRB Coordinates Network}
\bvolume{20296},
\bfpage{1}
(\byear{2016})
\end{barticle}
\endbibitem

\bibitem[\protect\citeauthoryear{D'Avanzo et~al.}{2013}]{Davanzo2013_130702A}
\begin{barticle}
\bauthor{\bsnm{D'Avanzo}, \binits{P.}},
\bauthor{\bsnm{Porterfield}, \binits{B.}},
\bauthor{\bsnm{Burrows}, \binits{D.N.}},
\bauthor{\bsnm{Siegel}, \binits{M.}},
\bauthor{\bsnm{Melandri}, \binits{A.}},
\bauthor{\bsnm{Evans}, \binits{P.A.}}:
\bjtitle{GRB Coordinates Network}
\bvolume{14973},
\bfpage{1}
(\byear{2013})
\end{barticle}
\endbibitem

\bibitem[\protect\citeauthoryear{{D'Avanzo} et~al.}{2014}]{DAvanzo2014}
\begin{barticle}
\bauthor{\bsnm{{D'Avanzo}}, \binits{P.}},
\bauthor{\bsnm{{Salvaterra}}, \binits{R.}},
\bauthor{\bsnm{{Bernardini}}, \binits{M.G.}},
\bauthor{\bsnm{{Nava}}, \binits{L.}},
\bauthor{\bsnm{{Campana}}, \binits{S.}},
\bauthor{\bsnm{{Covino}}, \binits{S.}},
\bauthor{\bsnm{{D'Elia}}, \binits{V.}},
\bauthor{\bsnm{{Ghirlanda}}, \binits{G.}},
\bauthor{\bsnm{{Ghisellini}}, \binits{G.}},
\bauthor{\bsnm{{Melandri}}, \binits{A.}},
\bauthor{\bsnm{{Sbarufatti}}, \binits{B.}},
\bauthor{\bsnm{{Vergani}}, \binits{S.D.}},
\bauthor{\bsnm{{Tagliaferri}}, \binits{G.}}:
\bjtitle{\mnras}
\bvolume{442}(\bissue{3}),
\bfpage{2342}
(\byear{2014}).
\arxivurl{1405.5131}.
doi:\doiurl{10.1093/mnras/stu994}
\end{barticle}
\endbibitem

\bibitem[\protect\citeauthoryear{de~Pasquale
  et~al.}{2005}]{DePasquale2005_051109B}
\begin{barticle}
\bauthor{\bparticle{de} \bsnm{Pasquale}, \binits{M.}},
\bauthor{\bsnm{Tagliaferri}, \binits{G.}},
\bauthor{\bsnm{Blustin}, \binits{A.J.}},
\bauthor{\bsnm{Hinshaw}, \binits{D.}},
\bauthor{\bsnm{Carter}, \binits{M.}},
\bauthor{\bsnm{Gehrels}, \binits{N.}}:
\bjtitle{GRB Coordinates Network}
\bvolume{4233},
\bfpage{1}
(\byear{2005})
\end{barticle}
\endbibitem

\bibitem[\protect\citeauthoryear{de~Ugarte~Postigo
  et~al.}{2005}]{UgartePostigo2005_050219A}
\begin{barticle}
\bauthor{\bparticle{de} \bsnm{Ugarte~Postigo}, \binits{A.}},
\bauthor{\bsnm{Eguchi}, \binits{S.}},
\bauthor{\bsnm{Gorosabel}, \binits{J.}},
\bauthor{\bsnm{Yock}, \binits{P.}},
\bauthor{\bsnm{Castro-Tirado}, \binits{A.J.}}:
\bjtitle{GRB Coordinates Network}
\bvolume{3041},
\bfpage{1}
(\byear{2005})
\end{barticle}
\endbibitem

\bibitem[\protect\citeauthoryear{de~Ugarte~Postigo
  et~al.}{2008}]{UgartePostigo2008_080905A}
\begin{barticle}
\bauthor{\bparticle{de} \bsnm{Ugarte~Postigo}, \binits{A.}},
\bauthor{\bsnm{Malesani}, \binits{D.}},
\bauthor{\bsnm{Levan}, \binits{A.J.}},
\bauthor{\bsnm{Hjorth}, \binits{J.}},
\bauthor{\bsnm{Tanvir}, \binits{N.R.}}:
\bjtitle{GRB Coordinates Network}
\bvolume{8195},
\bfpage{1}
(\byear{2008})
\end{barticle}
\endbibitem

\bibitem[\protect\citeauthoryear{de~Ugarte~Postigo
  et~al.}{2016}]{UgartePostigo2016_161219B}
\begin{barticle}
\bauthor{\bparticle{de} \bsnm{Ugarte~Postigo}, \binits{A.}},
\bauthor{\bsnm{Cano}, \binits{Z.}},
\bauthor{\bsnm{Izzo}, \binits{L.}},
\bauthor{\bsnm{Thoene}, \binits{C.}},
\bauthor{\bsnm{Sanchez-Ramirez}, \binits{R.}},
\bauthor{\bsnm{Bensch}, \binits{K.}},
\bauthor{\bsnm{Kann}, \binits{D.A.}},
\bauthor{\bsnm{Tanvir}, \binits{N.}},
\bauthor{\bsnm{Schulze}, \binits{S.}},
\bauthor{\bsnm{Leloudas}, \binits{G.}},
\bauthor{\bsnm{Geier}, \binits{S.}},
\bauthor{\bsnm{Tejero}, \binits{A.}}:
\bjtitle{GRB Coordinates Network}
\bvolume{20342},
\bfpage{1}
(\byear{2016})
\end{barticle}
\endbibitem

\bibitem[\protect\citeauthoryear{de~Ugarte~Postigo
  et~al.}{2017}]{UgartePostigo2017_171205A}
\begin{barticle}
\bauthor{\bparticle{de} \bsnm{Ugarte~Postigo}, \binits{A.}},
\bauthor{\bsnm{Izzo}, \binits{L.}},
\bauthor{\bsnm{Kann}, \binits{D.A.}},
\bauthor{\bsnm{Thoene}, \binits{C.C.}},
\bauthor{\bsnm{Pesev}, \binits{P.}},
\bauthor{\bsnm{Scarpa}, \binits{R.}},
\bauthor{\bsnm{Perez}, \binits{D.}}:
\bjtitle{The Astronomer's Telegram}
\bvolume{11038},
\bfpage{1}
(\byear{2017})
\end{barticle}
\endbibitem

\bibitem[\protect\citeauthoryear{{D'Elia} et~al.}{2015}]{Delia2015_150818A}
\begin{barticle}
\bauthor{\bsnm{{D'Elia}}, \binits{V.}},
\bauthor{\bsnm{{Breeveld}}, \binits{A.A.}},
\bauthor{\bsnm{{Evans}}, \binits{P.A.}},
\bauthor{\bsnm{{Gehrels}}, \binits{N.}},
\bauthor{\bsnm{{Izzo}}, \binits{L.}},
\bauthor{\bsnm{{Marshall}}, \binits{F.E.}},
\bauthor{\bsnm{{Page}}, \binits{K.L.}},
\bauthor{\bsnm{{Sbarufatti}}, \binits{B.}}:
\bjtitle{GRB Coordinates Network}
\bvolume{18152},
\bfpage{1}
(\byear{2015})
\end{barticle}
\endbibitem

\bibitem[\protect\citeauthoryear{D'Elia et~al.}{2018}]{Delia2018_171205A}
\begin{barticle}
\bauthor{\bsnm{D'Elia}, \binits{V.}},
\bauthor{\bsnm{Campana}, \binits{S.}},
\bauthor{\bsnm{D'A$\backslash$`$\backslash$i}, \binits{A.}},
\bauthor{\bsnm{{De Pasquale}}, \binits{M.}},
\bauthor{\bsnm{Emery}, \binits{S.{\~{}}.{\~{}}.}},
\bauthor{\bsnm{Frederiks}, \binits{D.{\~{}}.}},
\bauthor{\bsnm{Lien}, \binits{A.}},
\bauthor{\bsnm{Melandri}, \binits{A.}},
\bauthor{\bsnm{Page}, \binits{K.{\~{}}.}},
\bauthor{\bsnm{Starling}, \binits{R.{\~{}}.{\~{}}.}},
\bauthor{\bsnm{Burrows}, \binits{D.{\~{}}.}},
\bauthor{\bsnm{Breeveld}, \binits{A.{\~{}}.}},
\bauthor{\bsnm{Oates}, \binits{S.{\~{}}.}},
\bauthor{\bsnm{O'Brien}, \binits{P.{\~{}}.}},
\bauthor{\bsnm{Osborne}, \binits{J.{\~{}}.}},
\bauthor{\bsnm{Siegel}, \binits{M.{\~{}}.}},
\bauthor{\bsnm{Tagliaferri}, \binits{G.}},
\bauthor{\bsnm{Brown}, \binits{P.{\~{}}.}},
\bauthor{\bsnm{Cenko}, \binits{S.{\~{}}.}},
\bauthor{\bsnm{Svinkin}, \binits{D.{\~{}}.}},
\bauthor{\bsnm{Tohuvavohu}, \binits{A.}},
\bauthor{\bsnm{Tsvetkova}, \binits{A.{\~{}}.}}:
\bjtitle{$\backslash$aap}
\bvolume{619},
\bfpage{66}
(\byear{2018}).
\arxivurl{1810.03339}.
doi:\doiurl{10.1051/0004-6361/201833847}
\end{barticle}
\endbibitem

\bibitem[\protect\citeauthoryear{Dichiara et~al.}{2019}]{Dichiara2019_190829A}
\begin{barticle}
\bauthor{\bsnm{Dichiara}, \binits{S.}},
\bauthor{\bsnm{Bernardini}, \binits{M.G.}},
\bauthor{\bsnm{Burrows}, \binits{D.N.}},
\bauthor{\bsnm{D'Avanzo}, \binits{P.}},
\bauthor{\bsnm{Gronwall}, \binits{C.}},
\bauthor{\bsnm{Gropp}, \binits{J.D.}},
\bauthor{\bsnm{Kennea}, \binits{J.A.}},
\bauthor{\bsnm{Klingler}, \binits{N.J.}},
\bauthor{\bsnm{Krimm}, \binits{H.A.}},
\bauthor{\bsnm{Kuin}, \binits{N.P.M.}},
\bauthor{\bsnm{LaPorte}, \binits{S.J.}},
\bauthor{\bsnm{Melandri}, \binits{A.}},
\bauthor{\bsnm{Page}, \binits{K.L.}},
\bauthor{\bsnm{Palmer}, \binits{D.M.}},
\bauthor{\bsnm{Siegel}, \binits{M.H.}},
\bauthor{\bsnm{Simpson}, \binits{K.K.}},
\bauthor{\bsnm{Tohuvavohu}, \binits{A.}},
\bauthor{\bsnm{Team}, \binits{N.G.S.O.}}:
\bjtitle{GRB Coordinates Network}
\bvolume{25552},
\bfpage{1}
(\byear{2019})
\end{barticle}
\endbibitem

\bibitem[\protect\citeauthoryear{{Dichiara} et~al.}{2020}]{Dichiara2020}
\begin{barticle}
\bauthor{\bsnm{{Dichiara}}, \binits{S.}},
\bauthor{\bsnm{{Troja}}, \binits{E.}},
\bauthor{\bsnm{{O'Connor}}, \binits{B.}},
\bauthor{\bsnm{{Marshall}}, \binits{F.E.}},
\bauthor{\bsnm{{Beniamini}}, \binits{P.}},
\bauthor{\bsnm{{Cannizzo}}, \binits{J.K.}},
\bauthor{\bsnm{{Lien}}, \binits{A.Y.}},
\bauthor{\bsnm{{Sakamoto}}, \binits{T.}}:
\bjtitle{\mnras}
\bvolume{492}(\bissue{4}),
\bfpage{5011}
(\byear{2020}).
\arxivurl{1912.08698}.
doi:\doiurl{10.1093/mnras/staa124}
\end{barticle}
\endbibitem

\bibitem[\protect\citeauthoryear{Duran and Giannios}{2015}]{Barniol2015}
\begin{botherref}
\oauthor{\bsnm{Duran}, \binits{R.B.}},
\oauthor{\bsnm{Giannios}, \binits{D.}}:
Radio rebrightening of the grb afterglow by the accompanying supernova
\textbf{454},
1711
(2015).
doi:\doiurl{10.1093/mnras/stv2004}
\end{botherref}
\endbibitem

\bibitem[\protect\citeauthoryear{Evans and Wilkinson}{2000}]{Evans2000_M31}
\begin{barticle}
\bauthor{\bsnm{Evans}, \binits{N.W.}},
\bauthor{\bsnm{Wilkinson}, \binits{M.I.}}:
\bjtitle{Monthly Notices of the Royal Astronomical Society}
\bvolume{316}(\bissue{4}),
\bfpage{929}
(\byear{2000}).
\arxivurl{https://academic.oup.com/mnras/article-pdf/316/4/929/3798653/316-4-929.pdf}.
doi:\doiurl{10.1046/j.1365-8711.2000.03645.x}
\end{barticle}
\endbibitem

\bibitem[\protect\citeauthoryear{{Evans} et~al.}{2010}]{BATlightcurves}
\begin{barticle}
\bauthor{\bsnm{{Evans}}, \binits{P.A.}},
\bauthor{\bsnm{{Willingale}}, \binits{R.}},
\bauthor{\bsnm{{Osborne}}, \binits{J.P.}},
\bauthor{\bsnm{{O'Brien}}, \binits{P.T.}},
\bauthor{\bsnm{{Page}}, \binits{K.L.}},
\bauthor{\bsnm{{Markwardt}}, \binits{C.B.}},
\bauthor{\bsnm{{Barthelmy}}, \binits{S.D.}},
\bauthor{\bsnm{{Beardmore}}, \binits{A.P.}},
\bauthor{\bsnm{{Burrows}}, \binits{D.N.}},
\bauthor{\bsnm{{Pagani}}, \binits{C.}},
\bauthor{\bsnm{{Starling}}, \binits{R.L.C.}},
\bauthor{\bsnm{{Gehrels}}, \binits{N.}},
\bauthor{\bsnm{{Romano}}, \binits{P.}}:
\bjtitle{\aap}
\bvolume{519},
\bfpage{102}
(\byear{2010}).
\arxivurl{1004.3208}.
doi:\doiurl{10.1051/0004-6361/201014819}
\end{barticle}
\endbibitem

\bibitem[\protect\citeauthoryear{{Evans} et~al.}{1980}]{Evans1980_790305}
\begin{barticle}
\bauthor{\bsnm{{Evans}}, \binits{W.D.}},
\bauthor{\bsnm{{Klebesadel}}, \binits{R.W.}},
\bauthor{\bsnm{{Laros}}, \binits{J.G.}},
\bauthor{\bsnm{{Cline}}, \binits{T.L.}},
\bauthor{\bsnm{{Desai}}, \binits{U.D.}},
\bauthor{\bsnm{{Teegarden}}, \binits{B.J.}},
\bauthor{\bsnm{{Pizzichini}}, \binits{G.}},
\bauthor{\bsnm{{Hurley}}, \binits{K.}},
\bauthor{\bsnm{{Niel}}, \binits{M.}},
\bauthor{\bsnm{{Vedrenne}}, \binits{G.}}:
\bjtitle{\apjl}
\bvolume{237},
\bfpage{7}
(\byear{1980}).
doi:\doiurl{10.1086/183222}
\end{barticle}
\endbibitem

\bibitem[\protect\citeauthoryear{Fargion}{2003}]{Fargion2003}
\begin{barticle}
\bauthor{\bsnm{Fargion}, \binits{D.}}:
\bjtitle{Chinese Journal of Astronomy and Astrophysics Supplement}
\bvolume{3},
\bfpage{472}
(\byear{2003}).
doi:\doiurl{10.1088/1009-9271/3/S1/472}
\end{barticle}
\endbibitem

\bibitem[\protect\citeauthoryear{Fenimore et~al.}{1996}]{Fenimore1996_790305}
\begin{barticle}
\bauthor{\bsnm{Fenimore}, \binits{E.E.}},
\bauthor{\bsnm{Klebesadel}, \binits{R.W.}},
\bauthor{\bsnm{Laros}, \binits{J.G.}}:
\bjtitle{\apj}
\bvolume{460},
\bfpage{964}
(\byear{1996}).
doi:\doiurl{10.1086/177024}
\end{barticle}
\endbibitem

\bibitem[\protect\citeauthoryear{Fong et~al.}{2016a}]{Fong2016_160821B}
\begin{barticle}
\bauthor{\bsnm{Fong}, \binits{W.}},
\bauthor{\bsnm{Alexander}, \binits{K.D.}},
\bauthor{\bsnm{Laskar}, \binits{T.}}:
\bjtitle{GRB Coordinates Network}
\bvolume{19854},
\bfpage{1}
(\byear{2016}a)
\end{barticle}
\endbibitem

\bibitem[\protect\citeauthoryear{Fong et~al.}{2016b}]{Fong2016_150101B}
\begin{barticle}
\bauthor{\bsnm{Fong}, \binits{W.}},
\bauthor{\bsnm{Margutti}, \binits{R.}},
\bauthor{\bsnm{Chornock}, \binits{R.}},
\bauthor{\bsnm{Berger}, \binits{E.}},
\bauthor{\bsnm{Shappee}, \binits{B.J.}},
\bauthor{\bsnm{Levan}, \binits{A.J.}},
\bauthor{\bsnm{Tanvir}, \binits{N.R.}},
\bauthor{\bsnm{Smith}, \binits{N.}},
\bauthor{\bsnm{Milne}, \binits{P.A.}},
\bauthor{\bsnm{Laskar}, \binits{T.}},
\bauthor{\bsnm{Fox}, \binits{D.B.}},
\bauthor{\bsnm{Lunnan}, \binits{R.}},
\bauthor{\bsnm{Blanchard}, \binits{P.K.}},
\bauthor{\bsnm{Hjorth}, \binits{J.}},
\bauthor{\bsnm{Wiersema}, \binits{K.}},
\bauthor{\bparticle{van~der} \bsnm{Horst}, \binits{A.J.}},
\bauthor{\bsnm{Zaritsky}, \binits{D.}}:
\bjtitle{\apj}
\bvolume{833},
\bfpage{151}
(\byear{2016}b).
doi:\doiurl{10.3847/1538-4357/833/2/151}
\end{barticle}
\endbibitem

\bibitem[\protect\citeauthoryear{{Fouqu{\'e}} et~al.}{2001}]{Fouque2001_Virgo}
\begin{barticle}
\bauthor{\bsnm{{Fouqu{\'e}}}, \binits{P.}},
\bauthor{\bsnm{{Solanes}}, \binits{J.M.}},
\bauthor{\bsnm{{Sanchis}}, \binits{T.}},
\bauthor{\bsnm{{Balkowski}}, \binits{C.}}:
\bjtitle{\aap}
\bvolume{375},
\bfpage{770}
(\byear{2001}).
\arxivurl{astro-ph/0106261}.
doi:\doiurl{10.1051/0004-6361:20010833}
\end{barticle}
\endbibitem

\bibitem[\protect\citeauthoryear{Fox et~al.}{2005}]{Fox2005_050709}
\begin{barticle}
\bauthor{\bsnm{Fox}, \binits{D.B.}},
\bauthor{\bsnm{Frail}, \binits{D.A.}},
\bauthor{\bsnm{Price}, \binits{P.A.}},
\bauthor{\bsnm{Kulkarni}, \binits{S.R.}},
\bauthor{\bsnm{Berger}, \binits{E.}},
\bauthor{\bsnm{Piran}, \binits{T.}},
\bauthor{\bsnm{Soderberg}, \binits{A.M.}},
\bauthor{\bsnm{Cenko}, \binits{S.B.}},
\bauthor{\bsnm{Cameron}, \binits{P.B.}},
\bauthor{\bsnm{Gal-Yam}, \binits{A.}},
\bauthor{\bsnm{Kasliwal}, \binits{M.M.}},
\bauthor{\bsnm{Moon}, \binits{D.-S.}},
\bauthor{\bsnm{Harrison}, \binits{F.A.}},
\bauthor{\bsnm{Nakar}, \binits{E.}},
\bauthor{\bsnm{Schmidt}, \binits{B.P.}},
\bauthor{\bsnm{Penprase}, \binits{B.}},
\bauthor{\bsnm{Chevalier}, \binits{R.A.}},
\bauthor{\bsnm{Kumar}, \binits{P.}},
\bauthor{\bsnm{Roth}, \binits{K.}},
\bauthor{\bsnm{Watson}, \binits{D.}},
\bauthor{\bsnm{Lee}, \binits{B.L.}},
\bauthor{\bsnm{Shectman}, \binits{S.}},
\bauthor{\bsnm{Phillips}, \binits{M.M.}},
\bauthor{\bsnm{Roth}, \binits{M.}},
\bauthor{\bsnm{McCarthy}, \binits{P.J.}},
\bauthor{\bsnm{Rauch}, \binits{M.}},
\bauthor{\bsnm{Cowie}, \binits{L.}},
\bauthor{\bsnm{Peterson}, \binits{B.A.}},
\bauthor{\bsnm{Rich}, \binits{J.}},
\bauthor{\bsnm{Kawai}, \binits{N.}},
\bauthor{\bsnm{Aoki}, \binits{K.}},
\bauthor{\bsnm{Kosugi}, \binits{G.}},
\bauthor{\bsnm{Totani}, \binits{T.}},
\bauthor{\bsnm{Park}, \binits{H.-S.}},
\bauthor{\bsnm{MacFadyen}, \binits{A.}},
\bauthor{\bsnm{Hurley}, \binits{K.C.}}:
\bjtitle{\nat}
\bvolume{437},
\bfpage{845}
(\byear{2005}).
doi:\doiurl{10.1038/nature04189}
\end{barticle}
\endbibitem

\bibitem[\protect\citeauthoryear{Frail}{2003}]{Frail2003_031203}
\begin{barticle}
\bauthor{\bsnm{Frail}, \binits{D.A.}}:
\bjtitle{GRB Coordinates Network}
\bvolume{2473},
\bfpage{1}
(\byear{2003})
\end{barticle}
\endbibitem

\bibitem[\protect\citeauthoryear{Frail et~al.}{2000}]{Frail2000_980519}
\begin{botherref}
\oauthor{\bsnm{Frail}, \binits{D.A.}},
\oauthor{\bsnm{Kulkarni}, \binits{S.R.}},
\oauthor{\bsnm{Sari}, \binits{R.}},
\oauthor{\bsnm{Taylor}, \binits{G.B.}},
\oauthor{\bsnm{Shepherd}, \binits{D.S.}},
\oauthor{\bsnm{Bloom}, \binits{J.S.}},
\oauthor{\bsnm{Young}, \binits{C.H.}},
\oauthor{\bsnm{Nicastro}, \binits{L.}},
\oauthor{\bsnm{Masetti}, \binits{N.}}:
The radio afterglow from grb 980519: A test of the jet and circumstellar models
\textbf{534},
559
(2000).
doi:\doiurl{10.1086/308802}
\end{botherref}
\endbibitem

\bibitem[\protect\citeauthoryear{Frail et~al.}{2001}]{Frail2001}
\begin{botherref}
\oauthor{\bsnm{Frail}, \binits{D.A.}},
\oauthor{\bsnm{Kulkarni}, \binits{S.R.}},
\oauthor{\bsnm{Sari}, \binits{R.}},
\oauthor{\bsnm{Djorgovski}, \binits{S.G.}},
\oauthor{\bsnm{Bloom}, \binits{J.S.}},
\oauthor{\bsnm{Galama}, \binits{T.J.}},
\oauthor{\bsnm{Reichart}, \binits{D.E.}},
\oauthor{\bsnm{Berger}, \binits{E.}},
\oauthor{\bsnm{Harrison}, \binits{F.A.}},
\oauthor{\bsnm{Price}, \binits{P.A.}},
\oauthor{\bsnm{Yost}, \binits{S.A.}},
\oauthor{\bsnm{Diercks}, \binits{A.}},
\oauthor{\bsnm{Goodrich}, \binits{R.W.}},
\oauthor{\bsnm{Chaffee}, \binits{F.}}:
Beaming in gamma-ray bursts: Evidence for a standard energy reservoir
\textbf{562},
55
(2001).
doi:\doiurl{10.1086/338119}
\end{botherref}
\endbibitem

\bibitem[\protect\citeauthoryear{Frederiks
  et~al.}{2007a}]{Frederiks2007_041227}
\begin{barticle}
\bauthor{\bsnm{Frederiks}, \binits{D.D.}},
\bauthor{\bsnm{Golenetskii}, \binits{S.V.}},
\bauthor{\bsnm{Palshin}, \binits{V.D.}},
\bauthor{\bsnm{Aptekar}, \binits{R.L.}},
\bauthor{\bsnm{Ilyinskii}, \binits{V.N.}},
\bauthor{\bsnm{Oleinik}, \binits{F.P.}},
\bauthor{\bsnm{Mazets}, \binits{E.P.}},
\bauthor{\bsnm{Cline}, \binits{T.L.}}:
\bjtitle{Astronomy Letters}
\bvolume{33},
\bfpage{1}
(\byear{2007}a).
doi:\doiurl{10.1134/S106377370701001X}
\end{barticle}
\endbibitem

\bibitem[\protect\citeauthoryear{Frederiks
  et~al.}{2007b}]{Frederiks2007_051103}
\begin{barticle}
\bauthor{\bsnm{Frederiks}, \binits{D.D.}},
\bauthor{\bsnm{Palshin}, \binits{V.D.}},
\bauthor{\bsnm{Aptekar}, \binits{R.L.}},
\bauthor{\bsnm{Golenetskii}, \binits{S.V.}},
\bauthor{\bsnm{Cline}, \binits{T.L.}},
\bauthor{\bsnm{Mazets}, \binits{E.P.}}:
\bjtitle{Astronomy Letters}
\bvolume{33},
\bfpage{19}
(\byear{2007}b).
doi:\doiurl{10.1134/S1063773707010021}
\end{barticle}
\endbibitem

\bibitem[\protect\citeauthoryear{{Frederiks}
  et~al.}{2016}]{Frederiks2016_161219B}
\begin{barticle}
\bauthor{\bsnm{{Frederiks}}, \binits{D.}},
\bauthor{\bsnm{{Golenetskii}}, \binits{S.}},
\bauthor{\bsnm{{Aptekar}}, \binits{R.}},
\bauthor{\bsnm{{Oleynik}}, \binits{P.}},
\bauthor{\bsnm{{Ulanov}}, \binits{M.}},
\bauthor{\bsnm{{Svinkin}}, \binits{D.}},
\bauthor{\bsnm{{Tsvetkova}}, \binits{A.}},
\bauthor{\bsnm{{Lysenko}}, \binits{A.}},
\bauthor{\bsnm{{Kozlova}}, \binits{A.}},
\bauthor{\bsnm{{Cline}}, \binits{T.}}:
\bjtitle{GRB Coordinates Network}
\bvolume{20323},
\bfpage{1}
(\byear{2016})
\end{barticle}
\endbibitem

\bibitem[\protect\citeauthoryear{{Frederiks}
  et~al.}{2020}]{Frederiks2020_200415A}
\begin{barticle}
\bauthor{\bsnm{{Frederiks}}, \binits{D.}},
\bauthor{\bsnm{{Golenetskii}}, \binits{S.}},
\bauthor{\bsnm{{Aptekar}}, \binits{R.}},
\bauthor{\bsnm{{Lysenko}}, \binits{A.}},
\bauthor{\bsnm{{Ridnaia}}, \binits{A.}},
\bauthor{\bsnm{{Svinkin}}, \binits{D.}},
\bauthor{\bsnm{{Tsvetkova}}, \binits{A.}},
\bauthor{\bsnm{{Ulanov}}, \binits{M.}},
\bauthor{\bsnm{{Cline}}, \binits{T.}},
\bauthor{\bsnm{{Konus-Wind Team}}}:
\bjtitle{GRB Coordinates Network}
\bvolume{27596},
\bfpage{1}
(\byear{2020})
\end{barticle}
\endbibitem

\bibitem[\protect\citeauthoryear{{Fuentes-Fern{\'a}ndez}
  et~al.}{2020}]{Fuentes2020_COLIBRI}
\begin{barticle}
\bauthor{\bsnm{{Fuentes-Fern{\'a}ndez}}, \binits{J.}},
\bauthor{\bsnm{{Watson}}, \binits{A.M.}},
\bauthor{\bsnm{{Cuevas}}, \binits{S.}},
\bauthor{\bsnm{{Basa}}, \binits{S.}},
\bauthor{\bsnm{{Floriot}}, \binits{J.}},
\bauthor{\bsnm{{Dolon}}, \binits{F.}},
\bauthor{\bsnm{{Valentin}}, \binits{H.}},
\bauthor{\bsnm{{Challita}}, \binits{Z.}},
\bauthor{\bsnm{{Vola}}, \binits{P.}}:
\bjtitle{Journal of Astronomical Instrumentation}
\bvolume{9}(\bissue{1}),
\bfpage{2050001}
(\byear{2020}).
doi:\doiurl{10.1142/S2251171720500014}
\end{barticle}
\endbibitem

\bibitem[\protect\citeauthoryear{Fynbo et~al.}{2006}]{Fynbo2006_060614}
\begin{barticle}
\bauthor{\bsnm{Fynbo}, \binits{J.P.U.}},
\bauthor{\bsnm{Thoene}, \binits{C.C.}},
\bauthor{\bsnm{Jensen}, \binits{B.L.}},
\bauthor{\bsnm{Hjorth}, \binits{J.}},
\bauthor{\bsnm{Sollerman}, \binits{J.}},
\bauthor{\bsnm{Watson}, \binits{D.}},
\bauthor{\bsnm{Xu}, \binits{D.}},
\bauthor{\bsnm{Ovaldsen}, \binits{J.-E.}},
\bauthor{\bsnm{Joergensen}, \binits{U.G.}},
\bauthor{\bsnm{Hinse}, \binits{T.}},
\bauthor{\bsnm{Woller}, \binits{K.}}:
\bjtitle{GRB Coordinates Network}
\bvolume{5277},
\bfpage{1}
(\byear{2006})
\end{barticle}
\endbibitem

\bibitem[\protect\citeauthoryear{Fynbo et~al.}{2019}]{Fynbo2019_191019A}
\begin{barticle}
\bauthor{\bsnm{Fynbo}, \binits{J.P.U.}},
\bauthor{\bsnm{Perley}, \binits{D.A.}},
\bauthor{\bsnm{{de Ugarte Postigo}}, \binits{A.}},
\bauthor{\bsnm{Malesani}, \binits{D.B.}},
\bauthor{\bsnm{Jannsen}, \binits{N.E.}},
\bauthor{\bsnm{Al.}, \binits{E.}}:
\bjtitle{GCN}
\bvolume{26041},
\bfpage{1}
(\byear{2019})
\end{barticle}
\endbibitem

\bibitem[\protect\citeauthoryear{{Fynbo} et~al.}{2006}]{Fynbo2006_060505}
\begin{barticle}
\bauthor{\bsnm{{Fynbo}}, \binits{J.P.U.}},
\bauthor{\bsnm{{Watson}}, \binits{D.}},
\bauthor{\bsnm{{Th{\"o}ne}}, \binits{C.C.}},
\bauthor{\bsnm{{Sollerman}}, \binits{J.}},
\bauthor{\bsnm{{Bloom}}, \binits{J.S.}},
\bauthor{\bsnm{{Davis}}, \binits{T.M.}},
\bauthor{\bsnm{{Hjorth}}, \binits{J.}},
\bauthor{\bsnm{{Jakobsson}}, \binits{P.}},
\bauthor{\bsnm{{J{\o}rgensen}}, \binits{U.G.}},
\bauthor{\bsnm{{Graham}}, \binits{J.F.}},
\bauthor{\bsnm{{Fruchter}}, \binits{A.S.}},
\bauthor{\bsnm{{Bersier}}, \binits{D.}},
\bauthor{\bsnm{{Kewley}}, \binits{L.}},
\bauthor{\bsnm{{Cassan}}, \binits{A.}},
\bauthor{\bsnm{{Castro Cer{\'o}n}}, \binits{J.M.}},
\bauthor{\bsnm{{Foley}}, \binits{S.}},
\bauthor{\bsnm{{Gorosabel}}, \binits{J.}},
\bauthor{\bsnm{{Hinse}}, \binits{T.C.}},
\bauthor{\bsnm{{Horne}}, \binits{K.D.}},
\bauthor{\bsnm{{Jensen}}, \binits{B.L.}},
\bauthor{\bsnm{{Klose}}, \binits{S.}},
\bauthor{\bsnm{{Kocevski}}, \binits{D.}},
\bauthor{\bsnm{{Marquette}}, \binits{J.-B.}},
\bauthor{\bsnm{{Perley}}, \binits{D.}},
\bauthor{\bsnm{{Ramirez-Ruiz}}, \binits{E.}},
\bauthor{\bsnm{{Stritzinger}}, \binits{M.D.}},
\bauthor{\bsnm{{Vreeswijk}}, \binits{P.M.}},
\bauthor{\bsnm{{Wijers}}, \binits{R.A.M.}},
\bauthor{\bsnm{{Woller}}, \binits{K.G.}},
\bauthor{\bsnm{{Xu}}, \binits{D.}},
\bauthor{\bsnm{{Zub}}, \binits{M.}}:
\bjtitle{\nat}
\bvolume{444}(\bissue{7122}),
\bfpage{1047}
(\byear{2006}).
\arxivurl{astro-ph/0608313}.
doi:\doiurl{10.1038/nature05375}
\end{barticle}
\endbibitem

\bibitem[\protect\citeauthoryear{Gal-Yam et~al.}{2006}]{GalYam2006_060614}
\begin{barticle}
\bauthor{\bsnm{Gal-Yam}, \binits{A.}},
\bauthor{\bsnm{Fox}, \binits{D.B.}},
\bauthor{\bsnm{Price}, \binits{P.A.}},
\bauthor{\bsnm{Ofek}, \binits{E.O.}},
\bauthor{\bsnm{Davis}, \binits{M.R.}},
\bauthor{\bsnm{Leonard}, \binits{D.C.}},
\bauthor{\bsnm{Soderberg}, \binits{A.M.}},
\bauthor{\bsnm{Schmidt}, \binits{B.P.}},
\bauthor{\bsnm{Lewis}, \binits{K.M.}},
\bauthor{\bsnm{Peterson}, \binits{B.A.}},
\bauthor{\bsnm{Kulkarni}, \binits{S.R.}},
\bauthor{\bsnm{Berger}, \binits{E.}},
\bauthor{\bsnm{Cenko}, \binits{S.B.}},
\bauthor{\bsnm{Sari}, \binits{R.}},
\bauthor{\bsnm{Sharon}, \binits{K.}},
\bauthor{\bsnm{Frail}, \binits{D.}},
\bauthor{\bsnm{Moon}, \binits{D.-S.}},
\bauthor{\bsnm{Brown}, \binits{P.J.}},
\bauthor{\bsnm{Cucchiara}, \binits{A.}},
\bauthor{\bsnm{Harrison}, \binits{F.}},
\bauthor{\bsnm{Piran}, \binits{T.}},
\bauthor{\bsnm{Persson}, \binits{S.E.}},
\bauthor{\bsnm{McCarthy}, \binits{P.J.}},
\bauthor{\bsnm{Penprase}, \binits{B.E.}},
\bauthor{\bsnm{Chevalier}, \binits{R.A.}},
\bauthor{\bsnm{MacFadyen}, \binits{A.I.}}:
\bjtitle{\nat}
\bvolume{444},
\bfpage{1053}
(\byear{2006}).
doi:\doiurl{10.1038/nature05373}
\end{barticle}
\endbibitem

\bibitem[\protect\citeauthoryear{Galama et~al.}{1998}]{Galama1998_980425}
\begin{barticle}
\bauthor{\bsnm{Galama}, \binits{T.J.}},
\bauthor{\bsnm{Vreeswijk}, \binits{P.M.}},
\bauthor{\bparticle{van} \bsnm{Paradijs}, \binits{J.}},
\bauthor{\bsnm{Kouveliotou}, \binits{C.}},
\bauthor{\bsnm{Augusteijn}, \binits{T.}},
\bauthor{\bsnm{B\"ohnhardt}, \binits{H.}},
\bauthor{\bsnm{Brewer}, \binits{J.P.}},
\bauthor{\bsnm{Doublier}, \binits{V.}},
\bauthor{\bsnm{Gonzalez}, \binits{J.-F.}},
\bauthor{\bsnm{Leibundgut}, \binits{B.}},
\bauthor{\bsnm{Lidman}, \binits{C.}},
\bauthor{\bsnm{Hainaut}, \binits{O.R.}},
\bauthor{\bsnm{Patat}, \binits{F.}},
\bauthor{\bsnm{Heise}, \binits{J.}},
\bauthor{\bparticle{in't} \bsnm{Zand}, \binits{J.}},
\bauthor{\bsnm{Hurley}, \binits{K.}},
\bauthor{\bsnm{Groot}, \binits{P.J.}},
\bauthor{\bsnm{Strom}, \binits{R.G.}},
\bauthor{\bsnm{Mazzali}, \binits{P.A.}},
\bauthor{\bsnm{Iwamoto}, \binits{K.}},
\bauthor{\bsnm{Nomoto}, \binits{K.}},
\bauthor{\bsnm{Umeda}, \binits{H.}},
\bauthor{\bsnm{Nakamura}, \binits{T.}},
\bauthor{\bsnm{Young}, \binits{T.R.}},
\bauthor{\bsnm{Suzuki}, \binits{T.}},
\bauthor{\bsnm{Shigeyama}, \binits{T.}},
\bauthor{\bsnm{Koshut}, \binits{T.}},
\bauthor{\bsnm{Kippen}, \binits{M.}},
\bauthor{\bsnm{Robinson}, \binits{C.}},
\bauthor{\bparticle{de} \bsnm{Wildt}, \binits{P.}},
\bauthor{\bsnm{Wijers}, \binits{R.A.M.J.}},
\bauthor{\bsnm{Tanvir}, \binits{N.}},
\bauthor{\bsnm{Greiner}, \binits{J.}},
\bauthor{\bsnm{Pian}, \binits{E.}},
\bauthor{\bsnm{Palazzi}, \binits{E.}},
\bauthor{\bsnm{Frontera}, \binits{F.}},
\bauthor{\bsnm{Masetti}, \binits{N.}},
\bauthor{\bsnm{Nicastro}, \binits{L.}},
\bauthor{\bsnm{Feroci}, \binits{M.}},
\bauthor{\bsnm{Costa}, \binits{E.}},
\bauthor{\bsnm{Piro}, \binits{L.}},
\bauthor{\bsnm{Peterson}, \binits{B.A.}},
\bauthor{\bsnm{Tinney}, \binits{C.}},
\bauthor{\bsnm{Boyle}, \binits{B.}},
\bauthor{\bsnm{Cannon}, \binits{R.}},
\bauthor{\bsnm{Stathakis}, \binits{R.}},
\bauthor{\bsnm{Sadler}, \binits{E.}},
\bauthor{\bsnm{Begam}, \binits{M.C.}},
\bauthor{\bsnm{Ianna}, \binits{P.}}:
\bjtitle{\nat}
\bvolume{395},
\bfpage{670}
(\byear{1998}).
doi:\doiurl{10.1038/27150}
\end{barticle}
\endbibitem

\bibitem[\protect\citeauthoryear{Gao and Huang}{2006}]{Gao2006}
\begin{botherref}
\oauthor{\bsnm{Gao}, \binits{T.-T.}},
\oauthor{\bsnm{Huang}, \binits{Y.-F.}}:
On the evolution of the apparent size of gamma-ray burst remnants
\textbf{6},
305
(2006).
doi:\doiurl{10.1088/1009-9271/6/3/05}
\end{botherref}
\endbibitem

\bibitem[\protect\citeauthoryear{Gehrels et~al.}{2005}]{Gehrels2005_050509B}
\begin{barticle}
\bauthor{\bsnm{Gehrels}, \binits{N.}},
\bauthor{\bsnm{Sarazin}, \binits{C.L.}},
\bauthor{\bsnm{O'Brien}, \binits{P.T.}},
\bauthor{\bsnm{Zhang}, \binits{B.}},
\bauthor{\bsnm{Barbier}, \binits{L.}},
\bauthor{\bsnm{Barthelmy}, \binits{S.D.}},
\bauthor{\bsnm{Blustin}, \binits{A.}},
\bauthor{\bsnm{Burrows}, \binits{D.N.}},
\bauthor{\bsnm{Cannizzo}, \binits{J.}},
\bauthor{\bsnm{Cummings}, \binits{J.R.}},
\bauthor{\bsnm{Goad}, \binits{M.}},
\bauthor{\bsnm{Holland}, \binits{S.T.}},
\bauthor{\bsnm{Hurkett}, \binits{C.P.}},
\bauthor{\bsnm{Kennea}, \binits{J.A.}},
\bauthor{\bsnm{Levan}, \binits{A.}},
\bauthor{\bsnm{Markwardt}, \binits{C.B.}},
\bauthor{\bsnm{Mason}, \binits{K.O.}},
\bauthor{\bsnm{Meszaros}, \binits{P.}},
\bauthor{\bsnm{Page}, \binits{M.}},
\bauthor{\bsnm{Palmer}, \binits{D.M.}},
\bauthor{\bsnm{Rol}, \binits{E.}},
\bauthor{\bsnm{Sakamoto}, \binits{T.}},
\bauthor{\bsnm{Willingale}, \binits{R.}},
\bauthor{\bsnm{Angelini}, \binits{L.}},
\bauthor{\bsnm{Beardmore}, \binits{A.}},
\bauthor{\bsnm{Boyd}, \binits{P.T.}},
\bauthor{\bsnm{Breeveld}, \binits{A.}},
\bauthor{\bsnm{Campana}, \binits{S.}},
\bauthor{\bsnm{Chester}, \binits{M.M.}},
\bauthor{\bsnm{Chincarini}, \binits{G.}},
\bauthor{\bsnm{Cominsky}, \binits{L.R.}},
\bauthor{\bsnm{Cusumano}, \binits{G.}},
\bauthor{\bparticle{de} \bsnm{Pasquale}, \binits{M.}},
\bauthor{\bsnm{Fenimore}, \binits{E.E.}},
\bauthor{\bsnm{Giommi}, \binits{P.}},
\bauthor{\bsnm{Gronwall}, \binits{C.}},
\bauthor{\bsnm{Grupe}, \binits{D.}},
\bauthor{\bsnm{Hill}, \binits{J.E.}},
\bauthor{\bsnm{Hinshaw}, \binits{D.}},
\bauthor{\bsnm{Hjorth}, \binits{J.}},
\bauthor{\bsnm{Hullinger}, \binits{D.}},
\bauthor{\bsnm{Hurley}, \binits{K.C.}},
\bauthor{\bsnm{Klose}, \binits{S.}},
\bauthor{\bsnm{Kobayashi}, \binits{S.}},
\bauthor{\bsnm{Kouveliotou}, \binits{C.}},
\bauthor{\bsnm{Krimm}, \binits{H.A.}},
\bauthor{\bsnm{Mangano}, \binits{V.}},
\bauthor{\bsnm{Marshall}, \binits{F.E.}},
\bauthor{\bsnm{McGowan}, \binits{K.}},
\bauthor{\bsnm{Moretti}, \binits{A.}},
\bauthor{\bsnm{Mushotzky}, \binits{R.F.}},
\bauthor{\bsnm{Nakazawa}, \binits{K.}},
\bauthor{\bsnm{Norris}, \binits{J.P.}},
\bauthor{\bsnm{Nousek}, \binits{J.A.}},
\bauthor{\bsnm{Osborne}, \binits{J.P.}},
\bauthor{\bsnm{Page}, \binits{K.}},
\bauthor{\bsnm{Parsons}, \binits{A.M.}},
\bauthor{\bsnm{Patel}, \binits{S.}},
\bauthor{\bsnm{Perri}, \binits{M.}},
\bauthor{\bsnm{Poole}, \binits{T.}},
\bauthor{\bsnm{Romano}, \binits{P.}},
\bauthor{\bsnm{Roming}, \binits{P.W.A.}},
\bauthor{\bsnm{Rosen}, \binits{S.}},
\bauthor{\bsnm{Sato}, \binits{G.}},
\bauthor{\bsnm{Schady}, \binits{P.}},
\bauthor{\bsnm{Smale}, \binits{A.P.}},
\bauthor{\bsnm{Sollerman}, \binits{J.}},
\bauthor{\bsnm{Starling}, \binits{R.}},
\bauthor{\bsnm{Still}, \binits{M.}},
\bauthor{\bsnm{Suzuki}, \binits{M.}},
\bauthor{\bsnm{Tagliaferri}, \binits{G.}},
\bauthor{\bsnm{Takahashi}, \binits{T.}},
\bauthor{\bsnm{Tashiro}, \binits{M.}},
\bauthor{\bsnm{Tueller}, \binits{J.}},
\bauthor{\bsnm{Wells}, \binits{A.A.}},
\bauthor{\bsnm{White}, \binits{N.E.}},
\bauthor{\bsnm{Wijers}, \binits{R.A.M.J.}}:
\bjtitle{\nat}
\bvolume{437},
\bfpage{851}
(\byear{2005}).
doi:\doiurl{10.1038/nature04142}
\end{barticle}
\endbibitem

\bibitem[\protect\citeauthoryear{Gehrels et~al.}{2006}]{Gehrels2006_060614}
\begin{barticle}
\bauthor{\bsnm{Gehrels}, \binits{N.}},
\bauthor{\bsnm{Norris}, \binits{J.P.}},
\bauthor{\bsnm{Barthelmy}, \binits{S.D.}},
\bauthor{\bsnm{Granot}, \binits{J.}},
\bauthor{\bsnm{Kaneko}, \binits{Y.}},
\bauthor{\bsnm{Kouveliotou}, \binits{C.}},
\bauthor{\bsnm{Markwardt}, \binits{C.B.}},
\bauthor{\bsnm{M{\'{e}}sz{\'{a}}ros}, \binits{P.}},
\bauthor{\bsnm{Nakar}, \binits{E.}},
\bauthor{\bsnm{Nousek}, \binits{J.A.}},
\bauthor{\bsnm{O'Brien}, \binits{P.T.}},
\bauthor{\bsnm{Page}, \binits{M.}},
\bauthor{\bsnm{Palmer}, \binits{D.M.}},
\bauthor{\bsnm{Parsons}, \binits{A.M.}},
\bauthor{\bsnm{Roming}, \binits{P.W.A.}},
\bauthor{\bsnm{Sakamoto}, \binits{T.}},
\bauthor{\bsnm{Sarazin}, \binits{C.L.}},
\bauthor{\bsnm{Schady}, \binits{P.}},
\bauthor{\bsnm{Stamatikos}, \binits{M.}},
\bauthor{\bsnm{Woosley}, \binits{S.E.}}:
\bjtitle{Nature}
\bvolume{444}(\bissue{7122}),
\bfpage{1044}
(\byear{2006}).
\arxivurl{0610635}.
doi:\doiurl{10.1038/nature05376}
\end{barticle}
\endbibitem

\bibitem[\protect\citeauthoryear{{Gendre} et~al.}{2013}]{Gendre2013}
\begin{barticle}
\bauthor{\bsnm{{Gendre}}, \binits{B.}},
\bauthor{\bsnm{{Stratta}}, \binits{G.}},
\bauthor{\bsnm{{Atteia}}, \binits{J.-L.}},
\bauthor{\bsnm{{Basa}}, \binits{S.}},
\bauthor{\bsnm{{Bo{\"e}r}}, \binits{M.}},
\bauthor{\bsnm{{Coward}}, \binits{D.M.}},
\bauthor{\bsnm{{Cutini}}, \binits{S.}},
\bauthor{\bsnm{{D'Elia}}, \binits{V.}},
\bauthor{\bsnm{{Howell}}, \binits{E.J.}},
\bauthor{\bsnm{{Klotz}}, \binits{A.}},
\bauthor{\bsnm{{Piro}}, \binits{L.}}:
\bjtitle{\apj}
\bvolume{766}(\bissue{1}),
\bfpage{30}
(\byear{2013}).
\arxivurl{1212.2392}.
doi:\doiurl{10.1088/0004-637X/766/1/30}
\end{barticle}
\endbibitem

\bibitem[\protect\citeauthoryear{Gendre et~al.}{2006}]{Gendre2006}
\begin{barticle}
\bauthor{\bsnm{Gendre}, \binits{B.}},
\bauthor{\bsnm{Piro}, \binits{L.}},
\bauthor{\bparticle{de} \bsnm{Pasquale}, \binits{M.}}:
\bjtitle{Advances in Space Research}
\bvolume{38},
\bfpage{1325}
(\byear{2006}).
doi:\doiurl{10.1016/j.asr.2004.12.044}
\end{barticle}
\endbibitem

\bibitem[\protect\citeauthoryear{{Gevin} et~al.}{2009}]{Gevin2009}
\begin{barticle}
\bauthor{\bsnm{{Gevin}}, \binits{O.}},
\bauthor{\bsnm{{Baron}}, \binits{P.}},
\bauthor{\bsnm{{Coppolani}}, \binits{X.}},
\bauthor{\bsnm{{Daly}}, \binits{F.}},
\bauthor{\bsnm{{Delagnes}}, \binits{E.}},
\bauthor{\bsnm{{Limousin}}, \binits{O.}},
\bauthor{\bsnm{{Lugiez}}, \binits{F.}},
\bauthor{\bsnm{{Meuris}}, \binits{A.}},
\bauthor{\bsnm{{Pinsard}}, \binits{F.}},
\bauthor{\bsnm{{Renaud}}, \binits{D.}}:
\bjtitle{IEEE Transactions on Nuclear Science}
\bvolume{56}(\bissue{4}),
\bfpage{2351}
(\byear{2009}).
doi:\doiurl{10.1109/TNS.2009.2023989}
\end{barticle}
\endbibitem

\bibitem[\protect\citeauthoryear{{Godet} et~al.}{2014}]{Godet2014}
\begin{bchapter}
\bauthor{\bsnm{{Godet}}, \binits{O.}},
\bauthor{\bsnm{{Nasser}}, \binits{G.}},
\bauthor{\bsnm{{Atteia}}, \binits{J.-L.}},
\bauthor{\bsnm{{Cordier}}, \binits{B.}},
\bauthor{\bsnm{{Mandrou}}, \binits{P.}},
\bauthor{\bsnm{{Barret}}, \binits{D.}},
\bauthor{\bsnm{{Triou}}, \binits{H.}},
\bauthor{\bsnm{{Pons}}, \binits{R.}},
\bauthor{\bsnm{{Amoros}}, \binits{C.}},
\bauthor{\bsnm{{Bordon}}, \binits{S.}},
\bauthor{\bsnm{{Gevin}}, \binits{O.}},
\bauthor{\bsnm{{Gonzalez}}, \binits{F.}},
\bauthor{\bsnm{{G{\"o}tz}}, \binits{D.}},
\bauthor{\bsnm{{Gros}}, \binits{A.}},
\bauthor{\bsnm{{Houret}}, \binits{B.}},
\bauthor{\bsnm{{Lachaud}}, \binits{C.}},
\bauthor{\bsnm{{Lacombe}}, \binits{K.}},
\bauthor{\bsnm{{Marty}}, \binits{W.}},
\bauthor{\bsnm{{Mercier}}, \binits{K.}},
\bauthor{\bsnm{{Rambaud}}, \binits{D.}},
\bauthor{\bsnm{{Ramon}}, \binits{P.}},
\bauthor{\bsnm{{Rouaix}}, \binits{G.}},
\bauthor{\bsnm{{Schanne}}, \binits{S.}},
\bauthor{\bsnm{{Waegebaert}}, \binits{V.}}:
In: \bbtitle{\procspie}.
\bsertitle{Society of Photo-Optical Instrumentation Engineers (SPIE) Conference
  Series},
vol. \bseriesno{9144},
p. \bfpage{914424}
(\byear{2014}).
\arxivurl{1406.7759}.
doi:\doiurl{10.1117/12.2055507}.
\burl{https://ui.adsabs.harvard.edu/abs/2014SPIE.9144E..24G}
\end{bchapter}
\endbibitem

\bibitem[\protect\citeauthoryear{{Goldstein}
  et~al.}{2017}]{Goldstein2017_170817A}
\begin{barticle}
\bauthor{\bsnm{{Goldstein}}, \binits{A.}},
\bauthor{\bsnm{{Veres}}, \binits{P.}},
\bauthor{\bsnm{{Burns}}, \binits{E.}},
\bauthor{\bsnm{{Briggs}}, \binits{M.S.}},
\bauthor{\bsnm{{Hamburg}}, \binits{R.}},
\bauthor{\bsnm{{Kocevski}}, \binits{D.}},
\bauthor{\bsnm{{Wilson-Hodge}}, \binits{C.A.}},
\bauthor{\bsnm{{Preece}}, \binits{R.D.}},
\bauthor{\bsnm{{Poolakkil}}, \binits{S.}},
\bauthor{\bsnm{{Roberts}}, \binits{O.J.}},
\bauthor{\bsnm{{Hui}}, \binits{C.M.}},
\bauthor{\bsnm{{Connaughton}}, \binits{V.}},
\bauthor{\bsnm{{Racusin}}, \binits{J.}},
\bauthor{\bsnm{{von Kienlin}}, \binits{A.}},
\bauthor{\bsnm{{Dal Canton}}, \binits{T.}},
\bauthor{\bsnm{{Christensen}}, \binits{N.}},
\bauthor{\bsnm{{Littenberg}}, \binits{T.}},
\bauthor{\bsnm{{Siellez}}, \binits{K.}},
\bauthor{\bsnm{{Blackburn}}, \binits{L.}},
\bauthor{\bsnm{{Broida}}, \binits{J.}},
\bauthor{\bsnm{{Bissaldi}}, \binits{E.}},
\bauthor{\bsnm{{Cleveland}}, \binits{W.H.}},
\bauthor{\bsnm{{Gibby}}, \binits{M.H.}},
\bauthor{\bsnm{{Giles}}, \binits{M.M.}},
\bauthor{\bsnm{{Kippen}}, \binits{R.M.}},
\bauthor{\bsnm{{McBreen}}, \binits{S.}},
\bauthor{\bsnm{{McEnery}}, \binits{J.}},
\bauthor{\bsnm{{Meegan}}, \binits{C.A.}},
\bauthor{\bsnm{{Paciesas}}, \binits{W.S.}},
\bauthor{\bsnm{{Stanbro}}, \binits{M.}}:
\bjtitle{\apjl}
\bvolume{848}(\bissue{2}),
\bfpage{14}
(\byear{2017}).
\arxivurl{1710.05446}.
doi:\doiurl{10.3847/2041-8213/aa8f41}
\end{barticle}
\endbibitem

\bibitem[\protect\citeauthoryear{{Golenetskii}
  et~al.}{2006a}]{Golenetskii2006_060614}
\begin{barticle}
\bauthor{\bsnm{{Golenetskii}}, \binits{S.}},
\bauthor{\bsnm{{Aptekar}}, \binits{R.}},
\bauthor{\bsnm{{Mazets}}, \binits{E.}},
\bauthor{\bsnm{{Pal'Shin}}, \binits{V.}},
\bauthor{\bsnm{{Frederiks}}, \binits{D.}},
\bauthor{\bsnm{{Cline}}, \binits{T.}}:
\bjtitle{GRB Coordinates Network}
\bvolume{5264},
\bfpage{1}
(\byear{2006}a)
\end{barticle}
\endbibitem

\bibitem[\protect\citeauthoryear{{Golenetskii}
  et~al.}{2006b}]{Golenetskii2006_061201}
\begin{barticle}
\bauthor{\bsnm{{Golenetskii}}, \binits{S.}},
\bauthor{\bsnm{{Aptekar}}, \binits{R.}},
\bauthor{\bsnm{{Mazets}}, \binits{E.}},
\bauthor{\bsnm{{Pal'Shin}}, \binits{V.}},
\bauthor{\bsnm{{Frederiks}}, \binits{D.}},
\bauthor{\bsnm{{Cline}}, \binits{T.}}:
\bjtitle{GRB Coordinates Network}
\bvolume{5890},
\bfpage{1}
(\byear{2006}b)
\end{barticle}
\endbibitem

\bibitem[\protect\citeauthoryear{{Golenetskii}
  et~al.}{2015a}]{Golenetskii2015_150518A}
\begin{barticle}
\bauthor{\bsnm{{Golenetskii}}, \binits{S.}},
\bauthor{\bsnm{{Aptekar}}, \binits{R.}},
\bauthor{\bsnm{{Frederiks}}, \binits{D.}},
\bauthor{\bsnm{{Pal'Shin}}, \binits{V.}},
\bauthor{\bsnm{{Oleynik}}, \binits{P.}},
\bauthor{\bsnm{{Ulanov}}, \binits{M.}},
\bauthor{\bsnm{{Svinkin}}, \binits{D.}},
\bauthor{\bsnm{{Tsvetkova}}, \binits{A.}},
\bauthor{\bsnm{{Lysenko}}, \binits{A.}},
\bauthor{\bsnm{{Cline}}, \binits{T.}}:
\bjtitle{GRB Coordinates Network}
\bvolume{17837},
\bfpage{1}
(\byear{2015}a)
\end{barticle}
\endbibitem

\bibitem[\protect\citeauthoryear{{Golenetskii}
  et~al.}{2015b}]{Golenetskii2015_150818A}
\begin{barticle}
\bauthor{\bsnm{{Golenetskii}}, \binits{S.}},
\bauthor{\bsnm{{Aptekar}}, \binits{R.}},
\bauthor{\bsnm{{Frederiks}}, \binits{D.}},
\bauthor{\bsnm{{Pal'Shin}}, \binits{V.}},
\bauthor{\bsnm{{Oleynik}}, \binits{P.}},
\bauthor{\bsnm{{Ulanov}}, \binits{M.}},
\bauthor{\bsnm{{Svinkin}}, \binits{D.}},
\bauthor{\bsnm{{Tsvetkova}}, \binits{A.}},
\bauthor{\bsnm{{Lysenko}}, \binits{A.}},
\bauthor{\bsnm{{Kozlova}}, \binits{A.}},
\bauthor{\bsnm{{Cline}}, \binits{T.}}:
\bjtitle{GRB Coordinates Network}
\bvolume{18198},
\bfpage{1}
(\byear{2015}b)
\end{barticle}
\endbibitem

\bibitem[\protect\citeauthoryear{{Gompertz} et~al.}{2020}]{Gompertz2020}
\begin{barticle}
\bauthor{\bsnm{{Gompertz}}, \binits{B.P.}},
\bauthor{\bsnm{{Levan}}, \binits{A.J.}},
\bauthor{\bsnm{{Tanvir}}, \binits{N.R.}}:
\bjtitle{\apj}
\bvolume{895}(\bissue{1}),
\bfpage{58}
(\byear{2020}).
\arxivurl{2001.08706}.
doi:\doiurl{10.3847/1538-4357/ab8d24}
\end{barticle}
\endbibitem

\bibitem[\protect\citeauthoryear{{Gotz} et~al.}{2003}]{Gotz2003_031203}
\begin{barticle}
\bauthor{\bsnm{{Gotz}}, \binits{D.}},
\bauthor{\bsnm{{Mereghetti}}, \binits{S.}},
\bauthor{\bsnm{{Beck}}, \binits{M.}},
\bauthor{\bsnm{{Borkowski}}, \binits{J.}},
\bauthor{\bsnm{{Mowlavi}}, \binits{N.}}:
\bjtitle{GRB Coordinates Network}
\bvolume{2459},
\bfpage{1}
(\byear{2003})
\end{barticle}
\endbibitem

\bibitem[\protect\citeauthoryear{G\"otz et~al.}{2011}]{Gotz2011_041219A}
\begin{botherref}
\oauthor{\bsnm{G\"otz}, \binits{D.}},
\oauthor{\bsnm{Covino}, \binits{S.}},
\oauthor{\bsnm{Hasco\"et}, \binits{R.}},
\oauthor{\bsnm{ez-Soto}, \binits{A.F.}},
\oauthor{\bsnm{Daigne}, \binits{F.}},
\oauthor{\bsnm{Mochkovitch}, \binits{R.}},
\oauthor{\bsnm{Esposito}, \binits{P.}}:
A detailed spectral study of grb 041219a and its host galaxy
\textbf{413},
2173
(2011).
doi:\doiurl{10.1111/j.1365-2966.2011.18290.x}
\end{botherref}
\endbibitem

\bibitem[\protect\citeauthoryear{{G{\"o}tz} et~al.}{2014}]{Gotz2014_MXT}
\begin{bchapter}
\bauthor{\bsnm{{G{\"o}tz}}, \binits{D.}},
\bauthor{\bsnm{{Osborne}}, \binits{J.}},
\bauthor{\bsnm{{Cordier}}, \binits{B.}},
\bauthor{\bsnm{{Paul}}, \binits{J.}},
\bauthor{\bsnm{{Evans}}, \binits{P.}},
\bauthor{\bsnm{{Beardmore}}, \binits{A.}},
\bauthor{\bsnm{{Martindale}}, \binits{A.}},
\bauthor{\bsnm{{Willingale}}, \binits{R.}},
\bauthor{\bsnm{{O'Brien}}, \binits{P.}},
\bauthor{\bsnm{{Basa}}, \binits{S.}},
\bauthor{\bsnm{{Rossin}}, \binits{C.}},
\bauthor{\bsnm{{Godet}}, \binits{O.}},
\bauthor{\bsnm{{Webb}}, \binits{N.}},
\bauthor{\bsnm{{Greiner}}, \binits{J.}},
\bauthor{\bsnm{{Nandra}}, \binits{K.}},
\bauthor{\bsnm{{Meidinger}}, \binits{N.}},
\bauthor{\bsnm{{Perinati}}, \binits{E.}},
\bauthor{\bsnm{{Santangelo}}, \binits{A.}},
\bauthor{\bsnm{{Mercier}}, \binits{K.}},
\bauthor{\bsnm{{Gonzalez}}, \binits{F.}}:
In: \bbtitle{\procspie}.
\bsertitle{Society of Photo-Optical Instrumentation Engineers (SPIE) Conference
  Series},
vol. \bseriesno{9144},
p. \bfpage{914423}
(\byear{2014}).
\arxivurl{1407.2406}.
doi:\doiurl{10.1117/12.2054898}.
\burl{https://ui.adsabs.harvard.edu/abs/2014SPIE.9144E..23G}
\end{bchapter}
\endbibitem

\bibitem[\protect\citeauthoryear{Granot and Loeb}{2003}]{Granot2003}
\begin{botherref}
\oauthor{\bsnm{Granot}, \binits{J.}},
\oauthor{\bsnm{Loeb}, \binits{A.}}:
Radio imaging of gamma-ray burst jets in nearby supernovae
\textbf{593},
81
(2003).
doi:\doiurl{10.1086/378262}
\end{botherref}
\endbibitem

\bibitem[\protect\citeauthoryear{{Greiner} et~al.}{2003}]{Greiner2003_030329}
\begin{barticle}
\bauthor{\bsnm{{Greiner}}, \binits{J.}},
\bauthor{\bsnm{{Peimbert}}, \binits{M.}},
\bauthor{\bsnm{{Esteban}}, \binits{C.}},
\bauthor{\bsnm{{Kaufer}}, \binits{A.}},
\bauthor{\bsnm{{Jaunsen}}, \binits{A.}},
\bauthor{\bsnm{{Smoke}}, \binits{J.}},
\bauthor{\bsnm{{Klose}}, \binits{S.}},
\bauthor{\bsnm{{Reimer}}, \binits{O.}}:
\bjtitle{GRB Coordinates Network}
\bvolume{2020},
\bfpage{1}
(\byear{2003})
\end{barticle}
\endbibitem

\bibitem[\protect\citeauthoryear{{Guidorzi} and
  {Virgili}}{2013}]{Guidorzi2013_130702A}
\begin{barticle}
\bauthor{\bsnm{{Guidorzi}}, \binits{C.}},
\bauthor{\bsnm{{Virgili}}, \binits{F.}}:
\bjtitle{GRB Coordinates Network}
\bvolume{15050},
\bfpage{1}
(\byear{2013})
\end{barticle}
\endbibitem

\bibitem[\protect\citeauthoryear{Hallinan et~al.}{2017}]{Hallinan2017_170817A}
\begin{barticle}
\bauthor{\bsnm{Hallinan}, \binits{G.}},
\bauthor{\bsnm{Corsi}, \binits{A.}},
\bauthor{\bsnm{Mooley}, \binits{K.P.}},
\bauthor{\bsnm{Hotokezaka}, \binits{K.}},
\bauthor{\bsnm{Nakar}, \binits{E.}},
\bauthor{\bsnm{Kasliwal}, \binits{M.M.}},
\bauthor{\bsnm{Kaplan}, \binits{D.L.}},
\bauthor{\bsnm{Frail}, \binits{D.A.}},
\bauthor{\bsnm{Myers}, \binits{S.T.}},
\bauthor{\bsnm{Murphy}, \binits{T.}},
\bauthor{\bsnm{De}, \binits{K.}},
\bauthor{\bsnm{Dobie}, \binits{D.}},
\bauthor{\bsnm{Allison}, \binits{J.R.}},
\bauthor{\bsnm{Bannister}, \binits{K.W.}},
\bauthor{\bsnm{Bhalerao}, \binits{V.}},
\bauthor{\bsnm{Chandra}, \binits{P.}},
\bauthor{\bsnm{Clarke}, \binits{T.E.}},
\bauthor{\bsnm{Giacintucci}, \binits{S.}},
\bauthor{\bsnm{Ho}, \binits{A.Y.Q.}},
\bauthor{\bsnm{Horesh}, \binits{A.}},
\bauthor{\bsnm{Kassim}, \binits{N.E.}},
\bauthor{\bsnm{Kulkarni}, \binits{S.R.}},
\bauthor{\bsnm{Lenc}, \binits{E.}},
\bauthor{\bsnm{Lockman}, \binits{F.J.}},
\bauthor{\bsnm{Lynch}, \binits{C.}},
\bauthor{\bsnm{Nichols}, \binits{D.}},
\bauthor{\bsnm{Nissanke}, \binits{S.}},
\bauthor{\bsnm{Palliyaguru}, \binits{N.}},
\bauthor{\bsnm{Peters}, \binits{W.M.}},
\bauthor{\bsnm{Piran}, \binits{T.}},
\bauthor{\bsnm{Rana}, \binits{J.}},
\bauthor{\bsnm{Sadler}, \binits{E.M.}},
\bauthor{\bsnm{Singer}, \binits{L.P.}}:
\bjtitle{Science}
\bvolume{358},
\bfpage{1579}
(\byear{2017}).
doi:\doiurl{10.1126/science.aap9855}
\end{barticle}
\endbibitem

\bibitem[\protect\citeauthoryear{Halpern et~al.}{2005}]{Halpern2005_050826}
\begin{barticle}
\bauthor{\bsnm{Halpern}, \binits{J.P.}},
\bauthor{\bsnm{Halpern}},
\bauthor{\bsnm{P.}, \binits{J.}}:
\bjtitle{GCN}
\bvolume{3891},
\bfpage{1}
(\byear{2005})
\end{barticle}
\endbibitem

\bibitem[\protect\citeauthoryear{Halpern et~al.}{2006}]{Halpern2006_050826}
\begin{barticle}
\bauthor{\bsnm{Halpern}, \binits{J.P.}},
\bauthor{\bsnm{Mirabal}, \binits{N.}},
\bauthor{\bsnm{Halpern}, \binits{J.P.}},
\bauthor{\bsnm{Mirabal}, \binits{N.}}:
\bjtitle{GCN}
\bvolume{5982},
\bfpage{1}
(\byear{2006})
\end{barticle}
\endbibitem

\bibitem[\protect\citeauthoryear{{Heise} et~al.}{2001}]{Heise2001}
\begin{bchapter}
\bauthor{\bsnm{{Heise}}, \binits{J.}},
\bauthor{\bsnm{{Zand}}, \binits{J.I.}},
\bauthor{\bsnm{{Kippen}}, \binits{R.M.}},
\bauthor{\bsnm{{Woods}}, \binits{P.M.}}:
\bctitle{{X-Ray Flashes and X-Ray Rich Gamma Ray Bursts}},
p. \bfpage{16}
(\byear{2001}).
doi:\doiurl{10.1007/10853853}
\end{bchapter}
\endbibitem

\bibitem[\protect\citeauthoryear{{Heussaff} et~al.}{2013}]{Heussaff2013}
\begin{barticle}
\bauthor{\bsnm{{Heussaff}}, \binits{V.}},
\bauthor{\bsnm{{Atteia}}, \binits{J.-L.}},
\bauthor{\bsnm{{Zolnierowski}}, \binits{Y.}}:
\bjtitle{\aap}
\bvolume{557},
\bfpage{100}
(\byear{2013}).
\arxivurl{1306.1757}.
doi:\doiurl{10.1051/0004-6361/201321528}
\end{barticle}
\endbibitem

\bibitem[\protect\citeauthoryear{Hjorth et~al.}{2003}]{Hjorth2003_030329}
\begin{barticle}
\bauthor{\bsnm{Hjorth}, \binits{J.}},
\bauthor{\bsnm{Sollerman}, \binits{J.}},
\bauthor{\bsnm{Møller}, \binits{P.}},
\bauthor{\bsnm{Fynbo}, \binits{J.P.U.}},
\bauthor{\bsnm{Woosley}, \binits{S.E.}},
\bauthor{\bsnm{Kouveliotou}, \binits{C.}},
\bauthor{\bsnm{Tanvir}, \binits{N.R.}},
\bauthor{\bsnm{Greiner}, \binits{J.}},
\bauthor{\bsnm{Andersen}, \binits{M.I.}},
\bauthor{\bsnm{Castro-Tirado}, \binits{A.J.}},
\bauthor{\bsnm{Cerón}, \binits{J.M.C.}},
\bauthor{\bsnm{Fruchter}, \binits{A.S.}},
\bauthor{\bsnm{Gorosabel}, \binits{J.}},
\bauthor{\bsnm{Jakobsson}, \binits{P.}},
\bauthor{\bsnm{Kaper}, \binits{L.}},
\bauthor{\bsnm{Klose}, \binits{S.}},
\bauthor{\bsnm{Masetti}, \binits{N.}},
\bauthor{\bsnm{Pedersen}, \binits{H.}},
\bauthor{\bsnm{Pedersen}, \binits{K.}},
\bauthor{\bsnm{Pian}, \binits{E.}},
\bauthor{\bsnm{Palazzi}, \binits{E.}},
\bauthor{\bsnm{Rhoads}, \binits{J.E.}},
\bauthor{\bsnm{Rol}, \binits{E.}},
\bauthor{\bparticle{van~den} \bsnm{Heuvel}, \binits{E.P.J.}},
\bauthor{\bsnm{Vreeswijk}, \binits{P.M.}},
\bauthor{\bsnm{Watson}, \binits{D.}},
\bauthor{\bsnm{Wijers}, \binits{R.A.M.J.}}:
\bjtitle{\nat}
\bvolume{423},
\bfpage{847}
(\byear{2003}).
doi:\doiurl{10.1038/nature01750}
\end{barticle}
\endbibitem

\bibitem[\protect\citeauthoryear{Hjorth et~al.}{2005}]{Hjorth2005_050709}
\begin{barticle}
\bauthor{\bsnm{Hjorth}, \binits{J.}},
\bauthor{\bsnm{Watson}, \binits{D.}},
\bauthor{\bsnm{Fynbo}, \binits{J.P.U.}},
\bauthor{\bsnm{Price}, \binits{P.A.}},
\bauthor{\bsnm{Jensen}, \binits{B.L.}},
\bauthor{\bsnm{Jørgensen}, \binits{U.G.}},
\bauthor{\bsnm{Kubas}, \binits{D.}},
\bauthor{\bsnm{Gorosabel}, \binits{J.}},
\bauthor{\bsnm{Jakobsson}, \binits{P.}},
\bauthor{\bsnm{Sollerman}, \binits{J.}},
\bauthor{\bsnm{Pedersen}, \binits{K.}},
\bauthor{\bsnm{Kouveliotou}, \binits{C.}}:
\bjtitle{\nat}
\bvolume{437},
\bfpage{859}
(\byear{2005}).
doi:\doiurl{10.1038/nature04174}
\end{barticle}
\endbibitem

\bibitem[\protect\citeauthoryear{Holland and
  Marshall}{2006}]{Holland2006_061201}
\begin{barticle}
\bauthor{\bsnm{Holland}, \binits{S.T.}},
\bauthor{\bsnm{Marshall}, \binits{F.E.}}:
\bjtitle{GRB Coordinates Network}
\bvolume{5898},
\bfpage{1}
(\byear{2006})
\end{barticle}
\endbibitem

\bibitem[\protect\citeauthoryear{Howell et~al.}{2019}]{Howell2019_170817}
\begin{botherref}
\oauthor{\bsnm{Howell}, \binits{E.J.}},
\oauthor{\bsnm{Ackley}, \binits{K.}},
\oauthor{\bsnm{Rowlinson}, \binits{A.}},
\oauthor{\bsnm{Coward}, \binits{D.}}:
Joint gravitational wave - gamma-ray burst detection rates in the aftermath of
  gw170817
\textbf{485},
1435
(2019).
doi:\doiurl{10.1093/mnras/stz455}
\end{botherref}
\endbibitem

\bibitem[\protect\citeauthoryear{{Hurley} et~al.}{1999}]{Hurley1999_980827}
\begin{barticle}
\bauthor{\bsnm{{Hurley}}, \binits{K.}},
\bauthor{\bsnm{{Cline}}, \binits{T.}},
\bauthor{\bsnm{{Mazets}}, \binits{E.}},
\bauthor{\bsnm{{Barthelmy}}, \binits{S.}},
\bauthor{\bsnm{{Butterworth}}, \binits{P.}},
\bauthor{\bsnm{{Marshall}}, \binits{F.}},
\bauthor{\bsnm{{Palmer}}, \binits{D.}},
\bauthor{\bsnm{{Aptekar}}, \binits{R.}},
\bauthor{\bsnm{{Golenetskii}}, \binits{S.}},
\bauthor{\bsnm{{Il'Inskii}}, \binits{V.}},
\bauthor{\bsnm{{Frederiks}}, \binits{D.}},
\bauthor{\bsnm{{McTiernan}}, \binits{J.}},
\bauthor{\bsnm{{Gold}}, \binits{R.}},
\bauthor{\bsnm{{Trombka}}, \binits{J.}}:
\bjtitle{\nat}
\bvolume{397}(\bissue{6714}),
\bfpage{41}
(\byear{1999}).
\arxivurl{astro-ph/9811443}.
doi:\doiurl{10.1038/16199}
\end{barticle}
\endbibitem

\bibitem[\protect\citeauthoryear{Hurley et~al.}{2005a}]{Hurley2005_180620}
\begin{barticle}
\bauthor{\bsnm{Hurley}, \binits{K.}},
\bauthor{\bsnm{Boggs}, \binits{S.E.}},
\bauthor{\bsnm{Smith}, \binits{D.M.}},
\bauthor{\bsnm{Duncan}, \binits{R.C.}},
\bauthor{\bsnm{Lin}, \binits{R.}},
\bauthor{\bsnm{Zoglauer}, \binits{A.}},
\bauthor{\bsnm{Krucker}, \binits{S.}},
\bauthor{\bsnm{Hurford}, \binits{G.}},
\bauthor{\bsnm{Hudson}, \binits{H.}},
\bauthor{\bsnm{Wigger}, \binits{C.}},
\bauthor{\bsnm{Hajdas}, \binits{W.}},
\bauthor{\bsnm{Thompson}, \binits{C.}},
\bauthor{\bsnm{Mitrofanov}, \binits{I.}},
\bauthor{\bsnm{Sanin}, \binits{A.}},
\bauthor{\bsnm{Boynton}, \binits{W.}},
\bauthor{\bsnm{Fellows}, \binits{C.}},
\bauthor{\bparticle{von} \bsnm{Kienlin}, \binits{A.}},
\bauthor{\bsnm{Lichti}, \binits{G.}},
\bauthor{\bsnm{Rau}, \binits{A.}},
\bauthor{\bsnm{Cline}, \binits{T.}}:
\bjtitle{Nature}
\bvolume{434}(\bissue{7037}),
\bfpage{1098}
(\byear{2005}a).
\arxivurl{0502329}.
doi:\doiurl{10.1038/nature03519}
\end{barticle}
\endbibitem

\bibitem[\protect\citeauthoryear{Hurley et~al.}{2005b}]{Hurley2005_041227}
\begin{barticle}
\bauthor{\bsnm{Hurley}, \binits{K.}},
\bauthor{\bsnm{Boggs}, \binits{S.E.}},
\bauthor{\bsnm{Smith}, \binits{D.M.}},
\bauthor{\bsnm{Duncan}, \binits{R.C.}},
\bauthor{\bsnm{Lin}, \binits{R.}},
\bauthor{\bsnm{Zoglauer}, \binits{A.}},
\bauthor{\bsnm{Krucker}, \binits{S.}},
\bauthor{\bsnm{Hurford}, \binits{G.}},
\bauthor{\bsnm{Hudson}, \binits{H.}},
\bauthor{\bsnm{Wigger}, \binits{C.}},
\bauthor{\bsnm{Hajdas}, \binits{W.}},
\bauthor{\bsnm{Thompson}, \binits{C.}},
\bauthor{\bsnm{Mitrofanov}, \binits{I.}},
\bauthor{\bsnm{Sanin}, \binits{A.}},
\bauthor{\bsnm{Boynton}, \binits{W.}},
\bauthor{\bsnm{Fellows}, \binits{C.}},
\bauthor{\bparticle{von} \bsnm{Kienlin}, \binits{A.}},
\bauthor{\bsnm{Lichti}, \binits{G.}},
\bauthor{\bsnm{Rau}, \binits{A.}},
\bauthor{\bsnm{Cline}, \binits{T.}}:
\bjtitle{\nat}
\bvolume{434},
\bfpage{1098}
(\byear{2005}b).
doi:\doiurl{10.1038/nature03519}
\end{barticle}
\endbibitem

\bibitem[\protect\citeauthoryear{{Hurley} et~al.}{2010}]{Hurley2010_051103}
\begin{barticle}
\bauthor{\bsnm{{Hurley}}, \binits{K.}},
\bauthor{\bsnm{{Rowlinson}}, \binits{A.}},
\bauthor{\bsnm{{Bellm}}, \binits{E.}},
\bauthor{\bsnm{{Perley}}, \binits{D.}},
\bauthor{\bsnm{{Mitrofanov}}, \binits{I.G.}},
\bauthor{\bsnm{{Golovin}}, \binits{D.V.}},
\bauthor{\bsnm{{Kozyrev}}, \binits{A.S.}},
\bauthor{\bsnm{{Litvak}}, \binits{M.L.}},
\bauthor{\bsnm{{Sanin}}, \binits{A.B.}},
\bauthor{\bsnm{{Boynton}}, \binits{W.}},
\bauthor{\bsnm{{Fellows}}, \binits{C.}},
\bauthor{\bsnm{{Harshmann}}, \binits{K.}},
\bauthor{\bsnm{{Ohno}}, \binits{M.}},
\bauthor{\bsnm{{Yamaoka}}, \binits{K.}},
\bauthor{\bsnm{{Nakagawa}}, \binits{Y.E.}},
\bauthor{\bsnm{{Smith}}, \binits{D.M.}},
\bauthor{\bsnm{{Cline}}, \binits{T.}},
\bauthor{\bsnm{{Tanvir}}, \binits{N.R.}},
\bauthor{\bsnm{{O'Brien}}, \binits{P.T.}},
\bauthor{\bsnm{{Wiersema}}, \binits{K.}},
\bauthor{\bsnm{{Rol}}, \binits{E.}},
\bauthor{\bsnm{{Levan}}, \binits{A.}},
\bauthor{\bsnm{{Rhoads}}, \binits{J.}},
\bauthor{\bsnm{{Fruchter}}, \binits{A.}},
\bauthor{\bsnm{{Bersier}}, \binits{D.}},
\bauthor{\bsnm{{Kavelaars}}, \binits{J.J.}},
\bauthor{\bsnm{{Gehrels}}, \binits{N.}},
\bauthor{\bsnm{{Krimm}}, \binits{H.}},
\bauthor{\bsnm{{Palmer}}, \binits{D.M.}},
\bauthor{\bsnm{{Duncan}}, \binits{R.C.}},
\bauthor{\bsnm{{Wigger}}, \binits{C.}},
\bauthor{\bsnm{{Hajdas}}, \binits{W.}},
\bauthor{\bsnm{{Atteia}}, \binits{J.-L.}},
\bauthor{\bsnm{{Ricker}}, \binits{G.}},
\bauthor{\bsnm{{Vanderspek}}, \binits{R.}},
\bauthor{\bsnm{{Rau}}, \binits{A.}},
\bauthor{\bsnm{{von Kienlin}}, \binits{A.}}:
\bjtitle{\mnras}
\bvolume{403}(\bissue{1}),
\bfpage{342}
(\byear{2010}).
\arxivurl{0907.2462}.
doi:\doiurl{10.1111/j.1365-2966.2009.16118.x}
\end{barticle}
\endbibitem

\bibitem[\protect\citeauthoryear{Hurley}{2011a}]{Hurley2011}
\begin{barticle}
\bauthor{\bsnm{Hurley}, \binits{K.}}:
\bjtitle{Advances in Space Research}
\bvolume{47},
\bfpage{1337}
(\byear{2011}a).
doi:\doiurl{10.1016/j.asr.2010.08.036}
\end{barticle}
\endbibitem

\bibitem[\protect\citeauthoryear{Hurley}{2011b}]{Hurley2011_SGR}
\begin{barticle}
\bauthor{\bsnm{Hurley}, \binits{K.}}:
\bjtitle{Advances in Space Research}
\bvolume{47},
\bfpage{1326}
(\byear{2011}b).
doi:\doiurl{10.1016/j.asr.2010.03.001}
\end{barticle}
\endbibitem

\bibitem[\protect\citeauthoryear{Izzo et~al.}{2017}]{Izzo2017_171205A}
\begin{barticle}
\bauthor{\bsnm{Izzo}, \binits{L.}},
\bauthor{\bsnm{Kann}, \binits{D.A.}},
\bauthor{\bsnm{Fynbo}, \binits{J.P.U.}},
\bauthor{\bsnm{Levan}, \binits{A.J.}},
\bauthor{\bsnm{Tanvir}, \binits{N.R.}},
\bauthor{\bsnm{Izzo}, \binits{L.}},
\bauthor{\bsnm{Kann}, \binits{D.A.}},
\bauthor{\bsnm{Fynbo}, \binits{J.P.U.}},
\bauthor{\bsnm{Levan}, \binits{A.J.}},
\bauthor{\bsnm{Tanvir}, \binits{N.R.}}:
\bjtitle{GCN}
\bvolume{22178},
\bfpage{1}
(\byear{2017})
\end{barticle}
\endbibitem

\bibitem[\protect\citeauthoryear{Izzo et~al.}{2018}]{Izzo2018_180728A}
\begin{barticle}
\bauthor{\bsnm{Izzo}, \binits{L.}},
\bauthor{\bsnm{Rossi}, \binits{A.}},
\bauthor{\bsnm{Malesani}, \binits{D.B.}},
\bauthor{\bsnm{Heintz}, \binits{K.E.}},
\bauthor{\bsnm{Selsing}, \binits{J.}},
\bauthor{\bsnm{Schady}, \binits{P.}},
\bauthor{\bsnm{Starling}, \binits{R.L.C.}},
\bauthor{\bsnm{Sollerman}, \binits{J.}},
\bauthor{\bsnm{Leloudas}, \binits{G.}},
\bauthor{\bsnm{Cano}, \binits{Z.}},
\bauthor{\bsnm{Fynbo}, \binits{J.P.U.}},
\bauthor{\bsnm{Valle}, \binits{M.D.}},
\bauthor{\bsnm{Pian}, \binits{E.}},
\bauthor{\bsnm{Kann}, \binits{D.A.}},
\bauthor{\bsnm{Perley}, \binits{D.A.}},
\bauthor{\bsnm{Palazzi}, \binits{E.}},
\bauthor{\bsnm{Klose}, \binits{S.}},
\bauthor{\bsnm{Hjorth}, \binits{J.}},
\bauthor{\bsnm{Covino}, \binits{S.}},
\bauthor{\bsnm{D'Elia}, \binits{V.}},
\bauthor{\bsnm{Tanvir}, \binits{N.R.}},
\bauthor{\bsnm{Levan}, \binits{A.J.}},
\bauthor{\bsnm{Hartmann}, \binits{D.}},
\bauthor{\bsnm{Kouveliotou}, \binits{C.}}:
\bjtitle{GRB Coordinates Network}
\bvolume{23142},
\bfpage{1}
(\byear{2018})
\end{barticle}
\endbibitem

\bibitem[\protect\citeauthoryear{{Jin} et~al.}{2016}]{Jin2016_050709}
\begin{barticle}
\bauthor{\bsnm{{Jin}}, \binits{Z.-P.}},
\bauthor{\bsnm{{Hotokezaka}}, \binits{K.}},
\bauthor{\bsnm{{Li}}, \binits{X.}},
\bauthor{\bsnm{{Tanaka}}, \binits{M.}},
\bauthor{\bsnm{{D'Avanzo}}, \binits{P.}},
\bauthor{\bsnm{{Fan}}, \binits{Y.-Z.}},
\bauthor{\bsnm{{Covino}}, \binits{S.}},
\bauthor{\bsnm{{Wei}}, \binits{D.-M.}},
\bauthor{\bsnm{{Piran}}, \binits{T.}}:
\bjtitle{Nature Communications}
\bvolume{7},
\bfpage{12898}
(\byear{2016}).
\arxivurl{1603.07869}.
doi:\doiurl{10.1038/ncomms12898}
\end{barticle}
\endbibitem

\bibitem[\protect\citeauthoryear{Jin et~al.}{2020}]{Jin2020_070809}
\begin{barticle}
\bauthor{\bsnm{Jin}, \binits{Z.-P.}},
\bauthor{\bsnm{Covino}, \binits{S.}},
\bauthor{\bsnm{Liao}, \binits{N.-H.}},
\bauthor{\bsnm{Li}, \binits{X.}},
\bauthor{\bsnm{D'Avanzo}, \binits{P.}},
\bauthor{\bsnm{Fan}, \binits{Y.-Z.}},
\bauthor{\bsnm{Wei}, \binits{D.-M.}}:
\bjtitle{Nature Astronomy}
\bvolume{4},
\bfpage{77}
(\byear{2020}).
doi:\doiurl{10.1038/s41550-019-0892-y}
\end{barticle}
\endbibitem

\bibitem[\protect\citeauthoryear{{Kamble}}{2015}]{Kamble2015_150518A}
\begin{barticle}
\bauthor{\bsnm{{Kamble}}, \binits{A.}}:
\bjtitle{GRB Coordinates Network}
\bvolume{17859},
\bfpage{1}
(\byear{2015})
\end{barticle}
\endbibitem

\bibitem[\protect\citeauthoryear{{Karachentsev} and
  {Kashibadze}}{2006}]{Karachentsev2006_M81}
\begin{barticle}
\bauthor{\bsnm{{Karachentsev}}, \binits{I.D.}},
\bauthor{\bsnm{{Kashibadze}}, \binits{O.G.}}:
\bjtitle{Astrophysics}
\bvolume{49}(\bissue{1}),
\bfpage{3}
(\byear{2006}).
doi:\doiurl{10.1007/s10511-006-0002-6}
\end{barticle}
\endbibitem

\bibitem[\protect\citeauthoryear{{Kelson} et~al.}{2004}]{Kelson2004_040701}
\begin{barticle}
\bauthor{\bsnm{{Kelson}}, \binits{D.D.}},
\bauthor{\bsnm{{Koviak}}, \binits{K.}},
\bauthor{\bsnm{{Berger}}, \binits{E.}},
\bauthor{\bsnm{{Fox}}, \binits{D.B.}}:
\bjtitle{GRB Coordinates Network}
\bvolume{2627},
\bfpage{1}
(\byear{2004})
\end{barticle}
\endbibitem

\bibitem[\protect\citeauthoryear{Kennea et~al.}{2017}]{Kennea2017_171205A}
\begin{barticle}
\bauthor{\bsnm{Kennea}, \binits{J.A.}},
\bauthor{\bsnm{Sbarufatti}, \binits{B.}},
\bauthor{\bsnm{Burrows}, \binits{D.N.}},
\bauthor{\bsnm{Evans}, \binits{P.A.}},
\bauthor{\bsnm{Gibson}, \binits{S.L.}},
\bauthor{\bsnm{Osborne}, \binits{J.P.}},
\bauthor{\bsnm{ri}, \binits{A.M.}},
\bauthor{\bsnm{D'Avanzo}, \binits{P.}},
\bauthor{\bsnm{D'Elia}, \binits{V.}}:
\bjtitle{GRB Coordinates Network}
\bvolume{22183},
\bfpage{1}
(\byear{2017})
\end{barticle}
\endbibitem

\bibitem[\protect\citeauthoryear{{Kippen} et~al.}{2003}]{Kippen2003}
\begin{bchapter}
\bauthor{\bsnm{{Kippen}}, \binits{R.M.}},
\bauthor{\bsnm{{Woods}}, \binits{P.M.}},
\bauthor{\bsnm{{Heise}}, \binits{J.}},
\bauthor{\bsnm{{in't Zand }}, \binits{J.J.M.}},
\bauthor{\bsnm{{Briggs}}, \binits{M.S.}},
\bauthor{\bsnm{{Preece}}, \binits{R.D.}}:
In: \beditor{\bsnm{{Ricker}}, \binits{G.R.}},
\beditor{\bsnm{{Vanderspek}}, \binits{R.K.}} (eds.)
\bbtitle{Gamma-Ray Burst and Afterglow Astronomy 2001: A Workshop Celebrating
  the First Year of the HETE Mission}.
\bsertitle{American Institute of Physics Conference Series},
vol. \bseriesno{662},
p. \bfpage{244}
(\byear{2003}).
\arxivurl{astro-ph/0203114}.
doi:\doiurl{10.1063/1.1579349}.
\burl{https://ui.adsabs.harvard.edu/abs/2003AIPC..662..244K}
\end{bchapter}
\endbibitem

\bibitem[\protect\citeauthoryear{{Krimm} et~al.}{2009}]{Krimm2009_060505}
\begin{barticle}
\bauthor{\bsnm{{Krimm}}, \binits{H.A.}},
\bauthor{\bsnm{{Yamaoka}}, \binits{K.}},
\bauthor{\bsnm{{Sugita}}, \binits{S.}},
\bauthor{\bsnm{{Ohno}}, \binits{M.}},
\bauthor{\bsnm{{Sakamoto}}, \binits{T.}},
\bauthor{\bsnm{{Barthelmy}}, \binits{S.D.}},
\bauthor{\bsnm{{Gehrels}}, \binits{N.}},
\bauthor{\bsnm{{Hara}}, \binits{R.}},
\bauthor{\bsnm{{Norris}}, \binits{J.P.}},
\bauthor{\bsnm{{Ohmori}}, \binits{N.}},
\bauthor{\bsnm{{Onda}}, \binits{K.}},
\bauthor{\bsnm{{Sato}}, \binits{G.}},
\bauthor{\bsnm{{Tanaka}}, \binits{H.}},
\bauthor{\bsnm{{Tashiro}}, \binits{M.}},
\bauthor{\bsnm{{Yamauchi}}, \binits{M.}}:
\bjtitle{\apj}
\bvolume{704}(\bissue{2}),
\bfpage{1405}
(\byear{2009}).
\arxivurl{0908.1335}.
doi:\doiurl{10.1088/0004-637X/704/2/1405}
\end{barticle}
\endbibitem

\bibitem[\protect\citeauthoryear{Kruehler et~al.}{2016}]{Kruehler2016_161219B}
\begin{barticle}
\bauthor{\bsnm{Kruehler}, \binits{T.}},
\bauthor{\bsnm{Wiseman}, \binits{P.}},
\bauthor{\bsnm{Greiner}, \binits{J.}}:
\bjtitle{GRB Coordinates Network}
\bvolume{20299},
\bfpage{1}
(\byear{2016})
\end{barticle}
\endbibitem

\bibitem[\protect\citeauthoryear{Kuin and Troja}{2012}]{Kuin2012_120422A}
\begin{barticle}
\bauthor{\bsnm{Kuin}, \binits{N.P.M.}},
\bauthor{\bsnm{Troja}, \binits{E.}}:
\bjtitle{GRB Coordinates Network}
\bvolume{13248},
\bfpage{1}
(\byear{2012})
\end{barticle}
\endbibitem

\bibitem[\protect\citeauthoryear{{Lacombe} et~al.}{2018}]{Lacombe2018}
\begin{barticle}
\bauthor{\bsnm{{Lacombe}}, \binits{K.}},
\bauthor{\bsnm{{Dezalay}}, \binits{J.-P.}},
\bauthor{\bsnm{{Houret}}, \binits{B.}},
\bauthor{\bsnm{{Amoros}}, \binits{C.}},
\bauthor{\bsnm{{Atteia}}, \binits{J.-L.}},
\bauthor{\bsnm{{Bouchet}}, \binits{L.}},
\bauthor{\bsnm{{Cordier}}, \binits{B.}},
\bauthor{\bsnm{{Gevin}}, \binits{O.}},
\bauthor{\bsnm{{Godet}}, \binits{O.}},
\bauthor{\bsnm{{Gonzalez}}, \binits{F.}},
\bauthor{\bsnm{{Guillemot}}, \binits{P.}},
\bauthor{\bsnm{{Limousin}}, \binits{O.}},
\bauthor{\bsnm{{Pons}}, \binits{R.}},
\bauthor{\bsnm{{Ramon}}, \binits{P.}},
\bauthor{\bsnm{{Waegebaert}}, \binits{V.}}:
\bjtitle{Astroparticle Physics}
\bvolume{103},
\bfpage{131}
(\byear{2018}).
doi:\doiurl{10.1016/j.astropartphys.2018.08.002}
\end{barticle}
\endbibitem

\bibitem[\protect\citeauthoryear{{Lamb} et~al.}{2005}]{Lamb2005}
\begin{barticle}
\bauthor{\bsnm{{Lamb}}, \binits{D.Q.}},
\bauthor{\bsnm{{Donaghy}}, \binits{T.Q.}},
\bauthor{\bsnm{{Graziani}}, \binits{C.}}:
\bjtitle{\apj}
\bvolume{620}(\bissue{1}),
\bfpage{355}
(\byear{2005}).
\arxivurl{astro-ph/0312634}.
doi:\doiurl{10.1086/426099}
\end{barticle}
\endbibitem

\bibitem[\protect\citeauthoryear{Lamb et~al.}{2019}]{Lamb2019_160821B}
\begin{barticle}
\bauthor{\bsnm{Lamb}, \binits{G.P.}},
\bauthor{\bsnm{Tanvir}, \binits{N.R.}},
\bauthor{\bsnm{Levan}, \binits{A.J.}},
\bauthor{\bparticle{de} \bsnm{Ugarte~Postigo}, \binits{A.}},
\bauthor{\bsnm{Kawaguchi}, \binits{K.}},
\bauthor{\bsnm{Corsi}, \binits{A.}},
\bauthor{\bsnm{Evans}, \binits{P.A.}},
\bauthor{\bsnm{Gompertz}, \binits{B.}},
\bauthor{\bsnm{Malesani}, \binits{D.B.}},
\bauthor{\bsnm{Page}, \binits{K.L.}},
\bauthor{\bsnm{Wiersema}, \binits{K.}},
\bauthor{\bsnm{Rosswog}, \binits{S.}},
\bauthor{\bsnm{Shibata}, \binits{M.}},
\bauthor{\bsnm{Tanaka}, \binits{M.}},
\bauthor{\bparticle{van~der} \bsnm{Horst}, \binits{A.J.}},
\bauthor{\bsnm{Cano}, \binits{Z.}},
\bauthor{\bsnm{Fynbo}, \binits{J.P.U.}},
\bauthor{\bsnm{Fruchter}, \binits{A.S.}},
\bauthor{\bsnm{Greiner}, \binits{J.}},
\bauthor{\bsnm{Heintz}, \binits{K.E.}},
\bauthor{\bsnm{Higgins}, \binits{A.}},
\bauthor{\bsnm{Hjorth}, \binits{J.}},
\bauthor{\bsnm{Izzo}, \binits{L.}},
\bauthor{\bsnm{Jakobsson}, \binits{P.}},
\bauthor{\bsnm{Kann}, \binits{D.A.}},
\bauthor{\bsnm{O'Brien}, \binits{P.T.}},
\bauthor{\bsnm{Perley}, \binits{D.A.}},
\bauthor{\bsnm{Pian}, \binits{E.}},
\bauthor{\bsnm{Pugliese}, \binits{G.}},
\bauthor{\bsnm{Starling}, \binits{R.L.C.}},
\bauthor{\bsnm{Th\"one}, \binits{C.C.}},
\bauthor{\bsnm{Watson}, \binits{D.}},
\bauthor{\bsnm{Wijers}, \binits{R.A.M.J.}},
\bauthor{\bsnm{Xu}, \binits{D.}}:
\bjtitle{\apj}
\bvolume{883},
\bfpage{48}
(\byear{2019}).
doi:\doiurl{10.3847/1538-4357/ab38bb}
\end{barticle}
\endbibitem

\bibitem[\protect\citeauthoryear{Laporte and
  Starling}{2018}]{Laporte2018_180728A}
\begin{barticle}
\bauthor{\bsnm{Laporte}, \binits{S.J.}},
\bauthor{\bsnm{Starling}, \binits{R.L.C.}}:
\bjtitle{GRB Coordinates Network}
\bvolume{23064},
\bfpage{1}
(\byear{2018})
\end{barticle}
\endbibitem

\bibitem[\protect\citeauthoryear{Leloudas et~al.}{2013}]{Leloudas2013_130702A}
\begin{barticle}
\bauthor{\bsnm{Leloudas}, \binits{G.}},
\bauthor{\bsnm{Fynbo}, \binits{J.P.U.}},
\bauthor{\bsnm{Schulze}, \binits{S.}},
\bauthor{\bsnm{Xu}, \binits{D.}},
\bauthor{\bsnm{Malesani}, \binits{D.}},
\bauthor{\bsnm{Geier}, \binits{S.}},
\bauthor{\bsnm{Cano}, \binits{Z.}},
\bauthor{\bsnm{Jakobsson}, \binits{P.}},
\bauthor{\bsnm{Leloudas}, \binits{G.}},
\bauthor{\bsnm{Fynbo}, \binits{J.P.U.}},
\bauthor{\bsnm{Schulze}, \binits{S.}},
\bauthor{\bsnm{Xu}, \binits{D.}},
\bauthor{\bsnm{Malesani}, \binits{D.}},
\bauthor{\bsnm{Geier}, \binits{S.}},
\bauthor{\bsnm{Cano}, \binits{Z.}},
\bauthor{\bsnm{Jakobsson}, \binits{P.}}:
\bjtitle{GCN}
\bvolume{14983},
\bfpage{1}
(\byear{2013})
\end{barticle}
\endbibitem

\bibitem[\protect\citeauthoryear{Levan et~al.}{2007}]{Levan2007_050906}
\begin{barticle}
\bauthor{\bsnm{Levan}, \binits{A.J.}},
\bauthor{\bsnm{Tanvir}, \binits{N.R.}},
\bauthor{\bsnm{Jakobsson}, \binits{P.}},
\bauthor{\bsnm{Chapman}, \binits{R.}},
\bauthor{\bsnm{Hjorth}, \binits{J.}},
\bauthor{\bsnm{Priddey}, \binits{R.S.}},
\bauthor{\bsnm{Fynbo}, \binits{J.P.U.}},
\bauthor{\bsnm{Hurley}, \binits{K.}},
\bauthor{\bsnm{Jensen}, \binits{B.L.}},
\bauthor{\bsnm{Johnson}, \binits{R.}},
\bauthor{\bsnm{Gorosabel}, \binits{J.}},
\bauthor{\bsnm{Castro-Tirado}, \binits{A.J.}},
\bauthor{\bsnm{Jarvis}, \binits{M.}},
\bauthor{\bsnm{Watson}, \binits{D.}},
\bauthor{\bsnm{Wiersema}, \binits{K.}}:
\bjtitle{\mnras}
\bvolume{384}(\bissue{2}),
\bfpage{541}
(\byear{2007}).
\arxivurl{0705.1705}.
doi:\doiurl{10.1111/j.1365-2966.2007.11953.x}
\end{barticle}
\endbibitem

\bibitem[\protect\citeauthoryear{Levan et~al.}{2011}]{Levan2011_111005A}
\begin{barticle}
\bauthor{\bsnm{Levan}, \binits{A.J.}},
\bauthor{\bsnm{Tanvir}, \binits{N.R.}},
\bauthor{\bsnm{Wiersema}, \binits{K.}},
\bauthor{\bsnm{O'Brien}, \binits{P.T.}}:
\bjtitle{GRB Coordinates Network}
\bvolume{12414},
\bfpage{1}
(\byear{2011})
\end{barticle}
\endbibitem

\bibitem[\protect\citeauthoryear{{Levan} et~al.}{2014}]{Levan2014}
\begin{barticle}
\bauthor{\bsnm{{Levan}}, \binits{A.J.}},
\bauthor{\bsnm{{Tanvir}}, \binits{N.R.}},
\bauthor{\bsnm{{Starling}}, \binits{R.L.C.}},
\bauthor{\bsnm{{Wiersema}}, \binits{K.}},
\bauthor{\bsnm{{Page}}, \binits{K.L.}},
\bauthor{\bsnm{{Perley}}, \binits{D.A.}},
\bauthor{\bsnm{{Schulze}}, \binits{S.}},
\bauthor{\bsnm{{Wynn}}, \binits{G.A.}},
\bauthor{\bsnm{{Chornock}}, \binits{R.}},
\bauthor{\bsnm{{Hjorth}}, \binits{J.}},
\bauthor{\bsnm{{Cenko}}, \binits{S.B.}},
\bauthor{\bsnm{{Fruchter}}, \binits{A.S.}},
\bauthor{\bsnm{{O'Brien}}, \binits{P.T.}},
\bauthor{\bsnm{{Brown}}, \binits{G.C.}},
\bauthor{\bsnm{{Tunnicliffe}}, \binits{R.L.}},
\bauthor{\bsnm{{Malesani}}, \binits{D.}},
\bauthor{\bsnm{{Jakobsson}}, \binits{P.}},
\bauthor{\bsnm{{Watson}}, \binits{D.}},
\bauthor{\bsnm{{Berger}}, \binits{E.}},
\bauthor{\bsnm{{Bersier}}, \binits{D.}},
\bauthor{\bsnm{{Cobb}}, \binits{B.E.}},
\bauthor{\bsnm{{Covino}}, \binits{S.}},
\bauthor{\bsnm{{Cucchiara}}, \binits{A.}},
\bauthor{\bsnm{{de Ugarte Postigo}}, \binits{A.}},
\bauthor{\bsnm{{Fox}}, \binits{D.B.}},
\bauthor{\bsnm{{Gal-Yam}}, \binits{A.}},
\bauthor{\bsnm{{Goldoni}}, \binits{P.}},
\bauthor{\bsnm{{Gorosabel}}, \binits{J.}},
\bauthor{\bsnm{{Kaper}}, \binits{L.}},
\bauthor{\bsnm{{Kr{\"u}hler}}, \binits{T.}},
\bauthor{\bsnm{{Karjalainen}}, \binits{R.}},
\bauthor{\bsnm{{Osborne}}, \binits{J.P.}},
\bauthor{\bsnm{{Pian}}, \binits{E.}},
\bauthor{\bsnm{{S{\'a}nchez-Ram{\'\i}rez}}, \binits{R.}},
\bauthor{\bsnm{{Schmidt}}, \binits{B.}},
\bauthor{\bsnm{{Skillen}}, \binits{I.}},
\bauthor{\bsnm{{Tagliaferri}}, \binits{G.}},
\bauthor{\bsnm{{Th{\"o}ne}}, \binits{C.}},
\bauthor{\bsnm{{Vaduvescu}}, \binits{O.}},
\bauthor{\bsnm{{Wijers}}, \binits{R.A.M.J.}},
\bauthor{\bsnm{{Zauderer}}, \binits{B.A.}}:
\bjtitle{\apj}
\bvolume{781}(\bissue{1}),
\bfpage{13}
(\byear{2014}).
\arxivurl{1302.2352}.
doi:\doiurl{10.1088/0004-637X/781/1/13}
\end{barticle}
\endbibitem

\bibitem[\protect\citeauthoryear{Levan et~al.}{2015}]{Levan2015_150101B}
\begin{barticle}
\bauthor{\bsnm{Levan}, \binits{A.J.}},
\bauthor{\bsnm{Hjorth}, \binits{J.}},
\bauthor{\bsnm{Wiersema}, \binits{K.}},
\bauthor{\bsnm{Tanvir}, \binits{N.R.}},
\bauthor{\bsnm{Levan}, \binits{A.J.}},
\bauthor{\bsnm{Hjorth}, \binits{J.}},
\bauthor{\bsnm{Wiersema}, \binits{K.}},
\bauthor{\bsnm{Tanvir}, \binits{N.R.}}:
\bjtitle{GCN}
\bvolume{17281},
\bfpage{1}
(\byear{2015})
\end{barticle}
\endbibitem

\bibitem[\protect\citeauthoryear{Levan et~al.}{2016}]{Levan2016_160821B}
\begin{barticle}
\bauthor{\bsnm{Levan}, \binits{A.J.}},
\bauthor{\bsnm{Wiersema}, \binits{K.}},
\bauthor{\bsnm{Tanvir}, \binits{N.R.}},
\bauthor{\bsnm{Malesani}, \binits{D.}},
\bauthor{\bsnm{Xu}, \binits{D.}},
\bauthor{\bsnm{{de Ugarte Postigo}}, \binits{A.}},
\bauthor{\bsnm{Levan}, \binits{A.J.}},
\bauthor{\bsnm{Wiersema}, \binits{K.}},
\bauthor{\bsnm{Tanvir}, \binits{N.R.}},
\bauthor{\bsnm{Malesani}, \binits{D.}},
\bauthor{\bsnm{Xu}, \binits{D.}},
\bauthor{\bsnm{{de Ugarte Postigo}}, \binits{A.}}:
\bjtitle{GCN}
\bvolume{19846},
\bfpage{1}
(\byear{2016})
\end{barticle}
\endbibitem

\bibitem[\protect\citeauthoryear{Levan}{2012}]{Levan2012}
\begin{bchapter}
\bauthor{\bsnm{Levan}, \binits{A.}}:
\bctitle{Constraining gamma-ray burst progenitors},
vol. \bseriesno{279},
p. \bfpage{95}
(\byear{2012}).
doi:\doiurl{10.1017/S1743921312012756}
\end{bchapter}
\endbibitem

\bibitem[\protect\citeauthoryear{{Lewis} et~al.}{2015}]{Lewis2015}
\begin{barticle}
\bauthor{\bsnm{{Lewis}}, \binits{A.R.}},
\bauthor{\bsnm{{Dolphin}}, \binits{A.E.}},
\bauthor{\bsnm{{Dalcanton}}, \binits{J.J.}},
\bauthor{\bsnm{{Weisz}}, \binits{D.R.}},
\bauthor{\bsnm{{Williams}}, \binits{B.F.}},
\bauthor{\bsnm{{Bell}}, \binits{E.F.}},
\bauthor{\bsnm{{Seth}}, \binits{A.C.}},
\bauthor{\bsnm{{Simones}}, \binits{J.E.}},
\bauthor{\bsnm{{Skillman}}, \binits{E.D.}},
\bauthor{\bsnm{{Choi}}, \binits{Y.}},
\bauthor{\bsnm{{Fouesneau}}, \binits{M.}},
\bauthor{\bsnm{{Guhathakurta}}, \binits{P.}},
\bauthor{\bsnm{{Johnson}}, \binits{L.C.}},
\bauthor{\bsnm{{Kalirai}}, \binits{J.S.}},
\bauthor{\bsnm{{Leroy}}, \binits{A.K.}},
\bauthor{\bsnm{{Monachesi}}, \binits{A.}},
\bauthor{\bsnm{{Rix}}, \binits{H.-W.}},
\bauthor{\bsnm{{Schruba}}, \binits{A.}}:
\bjtitle{\apj}
\bvolume{805}(\bissue{2}),
\bfpage{183}
(\byear{2015}).
\arxivurl{1504.03338}.
doi:\doiurl{10.1088/0004-637X/805/2/183}
\end{barticle}
\endbibitem

\bibitem[\protect\citeauthoryear{Li and Paczyński}{2006}]{Li2006}
\begin{botherref}
\oauthor{\bsnm{Li}, \binits{L.-X.}},
\oauthor{\bsnm{Paczyński}, \binits{B.}}:
Improved correlation between the variability and peak luminosity of gamma-ray
  bursts
\textbf{366},
219
(2006).
doi:\doiurl{10.1111/j.1365-2966.2005.09836.x}
\end{botherref}
\endbibitem

\bibitem[\protect\citeauthoryear{Li and Song}{2004}]{Li2004}
\begin{botherref}
\oauthor{\bsnm{Li}, \binits{Z.}},
\oauthor{\bsnm{Song}, \binits{L.M.}}:
Late-time radio rebrightening of gamma-ray burst afterglows: Evidence for
  double-sided jets
\textbf{614},
17
(2004).
doi:\doiurl{10.1086/425498}
\end{botherref}
\endbibitem

\bibitem[\protect\citeauthoryear{{Lien} et~al.}{2016}]{BATcat2016}
\begin{barticle}
\bauthor{\bsnm{{Lien}}, \binits{A.}},
\bauthor{\bsnm{{Sakamoto}}, \binits{T.}},
\bauthor{\bsnm{{Barthelmy}}, \binits{S.D.}},
\bauthor{\bsnm{{Baumgartner}}, \binits{W.H.}},
\bauthor{\bsnm{{Cannizzo}}, \binits{J.K.}},
\bauthor{\bsnm{{Chen}}, \binits{K.}},
\bauthor{\bsnm{{Collins}}, \binits{N.R.}},
\bauthor{\bsnm{{Cummings}}, \binits{J.R.}},
\bauthor{\bsnm{{Gehrels}}, \binits{N.}},
\bauthor{\bsnm{{Krimm}}, \binits{H.A.}},
\bauthor{\bsnm{{Markwardt}}, \binits{C.B.}},
\bauthor{\bsnm{{Palmer}}, \binits{D.M.}},
\bauthor{\bsnm{{Stamatikos}}, \binits{M.}},
\bauthor{\bsnm{{Troja}}, \binits{E.}},
\bauthor{\bsnm{{Ukwatta}}, \binits{T.N.}}:
\bjtitle{\apj}
\bvolume{829}(\bissue{1}),
\bfpage{7}
(\byear{2016}).
\arxivurl{1606.01956}.
doi:\doiurl{10.3847/0004-637X/829/1/7}
\end{barticle}
\endbibitem

\bibitem[\protect\citeauthoryear{{Lin} et~al.}{2002}]{Lin2002_rhessi}
\begin{barticle}
\bauthor{\bsnm{{Lin}}, \binits{R.P.}},
\bauthor{\bsnm{{Dennis}}, \binits{B.R.}},
\bauthor{\bsnm{{Hurford}}, \binits{G.J.}},
\bauthor{\bsnm{{Smith}}, \binits{D.M.}},
\bauthor{\bsnm{{Zehnder}}, \binits{A.}},
\bauthor{\bsnm{{Harvey}}, \binits{P.R.}},
\bauthor{\bsnm{{Curtis}}, \binits{D.W.}},
\bauthor{\bsnm{{Pankow}}, \binits{D.}},
\bauthor{\bsnm{{Turin}}, \binits{P.}},
\bauthor{\bsnm{{Bester}}, \binits{M.}},
\bauthor{\bsnm{{Csillaghy}}, \binits{A.}},
\bauthor{\bsnm{{Lewis}}, \binits{M.}},
\bauthor{\bsnm{{Madden}}, \binits{N.}},
\bauthor{\bsnm{{van Beek}}, \binits{H.F.}},
\bauthor{\bsnm{{Appleby}}, \binits{M.}},
\bauthor{\bsnm{{Raudorf}}, \binits{T.}},
\bauthor{\bsnm{{McTiernan}}, \binits{J.}},
\bauthor{\bsnm{{Ramaty}}, \binits{R.}},
\bauthor{\bsnm{{Schmahl}}, \binits{E.}},
\bauthor{\bsnm{{Schwartz}}, \binits{R.}},
\bauthor{\bsnm{{Krucker}}, \binits{S.}},
\bauthor{\bsnm{{Abiad}}, \binits{R.}},
\bauthor{\bsnm{{Quinn}}, \binits{T.}},
\bauthor{\bsnm{{Berg}}, \binits{P.}},
\bauthor{\bsnm{{Hashii}}, \binits{M.}},
\bauthor{\bsnm{{Sterling}}, \binits{R.}},
\bauthor{\bsnm{{Jackson}}, \binits{R.}},
\bauthor{\bsnm{{Pratt}}, \binits{R.}},
\bauthor{\bsnm{{Campbell}}, \binits{R.D.}},
\bauthor{\bsnm{{Malone}}, \binits{D.}},
\bauthor{\bsnm{{Landis}}, \binits{D.}},
\bauthor{\bsnm{{Barrington-Leigh}}, \binits{C.P.}},
\bauthor{\bsnm{{Slassi-Sennou}}, \binits{S.}},
\bauthor{\bsnm{{Cork}}, \binits{C.}},
\bauthor{\bsnm{{Clark}}, \binits{D.}},
\bauthor{\bsnm{{Amato}}, \binits{D.}},
\bauthor{\bsnm{{Orwig}}, \binits{L.}},
\bauthor{\bsnm{{Boyle}}, \binits{R.}},
\bauthor{\bsnm{{Banks}}, \binits{I.S.}},
\bauthor{\bsnm{{Shirey}}, \binits{K.}},
\bauthor{\bsnm{{Tolbert}}, \binits{A.K.}},
\bauthor{\bsnm{{Zarro}}, \binits{D.}},
\bauthor{\bsnm{{Snow}}, \binits{F.}},
\bauthor{\bsnm{{Thomsen}}, \binits{K.}},
\bauthor{\bsnm{{Henneck}}, \binits{R.}},
\bauthor{\bsnm{{McHedlishvili}}, \binits{A.}},
\bauthor{\bsnm{{Ming}}, \binits{P.}},
\bauthor{\bsnm{{Fivian}}, \binits{M.}},
\bauthor{\bsnm{{Jordan}}, \binits{J.}},
\bauthor{\bsnm{{Wanner}}, \binits{R.}},
\bauthor{\bsnm{{Crubb}}, \binits{J.}},
\bauthor{\bsnm{{Preble}}, \binits{J.}},
\bauthor{\bsnm{{Matranga}}, \binits{M.}},
\bauthor{\bsnm{{Benz}}, \binits{A.}},
\bauthor{\bsnm{{Hudson}}, \binits{H.}},
\bauthor{\bsnm{{Canfield}}, \binits{R.C.}},
\bauthor{\bsnm{{Holman}}, \binits{G.D.}},
\bauthor{\bsnm{{Crannell}}, \binits{C.}},
\bauthor{\bsnm{{Kosugi}}, \binits{T.}},
\bauthor{\bsnm{{Emslie}}, \binits{A.G.}},
\bauthor{\bsnm{{Vilmer}}, \binits{N.}},
\bauthor{\bsnm{{Brown}}, \binits{J.C.}},
\bauthor{\bsnm{{Johns-Krull}}, \binits{C.}},
\bauthor{\bsnm{{Aschwanden}}, \binits{M.}},
\bauthor{\bsnm{{Metcalf}}, \binits{T.}},
\bauthor{\bsnm{{Conway}}, \binits{A.}}:
\bjtitle{\solphys}
\bvolume{210}(\bissue{1}),
\bfpage{3}
(\byear{2002}).
doi:\doiurl{10.1023/A:1022428818870}
\end{barticle}
\endbibitem

\bibitem[\protect\citeauthoryear{Lipunov et~al.}{2005}]{Lipunov2005_051103}
\begin{barticle}
\bauthor{\bsnm{Lipunov}, \binits{V.}},
\bauthor{\bsnm{Kornilov}, \binits{V.}},
\bauthor{\bsnm{Kuvshinov}, \binits{D.}},
\bauthor{\bsnm{Tyurina}, \binits{N.}},
\bauthor{\bsnm{Belinski}, \binits{A.}},
\bauthor{\bsnm{Gorbovskoy}, \binits{E.}},
\bauthor{\bsnm{Krylov}, \binits{A.}},
\bauthor{\bsnm{Borisov}, \binits{G.}},
\bauthor{\bsnm{Sankovich}, \binits{A.}},
\bauthor{\bsnm{Antipov}, \binits{G.}},
\bauthor{\bsnm{Vladimirov}, \binits{V.}}:
\bjtitle{GRB Coordinates Network}
\bvolume{4206},
\bfpage{1}
(\byear{2005})
\end{barticle}
\endbibitem

\bibitem[\protect\citeauthoryear{{L{\"u}} et~al.}{2017}]{Lu2017_160821B}
\begin{barticle}
\bauthor{\bsnm{{L{\"u}}}, \binits{H.-J.}},
\bauthor{\bsnm{{Zhang}}, \binits{H.-M.}},
\bauthor{\bsnm{{Zhong}}, \binits{S.-Q.}},
\bauthor{\bsnm{{Hou}}, \binits{S.-J.}},
\bauthor{\bsnm{{Sun}}, \binits{H.}},
\bauthor{\bsnm{{Rice}}, \binits{J.}},
\bauthor{\bsnm{{Liang}}, \binits{E.-W.}}:
\bjtitle{\apj}
\bvolume{835}(\bissue{2}),
\bfpage{181}
(\byear{2017}).
\arxivurl{1612.05691}.
doi:\doiurl{10.3847/1538-4357/835/2/181}
\end{barticle}
\endbibitem

\bibitem[\protect\citeauthoryear{Malesani et~al.}{2008}]{Malesani2008_080905A}
\begin{barticle}
\bauthor{\bsnm{Malesani}, \binits{D.}},
\bauthor{\bparticle{de} \bsnm{Ugarte~Postigo}, \binits{A.}},
\bauthor{\bsnm{Fynbo}, \binits{J.P.U.}},
\bauthor{\bsnm{Levan}, \binits{A.J.}},
\bauthor{\bsnm{Rol}, \binits{E.}},
\bauthor{\bsnm{Tanvir}, \binits{N.R.}},
\bauthor{\bsnm{Thoene}, \binits{C.C.}},
\bauthor{\bsnm{Telting}, \binits{J.}},
\bauthor{\bsnm{Baran}, \binits{A.}}:
\bjtitle{GRB Coordinates Network}
\bvolume{8190},
\bfpage{1}
(\byear{2008})
\end{barticle}
\endbibitem

\bibitem[\protect\citeauthoryear{Malesani et~al.}{2011}]{Malesani2011_111005A}
\begin{barticle}
\bauthor{\bsnm{Malesani}, \binits{D.}},
\bauthor{\bsnm{Levan}, \binits{A.J.}},
\bauthor{\bsnm{Tanvir}, \binits{N.R.}},
\bauthor{\bsnm{Wiersema}, \binits{K.}},
\bauthor{\bsnm{Hjorth}, \binits{J.}},
\bauthor{\bsnm{Fynbo}, \binits{J.P.U.}}:
\bjtitle{GRB Coordinates Network}
\bvolume{12418},
\bfpage{1}
(\byear{2011})
\end{barticle}
\endbibitem

\bibitem[\protect\citeauthoryear{Malesani et~al.}{2012}]{Malesani2012_120422A}
\begin{barticle}
\bauthor{\bsnm{Malesani}, \binits{D.}},
\bauthor{\bsnm{Schulze}, \binits{S.}},
\bauthor{\bsnm{Kruehler}, \binits{T.}},
\bauthor{\bsnm{Leloudas}, \binits{G.}},
\bauthor{\bparticle{de} \bsnm{Ugarte~Postigo}, \binits{A.}},
\bauthor{\bsnm{Xu}, \binits{D.}},
\bauthor{\bsnm{Fynbo}, \binits{J.P.U.}},
\bauthor{\bsnm{Hjorth}, \binits{J.}},
\bauthor{\bsnm{Jakobsson}, \binits{P.}},
\bauthor{\bsnm{Kotak}, \binits{R.}},
\bauthor{\bsnm{Wright}, \binits{D.}},
\bauthor{\bsnm{Barisevicius}, \binits{G.}},
\bauthor{\bsnm{Geier}, \binits{S.}}:
\bjtitle{GRB Coordinates Network}
\bvolume{13275},
\bfpage{1}
(\byear{2012})
\end{barticle}
\endbibitem

\bibitem[\protect\citeauthoryear{Mangano et~al.}{2005}]{Mangano2005_050826}
\begin{barticle}
\bauthor{\bsnm{Mangano}, \binits{V.}},
\bauthor{\bsnm{Racusin}, \binits{J.}},
\bauthor{\bsnm{Morris}, \binits{D.}},
\bauthor{\bsnm{Burrows}, \binits{D.N.}}:
\bjtitle{GRB Coordinates Network}
\bvolume{3885},
\bfpage{1}
(\byear{2005})
\end{barticle}
\endbibitem

\bibitem[\protect\citeauthoryear{Markwardt et~al.}{2005}]{Markwardt2005_050826}
\begin{barticle}
\bauthor{\bsnm{Markwardt}, \binits{C.}},
\bauthor{\bsnm{Barbier}, \binits{L.}},
\bauthor{\bsnm{Barthelmy}, \binits{S.}},
\bauthor{\bsnm{Cummings}, \binits{J.}},
\bauthor{\bsnm{Hullinger}, \binits{D.}},
\bauthor{\bsnm{Fenimore}, \binits{E.}},
\bauthor{\bsnm{Gehrels}, \binits{N.}},
\bauthor{\bsnm{Krimm}, \binits{H.}},
\bauthor{\bsnm{McMahon}, \binits{T.}},
\bauthor{\bsnm{Palmer}, \binits{D.}},
\bauthor{\bsnm{Parsons}, \binits{A.}},
\bauthor{\bsnm{Sakamoto}, \binits{T.}},
\bauthor{\bsnm{Sato}, \binits{G.}},
\bauthor{\bsnm{Tueller}, \binits{J.}},
\bauthor{\bsnm{White}, \binits{N.}}:
\bjtitle{GRB Coordinates Network}
\bvolume{3888},
\bfpage{1}
(\byear{2005})
\end{barticle}
\endbibitem

\bibitem[\protect\citeauthoryear{Markwardt et~al.}{2008}]{Markwardt2008_080517}
\begin{barticle}
\bauthor{\bsnm{Markwardt}, \binits{C.}},
\bauthor{\bsnm{Barthelmy}, \binits{S.D.}},
\bauthor{\bsnm{Baumgartner}, \binits{W.}},
\bauthor{\bsnm{Cummings}, \binits{J.}},
\bauthor{\bsnm{Fenimore}, \binits{E.}},
\bauthor{\bsnm{Gehrels}, \binits{N.}},
\bauthor{\bsnm{Krimm}, \binits{H.}},
\bauthor{\bsnm{McLean}, \binits{K.}},
\bauthor{\bsnm{Palmer}, \binits{D.}},
\bauthor{\bsnm{Parsons}, \binits{A.}},
\bauthor{\bsnm{Sakamoto}, \binits{T.}},
\bauthor{\bsnm{Sato}, \binits{G.}},
\bauthor{\bsnm{Stamatikos}, \binits{M.}},
\bauthor{\bsnm{Tueller}, \binits{J.}},
\bauthor{\bsnm{Ukwatta}, \binits{T.}}:
\bjtitle{GRB Coordinates Network}
\bvolume{7748},
\bfpage{1}
(\byear{2008})
\end{barticle}
\endbibitem

\bibitem[\protect\citeauthoryear{{Marshall} and
  {D'Elia}}{2015}]{Marshall2015_150818A}
\begin{barticle}
\bauthor{\bsnm{{Marshall}}, \binits{F.E.}},
\bauthor{\bsnm{{D'Elia}}, \binits{V.}}:
\bjtitle{GRB Coordinates Network}
\bvolume{18155},
\bfpage{1}
(\byear{2015})
\end{barticle}
\endbibitem

\bibitem[\protect\citeauthoryear{Marshall et~al.}{2003}]{Marshall2003_030329}
\begin{barticle}
\bauthor{\bsnm{Marshall}, \binits{F.E.}},
\bauthor{\bsnm{Markwardt}, \binits{C.}},
\bauthor{\bsnm{Swank}, \binits{J.H.}}:
\bjtitle{GRB Coordinates Network}
\bvolume{2052},
\bfpage{1}
(\byear{2003})
\end{barticle}
\endbibitem

\bibitem[\protect\citeauthoryear{Marshall et~al.}{2006}]{Marshall2006_061201}
\begin{barticle}
\bauthor{\bsnm{Marshall}, \binits{F.}},
\bauthor{\bsnm{Perri}, \binits{M.}},
\bauthor{\bsnm{Stratta}, \binits{G.}},
\bauthor{\bsnm{Barthelmy}, \binits{S.D.}},
\bauthor{\bsnm{Krimm}, \binits{H.}},
\bauthor{\bsnm{Burrows}, \binits{D.N.}},
\bauthor{\bsnm{Roming}, \binits{P.}},
\bauthor{\bsnm{Gehrels}, \binits{N.}}:
\bjtitle{GCN Report}
\bvolume{18},
\bfpage{1}
(\byear{2006})
\end{barticle}
\endbibitem

\bibitem[\protect\citeauthoryear{Masetti et~al.}{2006}]{Masetti2006_060218}
\begin{barticle}
\bauthor{\bsnm{Masetti}, \binits{N.}},
\bauthor{\bsnm{Palazzi}, \binits{E.}},
\bauthor{\bsnm{Pian}, \binits{E.}},
\bauthor{\bsnm{Patat}, \binits{F.}}:
\bjtitle{GRB Coordinates Network}
\bvolume{4803},
\bfpage{1}
(\byear{2006})
\end{barticle}
\endbibitem

\bibitem[\protect\citeauthoryear{{Mate} et~al.}{2019}]{Mate2019}
\begin{barticle}
\bauthor{\bsnm{{Mate}}, \binits{S.}},
\bauthor{\bsnm{{Bouchet}}, \binits{L.}},
\bauthor{\bsnm{{Atteia}}, \binits{J.-L.}},
\bauthor{\bsnm{{Claret}}, \binits{A.}},
\bauthor{\bsnm{{Cordier}}, \binits{B.}},
\bauthor{\bsnm{{Dagoneau}}, \binits{N.}},
\bauthor{\bsnm{{Godet}}, \binits{O.}},
\bauthor{\bsnm{{Gros}}, \binits{A.}},
\bauthor{\bsnm{{Schanne}}, \binits{S.}},
\bauthor{\bsnm{{Triou}}, \binits{H.}}:
\bjtitle{Experimental Astronomy}
\bvolume{48}(\bissue{2-3}),
\bfpage{171}
(\byear{2019}).
\arxivurl{1910.05061}.
doi:\doiurl{10.1007/s10686-019-09643-x}
\end{barticle}
\endbibitem

\bibitem[\protect\citeauthoryear{Matsuoka et~al.}{2004}]{Matsuoka2004}
\begin{barticle}
\bauthor{\bsnm{Matsuoka}, \binits{M.}},
\bauthor{\bsnm{Kawai}, \binits{N.}},
\bauthor{\bsnm{Yoshida}, \binits{A.}},
\bauthor{\bsnm{Tamagawa}, \binits{T.}},
\bauthor{\bsnm{Torii}, \binits{K.}},
\bauthor{\bsnm{Shirasaki}, \binits{Y.}},
\bauthor{\bsnm{Ricker}, \binits{G.}},
\bauthor{\bsnm{Doty}, \binits{J.}},
\bauthor{\bsnm{erspek}, \binits{R.V.}},
\bauthor{\bsnm{Crew}, \binits{G.}},
\bauthor{\bsnm{Villasenor}, \binits{J.}},
\bauthor{\bsnm{Atteia}, \binits{J.-L.}},
\bauthor{\bsnm{Fenimore}, \binits{E.E.}},
\bauthor{\bsnm{Galassi}, \binits{M.}},
\bauthor{\bsnm{Lamb}, \binits{D.Q.}},
\bauthor{\bsnm{Graziani}, \binits{C.}},
\bauthor{\bsnm{Hurley}, \binits{K.}},
\bauthor{\bsnm{Jernigan}, \binits{J.G.}},
\bauthor{\bsnm{Woosley}, \binits{S.}},
\bauthor{\bsnm{Martel}, \binits{F.}},
\bauthor{\bsnm{Prigozhin}, \binits{G.}},
\bauthor{\bsnm{Olive}, \binits{J.-F.}},
\bauthor{\bsnm{Dezalay}, \binits{J.-P.}},
\bauthor{\bsnm{Boer}, \binits{M.}},
\bauthor{\bsnm{Cline}, \binits{T.}},
\bauthor{\bsnm{Braga}, \binits{J.}},
\bauthor{\bsnm{Manchanda}, \binits{R.}},
\bauthor{\bsnm{Pizzichini}, \binits{G.}},
\bauthor{\bsnm{Levine}, \binits{A.}},
\bauthor{\bsnm{Morgan}, \binits{E.}},
\bauthor{\bsnm{Butler}, \binits{N.}},
\bauthor{\bsnm{Sakamoto}, \binits{T.}},
\bauthor{\bsnm{Urata}, \binits{Y.}},
\bauthor{\bsnm{Suzuki}, \binits{M.}},
\bauthor{\bsnm{Sato}, \binits{R.}},
\bauthor{\bsnm{Nakagawa}, \binits{Y.}},
\bauthor{\bsnm{Takagishi}, \binits{K.}},
\bauthor{\bsnm{Yamauchi}, \binits{M.}},
\bauthor{\bsnm{Hatsukade}, \binits{I.}}:
\bjtitle{Baltic Astronomy}
\bvolume{13},
\bfpage{201}
(\byear{2004})
\end{barticle}
\endbibitem

\bibitem[\protect\citeauthoryear{{Matsuoka} et~al.}{2009a}]{Matsuoka2009_maxi}
\begin{barticle}
\bauthor{\bsnm{{Matsuoka}}, \binits{M.}},
\bauthor{\bsnm{{Kawasaki}}, \binits{K.}},
\bauthor{\bsnm{{Ueno}}, \binits{S.}},
\bauthor{\bsnm{{Tomida}}, \binits{H.}},
\bauthor{\bsnm{{Kohama}}, \binits{M.}},
\bauthor{\bsnm{{Suzuki}}, \binits{M.}},
\bauthor{\bsnm{{Adachi}}, \binits{Y.}},
\bauthor{\bsnm{{Ishikawa}}, \binits{M.}},
\bauthor{\bsnm{{Mihara}}, \binits{T.}},
\bauthor{\bsnm{{Sugizaki}}, \binits{M.}},
\bauthor{\bsnm{{Isobe}}, \binits{N.}},
\bauthor{\bsnm{{Nakagawa}}, \binits{Y.}},
\bauthor{\bsnm{{Tsunemi}}, \binits{H.}},
\bauthor{\bsnm{{Miyata}}, \binits{E.}},
\bauthor{\bsnm{{Kawai}}, \binits{N.}},
\bauthor{\bsnm{{Kataoka}}, \binits{J.}},
\bauthor{\bsnm{{Morii}}, \binits{M.}},
\bauthor{\bsnm{{Yoshida}}, \binits{A.}},
\bauthor{\bsnm{{Negoro}}, \binits{H.}},
\bauthor{\bsnm{{Nakajima}}, \binits{M.}},
\bauthor{\bsnm{{Ueda}}, \binits{Y.}},
\bauthor{\bsnm{{Chujo}}, \binits{H.}},
\bauthor{\bsnm{{Yamaoka}}, \binits{K.}},
\bauthor{\bsnm{{Yamazaki}}, \binits{O.}},
\bauthor{\bsnm{{Nakahira}}, \binits{S.}},
\bauthor{\bsnm{{You}}, \binits{T.}},
\bauthor{\bsnm{{Ishiwata}}, \binits{R.}},
\bauthor{\bsnm{{Miyoshi}}, \binits{S.}},
\bauthor{\bsnm{{Eguchi}}, \binits{S.}},
\bauthor{\bsnm{{Hiroi}}, \binits{K.}},
\bauthor{\bsnm{{Katayama}}, \binits{H.}},
\bauthor{\bsnm{{Ebisawa}}, \binits{K.}}:
\bjtitle{\pasj}
\bvolume{61},
\bfpage{999}
(\byear{2009}a).
\arxivurl{0906.0631}.
doi:\doiurl{10.1093/pasj/61.5.999}
\end{barticle}
\endbibitem

\bibitem[\protect\citeauthoryear{{Matsuoka} et~al.}{2009b}]{Matsuoka2009}
\begin{barticle}
\bauthor{\bsnm{{Matsuoka}}, \binits{M.}},
\bauthor{\bsnm{{Kawasaki}}, \binits{K.}},
\bauthor{\bsnm{{Ueno}}, \binits{S.}},
\bauthor{\bsnm{{Tomida}}, \binits{H.}},
\bauthor{\bsnm{{Kohama}}, \binits{M.}},
\bauthor{\bsnm{{Suzuki}}, \binits{M.}},
\bauthor{\bsnm{{Adachi}}, \binits{Y.}},
\bauthor{\bsnm{{Ishikawa}}, \binits{M.}},
\bauthor{\bsnm{{Mihara}}, \binits{T.}},
\bauthor{\bsnm{{Sugizaki}}, \binits{M.}},
\bauthor{\bsnm{{Isobe}}, \binits{N.}},
\bauthor{\bsnm{{Nakagawa}}, \binits{Y.}},
\bauthor{\bsnm{{Tsunemi}}, \binits{H.}},
\bauthor{\bsnm{{Miyata}}, \binits{E.}},
\bauthor{\bsnm{{Kawai}}, \binits{N.}},
\bauthor{\bsnm{{Kataoka}}, \binits{J.}},
\bauthor{\bsnm{{Morii}}, \binits{M.}},
\bauthor{\bsnm{{Yoshida}}, \binits{A.}},
\bauthor{\bsnm{{Negoro}}, \binits{H.}},
\bauthor{\bsnm{{Nakajima}}, \binits{M.}},
\bauthor{\bsnm{{Ueda}}, \binits{Y.}},
\bauthor{\bsnm{{Chujo}}, \binits{H.}},
\bauthor{\bsnm{{Yamaoka}}, \binits{K.}},
\bauthor{\bsnm{{Yamazaki}}, \binits{O.}},
\bauthor{\bsnm{{Nakahira}}, \binits{S.}},
\bauthor{\bsnm{{You}}, \binits{T.}},
\bauthor{\bsnm{{Ishiwata}}, \binits{R.}},
\bauthor{\bsnm{{Miyoshi}}, \binits{S.}},
\bauthor{\bsnm{{Eguchi}}, \binits{S.}},
\bauthor{\bsnm{{Hiroi}}, \binits{K.}},
\bauthor{\bsnm{{Katayama}}, \binits{H.}},
\bauthor{\bsnm{{Ebisawa}}, \binits{K.}}:
\bjtitle{\pasj}
\bvolume{61},
\bfpage{999}
(\byear{2009}b).
\arxivurl{0906.0631}.
doi:\doiurl{10.1093/pasj/61.5.999}
\end{barticle}
\endbibitem

\bibitem[\protect\citeauthoryear{Mazaeva et~al.}{2015}]{Mazaeva2015_150818A}
\begin{barticle}
\bauthor{\bsnm{Mazaeva}, \binits{E.}},
\bauthor{\bsnm{Klunko}, \binits{E.}},
\bauthor{\bsnm{Volnova}, \binits{A.}},
\bauthor{\bsnm{Korobtsev}, \binits{I.}},
\bauthor{\bsnm{Pozanenko}, \binits{A.}}:
\bjtitle{GRB Coordinates Network}
\bvolume{18205},
\bfpage{1}
(\byear{2015})
\end{barticle}
\endbibitem

\bibitem[\protect\citeauthoryear{Mazets et~al.}{2005}]{Mazets2005_041227}
\begin{botherref}
\oauthor{\bsnm{Mazets}, \binits{E.P.}},
\oauthor{\bsnm{Cline}, \binits{T.L.}},
\oauthor{\bsnm{Aptekar}, \binits{R.L.}},
\oauthor{\bsnm{Frederiks}, \binits{D.D.}},
\oauthor{\bsnm{Golenetskii}, \binits{S.V.}},
\oauthor{\bsnm{Il'inskii}, \binits{V.N.}},
\oauthor{\bsnm{Pal'shin}, \binits{V.D.}}:
arXiv e-prints,
0502541
(2005)
\end{botherref}
\endbibitem

\bibitem[\protect\citeauthoryear{Mazets et~al.}{2008}]{Mazets2008}
\begin{barticle}
\bauthor{\bsnm{Mazets}, \binits{E.P.}},
\bauthor{\bsnm{Aptekar}, \binits{R.L.}},
\bauthor{\bsnm{Cline}, \binits{T.L.}},
\bauthor{\bsnm{Frederiks}, \binits{D.D.}},
\bauthor{\bsnm{Goldsten}, \binits{J.O.}},
\bauthor{\bsnm{Golenetskii}, \binits{S.V.}},
\bauthor{\bsnm{Hurley}, \binits{K.}},
\bauthor{\bparticle{von} \bsnm{Kienlin}, \binits{A.}},
\bauthor{\bsnm{Pal'shin}, \binits{V.D.}}:
\bjtitle{\apj}
\bvolume{680},
\bfpage{545}
(\byear{2008}).
doi:\doiurl{10.1086/587955}
\end{barticle}
\endbibitem

\bibitem[\protect\citeauthoryear{{Meegan} et~al.}{2009a}]{Meegan2009_gbm}
\begin{barticle}
\bauthor{\bsnm{{Meegan}}, \binits{C.}},
\bauthor{\bsnm{{Lichti}}, \binits{G.}},
\bauthor{\bsnm{{Bhat}}, \binits{P.N.}},
\bauthor{\bsnm{{Bissaldi}}, \binits{E.}},
\bauthor{\bsnm{{Briggs}}, \binits{M.S.}},
\bauthor{\bsnm{{Connaughton}}, \binits{V.}},
\bauthor{\bsnm{{Diehl}}, \binits{R.}},
\bauthor{\bsnm{{Fishman}}, \binits{G.}},
\bauthor{\bsnm{{Greiner}}, \binits{J.}},
\bauthor{\bsnm{{Hoover}}, \binits{A.S.}},
\bauthor{\bsnm{{van der Horst}}, \binits{A.e.J.}},
\bauthor{\bsnm{{von Kienlin}}, \binits{A.}},
\bauthor{\bsnm{{Kippen}}, \binits{R.M.}},
\bauthor{\bsnm{{Kouveliotou}}, \binits{C.}},
\bauthor{\bsnm{{McBreen}}, \binits{S.}},
\bauthor{\bsnm{{Paciesas}}, \binits{W.S.}},
\bauthor{\bsnm{{Preece}}, \binits{R.}},
\bauthor{\bsnm{{Steinle}}, \binits{H.}},
\bauthor{\bsnm{{Wallace}}, \binits{M.S.}},
\bauthor{\bsnm{{Wilson}}, \binits{R.B.}},
\bauthor{\bsnm{{Wilson-Hodge}}, \binits{C.}}:
\bjtitle{\apj}
\bvolume{702}(\bissue{1}),
\bfpage{791}
(\byear{2009}a).
\arxivurl{0908.0450}.
doi:\doiurl{10.1088/0004-637X/702/1/791}
\end{barticle}
\endbibitem

\bibitem[\protect\citeauthoryear{{Meegan} et~al.}{2009b}]{Meegan2009}
\begin{barticle}
\bauthor{\bsnm{{Meegan}}, \binits{C.}},
\bauthor{\bsnm{{Lichti}}, \binits{G.}},
\bauthor{\bsnm{{Bhat}}, \binits{P.N.}},
\bauthor{\bsnm{{Bissaldi}}, \binits{E.}},
\bauthor{\bsnm{{Briggs}}, \binits{M.S.}},
\bauthor{\bsnm{{Connaughton}}, \binits{V.}},
\bauthor{\bsnm{{Diehl}}, \binits{R.}},
\bauthor{\bsnm{{Fishman}}, \binits{G.}},
\bauthor{\bsnm{{Greiner}}, \binits{J.}},
\bauthor{\bsnm{{Hoover}}, \binits{A.S.}},
\bauthor{\bsnm{{van der Horst}}, \binits{A.e.J.}},
\bauthor{\bsnm{{von Kienlin}}, \binits{A.}},
\bauthor{\bsnm{{Kippen}}, \binits{R.M.}},
\bauthor{\bsnm{{Kouveliotou}}, \binits{C.}},
\bauthor{\bsnm{{McBreen}}, \binits{S.}},
\bauthor{\bsnm{{Paciesas}}, \binits{W.S.}},
\bauthor{\bsnm{{Preece}}, \binits{R.}},
\bauthor{\bsnm{{Steinle}}, \binits{H.}},
\bauthor{\bsnm{{Wallace}}, \binits{M.S.}},
\bauthor{\bsnm{{Wilson}}, \binits{R.B.}},
\bauthor{\bsnm{{Wilson-Hodge}}, \binits{C.}}:
\bjtitle{\apj}
\bvolume{702}(\bissue{1}),
\bfpage{791}
(\byear{2009}b).
\arxivurl{0908.0450}.
doi:\doiurl{10.1088/0004-637X/702/1/791}
\end{barticle}
\endbibitem

\bibitem[\protect\citeauthoryear{Melandri et~al.}{2012}]{Melandri2012_120422A}
\begin{barticle}
\bauthor{\bsnm{Melandri}, \binits{A.}},
\bauthor{\bsnm{Pian}, \binits{E.}},
\bauthor{\bsnm{Ferrero}, \binits{P.}},
\bauthor{\bsnm{D'Elia}, \binits{V.}},
\bauthor{\bsnm{Walker}, \binits{E.s.}},
\bauthor{\bsnm{Ghirlanda}, \binits{G.}},
\bauthor{\bsnm{Covino}, \binits{S.}},
\bauthor{\bsnm{Amati}, \binits{L.}},
\bauthor{\bsnm{D'Avanzo}, \binits{P.}},
\bauthor{\bsnm{Mazzali}, \binits{P.s.}},
\bauthor{\bsnm{{Della Valle}}, \binits{M.}},
\bauthor{\bsnm{Guidorzi}, \binits{C.}},
\bauthor{\bsnm{Antonelli}, \binits{L.s.}},
\bauthor{\bsnm{Bernardini}, \binits{M.s.}},
\bauthor{\bsnm{Bersier}, \binits{D.}},
\bauthor{\bsnm{Bufano}, \binits{F.}},
\bauthor{\bsnm{Campana}, \binits{S.}},
\bauthor{\bsnm{Castro-Tirado}, \binits{A.s.}},
\bauthor{\bsnm{Chincarini}, \binits{G.}},
\bauthor{\bsnm{Deng}, \binits{J.}},
\bauthor{\bsnm{Filippenko}, \binits{A.s.}},
\bauthor{\bsnm{Fugazza}, \binits{D.}},
\bauthor{\bsnm{Ghisellini}, \binits{G.}},
\bauthor{\bsnm{Kouveliotou}, \binits{C.}},
\bauthor{\bsnm{Maeda}, \binits{K.}},
\bauthor{\bsnm{Marconi}, \binits{G.}},
\bauthor{\bsnm{Masetti}, \binits{N.}},
\bauthor{\bsnm{Nomoto}, \binits{K.}},
\bauthor{\bsnm{Palazzi}, \binits{E.}},
\bauthor{\bsnm{Patat}, \binits{F.}},
\bauthor{\bsnm{Piranomonte}, \binits{S.}},
\bauthor{\bsnm{Salvaterra}, \binits{R.}},
\bauthor{\bsnm{Saviane}, \binits{I.}},
\bauthor{\bsnm{Starling}, \binits{R.s.s.}},
\bauthor{\bsnm{Tagliaferri}, \binits{G.}},
\bauthor{\bsnm{Tanaka}, \binits{M.}},
\bauthor{\bsnm{Vergani}, \binits{S.s.}}:
\bjtitle{\aap}
\bvolume{547},
\bfpage{82}
(\byear{2012}).
\arxivurl{1206.5532}.
doi:\doiurl{10.1051/0004-6361/201219879}
\end{barticle}
\endbibitem

\bibitem[\protect\citeauthoryear{Mereghetti and
  Gotz}{2003}]{Mereghetti2003_031203}
\begin{barticle}
\bauthor{\bsnm{Mereghetti}, \binits{S.}},
\bauthor{\bsnm{Gotz}, \binits{D.}}:
\bjtitle{GRB Coordinates Network}
\bvolume{2460},
\bfpage{1}
(\byear{2003})
\end{barticle}
\endbibitem

\bibitem[\protect\citeauthoryear{{Mesler}}{2013}]{2013PhDT.......497M}
\begin{botherref}
\oauthor{\bsnm{{Mesler}}, \binits{I.} \bsuffix{Robert~A.}}:
{Searching for the Long-Duration Gamma Ray Burst Progenitor}.
PhD thesis,
The University of New Mexico
(January 2013)
\end{botherref}
\endbibitem

\bibitem[\protect\citeauthoryear{{Micha{\l}owski}
  et~al.}{2018}]{Michalowski2018_111005A}
\begin{barticle}
\bauthor{\bsnm{{Micha{\l}owski}}, \binits{M.J.}},
\bauthor{\bsnm{{Xu}}, \binits{D.}},
\bauthor{\bsnm{{Stevens}}, \binits{J.}},
\bauthor{\bsnm{{Levan}}, \binits{A.}},
\bauthor{\bsnm{{Yang}}, \binits{J.}},
\bauthor{\bsnm{{Paragi}}, \binits{Z.}},
\bauthor{\bsnm{{Kamble}}, \binits{A.}},
\bauthor{\bsnm{{Tsai}}, \binits{A.-L.}},
\bauthor{\bsnm{{Dannerbauer}}, \binits{H.}},
\bauthor{\bsnm{{van der Horst}}, \binits{A.J.}},
\bauthor{\bsnm{{Shao}}, \binits{L.}},
\bauthor{\bsnm{{Crosby}}, \binits{D.}},
\bauthor{\bsnm{{Gentile}}, \binits{G.}},
\bauthor{\bsnm{{Stanway}}, \binits{E.}},
\bauthor{\bsnm{{Wiersema}}, \binits{K.}},
\bauthor{\bsnm{{Fynbo}}, \binits{J.P.U.}},
\bauthor{\bsnm{{Tanvir}}, \binits{N.R.}},
\bauthor{\bsnm{{Kamphuis}}, \binits{P.}},
\bauthor{\bsnm{{Garrett}}, \binits{M.}},
\bauthor{\bsnm{{Bartczak}}, \binits{P.}}:
\bjtitle{\aap}
\bvolume{616},
\bfpage{169}
(\byear{2018}).
\arxivurl{1610.06928}.
doi:\doiurl{10.1051/0004-6361/201629942}
\end{barticle}
\endbibitem

\bibitem[\protect\citeauthoryear{{Minaev} and
  {Pozanenko}}{2020}]{Minaev2020_200415A}
\begin{botherref}
\oauthor{\bsnm{{Minaev}}, \binits{P.}},
\oauthor{\bsnm{{Pozanenko}}, \binits{A.}}:
arXiv e-prints,
2008
(2020).
\arxivurl{2008.12752}
\end{botherref}
\endbibitem

\bibitem[\protect\citeauthoryear{Mirabal et~al.}{2007}]{Mirabal2007_050826}
\begin{barticle}
\bauthor{\bsnm{Mirabal}, \binits{N.}},
\bauthor{\bsnm{Halpern}, \binits{J.P.}},
\bauthor{\bsnm{O'Brien}, \binits{P.T.}}:
\bjtitle{Astrophys. J.}
\bvolume{661}(\bissue{2}),
\bfpage{127}
(\byear{2007}).
\arxivurl{0704.3069}.
doi:\doiurl{10.1086/519006}
\end{barticle}
\endbibitem

\bibitem[\protect\citeauthoryear{Mirabal et~al.}{2006}]{Mirabal2006_060218}
\begin{barticle}
\bauthor{\bsnm{Mirabal}, \binits{N.}},
\bauthor{\bsnm{Halpern}, \binits{J.P.}},
\bauthor{\bsnm{Mirabal}, \binits{N.}},
\bauthor{\bsnm{Halpern}, \binits{J.P.}}:
\bjtitle{GCN}
\bvolume{4792},
\bfpage{1}
(\byear{2006})
\end{barticle}
\endbibitem

\bibitem[\protect\citeauthoryear{{Mirabal}}{2004}]{2004PhDT........13M}
\begin{botherref}
\oauthor{\bsnm{{Mirabal}}, \binits{N.R.}}:
{On gamma-ray burst progenitors and environments}.
PhD thesis,
COLUMBIA UNIVERSITY
(January 2004)
\end{botherref}
\endbibitem

\bibitem[\protect\citeauthoryear{Misra and Fruchte}{2018}]{Misra2018}
\begin{barticle}
\bauthor{\bsnm{Misra}, \binits{K.}},
\bauthor{\bsnm{Fruchte}, \binits{A.S.}}:
\bjtitle{Bulletin de la Societe Royale des Sciences de Liege}
\bvolume{87},
\bfpage{347}
(\byear{2018})
\end{barticle}
\endbibitem

\bibitem[\protect\citeauthoryear{Moran and Reichart}{2005}]{Moran2005}
\begin{botherref}
\oauthor{\bsnm{Moran}, \binits{J.A.}},
\oauthor{\bsnm{Reichart}, \binits{D.E.}}:
Gamma-ray burst dust echoes revisited: Expectations at early times
\textbf{632},
438
(2005).
doi:\doiurl{10.1086/432634}
\end{botherref}
\endbibitem

\bibitem[\protect\citeauthoryear{{Narayana Bhat} et~al.}{2016}]{GBMcat2016}
\begin{barticle}
\bauthor{\bsnm{{Narayana Bhat}}, \binits{P.}},
\bauthor{\bsnm{{Meegan}}, \binits{C.A.}},
\bauthor{\bsnm{{von Kienlin}}, \binits{A.}},
\bauthor{\bsnm{{Paciesas}}, \binits{W.S.}},
\bauthor{\bsnm{{Briggs}}, \binits{M.S.}},
\bauthor{\bsnm{{Burgess}}, \binits{J.M.}},
\bauthor{\bsnm{{Burns}}, \binits{E.}},
\bauthor{\bsnm{{Chaplin}}, \binits{V.}},
\bauthor{\bsnm{{Cleveland}}, \binits{W.H.}},
\bauthor{\bsnm{{Collazzi}}, \binits{A.C.}},
\bauthor{\bsnm{{Connaughton}}, \binits{V.}},
\bauthor{\bsnm{{Diekmann}}, \binits{A.M.}},
\bauthor{\bsnm{{Fitzpatrick}}, \binits{G.}},
\bauthor{\bsnm{{Gibby}}, \binits{M.H.}},
\bauthor{\bsnm{{Giles}}, \binits{M.M.}},
\bauthor{\bsnm{{Goldstein}}, \binits{A.M.}},
\bauthor{\bsnm{{Greiner}}, \binits{J.}},
\bauthor{\bsnm{{Jenke}}, \binits{P.A.}},
\bauthor{\bsnm{{Kippen}}, \binits{R.M.}},
\bauthor{\bsnm{{Kouveliotou}}, \binits{C.}},
\bauthor{\bsnm{{Mailyan}}, \binits{B.}},
\bauthor{\bsnm{{McBreen}}, \binits{S.}},
\bauthor{\bsnm{{Pelassa}}, \binits{V.}},
\bauthor{\bsnm{{Preece}}, \binits{R.D.}},
\bauthor{\bsnm{{Roberts}}, \binits{O.J.}},
\bauthor{\bsnm{{Sparke}}, \binits{L.S.}},
\bauthor{\bsnm{{Stanbro}}, \binits{M.}},
\bauthor{\bsnm{{Veres}}, \binits{P.}},
\bauthor{\bsnm{{Wilson-Hodge}}, \binits{C.A.}},
\bauthor{\bsnm{{Xiong}}, \binits{S.}},
\bauthor{\bsnm{{Younes}}, \binits{G.}},
\bauthor{\bsnm{{Yu}}, \binits{H.-F.}},
\bauthor{\bsnm{{Zhang}}, \binits{B.}}:
\bjtitle{\apjs}
\bvolume{223}(\bissue{2}),
\bfpage{28}
(\byear{2016}).
\arxivurl{1603.07612}.
doi:\doiurl{10.3847/0067-0049/223/2/28}
\end{barticle}
\endbibitem

\bibitem[\protect\citeauthoryear{{Nicuesa Guelbenzu}
  et~al.}{2015}]{Guelbenzu2015_100628A}
\begin{barticle}
\bauthor{\bsnm{{Nicuesa Guelbenzu}}, \binits{A.}},
\bauthor{\bsnm{{Klose}}, \binits{S.}},
\bauthor{\bsnm{{Palazzi}}, \binits{E.}},
\bauthor{\bsnm{{Greiner}}, \binits{J.}},
\bauthor{\bsnm{{Micha{\l}owski}}, \binits{M.J.}},
\bauthor{\bsnm{{Kann}}, \binits{D.A.}},
\bauthor{\bsnm{{Hunt}}, \binits{L.K.}},
\bauthor{\bsnm{{Malesani}}, \binits{D.}},
\bauthor{\bsnm{{Rossi}}, \binits{A.}},
\bauthor{\bsnm{{Savaglio}}, \binits{S.}},
\bauthor{\bsnm{{Schulze}}, \binits{S.}},
\bauthor{\bsnm{{Xu}}, \binits{D.}},
\bauthor{\bsnm{{Afonso}}, \binits{P.M.J.}},
\bauthor{\bsnm{{Elliott}}, \binits{J.}},
\bauthor{\bsnm{{Ferrero}}, \binits{P.}},
\bauthor{\bsnm{{Filgas}}, \binits{R.}},
\bauthor{\bsnm{{Hartmann}}, \binits{D.H.}},
\bauthor{\bsnm{{Kr{\"u}hler}}, \binits{T.}},
\bauthor{\bsnm{{Knust}}, \binits{F.}},
\bauthor{\bsnm{{Masetti}}, \binits{N.}},
\bauthor{\bsnm{{Olivares E.}}, \binits{F.}},
\bauthor{\bsnm{{Rau}}, \binits{A.}},
\bauthor{\bsnm{{Schady}}, \binits{P.}},
\bauthor{\bsnm{{Schmidl}}, \binits{S.}},
\bauthor{\bsnm{{Tanga}}, \binits{M.}},
\bauthor{\bsnm{{Updike}}, \binits{A.C.}},
\bauthor{\bsnm{{Varela}}, \binits{K.}}:
\bjtitle{\aap}
\bvolume{583},
\bfpage{88}
(\byear{2015}).
doi:\doiurl{10.1051/0004-6361/201425160}
\end{barticle}
\endbibitem

\bibitem[\protect\citeauthoryear{Ofek et~al.}{2006}]{Ofek2006_060505}
\begin{barticle}
\bauthor{\bsnm{Ofek}, \binits{E.O.}},
\bauthor{\bsnm{Cenko}, \binits{S.B.}},
\bauthor{\bsnm{Gal-Yam}, \binits{A.}},
\bauthor{\bsnm{Peterson}, \binits{B.}},
\bauthor{\bsnm{Schmidt}, \binits{B.P.}},
\bauthor{\bsnm{Fox}, \binits{D.B.}},
\bauthor{\bsnm{Price}, \binits{P.A.}}:
\bjtitle{GRB Coordinates Network}
\bvolume{5123},
\bfpage{1}
(\byear{2006})
\end{barticle}
\endbibitem

\bibitem[\protect\citeauthoryear{Ofek et~al.}{2007}]{Ofek2007_060505}
\begin{barticle}
\bauthor{\bsnm{Ofek}, \binits{E.O.}},
\bauthor{\bsnm{Cenko}, \binits{S.B.}},
\bauthor{\bsnm{Gal-Yam}, \binits{A.}},
\bauthor{\bsnm{Fox}, \binits{D.B.}},
\bauthor{\bsnm{Nakar}, \binits{E.}},
\bauthor{\bsnm{Rau}, \binits{A.}},
\bauthor{\bsnm{Frail}, \binits{D.A.}},
\bauthor{\bsnm{Kulkarni}, \binits{S.R.}},
\bauthor{\bsnm{Price}, \binits{P.A.}},
\bauthor{\bsnm{Schmidt}, \binits{B.P.}},
\bauthor{\bsnm{Soderberg}, \binits{A.M.}},
\bauthor{\bsnm{Peterson}, \binits{B.}},
\bauthor{\bsnm{Berger}, \binits{E.}},
\bauthor{\bsnm{Sharon}, \binits{K.}},
\bauthor{\bsnm{Shemmer}, \binits{O.}},
\bauthor{\bsnm{Penprase}, \binits{B.E.}},
\bauthor{\bsnm{Chevalier}, \binits{R.A.}},
\bauthor{\bsnm{Brown}, \binits{P.J.}},
\bauthor{\bsnm{Burrows}, \binits{D.N.}},
\bauthor{\bsnm{Gehrels}, \binits{N.}},
\bauthor{\bsnm{Harrison}, \binits{F.}},
\bauthor{\bsnm{Holland}, \binits{S.T.}},
\bauthor{\bsnm{Mangano}, \binits{V.}},
\bauthor{\bsnm{McCarthy}, \binits{P.J.}},
\bauthor{\bsnm{Moon}, \binits{D.-S.}},
\bauthor{\bsnm{Nousek}, \binits{J.A.}},
\bauthor{\bsnm{Persson}, \binits{S.E.}},
\bauthor{\bsnm{Piran}, \binits{T.}},
\bauthor{\bsnm{Sari}, \binits{R.}}:
\bjtitle{Astrophys. J.}
\bvolume{662}(\bissue{2}),
\bfpage{1129}
(\byear{2007}).
\arxivurl{0703192}.
doi:\doiurl{10.1086/518082}
\end{barticle}
\endbibitem

\bibitem[\protect\citeauthoryear{Ofek et~al.}{2008}]{Ofek2008_070201}
\begin{barticle}
\bauthor{\bsnm{Ofek}, \binits{E.O.}},
\bauthor{\bsnm{Muno}, \binits{M.}},
\bauthor{\bsnm{Quimby}, \binits{R.}},
\bauthor{\bsnm{Kulkarni}, \binits{S.R.}},
\bauthor{\bsnm{Stiele}, \binits{H.}},
\bauthor{\bsnm{Pietsch}, \binits{W.}},
\bauthor{\bsnm{Nakar}, \binits{E.}},
\bauthor{\bsnm{Gal-Yam}, \binits{A.}},
\bauthor{\bsnm{Rau}, \binits{A.}},
\bauthor{\bsnm{Cameron}, \binits{P.B.}},
\bauthor{\bsnm{Cenko}, \binits{S.B.}},
\bauthor{\bsnm{Kasliwal}, \binits{M.M.}},
\bauthor{\bsnm{Fox}, \binits{D.B.}},
\bauthor{\bsnm{Chandra}, \binits{P.}},
\bauthor{\bsnm{Kong}, \binits{A.K.H.}},
\bauthor{\bsnm{Barnard}, \binits{R.}}:
\bjtitle{\apj}
\bvolume{681},
\bfpage{1464}
(\byear{2008}).
doi:\doiurl{10.1086/587686}
\end{barticle}
\endbibitem

\bibitem[\protect\citeauthoryear{Osborne et~al.}{2017}]{Osborne2017_171205A}
\begin{barticle}
\bauthor{\bsnm{Osborne}, \binits{J.P.}},
\bauthor{\bsnm{Beardmore}, \binits{A.P.}},
\bauthor{\bsnm{Evans}, \binits{P.A.}},
\bauthor{\bsnm{Goad}, \binits{M.R.}}:
\bjtitle{GRB Coordinates Network}
\bvolume{22179},
\bfpage{1}
(\byear{2017})
\end{barticle}
\endbibitem

\bibitem[\protect\citeauthoryear{Pagani et~al.}{2008}]{Pagani2008_080905A}
\begin{barticle}
\bauthor{\bsnm{Pagani}, \binits{C.}},
\bauthor{\bsnm{Baumgartner}, \binits{W.H.}},
\bauthor{\bsnm{Beardmore}, \binits{A.P.}},
\bauthor{\bsnm{Chester}, \binits{M.M.}},
\bauthor{\bsnm{Cummings}, \binits{J.R.}},
\bauthor{\bsnm{Evans}, \binits{P.A.}},
\bauthor{\bsnm{Gehrels}, \binits{N.}},
\bauthor{\bsnm{Grupe}, \binits{D.}},
\bauthor{\bsnm{Holland}, \binits{S.T.}},
\bauthor{\bsnm{Krimm}, \binits{H.A.}},
\bauthor{\bsnm{Mao}, \binits{J.}},
\bauthor{\bsnm{Markwardt}, \binits{C.B.}},
\bauthor{\bsnm{Marshall}, \binits{F.E.}},
\bauthor{\bsnm{Page}, \binits{K.L.}},
\bauthor{\bsnm{Palmer}, \binits{D.M.}},
\bauthor{\bsnm{Preger}, \binits{B.}},
\bauthor{\bsnm{Racusin}, \binits{J.L.}},
\bauthor{\bsnm{Romano}, \binits{P.}},
\bauthor{\bsnm{Sbarufatti}, \binits{B.}},
\bauthor{\bsnm{Schady}, \binits{P.}},
\bauthor{\bsnm{Stamatikos}, \binits{M.}},
\bauthor{\bsnm{Starling}, \binits{R.L.C.}},
\bauthor{\bsnm{Stroh}, \binits{M.C.}},
\bauthor{\bsnm{Troja}, \binits{E.}},
\bauthor{\bsnm{Ukwatta}, \binits{T.N.}},
\bauthor{\bsnm{Vetere}, \binits{L.}}:
\bjtitle{GRB Coordinates Network}
\bvolume{8180},
\bfpage{1}
(\byear{2008})
\end{barticle}
\endbibitem

\bibitem[\protect\citeauthoryear{Palmer et~al.}{2005}]{Palmer2005}
\begin{botherref}
\oauthor{\bsnm{Palmer}, \binits{D.M.}},
\oauthor{\bsnm{Barthelmy}, \binits{S.}},
\oauthor{\bsnm{Gehrels}, \binits{N.}},
\oauthor{\bsnm{Kippen}, \binits{R.M.}},
\oauthor{\bsnm{Cayton}, \binits{T.}},
\oauthor{\bsnm{Kouveliotou}, \binits{C.}},
\oauthor{\bsnm{Eichler}, \binits{D.}},
\oauthor{\bsnm{Wijers}, \binits{R.A.M.J.}},
\oauthor{\bsnm{Woods}, \binits{P.M.}},
\oauthor{\bsnm{Granot}, \binits{J.}},
\oauthor{\bsnm{Lyubarsky}, \binits{Y.E.}},
\oauthor{\bsnm{Ramirez-Ruiz}, \binits{E.}},
\oauthor{\bsnm{Barbier}, \binits{L.}},
\oauthor{\bsnm{Chester}, \binits{M.}},
\oauthor{\bsnm{Cummings}, \binits{J.}},
\oauthor{\bsnm{Fenimore}, \binits{E.E.}},
\oauthor{\bsnm{Finger}, \binits{M.H.}},
\oauthor{\bsnm{Gaensler}, \binits{B.M.}},
\oauthor{\bsnm{Hullinger}, \binits{D.}},
\oauthor{\bsnm{Krimm}, \binits{H.}},
\oauthor{\bsnm{Markwardt}, \binits{C.B.}},
\oauthor{\bsnm{Nousek}, \binits{J.A.}},
\oauthor{\bsnm{Parsons}, \binits{A.}},
\oauthor{\bsnm{Patel}, \binits{S.}},
\oauthor{\bsnm{Sakamoto}, \binits{T.}},
\oauthor{\bsnm{Sato}, \binits{G.}},
\oauthor{\bsnm{Suzuki}, \binits{M.}},
\oauthor{\bsnm{Tueller}, \binits{J.}}:
A giant gamma-ray flare from the magnetar sgr 1806 - 20
\textbf{434},
1107
(2005).
doi:\doiurl{10.1038/nature03525}
\end{botherref}
\endbibitem

\bibitem[\protect\citeauthoryear{Palmer et~al.}{2015}]{Palmer2015_150818A}
\begin{barticle}
\bauthor{\bsnm{Palmer}, \binits{D.M.}},
\bauthor{\bsnm{Barthelmy}, \binits{S.D.}},
\bauthor{\bsnm{Cummings}, \binits{J.R.}},
\bauthor{\bsnm{D'Elia}, \binits{V.}},
\bauthor{\bsnm{Gehrels}, \binits{N.}},
\bauthor{\bsnm{Krimm}, \binits{H.A.}},
\bauthor{\bsnm{Lien}, \binits{A.Y.}},
\bauthor{\bsnm{Markwardt}, \binits{C.B.}},
\bauthor{\bsnm{Sakamoto}, \binits{T.}},
\bauthor{\bsnm{Stamatikos}, \binits{M.}},
\bauthor{\bsnm{Ukwatta}, \binits{T.N.}}:
\bjtitle{GRB Coordinates Network}
\bvolume{18157},
\bfpage{1}
(\byear{2015})
\end{barticle}
\endbibitem

\bibitem[\protect\citeauthoryear{Parsons et~al.}{2006}]{Parsons2006_060614}
\begin{barticle}
\bauthor{\bsnm{Parsons}, \binits{A.M.}},
\bauthor{\bsnm{Cummings}, \binits{J.R.}},
\bauthor{\bsnm{Gehrels}, \binits{N.}},
\bauthor{\bsnm{Goad}, \binits{M.R.}},
\bauthor{\bsnm{Gronwall}, \binits{C.}},
\bauthor{\bsnm{Holland}, \binits{S.T.}},
\bauthor{\bsnm{Kennea}, \binits{J.A.}},
\bauthor{\bsnm{Parola}, \binits{V.L.}},
\bauthor{\bsnm{Mangano}, \binits{V.}},
\bauthor{\bsnm{Marshall}, \binits{F.E.}},
\bauthor{\bsnm{McLean}, \binits{K.M.}},
\bauthor{\bsnm{Pagani}, \binits{C.}},
\bauthor{\bsnm{Palmer}, \binits{D.M.}},
\bauthor{\bsnm{Romano}, \binits{P.}},
\bauthor{\bsnm{Stamatikos}, \binits{M.}}:
\bjtitle{GRB Coordinates Network}
\bvolume{5252},
\bfpage{1}
(\byear{2006})
\end{barticle}
\endbibitem

\bibitem[\protect\citeauthoryear{Parsons et~al.}{2008}]{Parsons2008_080517}
\begin{barticle}
\bauthor{\bsnm{Parsons}, \binits{A.M.}},
\bauthor{\bsnm{Beardmore}, \binits{A.P.}},
\bauthor{\bsnm{Burrows}, \binits{D.N.}},
\bauthor{\bsnm{Cummings}, \binits{J.R.}},
\bauthor{\bsnm{Evans}, \binits{P.A.}},
\bauthor{\bsnm{Gehrels}, \binits{N.}},
\bauthor{\bsnm{Holland}, \binits{S.T.}},
\bauthor{\bsnm{Hoversten}, \binits{E.A.}},
\bauthor{\bsnm{Markwardt}, \binits{C.B.}},
\bauthor{\bsnm{Marshall}, \binits{F.E.}},
\bauthor{\bsnm{O'Brien}, \binits{P.T.}},
\bauthor{\bsnm{Page}, \binits{K.L.}},
\bauthor{\bsnm{Palmer}, \binits{D.M.}},
\bauthor{\bsnm{Troja}, \binits{E.}}:
\bjtitle{GRB Coordinates Network}
\bvolume{7742},
\bfpage{1}
(\byear{2008})
\end{barticle}
\endbibitem

\bibitem[\protect\citeauthoryear{{Parsons} et~al.}{2005}]{Parsons2005_050906}
\begin{barticle}
\bauthor{\bsnm{{Parsons}}, \binits{A.}},
\bauthor{\bsnm{{Sarazin}}, \binits{C.}},
\bauthor{\bsnm{{Barbier}}, \binits{L.}},
\bauthor{\bsnm{{Barthelmy}}, \binits{S.}},
\bauthor{\bsnm{{Cummings}}, \binits{J.}},
\bauthor{\bsnm{{Hullinger}}, \binits{D.}},
\bauthor{\bsnm{{Fenimore}}, \binits{E.}},
\bauthor{\bsnm{{Gehrels}}, \binits{N.}},
\bauthor{\bsnm{{Krimm}}, \binits{H.}},
\bauthor{\bsnm{{Markwardt}}, \binits{C.}},
\bauthor{\bsnm{{Marshall}}, \binits{F.}},
\bauthor{\bsnm{{Palmer}}, \binits{D.}},
\bauthor{\bsnm{{Sakamoto}}, \binits{T.}},
\bauthor{\bsnm{{Sato}}, \binits{G.}},
\bauthor{\bsnm{{Takahashi}}, \binits{T.}},
\bauthor{\bsnm{{Tueller}}, \binits{J.}}:
\bjtitle{GRB Coordinates Network}
\bvolume{3935},
\bfpage{1}
(\byear{2005})
\end{barticle}
\endbibitem

\bibitem[\protect\citeauthoryear{{P{\'e}langeon}
  et~al.}{2008}]{Pelangeon2008_040701}
\begin{barticle}
\bauthor{\bsnm{{P{\'e}langeon}}, \binits{A.}},
\bauthor{\bsnm{{Atteia}}, \binits{J.-L.}},
\bauthor{\bsnm{{Nakagawa}}, \binits{Y.E.}},
\bauthor{\bsnm{{Hurley}}, \binits{K.}},
\bauthor{\bsnm{{Yoshida}}, \binits{A.}},
\bauthor{\bsnm{{Vanderspek}}, \binits{R.}},
\bauthor{\bsnm{{Suzuki}}, \binits{M.}},
\bauthor{\bsnm{{Kawai}}, \binits{N.}},
\bauthor{\bsnm{{Pizzichini}}, \binits{G.}},
\bauthor{\bsnm{{Bo{\"e}r}}, \binits{M.}},
\bauthor{\bsnm{{Braga}}, \binits{J.}},
\bauthor{\bsnm{{Crew}}, \binits{G.}},
\bauthor{\bsnm{{Donaghy}}, \binits{T.Q.}},
\bauthor{\bsnm{{Dezalay}}, \binits{J.-P.}},
\bauthor{\bsnm{{Doty}}, \binits{J.}},
\bauthor{\bsnm{{Fenimore}}, \binits{E.E.}},
\bauthor{\bsnm{{Galassi}}, \binits{M.}},
\bauthor{\bsnm{{Graziani}}, \binits{C.}},
\bauthor{\bsnm{{Jernigan}}, \binits{J.G.}},
\bauthor{\bsnm{{Lamb}}, \binits{D.Q.}},
\bauthor{\bsnm{{Levine}}, \binits{A.}},
\bauthor{\bsnm{{Manchand a}}, \binits{J.}},
\bauthor{\bsnm{{Martel}}, \binits{F.}},
\bauthor{\bsnm{{Matsuoka}}, \binits{M.}},
\bauthor{\bsnm{{Olive}}, \binits{J.-F.}},
\bauthor{\bsnm{{Prigozhin}}, \binits{G.}},
\bauthor{\bsnm{{Ricker}}, \binits{G.R.}},
\bauthor{\bsnm{{Sakamoto}}, \binits{T.}},
\bauthor{\bsnm{{Shirasaki}}, \binits{Y.}},
\bauthor{\bsnm{{Sugita}}, \binits{S.}},
\bauthor{\bsnm{{Takagishi}}, \binits{K.}},
\bauthor{\bsnm{{Tamagawa}}, \binits{T.}},
\bauthor{\bsnm{{Villasenor}}, \binits{J.}},
\bauthor{\bsnm{{Woosley}}, \binits{S.E.}},
\bauthor{\bsnm{{Yamauchi}}, \binits{M.}}:
\bjtitle{\aap}
\bvolume{491}(\bissue{1}),
\bfpage{157}
(\byear{2008}).
\arxivurl{0811.3304}.
doi:\doiurl{10.1051/0004-6361:200809709}
\end{barticle}
\endbibitem

\bibitem[\protect\citeauthoryear{Peng et~al.}{2005}]{Peng2005}
\begin{botherref}
\oauthor{\bsnm{Peng}, \binits{F.}},
\oauthor{\bsnm{K\"onigl}, \binits{A.}},
\oauthor{\bsnm{Granot}, \binits{J.}}:
Two-component jet models of gamma-ray burst sources
\textbf{626},
966
(2005).
doi:\doiurl{10.1086/430045}
\end{botherref}
\endbibitem

\bibitem[\protect\citeauthoryear{Perley and
  Cockeram}{2019}]{Perley2019_190829A}
\begin{barticle}
\bauthor{\bsnm{Perley}, \binits{D.A.}},
\bauthor{\bsnm{Cockeram}, \binits{A.M.}}:
\bjtitle{GRB Coordinates Network}
\bvolume{25657},
\bfpage{1}
(\byear{2019})
\end{barticle}
\endbibitem

\bibitem[\protect\citeauthoryear{Perley et~al.}{2006}]{Perley2006_051109B}
\begin{barticle}
\bauthor{\bsnm{Perley}, \binits{D.A.}},
\bauthor{\bsnm{Foley}, \binits{R.J.}},
\bauthor{\bsnm{Bloom}, \binits{J.S.}},
\bauthor{\bsnm{Butler}, \binits{N.R.}}:
\bjtitle{GRB Coordinates Network}
\bvolume{5387},
\bfpage{1}
(\byear{2006})
\end{barticle}
\endbibitem

\bibitem[\protect\citeauthoryear{Perley et~al.}{2008}]{Perley2008_070809}
\begin{barticle}
\bauthor{\bsnm{Perley}, \binits{D.A.}},
\bauthor{\bsnm{Bloom}, \binits{J.S.}},
\bauthor{\bsnm{Modjaz}, \binits{M.}},
\bauthor{\bsnm{Miller}, \binits{A.A.}},
\bauthor{\bsnm{Shiode}, \binits{J.}},
\bauthor{\bsnm{Brewer}, \binits{J.}},
\bauthor{\bsnm{Starr}, \binits{D.}},
\bauthor{\bsnm{Kennedy}, \binits{R.}}:
\bjtitle{GRB Coordinates Network}
\bvolume{7889},
\bfpage{1}
(\byear{2008})
\end{barticle}
\endbibitem

\bibitem[\protect\citeauthoryear{Perley et~al.}{2019}]{Perley2019_191019A}
\begin{barticle}
\bauthor{\bsnm{Perley}, \binits{D.A.}},
\bauthor{\bsnm{Malesani}, \binits{D.B.}},
\bauthor{\bsnm{Levan}, \binits{A.J.}},
\bauthor{\bsnm{Fynbo}, \binits{J.P.U.}},
\bauthor{\bsnm{Djupvik}, \binits{A.A.}},
\bauthor{\bparticle{et} \bsnm{al.}}:
\bjtitle{GRB Coordinates Network}
\bvolume{26062},
\bfpage{1}
(\byear{2019})
\end{barticle}
\endbibitem

\bibitem[\protect\citeauthoryear{Perri et~al.}{2018}]{Perri2018_180728A}
\begin{barticle}
\bauthor{\bsnm{Perri}, \binits{M.}},
\bauthor{\bsnm{Evans}, \binits{P.A.}},
\bauthor{\bsnm{Osborne}, \binits{J.P.}},
\bauthor{\bsnm{Burrows}, \binits{D.N.}},
\bauthor{\bsnm{Kennea}, \binits{J.A.}},
\bauthor{\bsnm{Cusumano}, \binits{G.}}:
\bjtitle{GRB Coordinates Network}
\bvolume{23049},
\bfpage{1}
(\byear{2018})
\end{barticle}
\endbibitem

\bibitem[\protect\citeauthoryear{Peters et~al.}{2019}]{Peters2019}
\begin{botherref}
\oauthor{\bsnm{Peters}, \binits{C.}},
\oauthor{\bparticle{van~der} \bsnm{Horst}, \binits{A.J.}},
\oauthor{\bsnm{Chomiuk}, \binits{L.}},
\oauthor{\bsnm{Kathirgamaraju}, \binits{A.}},
\oauthor{\bsnm{Duran}, \binits{R.B.}},
\oauthor{\bsnm{Giannios}, \binits{D.}},
\oauthor{\bsnm{Reynolds}, \binits{C.}},
\oauthor{\bsnm{Paragi}, \binits{Z.}},
\oauthor{\bsnm{Wilcots}, \binits{E.}}:
Observational constraints on late-time radio rebrightening of grb/supernovae
\textbf{872},
28
(2019).
doi:\doiurl{10.3847/1538-4357/aafb3c}
\end{botherref}
\endbibitem

\bibitem[\protect\citeauthoryear{Peterson and
  Price}{2003}]{Peterson2003_030329}
\begin{barticle}
\bauthor{\bsnm{Peterson}, \binits{B.A.}},
\bauthor{\bsnm{Price}, \binits{P.A.}}:
\bjtitle{GRB Coordinates Network}
\bvolume{1985},
\bfpage{1}
(\byear{2003})
\end{barticle}
\endbibitem

\bibitem[\protect\citeauthoryear{Pian et~al.}{1998}]{Pian1998_980425}
\begin{barticle}
\bauthor{\bsnm{Pian}, \binits{E.}},
\bauthor{\bsnm{Antonelli}, \binits{L.A.}},
\bauthor{\bsnm{Piro}, \binits{L.}},
\bauthor{\bsnm{Feroci}, \binits{M.}}:
\bjtitle{GRB Coordinates Network}
\bvolume{158},
\bfpage{1}
(\byear{1998})
\end{barticle}
\endbibitem

\bibitem[\protect\citeauthoryear{{Pian} et~al.}{2000}]{Pian2000_980425}
\begin{barticle}
\bauthor{\bsnm{{Pian}}, \binits{E.}},
\bauthor{\bsnm{{Amati}}, \binits{L.}},
\bauthor{\bsnm{{Antonelli}}, \binits{L.A.}},
\bauthor{\bsnm{{Butler}}, \binits{R.C.}},
\bauthor{\bsnm{{Costa}}, \binits{E.}},
\bauthor{\bsnm{{Cusumano}}, \binits{G.}},
\bauthor{\bsnm{{Danziger}}, \binits{J.}},
\bauthor{\bsnm{{Feroci}}, \binits{M.}},
\bauthor{\bsnm{{Fiore}}, \binits{F.}},
\bauthor{\bsnm{{Frontera}}, \binits{F.}},
\bauthor{\bsnm{{Giommi}}, \binits{P.}},
\bauthor{\bsnm{{Masetti}}, \binits{N.}},
\bauthor{\bsnm{{Muller}}, \binits{J.M.}},
\bauthor{\bsnm{{Nicastro}}, \binits{L.}},
\bauthor{\bsnm{{Oosterbroek}}, \binits{T.}},
\bauthor{\bsnm{{Orland ini}}, \binits{M.}},
\bauthor{\bsnm{{Owens}}, \binits{A.}},
\bauthor{\bsnm{{Palazzi}}, \binits{E.}},
\bauthor{\bsnm{{Parmar}}, \binits{A.}},
\bauthor{\bsnm{{Piro}}, \binits{L.}},
\bauthor{\bsnm{{in't Zand}}, \binits{J.J.M.}},
\bauthor{\bsnm{{Castro-Tirado}}, \binits{A.}},
\bauthor{\bsnm{{Coletta}}, \binits{A.}},
\bauthor{\bsnm{{Dal Fiume}}, \binits{D.}},
\bauthor{\bsnm{{Del Sordo}}, \binits{S.}},
\bauthor{\bsnm{{Heise}}, \binits{J.}},
\bauthor{\bsnm{{Soffitta}}, \binits{P.}},
\bauthor{\bsnm{{Torroni}}, \binits{V.}}:
\bjtitle{\apj}
\bvolume{536}(\bissue{2}),
\bfpage{778}
(\byear{2000}).
\arxivurl{astro-ph/9910235}.
doi:\doiurl{10.1086/308978}
\end{barticle}
\endbibitem

\bibitem[\protect\citeauthoryear{{Pian} et~al.}{2017}]{Pian2017}
\begin{barticle}
\bauthor{\bsnm{{Pian}}, \binits{E.}},
\bauthor{\bsnm{{D'Avanzo}}, \binits{P.}},
\bauthor{\bsnm{{Benetti}}, \binits{S.}},
\bauthor{\bsnm{{Branchesi}}, \binits{M.}},
\bauthor{\bsnm{{Brocato}}, \binits{E.}},
\bauthor{\bsnm{{Campana}}, \binits{S.}},
\bauthor{\bsnm{{Cappellaro}}, \binits{E.}},
\bauthor{\bsnm{{Covino}}, \binits{S.}},
\bauthor{\bsnm{{D'Elia}}, \binits{V.}},
\bauthor{\bsnm{{Fynbo}}, \binits{J.P.U.}},
\bauthor{\bsnm{{Getman}}, \binits{F.}},
\bauthor{\bsnm{{Ghirland a}}, \binits{G.}},
\bauthor{\bsnm{{Ghisellini}}, \binits{G.}},
\bauthor{\bsnm{{Grado}}, \binits{A.}},
\bauthor{\bsnm{{Greco}}, \binits{G.}},
\bauthor{\bsnm{{Hjorth}}, \binits{J.}},
\bauthor{\bsnm{{Kouveliotou}}, \binits{C.}},
\bauthor{\bsnm{{Levan}}, \binits{A.}},
\bauthor{\bsnm{{Limatola}}, \binits{L.}},
\bauthor{\bsnm{{Malesani}}, \binits{D.}},
\bauthor{\bsnm{{Mazzali}}, \binits{P.A.}},
\bauthor{\bsnm{{Melandri}}, \binits{A.}},
\bauthor{\bsnm{{M{\o}ller}}, \binits{P.}},
\bauthor{\bsnm{{Nicastro}}, \binits{L.}},
\bauthor{\bsnm{{Palazzi}}, \binits{E.}},
\bauthor{\bsnm{{Piranomonte}}, \binits{S.}},
\bauthor{\bsnm{{Rossi}}, \binits{A.}},
\bauthor{\bsnm{{Salafia}}, \binits{O.S.}},
\bauthor{\bsnm{{Selsing}}, \binits{J.}},
\bauthor{\bsnm{{Stratta}}, \binits{G.}},
\bauthor{\bsnm{{Tanaka}}, \binits{M.}},
\bauthor{\bsnm{{Tanvir}}, \binits{N.R.}},
\bauthor{\bsnm{{Tomasella}}, \binits{L.}},
\bauthor{\bsnm{{Watson}}, \binits{D.}},
\bauthor{\bsnm{{Yang}}, \binits{S.}},
\bauthor{\bsnm{{Amati}}, \binits{L.}},
\bauthor{\bsnm{{Antonelli}}, \binits{L.A.}},
\bauthor{\bsnm{{Ascenzi}}, \binits{S.}},
\bauthor{\bsnm{{Bernardini}}, \binits{M.G.}},
\bauthor{\bsnm{{Bo{\"e}r}}, \binits{M.}},
\bauthor{\bsnm{{Bufano}}, \binits{F.}},
\bauthor{\bsnm{{Bulgarelli}}, \binits{A.}},
\bauthor{\bsnm{{Capaccioli}}, \binits{M.}},
\bauthor{\bsnm{{Casella}}, \binits{P.}},
\bauthor{\bsnm{{Castro-Tirado}}, \binits{A.J.}},
\bauthor{\bsnm{{Chassande-Mottin}}, \binits{E.}},
\bauthor{\bsnm{{Ciolfi}}, \binits{R.}},
\bauthor{\bsnm{{Copperwheat}}, \binits{C.M.}},
\bauthor{\bsnm{{Dadina}}, \binits{M.}},
\bauthor{\bsnm{{De Cesare}}, \binits{G.}},
\bauthor{\bsnm{{di Paola}}, \binits{A.}},
\bauthor{\bsnm{{Fan}}, \binits{Y.Z.}},
\bauthor{\bsnm{{Gendre}}, \binits{B.}},
\bauthor{\bsnm{{Giuffrida}}, \binits{G.}},
\bauthor{\bsnm{{Giunta}}, \binits{A.}},
\bauthor{\bsnm{{Hunt}}, \binits{L.K.}},
\bauthor{\bsnm{{Israel}}, \binits{G.L.}},
\bauthor{\bsnm{{Jin}}, \binits{Z.-P.}},
\bauthor{\bsnm{{Kasliwal}}, \binits{M.M.}},
\bauthor{\bsnm{{Klose}}, \binits{S.}},
\bauthor{\bsnm{{Lisi}}, \binits{M.}},
\bauthor{\bsnm{{Longo}}, \binits{F.}},
\bauthor{\bsnm{{Maiorano}}, \binits{E.}},
\bauthor{\bsnm{{Mapelli}}, \binits{M.}},
\bauthor{\bsnm{{Masetti}}, \binits{N.}},
\bauthor{\bsnm{{Nava}}, \binits{L.}},
\bauthor{\bsnm{{Patricelli}}, \binits{B.}},
\bauthor{\bsnm{{Perley}}, \binits{D.}},
\bauthor{\bsnm{{Pescalli}}, \binits{A.}},
\bauthor{\bsnm{{Piran}}, \binits{T.}},
\bauthor{\bsnm{{Possenti}}, \binits{A.}},
\bauthor{\bsnm{{Pulone}}, \binits{L.}},
\bauthor{\bsnm{{Razzano}}, \binits{M.}},
\bauthor{\bsnm{{Salvaterra}}, \binits{R.}},
\bauthor{\bsnm{{Schipani}}, \binits{P.}},
\bauthor{\bsnm{{Spera}}, \binits{M.}},
\bauthor{\bsnm{{Stamerra}}, \binits{A.}},
\bauthor{\bsnm{{Stella}}, \binits{L.}},
\bauthor{\bsnm{{Tagliaferri}}, \binits{G.}},
\bauthor{\bsnm{{Testa}}, \binits{V.}},
\bauthor{\bsnm{{Troja}}, \binits{E.}},
\bauthor{\bsnm{{Turatto}}, \binits{M.}},
\bauthor{\bsnm{{Vergani}}, \binits{S.D.}},
\bauthor{\bsnm{{Vergani}}, \binits{D.}}:
\bjtitle{\nat}
\bvolume{551}(\bissue{7678}),
\bfpage{67}
(\byear{2017}).
\arxivurl{1710.05858}.
doi:\doiurl{10.1038/nature24298}
\end{barticle}
\endbibitem

\bibitem[\protect\citeauthoryear{{Planck Collaboration}
  et~al.}{2018}]{PlanckCollaboration2018}
\begin{botherref}
\oauthor{\bsnm{{Planck Collaboration}}},
\oauthor{\bsnm{{Aghanim}}, \binits{N.}},
\oauthor{\bsnm{{Akrami}}, \binits{Y.}},
\oauthor{\bsnm{{Ashdown}}, \binits{M.}},
\oauthor{\bsnm{{Aumont}}, \binits{J.}},
\oauthor{\bsnm{{Baccigalupi}}, \binits{C.}},
\oauthor{\bsnm{{Ballardini}}, \binits{M.}},
\oauthor{\bsnm{{Banday}}, \binits{A.J.}},
\oauthor{\bsnm{{Barreiro}}, \binits{R.B.}},
\oauthor{\bsnm{{Bartolo}}, \binits{N.}},
\oauthor{\bsnm{{Basak}}, \binits{S.}},
\oauthor{\bsnm{{Battye}}, \binits{R.}},
\oauthor{\bsnm{{Benabed}}, \binits{K.}},
\oauthor{\bsnm{{Bernard}}, \binits{J.-P.}},
\oauthor{\bsnm{{Bersanelli}}, \binits{M.}},
\oauthor{\bsnm{{Bielewicz}}, \binits{P.}},
\oauthor{\bsnm{{Bock}}, \binits{J.J.}},
\oauthor{\bsnm{{Bond}}, \binits{J.R.}},
\oauthor{\bsnm{{Borrill}}, \binits{J.}},
\oauthor{\bsnm{{Bouchet}}, \binits{F.R.}},
\oauthor{\bsnm{{Boulanger}}, \binits{F.}},
\oauthor{\bsnm{{Bucher}}, \binits{M.}},
\oauthor{\bsnm{{Burigana}}, \binits{C.}},
\oauthor{\bsnm{{Butler}}, \binits{R.C.}},
\oauthor{\bsnm{{Calabrese}}, \binits{E.}},
\oauthor{\bsnm{{Cardoso}}, \binits{J.-F.}},
\oauthor{\bsnm{{Carron}}, \binits{J.}},
\oauthor{\bsnm{{Challinor}}, \binits{A.}},
\oauthor{\bsnm{{Chiang}}, \binits{H.C.}},
\oauthor{\bsnm{{Chluba}}, \binits{J.}},
\oauthor{\bsnm{{Colombo}}, \binits{L.P.L.}},
\oauthor{\bsnm{{Combet}}, \binits{C.}},
\oauthor{\bsnm{{Contreras}}, \binits{D.}},
\oauthor{\bsnm{{Crill}}, \binits{B.P.}},
\oauthor{\bsnm{{Cuttaia}}, \binits{F.}},
\oauthor{\bsnm{{de Bernardis}}, \binits{P.}},
\oauthor{\bsnm{{de Zotti}}, \binits{G.}},
\oauthor{\bsnm{{Delabrouille}}, \binits{J.}},
\oauthor{\bsnm{{Delouis}}, \binits{J.-M.}},
\oauthor{\bsnm{{Di Valentino}}, \binits{E.}},
\oauthor{\bsnm{{Diego}}, \binits{J.M.}},
\oauthor{\bsnm{{Dor{\'e}}}, \binits{O.}},
\oauthor{\bsnm{{Douspis}}, \binits{M.}},
\oauthor{\bsnm{{Ducout}}, \binits{A.}},
\oauthor{\bsnm{{Dupac}}, \binits{X.}},
\oauthor{\bsnm{{Dusini}}, \binits{S.}},
\oauthor{\bsnm{{Efstathiou}}, \binits{G.}},
\oauthor{\bsnm{{Elsner}}, \binits{F.}},
\oauthor{\bsnm{{En{\ss}lin}}, \binits{T.A.}},
\oauthor{\bsnm{{Eriksen}}, \binits{H.K.}},
\oauthor{\bsnm{{Fantaye}}, \binits{Y.}},
\oauthor{\bsnm{{Farhang}}, \binits{M.}},
\oauthor{\bsnm{{Fergusson}}, \binits{J.}},
\oauthor{\bsnm{{Fernandez-Cobos}}, \binits{R.}},
\oauthor{\bsnm{{Finelli}}, \binits{F.}},
\oauthor{\bsnm{{Forastieri}}, \binits{F.}},
\oauthor{\bsnm{{Frailis}}, \binits{M.}},
\oauthor{\bsnm{{Fraisse}}, \binits{A.A.}},
\oauthor{\bsnm{{Franceschi}}, \binits{E.}},
\oauthor{\bsnm{{Frolov}}, \binits{A.}},
\oauthor{\bsnm{{Galeotta}}, \binits{S.}},
\oauthor{\bsnm{{Galli}}, \binits{S.}},
\oauthor{\bsnm{{Ganga}}, \binits{K.}},
\oauthor{\bsnm{{G{\'e}nova-Santos}}, \binits{R.T.}},
\oauthor{\bsnm{{Gerbino}}, \binits{M.}},
\oauthor{\bsnm{{Ghosh}}, \binits{T.}},
\oauthor{\bsnm{{Gonz{\'a}lez-Nuevo}}, \binits{J.}},
\oauthor{\bsnm{{G{\'o}rski}}, \binits{K.M.}},
\oauthor{\bsnm{{Gratton}}, \binits{S.}},
\oauthor{\bsnm{{Gruppuso}}, \binits{A.}},
\oauthor{\bsnm{{Gudmundsson}}, \binits{J.E.}},
\oauthor{\bsnm{{Hamann}}, \binits{J.}},
\oauthor{\bsnm{{Handley}}, \binits{W.}},
\oauthor{\bsnm{{Hansen}}, \binits{F.K.}},
\oauthor{\bsnm{{Herranz}}, \binits{D.}},
\oauthor{\bsnm{{Hildebrandt}}, \binits{S.R.}},
\oauthor{\bsnm{{Hivon}}, \binits{E.}},
\oauthor{\bsnm{{Huang}}, \binits{Z.}},
\oauthor{\bsnm{{Jaffe}}, \binits{A.H.}},
\oauthor{\bsnm{{Jones}}, \binits{W.C.}},
\oauthor{\bsnm{{Karakci}}, \binits{A.}},
\oauthor{\bsnm{{Keih{\"a}nen}}, \binits{E.}},
\oauthor{\bsnm{{Keskitalo}}, \binits{R.}},
\oauthor{\bsnm{{Kiiveri}}, \binits{K.}},
\oauthor{\bsnm{{Kim}}, \binits{J.}},
\oauthor{\bsnm{{Kisner}}, \binits{T.S.}},
\oauthor{\bsnm{{Knox}}, \binits{L.}},
\oauthor{\bsnm{{Krachmalnicoff}}, \binits{N.}},
\oauthor{\bsnm{{Kunz}}, \binits{M.}},
\oauthor{\bsnm{{Kurki-Suonio}}, \binits{H.}},
\oauthor{\bsnm{{Lagache}}, \binits{G.}},
\oauthor{\bsnm{{Lamarre}}, \binits{J.-M.}},
\oauthor{\bsnm{{Lasenby}}, \binits{A.}},
\oauthor{\bsnm{{Lattanzi}}, \binits{M.}},
\oauthor{\bsnm{{Lawrence}}, \binits{C.R.}},
\oauthor{\bsnm{{Le Jeune}}, \binits{M.}},
\oauthor{\bsnm{{Lemos}}, \binits{P.}},
\oauthor{\bsnm{{Lesgourgues}}, \binits{J.}},
\oauthor{\bsnm{{Levrier}}, \binits{F.}},
\oauthor{\bsnm{{Lewis}}, \binits{A.}},
\oauthor{\bsnm{{Liguori}}, \binits{M.}},
\oauthor{\bsnm{{Lilje}}, \binits{P.B.}},
\oauthor{\bsnm{{Lilley}}, \binits{M.}},
\oauthor{\bsnm{{Lindholm}}, \binits{V.}},
\oauthor{\bsnm{{L{\'o}pez-Caniego}}, \binits{M.}},
\oauthor{\bsnm{{Lubin}}, \binits{P.M.}},
\oauthor{\bsnm{{Ma}}, \binits{Y.-Z.}},
\oauthor{\bsnm{{Mac{\'\i}as-P{\'e}rez}}, \binits{J.F.}},
\oauthor{\bsnm{{Maggio}}, \binits{G.}},
\oauthor{\bsnm{{Maino}}, \binits{D.}},
\oauthor{\bsnm{{Mandolesi}}, \binits{N.}},
\oauthor{\bsnm{{Mangilli}}, \binits{A.}},
\oauthor{\bsnm{{Marcos-Caballero}}, \binits{A.}},
\oauthor{\bsnm{{Maris}}, \binits{M.}},
\oauthor{\bsnm{{Martin}}, \binits{P.G.}},
\oauthor{\bsnm{{Martinelli}}, \binits{M.}},
\oauthor{\bsnm{{Mart{\'\i}nez-Gonz{\'a}lez}}, \binits{E.}},
\oauthor{\bsnm{{Matarrese}}, \binits{S.}},
\oauthor{\bsnm{{Mauri}}, \binits{N.}},
\oauthor{\bsnm{{McEwen}}, \binits{J.D.}},
\oauthor{\bsnm{{Meinhold}}, \binits{P.R.}},
\oauthor{\bsnm{{Melchiorri}}, \binits{A.}},
\oauthor{\bsnm{{Mennella}}, \binits{A.}},
\oauthor{\bsnm{{Migliaccio}}, \binits{M.}},
\oauthor{\bsnm{{Millea}}, \binits{M.}},
\oauthor{\bsnm{{Mitra}}, \binits{S.}},
\oauthor{\bsnm{{Miville-Desch{\^e}nes}}, \binits{M.-A.}},
\oauthor{\bsnm{{Molinari}}, \binits{D.}},
\oauthor{\bsnm{{Montier}}, \binits{L.}},
\oauthor{\bsnm{{Morgante}}, \binits{G.}},
\oauthor{\bsnm{{Moss}}, \binits{A.}},
\oauthor{\bsnm{{Natoli}}, \binits{P.}},
\oauthor{\bsnm{{N{\o}rgaard-Nielsen}}, \binits{H.U.}},
\oauthor{\bsnm{{Pagano}}, \binits{L.}},
\oauthor{\bsnm{{Paoletti}}, \binits{D.}},
\oauthor{\bsnm{{Partridge}}, \binits{B.}},
\oauthor{\bsnm{{Patanchon}}, \binits{G.}},
\oauthor{\bsnm{{Peiris}}, \binits{H.V.}},
\oauthor{\bsnm{{Perrotta}}, \binits{F.}},
\oauthor{\bsnm{{Pettorino}}, \binits{V.}},
\oauthor{\bsnm{{Piacentini}}, \binits{F.}},
\oauthor{\bsnm{{Polastri}}, \binits{L.}},
\oauthor{\bsnm{{Polenta}}, \binits{G.}},
\oauthor{\bsnm{{Puget}}, \binits{J.-L.}},
\oauthor{\bsnm{{Rachen}}, \binits{J.P.}},
\oauthor{\bsnm{{Reinecke}}, \binits{M.}},
\oauthor{\bsnm{{Remazeilles}}, \binits{M.}},
\oauthor{\bsnm{{Renzi}}, \binits{A.}},
\oauthor{\bsnm{{Rocha}}, \binits{G.}},
\oauthor{\bsnm{{Rosset}}, \binits{C.}},
\oauthor{\bsnm{{Roudier}}, \binits{G.}},
\oauthor{\bsnm{{Rubi{\~n}o-Mart{\'\i}n}}, \binits{J.A.}},
\oauthor{\bsnm{{Ruiz-Granados}}, \binits{B.}},
\oauthor{\bsnm{{Salvati}}, \binits{L.}},
\oauthor{\bsnm{{Sandri}}, \binits{M.}},
\oauthor{\bsnm{{Savelainen}}, \binits{M.}},
\oauthor{\bsnm{{Scott}}, \binits{D.}},
\oauthor{\bsnm{{Shellard}}, \binits{E.P.S.}},
\oauthor{\bsnm{{Sirignano}}, \binits{C.}},
\oauthor{\bsnm{{Sirri}}, \binits{G.}},
\oauthor{\bsnm{{Spencer}}, \binits{L.D.}},
\oauthor{\bsnm{{Sunyaev}}, \binits{R.}},
\oauthor{\bsnm{{Suur-Uski}}, \binits{A.-S.}},
\oauthor{\bsnm{{Tauber}}, \binits{J.A.}},
\oauthor{\bsnm{{Tavagnacco}}, \binits{D.}},
\oauthor{\bsnm{{Tenti}}, \binits{M.}},
\oauthor{\bsnm{{Toffolatti}}, \binits{L.}},
\oauthor{\bsnm{{Tomasi}}, \binits{M.}},
\oauthor{\bsnm{{Trombetti}}, \binits{T.}},
\oauthor{\bsnm{{Valenziano}}, \binits{L.}},
\oauthor{\bsnm{{Valiviita}}, \binits{J.}},
\oauthor{\bsnm{{Van Tent}}, \binits{B.}},
\oauthor{\bsnm{{Vibert}}, \binits{L.}},
\oauthor{\bsnm{{Vielva}}, \binits{P.}},
\oauthor{\bsnm{{Villa}}, \binits{F.}},
\oauthor{\bsnm{{Vittorio}}, \binits{N.}},
\oauthor{\bsnm{{Wand elt}}, \binits{B.D.}},
\oauthor{\bsnm{{Wehus}}, \binits{I.K.}},
\oauthor{\bsnm{{White}}, \binits{M.}},
\oauthor{\bsnm{{White}}, \binits{S.D.M.}},
\oauthor{\bsnm{{Zacchei}}, \binits{A.}},
\oauthor{\bsnm{{Zonca}}, \binits{A.}}:
arXiv e-prints,
1807
(2018).
\arxivurl{1807.06209}
\end{botherref}
\endbibitem

\bibitem[\protect\citeauthoryear{Poole and Troja}{2006}]{Poole2006_060502B}
\begin{barticle}
\bauthor{\bsnm{Poole}, \binits{T.S.}},
\bauthor{\bsnm{Troja}, \binits{E.}}:
\bjtitle{GRB Coordinates Network}
\bvolume{5069},
\bfpage{1}
(\byear{2006})
\end{barticle}
\endbibitem

\bibitem[\protect\citeauthoryear{Pozanenko
  et~al.}{2015}]{Pozanenko2015_150518A}
\begin{barticle}
\bauthor{\bsnm{Pozanenko}, \binits{A.}},
\bauthor{\bsnm{Mazaeva}, \binits{E.}},
\bauthor{\bsnm{Sergeev}, \binits{A.}},
\bauthor{\bsnm{Reva}, \binits{I.}},
\bauthor{\bsnm{Volnova}, \binits{A.}},
\bauthor{\bsnm{Klunko}, \binits{E.}},
\bauthor{\bsnm{Korobtsev}, \binits{I.}}:
\bjtitle{GRB Coordinates Network}
\bvolume{17903},
\bfpage{1}
(\byear{2015})
\end{barticle}
\endbibitem

\bibitem[\protect\citeauthoryear{Price et~al.}{2005}]{Price2005_050709}
\begin{barticle}
\bauthor{\bsnm{Price}, \binits{P.A.}},
\bauthor{\bsnm{Roth}, \binits{K.}},
\bauthor{\bsnm{Fox}, \binits{D.W.}},
\bauthor{\bsnm{Price}, \binits{P.A.}},
\bauthor{\bsnm{Roth}, \binits{K.}},
\bauthor{\bsnm{Fox}, \binits{D.W.}}:
\bjtitle{GCN}
\bvolume{3605},
\bfpage{1}
(\byear{2005})
\end{barticle}
\endbibitem

\bibitem[\protect\citeauthoryear{Price et~al.}{2006}]{Price2006_060614}
\begin{barticle}
\bauthor{\bsnm{Price}, \binits{P.A.}},
\bauthor{\bsnm{Berger}, \binits{E.}},
\bauthor{\bsnm{Fox}, \binits{D.B.}},
\bauthor{\bsnm{Price}, \binits{P.A.}},
\bauthor{\bsnm{Berger}, \binits{E.}},
\bauthor{\bsnm{Fox}, \binits{D.B.}}:
\bjtitle{GCN}
\bvolume{5275},
\bfpage{1}
(\byear{2006})
\end{barticle}
\endbibitem

\bibitem[\protect\citeauthoryear{Prochaska et~al.}{2003}]{Prochaska2003_031203}
\begin{barticle}
\bauthor{\bsnm{Prochaska}, \binits{J.X.}},
\bauthor{\bsnm{Bloom}, \binits{J.S.}},
\bauthor{\bsnm{Chen}, \binits{H.W.}},
\bauthor{\bsnm{Hurley}, \binits{K.}},
\bauthor{\bsnm{Dressler}, \binits{A.}},
\bauthor{\bsnm{Osip}, \binits{D.}},
\bauthor{\bsnm{Prochaska}, \binits{J.X.}},
\bauthor{\bsnm{Bloom}, \binits{J.S.}},
\bauthor{\bsnm{Chen}, \binits{H.W.}},
\bauthor{\bsnm{Hurley}, \binits{K.}},
\bauthor{\bsnm{Dressler}, \binits{A.}},
\bauthor{\bsnm{Osip}, \binits{D.}}:
\bjtitle{GCN}
\bvolume{2482},
\bfpage{1}
(\byear{2003})
\end{barticle}
\endbibitem

\bibitem[\protect\citeauthoryear{Prochaska
  et~al.}{2005a}]{Prochaska2005_050724}
\begin{barticle}
\bauthor{\bsnm{Prochaska}, \binits{J.X.}},
\bauthor{\bsnm{Bloom}, \binits{J.S.}},
\bauthor{\bsnm{Chen}, \binits{H.-W.}},
\bauthor{\bsnm{Hansen}, \binits{B.}},
\bauthor{\bsnm{Kalirai}, \binits{J.}},
\bauthor{\bsnm{Rich}, \binits{M.}},
\bauthor{\bsnm{Richer}, \binits{H.}},
\bauthor{\bsnm{Prochaska}, \binits{J.X.}},
\bauthor{\bsnm{Bloom}, \binits{J.S.}},
\bauthor{\bsnm{Chen}, \binits{H.-W.}},
\bauthor{\bsnm{Hansen}, \binits{B.}},
\bauthor{\bsnm{Kalirai}, \binits{J.}},
\bauthor{\bsnm{Rich}, \binits{M.}},
\bauthor{\bsnm{Richer}, \binits{H.}}:
\bjtitle{GCN}
\bvolume{3700},
\bfpage{1}
(\byear{2005}a)
\end{barticle}
\endbibitem

\bibitem[\protect\citeauthoryear{Prochaska
  et~al.}{2005b}]{Prochaska2005_050509B}
\begin{barticle}
\bauthor{\bsnm{Prochaska}, \binits{J.X.}},
\bauthor{\bsnm{Cooper}, \binits{M.}},
\bauthor{\bsnm{Newman}, \binits{J.}},
\bauthor{\bsnm{Bloom}, \binits{J.S.}},
\bauthor{\bsnm{Hurley}, \binits{K.}},
\bauthor{\bsnm{Blake}, \binits{C.}},
\bauthor{\bsnm{Gerke}, \binits{B.}},
\bauthor{\bsnm{Chen}, \binits{H.W.}}:
\bjtitle{GRB Coordinates Network}
\bvolume{3390},
\bfpage{1}
(\byear{2005}b)
\end{barticle}
\endbibitem

\bibitem[\protect\citeauthoryear{{Punturo}
  et~al.}{2010}]{Punturo2010_einsteintelescope}
\begin{barticle}
\bauthor{\bsnm{{Punturo}}, \binits{M.}},
\bauthor{\bsnm{{Abernathy}}, \binits{M.}},
\bauthor{\bsnm{{Acernese}}, \binits{F.}},
\bauthor{\bsnm{{Allen}}, \binits{B.}},
\bauthor{\bsnm{{Andersson}}, \binits{N.}},
\bauthor{\bsnm{{Arun}}, \binits{K.}},
\bauthor{\bsnm{{Barone}}, \binits{F.}},
\bauthor{\bsnm{{Barr}}, \binits{B.}},
\bauthor{\bsnm{{Barsuglia}}, \binits{M.}},
\bauthor{\bsnm{{Beker}}, \binits{M.}},
\bauthor{\bsnm{{Beveridge}}, \binits{N.}},
\bauthor{\bsnm{{Birindelli}}, \binits{S.}},
\bauthor{\bsnm{{Bose}}, \binits{S.}},
\bauthor{\bsnm{{Bosi}}, \binits{L.}},
\bauthor{\bsnm{{Braccini}}, \binits{S.}},
\bauthor{\bsnm{{Bradaschia}}, \binits{C.}},
\bauthor{\bsnm{{Bulik}}, \binits{T.}},
\bauthor{\bsnm{{Calloni}}, \binits{E.}},
\bauthor{\bsnm{{Cella}}, \binits{G.}},
\bauthor{\bsnm{{Chassande Mottin}}, \binits{E.}},
\bauthor{\bsnm{{Chelkowski}}, \binits{S.}},
\bauthor{\bsnm{{Chincarini}}, \binits{A.}},
\bauthor{\bsnm{{Clark}}, \binits{J.}},
\bauthor{\bsnm{{Coccia}}, \binits{E.}},
\bauthor{\bsnm{{Colacino}}, \binits{C.}},
\bauthor{\bsnm{{Colas}}, \binits{J.}},
\bauthor{\bsnm{{Cumming}}, \binits{A.}},
\bauthor{\bsnm{{Cunningham}}, \binits{L.}},
\bauthor{\bsnm{{Cuoco}}, \binits{E.}},
\bauthor{\bsnm{{Danilishin}}, \binits{S.}},
\bauthor{\bsnm{{Danzmann}}, \binits{K.}},
\bauthor{\bsnm{{De Luca}}, \binits{G.}},
\bauthor{\bsnm{{De Salvo}}, \binits{R.}},
\bauthor{\bsnm{{Dent}}, \binits{T.}},
\bauthor{\bsnm{{De Rosa}}, \binits{R.}},
\bauthor{\bsnm{{Di Fiore}}, \binits{L.}},
\bauthor{\bsnm{{Di Virgilio}}, \binits{A.}},
\bauthor{\bsnm{{Doets}}, \binits{M.}},
\bauthor{\bsnm{{Fafone}}, \binits{V.}},
\bauthor{\bsnm{{Falferi}}, \binits{P.}},
\bauthor{\bsnm{{Flaminio}}, \binits{R.}},
\bauthor{\bsnm{{Franc}}, \binits{J.}},
\bauthor{\bsnm{{Frasconi}}, \binits{F.}},
\bauthor{\bsnm{{Freise}}, \binits{A.}},
\bauthor{\bsnm{{Fulda}}, \binits{P.}},
\bauthor{\bsnm{{Gair}}, \binits{J.}},
\bauthor{\bsnm{{Gemme}}, \binits{G.}},
\bauthor{\bsnm{{Gennai}}, \binits{A.}},
\bauthor{\bsnm{{Giazotto}}, \binits{A.}},
\bauthor{\bsnm{{Glampedakis}}, \binits{K.}},
\bauthor{\bsnm{{Granata}}, \binits{M.}},
\bauthor{\bsnm{{Grote}}, \binits{H.}},
\bauthor{\bsnm{{Guidi}}, \binits{G.}},
\bauthor{\bsnm{{Hammond}}, \binits{G.}},
\bauthor{\bsnm{{Hannam}}, \binits{M.}},
\bauthor{\bsnm{{Harms}}, \binits{J.}},
\bauthor{\bsnm{{Heinert}}, \binits{D.}},
\bauthor{\bsnm{{Hendry}}, \binits{M.}},
\bauthor{\bsnm{{Heng}}, \binits{I.}},
\bauthor{\bsnm{{Hennes}}, \binits{E.}},
\bauthor{\bsnm{{Hild}}, \binits{S.}},
\bauthor{\bsnm{{Hough}}, \binits{J.}},
\bauthor{\bsnm{{Husa}}, \binits{S.}},
\bauthor{\bsnm{{Huttner}}, \binits{S.}},
\bauthor{\bsnm{{Jones}}, \binits{G.}},
\bauthor{\bsnm{{Khalili}}, \binits{F.}},
\bauthor{\bsnm{{Kokeyama}}, \binits{K.}},
\bauthor{\bsnm{{Kokkotas}}, \binits{K.}},
\bauthor{\bsnm{{Krishnan}}, \binits{B.}},
\bauthor{\bsnm{{Lorenzini}}, \binits{M.}},
\bauthor{\bsnm{{L{\"u}ck}}, \binits{H.}},
\bauthor{\bsnm{{Majorana}}, \binits{E.}},
\bauthor{\bsnm{{Mandel}}, \binits{I.}},
\bauthor{\bsnm{{Mandic}}, \binits{V.}},
\bauthor{\bsnm{{Martin}}, \binits{I.}},
\bauthor{\bsnm{{Michel}}, \binits{C.}},
\bauthor{\bsnm{{Minenkov}}, \binits{Y.}},
\bauthor{\bsnm{{Morgado}}, \binits{N.}},
\bauthor{\bsnm{{Mosca}}, \binits{S.}},
\bauthor{\bsnm{{Mours}}, \binits{B.}},
\bauthor{\bsnm{{M{\"u}ller-Ebhardt}}, \binits{H.}},
\bauthor{\bsnm{{Murray}}, \binits{P.}},
\bauthor{\bsnm{{Nawrodt}}, \binits{R.}},
\bauthor{\bsnm{{Nelson}}, \binits{J.}},
\bauthor{\bsnm{{Oshaughnessy}}, \binits{R.}},
\bauthor{\bsnm{{Ott}}, \binits{C.D.}},
\bauthor{\bsnm{{Palomba}}, \binits{C.}},
\bauthor{\bsnm{{Paoli}}, \binits{A.}},
\bauthor{\bsnm{{Parguez}}, \binits{G.}},
\bauthor{\bsnm{{Pasqualetti}}, \binits{A.}},
\bauthor{\bsnm{{Passaquieti}}, \binits{R.}},
\bauthor{\bsnm{{Passuello}}, \binits{D.}},
\bauthor{\bsnm{{Pinard}}, \binits{L.}},
\bauthor{\bsnm{{Poggiani}}, \binits{R.}},
\bauthor{\bsnm{{Popolizio}}, \binits{P.}},
\bauthor{\bsnm{{Prato}}, \binits{M.}},
\bauthor{\bsnm{{Puppo}}, \binits{P.}},
\bauthor{\bsnm{{Rabeling}}, \binits{D.}},
\bauthor{\bsnm{{Rapagnani}}, \binits{P.}},
\bauthor{\bsnm{{Read}}, \binits{J.}},
\bauthor{\bsnm{{Regimbau}}, \binits{T.}},
\bauthor{\bsnm{{Rehbein}}, \binits{H.}},
\bauthor{\bsnm{{Reid}}, \binits{S.}},
\bauthor{\bsnm{{Rezzolla}}, \binits{L.}},
\bauthor{\bsnm{{Ricci}}, \binits{F.}},
\bauthor{\bsnm{{Richard}}, \binits{F.}},
\bauthor{\bsnm{{Rocchi}}, \binits{A.}},
\bauthor{\bsnm{{Rowan}}, \binits{S.}},
\bauthor{\bsnm{{R{\"u}diger}}, \binits{A.}},
\bauthor{\bsnm{{Sassolas}}, \binits{B.}},
\bauthor{\bsnm{{Sathyaprakash}}, \binits{B.}},
\bauthor{\bsnm{{Schnabel}}, \binits{R.}},
\bauthor{\bsnm{{Schwarz}}, \binits{C.}},
\bauthor{\bsnm{{Seidel}}, \binits{P.}},
\bauthor{\bsnm{{Sintes}}, \binits{A.}},
\bauthor{\bsnm{{Somiya}}, \binits{K.}},
\bauthor{\bsnm{{Speirits}}, \binits{F.}},
\bauthor{\bsnm{{Strain}}, \binits{K.}},
\bauthor{\bsnm{{Strigin}}, \binits{S.}},
\bauthor{\bsnm{{Sutton}}, \binits{P.}},
\bauthor{\bsnm{{Tarabrin}}, \binits{S.}},
\bauthor{\bsnm{{Th{\"u}ring}}, \binits{A.}},
\bauthor{\bsnm{{van den Brand}}, \binits{J.}},
\bauthor{\bsnm{{van Leewen}}, \binits{C.}},
\bauthor{\bsnm{{van Veggel}}, \binits{M.}},
\bauthor{\bsnm{{van den Broeck}}, \binits{C.}},
\bauthor{\bsnm{{Vecchio}}, \binits{A.}},
\bauthor{\bsnm{{Veitch}}, \binits{J.}},
\bauthor{\bsnm{{Vetrano}}, \binits{F.}},
\bauthor{\bsnm{{Vicere}}, \binits{A.}},
\bauthor{\bsnm{{Vyatchanin}}, \binits{S.}},
\bauthor{\bsnm{{Willke}}, \binits{B.}},
\bauthor{\bsnm{{Woan}}, \binits{G.}},
\bauthor{\bsnm{{Wolfango}}, \binits{P.}},
\bauthor{\bsnm{{Yamamoto}}, \binits{K.}}:
\bjtitle{Classical and Quantum Gravity}
\bvolume{27}(\bissue{19}),
\bfpage{194002}
(\byear{2010}).
doi:\doiurl{10.1088/0264-9381/27/19/194002}
\end{barticle}
\endbibitem

\bibitem[\protect\citeauthoryear{Razzaque et~al.}{2004}]{Razzaque2004}
\begin{botherref}
\oauthor{\bsnm{Razzaque}, \binits{S.}},
\oauthor{\bsnm{M\'eszáros}, \binits{P.}},
\oauthor{\bsnm{Waxman}, \binits{E.}}:
Neutrino signatures of the supernova: Gamma ray burst relationship
\textbf{69},
23001
(2004).
doi:\doiurl{10.1103/PhysRevD.69.023001}
\end{botherref}
\endbibitem

\bibitem[\protect\citeauthoryear{Reichart}{2003}]{Reichart2003_031203}
\begin{barticle}
\bauthor{\bsnm{Reichart}, \binits{D.}}:
\bjtitle{GRB Coordinates Network}
\bvolume{2469},
\bfpage{1}
(\byear{2003})
\end{barticle}
\endbibitem

\bibitem[\protect\citeauthoryear{{Reitze}
  et~al.}{2019}]{Reitz2019_cosmis_explorer}
\begin{barticle}
\bauthor{\bsnm{{Reitze}}, \binits{D.}},
\bauthor{\bsnm{{LIGO Laboratory: California Institute of Technology}}},
\bauthor{\bsnm{{LIGO Laboratory: Massachusetts Institute of Technology}}},
\bauthor{\bsnm{{LIGO Hanford Observatory}}},
\bauthor{\bsnm{{LIGO Livingston Observatory}}}:
\bjtitle{\baas}
\bvolume{51}(\bissue{3}),
\bfpage{141}
(\byear{2019}).
\arxivurl{1903.04615}
\end{barticle}
\endbibitem

\bibitem[\protect\citeauthoryear{{Remou{\'e}} et~al.}{2010}]{Remoue2010}
\begin{barticle}
\bauthor{\bsnm{{Remou{\'e}}}, \binits{N.}},
\bauthor{\bsnm{{Barret}}, \binits{D.}},
\bauthor{\bsnm{{Godet}}, \binits{O.}},
\bauthor{\bsnm{{Mandrou}}, \binits{P.}}:
\bjtitle{Nuclear Instruments and Methods in Physics Research A}
\bvolume{618}(\bissue{1-3}),
\bfpage{199}
(\byear{2010}).
doi:\doiurl{10.1016/j.nima.2010.02.137}
\end{barticle}
\endbibitem

\bibitem[\protect\citeauthoryear{Reva et~al.}{2019}]{Reva2019_191019A}
\begin{barticle}
\bauthor{\bsnm{Reva}, \binits{I.}},
\bauthor{\bsnm{Pozanenko}, \binits{A.}},
\bauthor{\bsnm{Krugov}, \binits{M.}},
\bauthor{\bsnm{Belkin}, \binits{S.}},
\bauthor{\bsnm{Mazaeva}, \binits{E.}},
\bauthor{\bsnm{Volnova}, \binits{A.}},
\bauthor{\bsnm{Team}, \binits{I.-G.-F.}}:
\bjtitle{GRB Coordinates Network}
\bvolume{26036},
\bfpage{1}
(\byear{2019})
\end{barticle}
\endbibitem

\bibitem[\protect\citeauthoryear{Rhodes et~al.}{2020}]{Rhodes2020_190829A}
\begin{botherref}
\oauthor{\bsnm{Rhodes}, \binits{L.}},
\oauthor{\bparticle{van~der} \bsnm{Horst}, \binits{A.J.}},
\oauthor{\bsnm{Fender}, \binits{R.}},
\oauthor{\bsnm{Monageng}, \binits{I.}},
\oauthor{\bsnm{Anderson}, \binits{G.E.}},
\oauthor{\bsnm{Antoniadis}, \binits{J.}},
\oauthor{\bsnm{Bietenholz}, \binits{M.F.}},
\oauthor{\bsnm{Bottcher}, \binits{M.}},
\oauthor{\bsnm{Bright}, \binits{J.S.}},
\oauthor{\bsnm{Kouveliotou}, \binits{C.}},
\oauthor{\bsnm{Kramer}, \binits{M.}},
\oauthor{\bsnm{Motta}, \binits{S.E.}},
\oauthor{\bsnm{Williams}, \binits{D.R.A.}},
\oauthor{\bsnm{Woudt}, \binits{P.A.}},
\oauthor{\bsnm{.}}:
arXiv e-prints,
2004
(2020)
\end{botherref}
\endbibitem

\bibitem[\protect\citeauthoryear{Romano et~al.}{2005}]{Romano2005_050219A}
\begin{barticle}
\bauthor{\bsnm{Romano}, \binits{P.}},
\bauthor{\bsnm{Perri}, \binits{M.}},
\bauthor{\bsnm{Beardmore}, \binits{A.}},
\bauthor{\bsnm{Mangano}, \binits{V.}},
\bauthor{\bsnm{Burrows}, \binits{D.N.}},
\bauthor{\bsnm{Hill}, \binits{J.E.}},
\bauthor{\bsnm{Zhang}, \binits{B.}},
\bauthor{\bsnm{Gehrels}, \binits{N.}}:
\bjtitle{GRB Coordinates Network}
\bvolume{3036},
\bfpage{1}
(\byear{2005})
\end{barticle}
\endbibitem

\bibitem[\protect\citeauthoryear{Rossi et~al.}{2014}]{Rossi2014_050219A}
\begin{barticle}
\bauthor{\bsnm{Rossi}, \binits{A.}},
\bauthor{\bsnm{Piranomonte}, \binits{S.}},
\bauthor{\bsnm{Savaglio}, \binits{S.}},
\bauthor{\bsnm{Palazzi}, \binits{E.}},
\bauthor{\bsnm{Micha{\l}owski}, \binits{M.s.J.}},
\bauthor{\bsnm{Klose}, \binits{S.}},
\bauthor{\bsnm{Hunt}, \binits{L.s.K.}},
\bauthor{\bsnm{Amati}, \binits{L.}},
\bauthor{\bsnm{Elliott}, \binits{J.}},
\bauthor{\bsnm{Greiner}, \binits{J.}},
\bauthor{\bsnm{Guidorzi}, \binits{C.}},
\bauthor{\bsnm{Japelj}, \binits{J.}},
\bauthor{\bsnm{Kann}, \binits{D.s.A.}},
\bauthor{\bsnm{{Lo Faro}}, \binits{B.}},
\bauthor{\bsnm{{Nicuesa Guelbenzu}}, \binits{A.}},
\bauthor{\bsnm{Schulze}, \binits{S.}},
\bauthor{\bsnm{Vergani}, \binits{S.s.D.}},
\bauthor{\bsnm{Arnold}, \binits{L.s.A.}},
\bauthor{\bsnm{Covino}, \binits{S.}},
\bauthor{\bsnm{D'Elia}, \binits{V.}},
\bauthor{\bsnm{Ferrero}, \binits{P.}},
\bauthor{\bsnm{Filgas}, \binits{R.}},
\bauthor{\bsnm{Goldoni}, \binits{P.}},
\bauthor{\bsnm{{K{\"{u}}pc{\"{u}} Yolda\textcommabelow s}}, \binits{A.}},
\bauthor{\bsnm{{Le Borgne}}, \binits{D.}},
\bauthor{\bsnm{Pian}, \binits{E.}},
\bauthor{\bsnm{Schady}, \binits{P.}},
\bauthor{\bsnm{Stratta}, \binits{G.}},
\bauthor{\bsnm{Faro}, \binits{B.L.}},
\bauthor{\bsnm{Guelbenzu}, \binits{A.N.}},
\bauthor{\bsnm{Schulze}, \binits{S.}},
\bauthor{\bsnm{Vergani}, \binits{S.s.D.}},
\bauthor{\bsnm{Arnold}, \binits{L.s.A.}},
\bauthor{\bsnm{Covino}, \binits{S.}},
\bauthor{\bsnm{D'Elia}, \binits{V.}},
\bauthor{\bsnm{Ferrero}, \binits{P.}},
\bauthor{\bsnm{Filgas}, \binits{R.}},
\bauthor{\bsnm{Goldoni}, \binits{P.}},
\bauthor{\bsnm{Yoldaş}, \binits{A.K.}},
\bauthor{\bsnm{Borgne}, \binits{D.L.}},
\bauthor{\bsnm{Pian}, \binits{E.}},
\bauthor{\bsnm{Schady}, \binits{P.}},
\bauthor{\bsnm{Stratta}, \binits{G.}}:
\bjtitle{\aap}
\bvolume{572},
\bfpage{47}
(\byear{2014}).
\arxivurl{1409.0017}.
doi:\doiurl{10.1051/0004-6361/201423865}
\end{barticle}
\endbibitem

\bibitem[\protect\citeauthoryear{Rossi et~al.}{2018}]{Rossi2018_180728A}
\begin{barticle}
\bauthor{\bsnm{Rossi}, \binits{A.}},
\bauthor{\bsnm{Izzo}, \binits{L.}},
\bauthor{\bsnm{Milvang-Jensen}, \binits{B.}},
\bauthor{\bsnm{Perley}, \binits{D.A.}},
\bauthor{\bparticle{de} \bsnm{Ugarte~Postigo}, \binits{A.}},
\bauthor{\bsnm{Kann}, \binits{D.A.}},
\bauthor{\bsnm{Levan}, \binits{A.J.}},
\bauthor{\bsnm{Tanvir}, \binits{N.R.}},
\bauthor{\bsnm{Covino}, \binits{S.}},
\bauthor{\bsnm{Malesani}, \binits{D.B.}}:
\bjtitle{GRB Coordinates Network}
\bvolume{23055},
\bfpage{1}
(\byear{2018})
\end{barticle}
\endbibitem

\bibitem[\protect\citeauthoryear{Rowlinson
  et~al.}{2010}]{Rowlinson2010_080905A}
\begin{barticle}
\bauthor{\bsnm{Rowlinson}, \binits{A.}},
\bauthor{\bsnm{Wiersema}, \binits{K.}},
\bauthor{\bsnm{Levan}, \binits{A.J.}},
\bauthor{\bsnm{Tanvir}, \binits{N.R.}},
\bauthor{\bsnm{O'Brien}, \binits{P.T.}},
\bauthor{\bsnm{Rol}, \binits{E.}},
\bauthor{\bsnm{Hjorth}, \binits{J.}},
\bauthor{\bsnm{Th{\"{o}}ne}, \binits{C.C.}},
\bauthor{\bsnm{{De Ugarte Postigo}}, \binits{A.}},
\bauthor{\bsnm{Fynbo}, \binits{J.P.U.}},
\bauthor{\bsnm{Jakobsson}, \binits{P.}},
\bauthor{\bsnm{Pagani}, \binits{C.}},
\bauthor{\bsnm{Stamatikos}, \binits{M.}}:
\bjtitle{Mon. Not. R. Astron. Soc.}
\bvolume{408}(\bissue{1}),
\bfpage{383}
(\byear{2010}).
\arxivurl{1006.0487}.
doi:\doiurl{10.1111/j.1365-2966.2010.17115.x}
\end{barticle}
\endbibitem

\bibitem[\protect\citeauthoryear{Sakamoto et~al.}{2004}]{Sakamoto2004_020903}
\begin{barticle}
\bauthor{\bsnm{Sakamoto}, \binits{T.}},
\bauthor{\bsnm{Lamb}, \binits{D.Q.}},
\bauthor{\bsnm{Graziani}, \binits{C.}},
\bauthor{\bsnm{Donaghy}, \binits{T.Q.}},
\bauthor{\bsnm{Suzuki}, \binits{M.}},
\bauthor{\bsnm{Ricker}, \binits{G.}},
\bauthor{\bsnm{Atteia}, \binits{J.-L.}},
\bauthor{\bsnm{Kawai}, \binits{N.}},
\bauthor{\bsnm{Yoshida}, \binits{A.}},
\bauthor{\bsnm{Shirasaki}, \binits{Y.}},
\bauthor{\bsnm{Tamagawa}, \binits{T.}},
\bauthor{\bsnm{Torii}, \binits{K.}},
\bauthor{\bsnm{Matsuoka}, \binits{M.}},
\bauthor{\bsnm{Fenimore}, \binits{E.E.}},
\bauthor{\bsnm{Galassi}, \binits{M.}},
\bauthor{\bsnm{Tavenner}, \binits{T.}},
\bauthor{\bsnm{Doty}, \binits{J.}},
\bauthor{\bsnm{Vanderspek}, \binits{R.}},
\bauthor{\bsnm{Crew}, \binits{G.B.}},
\bauthor{\bsnm{Villasenor}, \binits{J.}},
\bauthor{\bsnm{Butler}, \binits{N.}},
\bauthor{\bsnm{Prigozhin}, \binits{G.}},
\bauthor{\bsnm{Jernigan}, \binits{J.G.}},
\bauthor{\bsnm{Barraud}, \binits{C.}},
\bauthor{\bsnm{Boer}, \binits{M.}},
\bauthor{\bsnm{Dezalay}, \binits{J.-P.}},
\bauthor{\bsnm{Olive}, \binits{J.-F.}},
\bauthor{\bsnm{Hurley}, \binits{K.}},
\bauthor{\bsnm{Levine}, \binits{A.}},
\bauthor{\bsnm{Monnelly}, \binits{G.}},
\bauthor{\bsnm{Martel}, \binits{F.}},
\bauthor{\bsnm{Morgan}, \binits{E.}},
\bauthor{\bsnm{Woosley}, \binits{S.E.}},
\bauthor{\bsnm{Cline}, \binits{T.}},
\bauthor{\bsnm{Braga}, \binits{J.}},
\bauthor{\bsnm{Manchanda}, \binits{R.}},
\bauthor{\bsnm{Pizzichini}, \binits{G.}},
\bauthor{\bsnm{Takagishi}, \binits{K.}},
\bauthor{\bsnm{Yamauchi}, \binits{M.}}:
\bjtitle{Astrophys. J.}
\bvolume{602}(\bissue{2}),
\bfpage{875}
(\byear{2004}).
doi:\doiurl{10.1086/381232}
\end{barticle}
\endbibitem

\bibitem[\protect\citeauthoryear{Sakamoto et~al.}{2005}]{Sakamoto2005}
\begin{barticle}
\bauthor{\bsnm{Sakamoto}, \binits{T.}},
\bauthor{\bsnm{Lamb}, \binits{D.Q.}},
\bauthor{\bsnm{Kawai}, \binits{N.}},
\bauthor{\bsnm{Yoshida}, \binits{A.}},
\bauthor{\bsnm{Graziani}, \binits{C.}},
\bauthor{\bsnm{Fenimore}, \binits{E.E.}},
\bauthor{\bsnm{Donaghy}, \binits{T.Q.}},
\bauthor{\bsnm{Matsuoka}, \binits{M.}},
\bauthor{\bsnm{Suzuki}, \binits{M.}},
\bauthor{\bsnm{Ricker}, \binits{G.}},
\bauthor{\bsnm{Atteia}, \binits{J.-L.}},
\bauthor{\bsnm{Shirasaki}, \binits{Y.}},
\bauthor{\bsnm{Tamagawa}, \binits{T.}},
\bauthor{\bsnm{Torii}, \binits{K.}},
\bauthor{\bsnm{Galassi}, \binits{M.}},
\bauthor{\bsnm{Doty}, \binits{J.}},
\bauthor{\bsnm{Vanderspek}, \binits{R.}},
\bauthor{\bsnm{Crew}, \binits{G.B.}},
\bauthor{\bsnm{Villasenor}, \binits{J.}},
\bauthor{\bsnm{Butler}, \binits{N.}},
\bauthor{\bsnm{Prigozhin}, \binits{G.}},
\bauthor{\bsnm{Jernigan}, \binits{J.G.}},
\bauthor{\bsnm{Barraud}, \binits{C.}},
\bauthor{\bsnm{Boer}, \binits{M.}},
\bauthor{\bsnm{Dezalay}, \binits{J.-P.}},
\bauthor{\bsnm{Olive}, \binits{J.-F.}},
\bauthor{\bsnm{Hurley}, \binits{K.}},
\bauthor{\bsnm{Levine}, \binits{A.}},
\bauthor{\bsnm{Monnelly}, \binits{G.}},
\bauthor{\bsnm{Martel}, \binits{F.}},
\bauthor{\bsnm{Morgan}, \binits{E.}},
\bauthor{\bsnm{Woosley}, \binits{S.E.}},
\bauthor{\bsnm{Cline}, \binits{T.}},
\bauthor{\bsnm{Braga}, \binits{J.}},
\bauthor{\bsnm{Manchanda}, \binits{R.}},
\bauthor{\bsnm{Pizzichini}, \binits{G.}},
\bauthor{\bsnm{Takagishi}, \binits{K.}},
\bauthor{\bsnm{Yamauchi}, \binits{M.}}:
\bjtitle{\apj}
\bvolume{629},
\bfpage{311}
(\byear{2005}).
doi:\doiurl{10.1086/431235}
\end{barticle}
\endbibitem

\bibitem[\protect\citeauthoryear{{Sakamoto} et~al.}{2008}]{Sakamoto2008}
\begin{barticle}
\bauthor{\bsnm{{Sakamoto}}, \binits{T.}},
\bauthor{\bsnm{{Hullinger}}, \binits{D.}},
\bauthor{\bsnm{{Sato}}, \binits{G.}},
\bauthor{\bsnm{{Yamazaki}}, \binits{R.}},
\bauthor{\bsnm{{Barbier}}, \binits{L.}},
\bauthor{\bsnm{{Barthelmy}}, \binits{S.D.}},
\bauthor{\bsnm{{Cummings}}, \binits{J.R.}},
\bauthor{\bsnm{{Fenimore}}, \binits{E.E.}},
\bauthor{\bsnm{{Gehrels}}, \binits{N.}},
\bauthor{\bsnm{{Krimm}}, \binits{H.A.}},
\bauthor{\bsnm{{Lamb}}, \binits{D.Q.}},
\bauthor{\bsnm{{Markwardt}}, \binits{C.B.}},
\bauthor{\bsnm{{Osborne}}, \binits{J.P.}},
\bauthor{\bsnm{{Palmer}}, \binits{D.M.}},
\bauthor{\bsnm{{Parsons}}, \binits{A.M.}},
\bauthor{\bsnm{{Stamatikos}}, \binits{M.}},
\bauthor{\bsnm{{Tueller}}, \binits{J.}}:
\bjtitle{\apj}
\bvolume{679}(\bissue{1}),
\bfpage{570}
(\byear{2008}).
\arxivurl{0801.4319}.
doi:\doiurl{10.1086/586884}
\end{barticle}
\endbibitem

\bibitem[\protect\citeauthoryear{{Sakamoto} et~al.}{2011}]{Sakamoto2011}
\begin{barticle}
\bauthor{\bsnm{{Sakamoto}}, \binits{T.}},
\bauthor{\bsnm{{Barthelmy}}, \binits{S.D.}},
\bauthor{\bsnm{{Baumgartner}}, \binits{W.H.}},
\bauthor{\bsnm{{Cummings}}, \binits{J.R.}},
\bauthor{\bsnm{{Fenimore}}, \binits{E.E.}},
\bauthor{\bsnm{{Gehrels}}, \binits{N.}},
\bauthor{\bsnm{{Krimm}}, \binits{H.A.}},
\bauthor{\bsnm{{Markwardt}}, \binits{C.B.}},
\bauthor{\bsnm{{Palmer}}, \binits{D.M.}},
\bauthor{\bsnm{{Parsons}}, \binits{A.M.}},
\bauthor{\bsnm{{Sato}}, \binits{G.}},
\bauthor{\bsnm{{Stamatikos}}, \binits{M.}},
\bauthor{\bsnm{{Tueller}}, \binits{J.}},
\bauthor{\bsnm{{Ukwatta}}, \binits{T.N.}},
\bauthor{\bsnm{{Zhang}}, \binits{B.}}:
\bjtitle{\apjs}
\bvolume{195}(\bissue{1}),
\bfpage{2}
(\byear{2011}).
\arxivurl{1104.4689}.
doi:\doiurl{10.1088/0067-0049/195/1/2}
\end{barticle}
\endbibitem

\bibitem[\protect\citeauthoryear{{Sakamoto}
  et~al.}{2015}]{Sakamoto2015_150518A}
\begin{barticle}
\bauthor{\bsnm{{Sakamoto}}, \binits{T.}},
\bauthor{\bsnm{{Serino}}, \binits{M.}},
\bauthor{\bsnm{{Nakahira}}, \binits{S.}},
\bauthor{\bsnm{{Negoro}}, \binits{H.}},
\bauthor{\bsnm{{Ueno}}, \binits{S.}},
\bauthor{\bsnm{{Tomida}}, \binits{H.}},
\bauthor{\bsnm{{Kimura}}, \binits{M.}},
\bauthor{\bsnm{{Ishikawa}}, \binits{M.}},
\bauthor{\bsnm{{Nakagawa}}, \binits{Y.E.}},
\bauthor{\bsnm{{Mihara}}, \binits{T.}},
\bauthor{\bsnm{{Sugizaki}}, \binits{M.}},
\bauthor{\bsnm{{Shidatsu}}, \binits{M.}},
\bauthor{\bsnm{{Sugimoto}}, \binits{J.}},
\bauthor{\bsnm{{Takagi}}, \binits{T.}},
\bauthor{\bsnm{{Matsuoka}}, \binits{M.}},
\bauthor{\bsnm{{Kawai}}, \binits{N.}},
\bauthor{\bsnm{{Yoshii}}, \binits{T.}},
\bauthor{\bsnm{{Tachibana}}, \binits{Y.}},
\bauthor{\bsnm{{Yoshida}}, \binits{A.}},
\bauthor{\bsnm{{Kawakubo}}, \binits{Y.}},
\bauthor{\bsnm{{Ohtsuki}}, \binits{H.}},
\bauthor{\bsnm{{Tsunemi}}, \binits{H.}},
\bauthor{\bsnm{{Imatani}}, \binits{R.}},
\bauthor{\bsnm{{Nakajima}}, \binits{M.}},
\bauthor{\bsnm{{Tanaka}}, \binits{K.}},
\bauthor{\bsnm{{Masumitsu}}, \binits{T.}},
\bauthor{\bsnm{{Ueda}}, \binits{Y.}},
\bauthor{\bsnm{{Hori}}, \binits{T.}},
\bauthor{\bsnm{{Kawamuro}}, \binits{T.}},
\bauthor{\bsnm{{Tsuboi}}, \binits{Y.}},
\bauthor{\bsnm{{Kanetou}}, \binits{S.}},
\bauthor{\bsnm{{Yamauchi}}, \binits{M.}},
\bauthor{\bsnm{{Itoh}}, \binits{D.}},
\bauthor{\bsnm{{Yamaoka}}, \binits{K.}},
\bauthor{\bsnm{{Morii}}, \binits{M.}}:
\bjtitle{GRB Coordinates Network}
\bvolume{17860},
\bfpage{1}
(\byear{2015})
\end{barticle}
\endbibitem

\bibitem[\protect\citeauthoryear{Sanchez-Ramirez
  et~al.}{2015}]{SanchezRamirez2015_150818A}
\begin{barticle}
\bauthor{\bsnm{Sanchez-Ramirez}, \binits{R.}},
\bauthor{\bsnm{Gorosabel}, \binits{J.}},
\bauthor{\bsnm{Perez-Ramirez}, \binits{D.}},
\bauthor{\bsnm{Jeong}, \binits{S.}},
\bauthor{\bsnm{Castro-Tirado}, \binits{A.J.}},
\bauthor{\bsnm{Aceituno}, \binits{F.J.}},
\bauthor{\bsnm{Cunniffe}, \binits{R.}},
\bauthor{\bsnm{Ferrero}, \binits{P.}},
\bauthor{\bsnm{Hu}, \binits{Y.}},
\bauthor{\bsnm{Oates}, \binits{S.R.}},
\bauthor{\bsnm{Tello}, \binits{J.C.}},
\bauthor{\bsnm{Zhang}, \binits{B.-B.}},
\bauthor{\bsnm{Jelinek}, \binits{M.}},
\bauthor{\bsnm{Guziy}, \binits{S.}},
\bauthor{\bsnm{Sokolov}, \binits{V.}},
\bauthor{\bsnm{{Castro Ceron}}, \binits{J.M.}},
\bauthor{\bsnm{Cepa}, \binits{J.}},
\bauthor{\bsnm{Garcia}, \binits{A.}},
\bauthor{\bsnm{Scarpa}, \binits{R.}},
\bauthor{\bsnm{Sanchez-Ramirez}, \binits{R.}},
\bauthor{\bsnm{Gorosabel}, \binits{J.}},
\bauthor{\bsnm{Perez-Ramirez}, \binits{D.}},
\bauthor{\bsnm{Jeong}, \binits{S.}},
\bauthor{\bsnm{Castro-Tirado}, \binits{A.J.}},
\bauthor{\bsnm{Aceituno}, \binits{F.J.}},
\bauthor{\bsnm{Cunniffe}, \binits{R.}},
\bauthor{\bsnm{Ferrero}, \binits{P.}},
\bauthor{\bsnm{Hu}, \binits{Y.}},
\bauthor{\bsnm{Oates}, \binits{S.R.}},
\bauthor{\bsnm{Tello}, \binits{J.C.}},
\bauthor{\bsnm{Zhang}, \binits{B.-B.}},
\bauthor{\bsnm{Jelinek}, \binits{M.}},
\bauthor{\bsnm{Guziy}, \binits{S.}},
\bauthor{\bsnm{Sokolov}, \binits{V.}},
\bauthor{\bsnm{{Castro Ceron}}, \binits{J.M.}},
\bauthor{\bsnm{Cepa}, \binits{J.}},
\bauthor{\bsnm{Garcia}, \binits{A.}},
\bauthor{\bsnm{Scarpa}, \binits{R.}}:
\bjtitle{GCN}
\bvolume{18177},
\bfpage{1}
(\byear{2015})
\end{barticle}
\endbibitem

\bibitem[\protect\citeauthoryear{Santos-Lleo
  et~al.}{2003}]{SantosLleo2003_031203}
\begin{barticle}
\bauthor{\bsnm{Santos-Lleo}, \binits{M.}},
\bauthor{\bsnm{Calderon}, \binits{P.}},
\bauthor{\bsnm{Gotz}, \binits{D.}}:
\bjtitle{GRB Coordinates Network}
\bvolume{2464},
\bfpage{1}
(\byear{2003})
\end{barticle}
\endbibitem

\bibitem[\protect\citeauthoryear{{Savchenko}
  et~al.}{2017}]{Savchenko2017_170817A}
\begin{barticle}
\bauthor{\bsnm{{Savchenko}}, \binits{V.}},
\bauthor{\bsnm{{Ferrigno}}, \binits{C.}},
\bauthor{\bsnm{{Kuulkers}}, \binits{E.}},
\bauthor{\bsnm{{Bazzano}}, \binits{A.}},
\bauthor{\bsnm{{Bozzo}}, \binits{E.}},
\bauthor{\bsnm{{Brandt}}, \binits{S.}},
\bauthor{\bsnm{{Chenevez}}, \binits{J.}},
\bauthor{\bsnm{{Courvoisier}}, \binits{T.J.-L.}},
\bauthor{\bsnm{{Diehl}}, \binits{R.}},
\bauthor{\bsnm{{Domingo}}, \binits{A.}},
\bauthor{\bsnm{{Hanlon}}, \binits{L.}},
\bauthor{\bsnm{{Jourdain}}, \binits{E.}},
\bauthor{\bsnm{{von Kienlin}}, \binits{A.}},
\bauthor{\bsnm{{Laurent}}, \binits{P.}},
\bauthor{\bsnm{{Lebrun}}, \binits{F.}},
\bauthor{\bsnm{{Lutovinov}}, \binits{A.}},
\bauthor{\bsnm{{Martin-Carrillo}}, \binits{A.}},
\bauthor{\bsnm{{Mereghetti}}, \binits{S.}},
\bauthor{\bsnm{{Natalucci}}, \binits{L.}},
\bauthor{\bsnm{{Rodi}}, \binits{J.}},
\bauthor{\bsnm{{Roques}}, \binits{J.-P.}},
\bauthor{\bsnm{{Sunyaev}}, \binits{R.}},
\bauthor{\bsnm{{Ubertini}}, \binits{P.}}:
\bjtitle{\apjl}
\bvolume{848}(\bissue{2}),
\bfpage{15}
(\byear{2017}).
\arxivurl{1710.05449}.
doi:\doiurl{10.3847/2041-8213/aa8f94}
\end{barticle}
\endbibitem

\bibitem[\protect\citeauthoryear{{Sazonov} et~al.}{2004}]{Sazonov2004_031203}
\begin{barticle}
\bauthor{\bsnm{{Sazonov}}, \binits{S.Y.}},
\bauthor{\bsnm{{Lutovinov}}, \binits{A.A.}},
\bauthor{\bsnm{{Sunyaev}}, \binits{R.A.}}:
\bjtitle{\nat}
\bvolume{430}(\bissue{7000}),
\bfpage{646}
(\byear{2004}).
\arxivurl{astro-ph/0408095}.
doi:\doiurl{10.1038/nature02748}
\end{barticle}
\endbibitem

\bibitem[\protect\citeauthoryear{{Sazonov} et~al.}{2007}]{Sazonov2007}
\begin{barticle}
\bauthor{\bsnm{{Sazonov}}, \binits{S.}},
\bauthor{\bsnm{{Churazov}}, \binits{E.}},
\bauthor{\bsnm{{Sunyaev}}, \binits{R.}},
\bauthor{\bsnm{{Revnivtsev}}, \binits{M.}}:
\bjtitle{\mnras}
\bvolume{377}(\bissue{4}),
\bfpage{1726}
(\byear{2007}).
\arxivurl{astro-ph/0608253}.
doi:\doiurl{10.1111/j.1365-2966.2007.11746.x}
\end{barticle}
\endbibitem

\bibitem[\protect\citeauthoryear{Sbarufatti
  et~al.}{2015}]{Sbarufatti2015_150518A}
\begin{barticle}
\bauthor{\bsnm{Sbarufatti}, \binits{B.}},
\bauthor{\bsnm{Pagani}, \binits{C.}},
\bauthor{\bsnm{Beardmore}, \binits{A.P.}},
\bauthor{\bsnm{ri}, \binits{A.M.}},
\bauthor{\bsnm{D'Avanzo}, \binits{P.}},
\bauthor{\bsnm{D'Elia}, \binits{V.}},
\bauthor{\bsnm{Burrows}, \binits{D.N.}},
\bauthor{\bsnm{Kennea}, \binits{J.A.}},
\bauthor{\bsnm{Evans}, \binits{P.A.}}:
\bjtitle{GRB Coordinates Network}
\bvolume{17827},
\bfpage{1}
(\byear{2015})
\end{barticle}
\endbibitem

\bibitem[\protect\citeauthoryear{Sbarufatti
  et~al.}{2019}]{Sbarufatti2019_191019A}
\begin{barticle}
\bauthor{\bsnm{Sbarufatti}, \binits{B.}},
\bauthor{\bsnm{Evans}, \binits{P.A.}},
\bauthor{\bsnm{Osborne}, \binits{J.P.}},
\bauthor{\bsnm{Simpson}, \binits{K.K.}},
\bauthor{\bsnm{Team}, \binits{S.-X.}}:
\bjtitle{GRB Coordinates Network}
\bvolume{26048},
\bfpage{1}
(\byear{2019})
\end{barticle}
\endbibitem

\bibitem[\protect\citeauthoryear{{Schanne} et~al.}{2019}]{Schanne2019}
\begin{barticle}
\bauthor{\bsnm{{Schanne}}, \binits{S.}},
\bauthor{\bsnm{{Dagoneau}}, \binits{N.}},
\bauthor{\bsnm{{Ch{\^a}teau}}, \binits{F.}},
\bauthor{\bsnm{{Le Provost}}, \binits{H.}},
\bauthor{\bsnm{{Daly}}, \binits{F.}},
\bauthor{\bsnm{{Anvar}}, \binits{S.}},
\bauthor{\bsnm{{Antier}}, \binits{S.}},
\bauthor{\bsnm{{Gros}}, \binits{A.}},
\bauthor{\bsnm{{Cordier}}, \binits{B.}}:
\bjtitle{\memsai}
\bvolume{90},
\bfpage{267}
(\byear{2019})
\end{barticle}
\endbibitem

\bibitem[\protect\citeauthoryear{Schulze et~al.}{2012}]{Schulze2012_120422A}
\begin{barticle}
\bauthor{\bsnm{Schulze}, \binits{S.}},
\bauthor{\bsnm{Levan}, \binits{A.J.}},
\bauthor{\bsnm{Malesani}, \binits{D.}},
\bauthor{\bsnm{Fynbo}, \binits{J.P.U.}},
\bauthor{\bsnm{Tanvir}, \binits{N.R.}},
\bauthor{\bsnm{Milvang-Jensen}, \binits{B.}},
\bauthor{\bsnm{{de Ugarte Postigo}}, \binits{A.}},
\bauthor{\bsnm{D'Elia}, \binits{V.}},
\bauthor{\bsnm{Sollerman}, \binits{J.}},
\bauthor{\bsnm{Hjorth}, \binits{J.}},
\bauthor{\bsnm{Schulze}, \binits{S.}},
\bauthor{\bsnm{Levan}, \binits{A.J.}},
\bauthor{\bsnm{Malesani}, \binits{D.}},
\bauthor{\bsnm{Fynbo}, \binits{J.P.U.}},
\bauthor{\bsnm{Tanvir}, \binits{N.R.}},
\bauthor{\bsnm{Milvang-Jensen}, \binits{B.}},
\bauthor{\bsnm{{de Ugarte Postigo}}, \binits{A.}},
\bauthor{\bsnm{D'Elia}, \binits{V.}},
\bauthor{\bsnm{Sollerman}, \binits{J.}},
\bauthor{\bsnm{Hjorth}, \binits{J.}}:
\bjtitle{GCN}
\bvolume{13257},
\bfpage{1}
(\byear{2012})
\end{barticle}
\endbibitem

\bibitem[\protect\citeauthoryear{Schulze et~al.}{2013}]{Schulze2013_130702A}
\begin{barticle}
\bauthor{\bsnm{Schulze}, \binits{S.}},
\bauthor{\bsnm{Leloudas}, \binits{G.}},
\bauthor{\bsnm{Xu}, \binits{D.}},
\bauthor{\bsnm{Fynbo}, \binits{J.P.U.}},
\bauthor{\bsnm{Geier}, \binits{S.}},
\bauthor{\bsnm{Jakobsson}, \binits{P.}}:
\bjtitle{GRB Coordinates Network}
\bvolume{14994},
\bfpage{1}
(\byear{2013})
\end{barticle}
\endbibitem

\bibitem[\protect\citeauthoryear{Semkov and Borisova}{2004}]{Semkov2004}
\begin{barticle}
\bauthor{\bsnm{Semkov}, \binits{E.H.}},
\bauthor{\bsnm{Borisova}, \binits{A.P.}}:
\bjtitle{Baltic Astronomy}
\bvolume{13},
\bfpage{257}
(\byear{2004})
\end{barticle}
\endbibitem

\bibitem[\protect\citeauthoryear{{Shirasaki} et~al.}{2003}]{Shirasaki2003}
\begin{barticle}
\bauthor{\bsnm{{Shirasaki}}, \binits{Y.}},
\bauthor{\bsnm{{Kawai}}, \binits{N.}},
\bauthor{\bsnm{{Yoshida}}, \binits{A.}},
\bauthor{\bsnm{{Matsuoka}}, \binits{M.}},
\bauthor{\bsnm{{Tamagawa}}, \binits{T.}},
\bauthor{\bsnm{{Torii}}, \binits{K.}},
\bauthor{\bsnm{{Sakamoto}}, \binits{T.}},
\bauthor{\bsnm{{Suzuki}}, \binits{M.}},
\bauthor{\bsnm{{Urata}}, \binits{Y.}},
\bauthor{\bsnm{{Sato}}, \binits{R.}},
\bauthor{\bsnm{{Nakagawa}}, \binits{Y.}},
\bauthor{\bsnm{{Takahashi}}, \binits{D.}},
\bauthor{\bsnm{{Fenimore}}, \binits{E.E.}},
\bauthor{\bsnm{{Galassi}}, \binits{M.}},
\bauthor{\bsnm{{Lamb}}, \binits{D.Q.}},
\bauthor{\bsnm{{Graziani}}, \binits{C.}},
\bauthor{\bsnm{{Donaghy}}, \binits{T.Q.}},
\bauthor{\bsnm{{Vanderspek}}, \binits{R.}},
\bauthor{\bsnm{{Yamauchi}}, \binits{M.}},
\bauthor{\bsnm{{Takagishi}}, \binits{K.}},
\bauthor{\bsnm{{Hatsukade}}, \binits{I.}}:
\bjtitle{\pasj}
\bvolume{55},
\bfpage{1033}
(\byear{2003}).
\arxivurl{astro-ph/0311067}.
doi:\doiurl{10.1093/pasj/55.5.1033}
\end{barticle}
\endbibitem

\bibitem[\protect\citeauthoryear{Siegel et~al.}{2011}]{Siegel2011_111225A}
\begin{barticle}
\bauthor{\bsnm{Siegel}, \binits{M.H.}},
\bauthor{\bsnm{Baumgartner}, \binits{W.H.}},
\bauthor{\bsnm{Chester}, \binits{M.M.}},
\bauthor{\bsnm{Gehrels}, \binits{N.}},
\bauthor{\bsnm{Grupe}, \binits{D.}},
\bauthor{\bsnm{Marshall}, \binits{F.E.}},
\bauthor{\bsnm{Melandri}, \binits{A.}},
\bauthor{\bsnm{Palmer}, \binits{D.M.}},
\bauthor{\bsnm{Sakamoto}, \binits{T.}},
\bauthor{\bsnm{Swenson}, \binits{C.A.}},
\bauthor{\bsnm{Ukwatta}, \binits{T.N.}},
\bauthor{\bsnm{Zhang}, \binits{B.-B.}}:
\bjtitle{GRB Coordinates Network}
\bvolume{12720},
\bfpage{1}
(\byear{2011})
\end{barticle}
\endbibitem

\bibitem[\protect\citeauthoryear{Singer et~al.}{2013}]{Singer2013_130702A}
\begin{barticle}
\bauthor{\bsnm{Singer}, \binits{L.P.}},
\bauthor{\bsnm{Cenko}, \binits{S.B.}},
\bauthor{\bsnm{Kasliwal}, \binits{M.M.}},
\bauthor{\bsnm{Perley}, \binits{D.A.}},
\bauthor{\bsnm{Ofek}, \binits{E.O.}},
\bauthor{\bsnm{Brown}, \binits{D.A.}},
\bauthor{\bsnm{Nugent}, \binits{P.E.}},
\bauthor{\bsnm{Kulkarni}, \binits{S.R.}},
\bauthor{\bsnm{Corsi}, \binits{A.}},
\bauthor{\bsnm{Frail}, \binits{D.A.}},
\bauthor{\bsnm{Bellm}, \binits{E.}},
\bauthor{\bsnm{Mulchaey}, \binits{J.}},
\bauthor{\bsnm{Arcavi}, \binits{I.}},
\bauthor{\bsnm{Barlow}, \binits{T.}},
\bauthor{\bsnm{Bloom}, \binits{J.S.}},
\bauthor{\bsnm{Cao}, \binits{Y.}},
\bauthor{\bsnm{Gehrels}, \binits{N.}},
\bauthor{\bsnm{Horesh}, \binits{A.}},
\bauthor{\bsnm{Masci}, \binits{F.J.}},
\bauthor{\bsnm{McEnery}, \binits{J.}},
\bauthor{\bsnm{Rau}, \binits{A.}},
\bauthor{\bsnm{Surace}, \binits{J.A.}},
\bauthor{\bsnm{Yaron}, \binits{O.}}:
\bjtitle{\apj}
\bvolume{776},
\bfpage{34}
(\byear{2013}).
doi:\doiurl{10.1088/2041-8205/776/2/L34}
\end{barticle}
\endbibitem

\bibitem[\protect\citeauthoryear{Soderberg and
  Frail}{2006}]{Soderberg2006_060218}
\begin{barticle}
\bauthor{\bsnm{Soderberg}, \binits{A.M.}},
\bauthor{\bsnm{Frail}, \binits{D.A.}}:
\bjtitle{GRB Coordinates Network}
\bvolume{4794},
\bfpage{1}
(\byear{2006})
\end{barticle}
\endbibitem

\bibitem[\protect\citeauthoryear{Soderberg et~al.}{2002}]{Soderberg2002_020903}
\begin{barticle}
\bauthor{\bsnm{Soderberg}, \binits{A.M.}},
\bauthor{\bsnm{Price}, \binits{P.A.}},
\bauthor{\bsnm{Fox}, \binits{D.W.}},
\bauthor{\bsnm{Kulkarni}, \binits{S.R.}},
\bauthor{\bsnm{Djorgovski}, \binits{S.G.}},
\bauthor{\bsnm{Berger}, \binits{E.}},
\bauthor{\bsnm{Harrison}, \binits{F.}},
\bauthor{\bsnm{Yost}, \binits{S.}},
\bauthor{\bsnm{Hamuy}, \binits{M.}},
\bauthor{\bsnm{Shectman}, \binits{S.}},
\bauthor{\bsnm{Mirabal}, \binits{N.}},
\bauthor{\bsnm{Halpern}, \binits{J.}}:
\bjtitle{GCN}
\bvolume{1554},
\bfpage{1}
(\byear{2002})
\end{barticle}
\endbibitem

\bibitem[\protect\citeauthoryear{Soderberg
  et~al.}{2004a}]{Soderberg2004_020903}
\begin{barticle}
\bauthor{\bsnm{Soderberg}, \binits{A.M.}},
\bauthor{\bsnm{Kulkarni}, \binits{S.R.}},
\bauthor{\bsnm{Berger}, \binits{E.}},
\bauthor{\bsnm{Fox}, \binits{D.B.}},
\bauthor{\bsnm{Price}, \binits{P.A.}},
\bauthor{\bsnm{Yost}, \binits{S.A.}},
\bauthor{\bsnm{Hunt}, \binits{M.P.}},
\bauthor{\bsnm{Frail}, \binits{D.A.}},
\bauthor{\bsnm{Walker}, \binits{R.C.}},
\bauthor{\bsnm{Hamuy}, \binits{M.}},
\bauthor{\bsnm{Shectman}, \binits{S.A.}},
\bauthor{\bsnm{Halpern}, \binits{J.P.}},
\bauthor{\bsnm{Mirabal}, \binits{N.}}:
\bjtitle{\apj}
\bvolume{606},
\bfpage{994}
(\byear{2004}a).
doi:\doiurl{10.1086/383082}
\end{barticle}
\endbibitem

\bibitem[\protect\citeauthoryear{Soderberg
  et~al.}{2004b}]{Soderberg2004_031203}
\begin{barticle}
\bauthor{\bsnm{Soderberg}, \binits{A.M.}},
\bauthor{\bsnm{Kulkarni}, \binits{S.R.}},
\bauthor{\bsnm{Berger}, \binits{E.}},
\bauthor{\bsnm{Fox}, \binits{D.W.}},
\bauthor{\bsnm{Sako}, \binits{M.}},
\bauthor{\bsnm{Frail}, \binits{D.A.}},
\bauthor{\bsnm{Gal-Yam}, \binits{A.}},
\bauthor{\bsnm{Moon}, \binits{D.S.}},
\bauthor{\bsnm{Cenko}, \binits{S.B.}},
\bauthor{\bsnm{Yost}, \binits{S.A.}},
\bauthor{\bsnm{Phillips}, \binits{M.M.}},
\bauthor{\bsnm{Persson}, \binits{S.E.}},
\bauthor{\bsnm{Freedman}, \binits{W.L.}},
\bauthor{\bsnm{Wyatt}, \binits{P.}},
\bauthor{\bsnm{Jayawardhana}, \binits{R.}},
\bauthor{\bsnm{Paulson}, \binits{D.}}:
\bjtitle{\nat}
\bvolume{430},
\bfpage{648}
(\byear{2004}b).
doi:\doiurl{10.1038/nature02757}
\end{barticle}
\endbibitem

\bibitem[\protect\citeauthoryear{{Soderberg}
  et~al.}{2005}]{Soderberg2005_040701}
\begin{barticle}
\bauthor{\bsnm{{Soderberg}}, \binits{A.M.}},
\bauthor{\bsnm{{Kulkarni}}, \binits{S.R.}},
\bauthor{\bsnm{{Fox}}, \binits{D.B.}},
\bauthor{\bsnm{{Berger}}, \binits{E.}},
\bauthor{\bsnm{{Price}}, \binits{P.A.}},
\bauthor{\bsnm{{Cenko}}, \binits{S.B.}},
\bauthor{\bsnm{{Howell}}, \binits{D.A.}},
\bauthor{\bsnm{{Gal-Yam}}, \binits{A.}},
\bauthor{\bsnm{{Leonard}}, \binits{D.C.}},
\bauthor{\bsnm{{Frail}}, \binits{D.A.}},
\bauthor{\bsnm{{Moon}}, \binits{D.}},
\bauthor{\bsnm{{Chevalier}}, \binits{R.A.}},
\bauthor{\bsnm{{Hamuy}}, \binits{M.}},
\bauthor{\bsnm{{Hurley}}, \binits{K.C.}},
\bauthor{\bsnm{{Kelson}}, \binits{D.}},
\bauthor{\bsnm{{Koviak}}, \binits{K.}},
\bauthor{\bsnm{{Krzeminski}}, \binits{W.}},
\bauthor{\bsnm{{Kumar}}, \binits{P.}},
\bauthor{\bsnm{{MacFadyen}}, \binits{A.}},
\bauthor{\bsnm{{McCarthy}}, \binits{P.J.}},
\bauthor{\bsnm{{Park}}, \binits{H.S.}},
\bauthor{\bsnm{{Peterson}}, \binits{B.A.}},
\bauthor{\bsnm{{Phillips}}, \binits{M.M.}},
\bauthor{\bsnm{{Rauch}}, \binits{M.}},
\bauthor{\bsnm{{Roth}}, \binits{M.}},
\bauthor{\bsnm{{Schmidt}}, \binits{B.P.}},
\bauthor{\bsnm{{Shectman}}, \binits{S.}}:
\bjtitle{\apj}
\bvolume{627}(\bissue{2}),
\bfpage{877}
(\byear{2005}).
\arxivurl{astro-ph/0502553}.
doi:\doiurl{10.1086/430405}
\end{barticle}
\endbibitem

\bibitem[\protect\citeauthoryear{{Soderberg}
  et~al.}{2006}]{Soderberg2006_060218_nature}
\begin{barticle}
\bauthor{\bsnm{{Soderberg}}, \binits{A.M.}},
\bauthor{\bsnm{{Kulkarni}}, \binits{S.R.}},
\bauthor{\bsnm{{Nakar}}, \binits{E.}},
\bauthor{\bsnm{{Berger}}, \binits{E.}},
\bauthor{\bsnm{{Cameron}}, \binits{P.B.}},
\bauthor{\bsnm{{Fox}}, \binits{D.B.}},
\bauthor{\bsnm{{Frail}}, \binits{D.}},
\bauthor{\bsnm{{Gal-Yam}}, \binits{A.}},
\bauthor{\bsnm{{Sari}}, \binits{R.}},
\bauthor{\bsnm{{Cenko}}, \binits{S.B.}},
\bauthor{\bsnm{{Kasliwal}}, \binits{M.}},
\bauthor{\bsnm{{Chevalier}}, \binits{R.A.}},
\bauthor{\bsnm{{Piran}}, \binits{T.}},
\bauthor{\bsnm{{Price}}, \binits{P.A.}},
\bauthor{\bsnm{{Schmidt}}, \binits{B.P.}},
\bauthor{\bsnm{{Pooley}}, \binits{G.}},
\bauthor{\bsnm{{Moon}}, \binits{D.-S.}},
\bauthor{\bsnm{{Penprase}}, \binits{B.E.}},
\bauthor{\bsnm{{Ofek}}, \binits{E.}},
\bauthor{\bsnm{{Rau}}, \binits{A.}},
\bauthor{\bsnm{{Gehrels}}, \binits{N.}},
\bauthor{\bsnm{{Nousek}}, \binits{J.A.}},
\bauthor{\bsnm{{Burrows}}, \binits{D.N.}},
\bauthor{\bsnm{{Persson}}, \binits{S.E.}},
\bauthor{\bsnm{{McCarthy}}, \binits{P.J.}}:
\bjtitle{\nat}
\bvolume{442}(\bissue{7106}),
\bfpage{1014}
(\byear{2006}).
\arxivurl{astro-ph/0604389}.
doi:\doiurl{10.1038/nature05087}
\end{barticle}
\endbibitem

\bibitem[\protect\citeauthoryear{Soffitta et~al.}{1998}]{Soffitta1998_980425}
\begin{barticle}
\bauthor{\bsnm{Soffitta}, \binits{P.}},
\bauthor{\bsnm{Feroci}, \binits{M.}},
\bauthor{\bsnm{Piro}, \binits{L.}},
\bauthor{\bparticle{in~'t} \bsnm{Zand}, \binits{J.}},
\bauthor{\bsnm{Heise}, \binits{J.}},
\bauthor{\bparticle{di} \bsnm{Ciolo}, \binits{L.}},
\bauthor{\bsnm{Muller}, \binits{J.M.}},
\bauthor{\bsnm{Palazzi}, \binits{E.}},
\bauthor{\bsnm{Frontera}, \binits{F.}}:
\bjtitle{\iaucirc}
\bvolume{6884},
\bfpage{1}
(\byear{1998})
\end{barticle}
\endbibitem

\bibitem[\protect\citeauthoryear{{Somiya}}{2012}]{Somiya2012}
\begin{barticle}
\bauthor{\bsnm{{Somiya}}, \binits{K.}}:
\bjtitle{Classical and Quantum Gravity}
\bvolume{29}(\bissue{12}),
\bfpage{124007}
(\byear{2012}).
\arxivurl{1111.7185}.
doi:\doiurl{10.1088/0264-9381/29/12/124007}
\end{barticle}
\endbibitem

\bibitem[\protect\citeauthoryear{Stamatikos
  et~al.}{2010}]{Stamatikos2010_100316D}
\begin{barticle}
\bauthor{\bsnm{Stamatikos}, \binits{M.}},
\bauthor{\bsnm{Barthelmy}, \binits{S.D.}},
\bauthor{\bsnm{Baumgartner}, \binits{W.H.}},
\bauthor{\bsnm{Beardmore}, \binits{A.P.}},
\bauthor{\bsnm{Campana}, \binits{S.}},
\bauthor{\bparticle{de} \bsnm{Ugarte~Postigo}, \binits{A.}},
\bauthor{\bsnm{Evans}, \binits{P.A.}},
\bauthor{\bsnm{Gelbord}, \binits{J.M.}},
\bauthor{\bsnm{Guidorzi}, \binits{C.}},
\bauthor{\bsnm{Holland}, \binits{S.T.}},
\bauthor{\bsnm{Kennea}, \binits{J.A.}},
\bauthor{\bsnm{Krimm}, \binits{H.A.}},
\bauthor{\bsnm{Margutti}, \binits{R.}},
\bauthor{\bsnm{Marshall}, \binits{F.E.}},
\bauthor{\bsnm{Oates}, \binits{S.R.}},
\bauthor{\bsnm{Pagani}, \binits{C.}},
\bauthor{\bsnm{Page}, \binits{K.L.}},
\bauthor{\bsnm{Palmer}, \binits{D.M.}},
\bauthor{\bsnm{Romano}, \binits{P.}},
\bauthor{\bsnm{Sbarufatti}, \binits{B.}},
\bauthor{\bsnm{Starling}, \binits{R.L.C.}},
\bauthor{\bsnm{Ukwatta}, \binits{T.N.}}:
\bjtitle{GRB Coordinates Network}
\bvolume{10496},
\bfpage{1}
(\byear{2010})
\end{barticle}
\endbibitem

\bibitem[\protect\citeauthoryear{{Stanbro} and
  {Meegan}}{2016}]{Stanbro2016_160821B}
\begin{barticle}
\bauthor{\bsnm{{Stanbro}}, \binits{M.}},
\bauthor{\bsnm{{Meegan}}, \binits{C.}}:
\bjtitle{GRB Coordinates Network}
\bvolume{19843},
\bfpage{1}
(\byear{2016})
\end{barticle}
\endbibitem

\bibitem[\protect\citeauthoryear{Stanek et~al.}{2003}]{Stanek2003_030329}
\begin{barticle}
\bauthor{\bsnm{Stanek}, \binits{K.Z.}},
\bauthor{\bsnm{Latham}, \binits{D.W.}},
\bauthor{\bsnm{Everett}, \binits{M.E.}}:
\bjtitle{GRB Coordinates Network}
\bvolume{2244},
\bfpage{1}
(\byear{2003})
\end{barticle}
\endbibitem

\bibitem[\protect\citeauthoryear{Stanway et~al.}{2014}]{Stanway2014_080517}
\begin{barticle}
\bauthor{\bsnm{Stanway}, \binits{E.R.}},
\bauthor{\bsnm{Levan}, \binits{A.J.}},
\bauthor{\bsnm{Tanvir}, \binits{N.}},
\bauthor{\bsnm{Wiersema}, \binits{K.}},
\bauthor{\bsnm{{Van der Horst}}, \binits{A.}},
\bauthor{\bsnm{Mundell}, \binits{C.G.}},
\bauthor{\bsnm{Guidorzi}, \binits{C.}}:
\bjtitle{Mon. Not. R. Astron. Soc.}
\bvolume{446}(\bissue{4}),
\bfpage{3911}
(\byear{2014}).
\arxivurl{1409.5791}.
doi:\doiurl{10.1093/mnras/stu2286}
\end{barticle}
\endbibitem

\bibitem[\protect\citeauthoryear{Stanway et~al.}{2015}]{Stanway2015_080517}
\begin{barticle}
\bauthor{\bsnm{Stanway}, \binits{E.R.}},
\bauthor{\bsnm{Levan}, \binits{A.J.}},
\bauthor{\bsnm{Tanvir}, \binits{N.}},
\bauthor{\bsnm{Wiersema}, \binits{K.}},
\bauthor{\bparticle{van~der} \bsnm{Horst}, \binits{A.}},
\bauthor{\bsnm{Mundell}, \binits{C.G.}},
\bauthor{\bsnm{Guidorzi}, \binits{C.}}:
\bjtitle{\mnras}
\bvolume{446},
\bfpage{3911}
(\byear{2015}).
doi:\doiurl{10.1093/mnras/stu2286}
\end{barticle}
\endbibitem

\bibitem[\protect\citeauthoryear{Starling et~al.}{2011}]{Starling2011_100316D}
\begin{barticle}
\bauthor{\bsnm{Starling}, \binits{R.L.C.}},
\bauthor{\bsnm{Wiersema}, \binits{K.}},
\bauthor{\bsnm{Levan}, \binits{A.J.}},
\bauthor{\bsnm{Sakamoto}, \binits{T.}},
\bauthor{\bsnm{Bersier}, \binits{D.}},
\bauthor{\bsnm{Goldoni}, \binits{P.}},
\bauthor{\bsnm{Oates}, \binits{S.R.}},
\bauthor{\bsnm{Rowlinson}, \binits{A.}},
\bauthor{\bsnm{Campana}, \binits{S.}},
\bauthor{\bsnm{Sollerman}, \binits{J.}},
\bauthor{\bsnm{Tanvir}, \binits{N.R.}},
\bauthor{\bsnm{Malesani}, \binits{D.}},
\bauthor{\bsnm{Fynbo}, \binits{J.P.U.}},
\bauthor{\bsnm{Covino}, \binits{S.}},
\bauthor{\bsnm{D'Avanzo}, \binits{P.}},
\bauthor{\bsnm{O'Brien}, \binits{P.T.}},
\bauthor{\bsnm{Page}, \binits{K.L.}},
\bauthor{\bsnm{Osborne}, \binits{J.P.}},
\bauthor{\bsnm{Vergani}, \binits{S.D.}},
\bauthor{\bsnm{Barthelmy}, \binits{S.}},
\bauthor{\bsnm{Burrows}, \binits{D.N.}},
\bauthor{\bsnm{Cano}, \binits{Z.}},
\bauthor{\bsnm{Curran}, \binits{P.A.}},
\bauthor{\bsnm{{De Pasquale}}, \binits{M.}},
\bauthor{\bsnm{D'Elia}, \binits{V.}},
\bauthor{\bsnm{Evans}, \binits{P.A.}},
\bauthor{\bsnm{Flores}, \binits{H.}},
\bauthor{\bsnm{Fruchter}, \binits{A.S.}},
\bauthor{\bsnm{Garnavich}, \binits{P.}},
\bauthor{\bsnm{Gehrels}, \binits{N.}},
\bauthor{\bsnm{Gorosabel}, \binits{J.}},
\bauthor{\bsnm{Hjorth}, \binits{J.}},
\bauthor{\bsnm{Holland}, \binits{S.T.}},
\bauthor{\bsnm{{Van Der Horst}}, \binits{A.J.}},
\bauthor{\bsnm{Hurkett}, \binits{C.P.}},
\bauthor{\bsnm{Jakobsson}, \binits{P.}},
\bauthor{\bsnm{Kamble}, \binits{A.P.}},
\bauthor{\bsnm{Kouveliotou}, \binits{C.}},
\bauthor{\bsnm{Kuin}, \binits{N.P.M.}},
\bauthor{\bsnm{Kaper}, \binits{L.}},
\bauthor{\bsnm{Mazzali}, \binits{P.A.}},
\bauthor{\bsnm{Nugent}, \binits{P.E.}},
\bauthor{\bsnm{Pian}, \binits{E.}},
\bauthor{\bsnm{Stamatikos}, \binits{M.}},
\bauthor{\bsnm{Th{\"{o}}ne}, \binits{C.C.}},
\bauthor{\bsnm{Woosley}, \binits{S.E.}}:
\bjtitle{Mon. Not. R. Astron. Soc.}
\bvolume{411}(\bissue{4}),
\bfpage{2792}
(\byear{2011}).
\arxivurl{1004.2919}.
doi:\doiurl{10.1111/j.1365-2966.2010.17879.x}
\end{barticle}
\endbibitem

\bibitem[\protect\citeauthoryear{Stratta et~al.}{2007}]{Stratta2007_061201}
\begin{barticle}
\bauthor{\bsnm{Stratta}, \binits{G.}},
\bauthor{\bsnm{D'Avanzo}, \binits{P.}},
\bauthor{\bsnm{Piranomonte}, \binits{S.}},
\bauthor{\bsnm{Cutini}, \binits{S.}},
\bauthor{\bsnm{Preger}, \binits{B.}},
\bauthor{\bsnm{Perri}, \binits{M.}},
\bauthor{\bsnm{Conciatore}, \binits{M.L.}},
\bauthor{\bsnm{Covino}, \binits{S.}},
\bauthor{\bsnm{Stella}, \binits{L.}},
\bauthor{\bsnm{Guetta}, \binits{D.}},
\bauthor{\bsnm{Marshall}, \binits{F.E.}},
\bauthor{\bsnm{Holland}, \binits{S.T.}},
\bauthor{\bsnm{Stamatikos}, \binits{M.}},
\bauthor{\bsnm{Guidorzi}, \binits{C.}},
\bauthor{\bsnm{Mangano}, \binits{V.}},
\bauthor{\bsnm{Antonelli}, \binits{L.A.}},
\bauthor{\bsnm{Burrows}, \binits{D.}},
\bauthor{\bsnm{Campana}, \binits{S.}},
\bauthor{\bsnm{Capalbi}, \binits{M.}},
\bauthor{\bsnm{Chincarini}, \binits{G.}},
\bauthor{\bsnm{Cusumano}, \binits{G.}},
\bauthor{\bsnm{D'Elia}, \binits{V.}},
\bauthor{\bsnm{Evans}, \binits{P.A.}},
\bauthor{\bsnm{Fiore}, \binits{F.}},
\bauthor{\bsnm{Fugazza}, \binits{D.}},
\bauthor{\bsnm{Giommi}, \binits{P.}},
\bauthor{\bsnm{Osborne}, \binits{J.P.}},
\bauthor{\bsnm{{La Parola}}, \binits{V.}},
\bauthor{\bsnm{Mineo}, \binits{T.}},
\bauthor{\bsnm{Moretti}, \binits{A.}},
\bauthor{\bsnm{Page}, \binits{K.L.}},
\bauthor{\bsnm{Romano}, \binits{P.}},
\bauthor{\bsnm{Tagliaferri}, \binits{G.}}:
\bjtitle{Astron. Astrophys.}
\bvolume{474}(\bissue{3}),
\bfpage{827}
(\byear{2007}).
doi:\doiurl{10.1051/0004-6361:20078006}
\end{barticle}
\endbibitem

\bibitem[\protect\citeauthoryear{{Svinkin} et~al.}{2020}]{Svinkin2020_200415A}
\begin{barticle}
\bauthor{\bsnm{{Svinkin}}, \binits{D.}},
\bauthor{\bsnm{{Hurley}}, \binits{K.}},
\bauthor{\bsnm{{Frederiks}}, \binits{D.}},
\bauthor{\bsnm{{Hurley}}, \binits{K.}},
\bauthor{\bsnm{{Mitrofanov}}, \binits{I.G.}},
\bauthor{\bsnm{{Golovin}}, \binits{D.}},
\bauthor{\bsnm{{Litvak}}, \binits{M.L.}},
\bauthor{\bsnm{{Sanin}}, \binits{A.B.}},
\bauthor{\bsnm{{Kozlova}}, \binits{A.}},
\bauthor{\bsnm{{Golenetskii}}, \binits{S.}},
\bauthor{\bsnm{{Aptekar}}, \binits{R.}},
\bauthor{\bsnm{{Frederiks}}, \binits{D.}},
\bauthor{\bsnm{{Svinkin}}, \binits{D.}},
\bauthor{\bsnm{{Cline}}, \binits{T.}},
\bauthor{\bsnm{{Goldstein}}, \binits{A.}},
\bauthor{\bsnm{{Briggs}}, \binits{M.S.}},
\bauthor{\bsnm{{Wilson-Hodge}}, \binits{C.}},
\bauthor{\bsnm{{von Kienlin}}, \binits{A.}},
\bauthor{\bsnm{{Zhang}}, \binits{X.}},
\bauthor{\bsnm{{Rau}}, \binits{A.}},
\bauthor{\bsnm{{Savchenko}}, \binits{V.}},
\bauthor{\bsnm{{E. Bozzo}}, \binits{E.}},
\bauthor{\bsnm{{Ferrigno}}, \binits{C.}},
\bauthor{\bsnm{{Barthelmy}}, \binits{S.}},
\bauthor{\bsnm{{Cummings}}, \binits{J.}},
\bauthor{\bsnm{{Krimm}}, \binits{H.}},
\bauthor{\bsnm{{Palmer}}, \binits{D.}},
\bauthor{\bsnm{{Boynton}}, \binits{W.}},
\bauthor{\bsnm{{Fellows}}, \binits{C.}},
\bauthor{\bsnm{{Harshman}}, \binits{K.}},
\bauthor{\bsnm{{Enos}}, \binits{H.}},
\bauthor{\bsnm{{Starr}}, \binits{R.}}:
\bjtitle{GRB Coordinates Network}
\bvolume{27585},
\bfpage{1}
(\byear{2020})
\end{barticle}
\endbibitem

\bibitem[\protect\citeauthoryear{{Tagliaferri}
  et~al.}{2005}]{Tagliaferri2005_050219A}
\begin{barticle}
\bauthor{\bsnm{{Tagliaferri}}, \binits{G.}},
\bauthor{\bsnm{{Goad}}, \binits{M.}},
\bauthor{\bsnm{{Chincarini}}, \binits{G.}},
\bauthor{\bsnm{{Moretti}}, \binits{A.}},
\bauthor{\bsnm{{Campana}}, \binits{S.}},
\bauthor{\bsnm{{Burrows}}, \binits{D.N.}},
\bauthor{\bsnm{{Perri}}, \binits{M.}},
\bauthor{\bsnm{{Barthelmy}}, \binits{S.D.}},
\bauthor{\bsnm{{Gehrels}}, \binits{N.}},
\bauthor{\bsnm{{Krimm}}, \binits{H.}},
\bauthor{\bsnm{{Sakamoto}}, \binits{T.}},
\bauthor{\bsnm{{Kumar}}, \binits{P.}},
\bauthor{\bsnm{{M{\'e}sz{\'a}ros}}, \binits{P.I.}},
\bauthor{\bsnm{{Kobayashi}}, \binits{S.}},
\bauthor{\bsnm{{Zhang}}, \binits{B.}},
\bauthor{\bsnm{{Angelini}}, \binits{L.}},
\bauthor{\bsnm{{Banat}}, \binits{P.}},
\bauthor{\bsnm{{Beardmore}}, \binits{A.P.}},
\bauthor{\bsnm{{Capalbi}}, \binits{M.}},
\bauthor{\bsnm{{Covino}}, \binits{S.}},
\bauthor{\bsnm{{Cusumano}}, \binits{G.}},
\bauthor{\bsnm{{Giommi}}, \binits{P.}},
\bauthor{\bsnm{{Godet}}, \binits{O.}},
\bauthor{\bsnm{{Hill}}, \binits{J.E.}},
\bauthor{\bsnm{{Kennea}}, \binits{J.A.}},
\bauthor{\bsnm{{Mangano}}, \binits{V.}},
\bauthor{\bsnm{{Morris}}, \binits{D.C.}},
\bauthor{\bsnm{{Nousek}}, \binits{J.A.}},
\bauthor{\bsnm{{O'Brien}}, \binits{P.T.}},
\bauthor{\bsnm{{Osborne}}, \binits{J.P.}},
\bauthor{\bsnm{{Pagani}}, \binits{C.}},
\bauthor{\bsnm{{Page}}, \binits{K.L.}},
\bauthor{\bsnm{{Romano}}, \binits{P.}},
\bauthor{\bsnm{{Stella}}, \binits{L.}},
\bauthor{\bsnm{{Wells}}, \binits{A.}}:
\bjtitle{\nat}
\bvolume{436}(\bissue{7053}),
\bfpage{985}
(\byear{2005}).
\arxivurl{astro-ph/0506355}.
doi:\doiurl{10.1038/nature03934}
\end{barticle}
\endbibitem

\bibitem[\protect\citeauthoryear{Tagliaferri
  et~al.}{2006}]{Tagliaferri2006_031203}
\begin{barticle}
\bauthor{\bsnm{Tagliaferri}, \binits{G.}},
\bauthor{\bsnm{Malesani}, \binits{D.}},
\bauthor{\bsnm{Chincarini}, \binits{G.}},
\bauthor{\bsnm{Covino}, \binits{S.}},
\bauthor{\bsnm{Valle}, \binits{M.D.}},
\bauthor{\bsnm{Fugazza}, \binits{D.}},
\bauthor{\bsnm{Mazzali}, \binits{P.}}:
\bjtitle{Advances in Space Research}
\bvolume{38},
\bfpage{1295}
(\byear{2006}).
doi:\doiurl{10.1016/j.asr.2005.05.040}
\end{barticle}
\endbibitem

\bibitem[\protect\citeauthoryear{Tanaka}{2007}]{Tanaka2007_980827}
\begin{botherref}
\oauthor{\bsnm{Tanaka}, \binits{Y.T.}}:
In: Becker, W., Huang, H.H. (eds.)
GEOTAIL observation of SGR 1900+14 giant flare on 27 August 1998,
p. 205
(2007)
\end{botherref}
\endbibitem

\bibitem[\protect\citeauthoryear{Tanaka et~al.}{2007}]{Tanaka2007b_980827}
\begin{barticle}
\bauthor{\bsnm{Tanaka}, \binits{Y.T.}},
\bauthor{\bsnm{Terasawa}, \binits{T.}},
\bauthor{\bsnm{Kawai}, \binits{N.}},
\bauthor{\bsnm{Yoshida}, \binits{A.}},
\bauthor{\bsnm{Yoshikawa}, \binits{I.}},
\bauthor{\bsnm{Saito}, \binits{Y.}},
\bauthor{\bsnm{Takashima}, \binits{T.}},
\bauthor{\bsnm{Mukai}, \binits{T.}}:
\bjtitle{Astrophys. J.}
\bvolume{665}(\bissue{1}),
\bfpage{55}
(\byear{2007}).
\arxivurl{0706.3123}.
doi:\doiurl{10.1086/521025}
\end{barticle}
\endbibitem

\bibitem[\protect\citeauthoryear{Tanga et~al.}{2018}]{Tanga2018_111005A}
\begin{barticle}
\bauthor{\bsnm{Tanga}, \binits{M.}},
\bauthor{\bsnm{Kr{\"{u}}hler}, \binits{T.}},
\bauthor{\bsnm{Schady}, \binits{P.}},
\bauthor{\bsnm{Klose}, \binits{S.}},
\bauthor{\bsnm{Graham}, \binits{J.s.}},
\bauthor{\bsnm{Greiner}, \binits{J.}},
\bauthor{\bsnm{Kann}, \binits{D.s.}},
\bauthor{\bsnm{Nardini}, \binits{M.}}:
\bjtitle{\aap}
\bvolume{615},
\bfpage{136}
(\byear{2018}).
\arxivurl{1708.06270}.
doi:\doiurl{10.1051/0004-6361/201731799}
\end{barticle}
\endbibitem

\bibitem[\protect\citeauthoryear{Tanvir et~al.}{2012}]{Tanvir2012_120422A}
\begin{barticle}
\bauthor{\bsnm{Tanvir}, \binits{N.R.}},
\bauthor{\bsnm{Levan}, \binits{A.J.}},
\bauthor{\bsnm{Cucchiara}, \binits{A.}},
\bauthor{\bsnm{Fox}, \binits{D.B.}}:
\bjtitle{GRB Coordinates Network}
\bvolume{13251},
\bfpage{1}
(\byear{2012})
\end{barticle}
\endbibitem

\bibitem[\protect\citeauthoryear{{Tanvir} et~al.}{2015}]{Tanvir2015_150424A}
\begin{barticle}
\bauthor{\bsnm{{Tanvir}}, \binits{N.R.}},
\bauthor{\bsnm{{Levan}}, \binits{A.J.}},
\bauthor{\bsnm{{Fruchter}}, \binits{A.S.}},
\bauthor{\bsnm{{Hjorth}}, \binits{J.}},
\bauthor{\bsnm{{Watson}}, \binits{D.}},
\bauthor{\bsnm{{Perley}}, \binits{D.}},
\bauthor{\bsnm{{Greiner}}, \binits{J.}},
\bauthor{\bsnm{{de Ugarte Postigo}}, \binits{A.}},
\bauthor{\bsnm{{Thoene}}, \binits{C.}},
\bauthor{\bsnm{{Hounsell}}, \binits{R.A.}},
\bauthor{\bsnm{{Rosswog}}, \binits{S.}}:
\bjtitle{GRB Coordinates Network}
\bvolume{18100},
\bfpage{1}
(\byear{2015})
\end{barticle}
\endbibitem

\bibitem[\protect\citeauthoryear{Tanvir et~al.}{2016}]{Tanvir2016_161219B}
\begin{barticle}
\bauthor{\bsnm{Tanvir}, \binits{N.R.}},
\bauthor{\bsnm{Kruehler}, \binits{T.}},
\bauthor{\bsnm{Wiersema}, \binits{K.}},
\bauthor{\bsnm{Xu}, \binits{D.}},
\bauthor{\bsnm{Malesani}, \binits{D.}},
\bauthor{\bsnm{Milvang-Jensen}, \binits{B.}},
\bauthor{\bsnm{Fynbo}, \binits{J.P.U.}},
\bauthor{\bsnm{Tanvir}, \binits{N.R.}},
\bauthor{\bsnm{Kruehler}, \binits{T.}},
\bauthor{\bsnm{Wiersema}, \binits{K.}},
\bauthor{\bsnm{Xu}, \binits{D.}},
\bauthor{\bsnm{Malesani}, \binits{D.}},
\bauthor{\bsnm{Milvang-Jensen}, \binits{B.}},
\bauthor{\bsnm{Fynbo}, \binits{J.P.U.}}:
\bjtitle{GCN}
\bvolume{20321},
\bfpage{1}
(\byear{2016})
\end{barticle}
\endbibitem

\bibitem[\protect\citeauthoryear{Tanvir et~al.}{2017}]{Tanvir2017_170817A}
\begin{barticle}
\bauthor{\bsnm{Tanvir}, \binits{N.R.}},
\bauthor{\bsnm{Levan}, \binits{A.J.}},
\bauthor{\bsnm{González-Fernández}, \binits{C.}},
\bauthor{\bsnm{Korobkin}, \binits{O.}},
\bauthor{\bsnm{Mandel}, \binits{I.}},
\bauthor{\bsnm{Rosswog}, \binits{S.}},
\bauthor{\bsnm{Hjorth}, \binits{J.}},
\bauthor{\bsnm{D'Avanzo}, \binits{P.}},
\bauthor{\bsnm{Fruchter}, \binits{A.S.}},
\bauthor{\bsnm{Fryer}, \binits{C.L.}},
\bauthor{\bsnm{Kangas}, \binits{T.}},
\bauthor{\bsnm{Milvang-Jensen}, \binits{B.}},
\bauthor{\bsnm{Rosetti}, \binits{S.}},
\bauthor{\bsnm{Steeghs}, \binits{D.}},
\bauthor{\bsnm{Wollaeger}, \binits{R.T.}},
\bauthor{\bsnm{Cano}, \binits{Z.}},
\bauthor{\bsnm{Copperwheat}, \binits{C.M.}},
\bauthor{\bsnm{Covino}, \binits{S.}},
\bauthor{\bsnm{D'Elia}, \binits{V.}},
\bauthor{\bparticle{de} \bsnm{Ugarte~Postigo}, \binits{A.}},
\bauthor{\bsnm{Evans}, \binits{P.A.}},
\bauthor{\bsnm{Even}, \binits{W.P.}},
\bauthor{\bsnm{Fairhurst}, \binits{S.}},
\bauthor{\bsnm{Jaimes}, \binits{R.F.}},
\bauthor{\bsnm{Fontes}, \binits{C.J.}},
\bauthor{\bsnm{Fujii}, \binits{Y.I.}},
\bauthor{\bsnm{Fynbo}, \binits{J.P.U.}},
\bauthor{\bsnm{Gompertz}, \binits{B.P.}},
\bauthor{\bsnm{Greiner}, \binits{J.}},
\bauthor{\bsnm{Hodosan}, \binits{G.}},
\bauthor{\bsnm{Irwin}, \binits{M.J.}},
\bauthor{\bsnm{Jakobsson}, \binits{P.}},
\bauthor{\bsnm{Jørgensen}, \binits{U.G.}},
\bauthor{\bsnm{Kann}, \binits{D.A.}},
\bauthor{\bsnm{Lyman}, \binits{J.D.}},
\bauthor{\bsnm{Malesani}, \binits{D.}},
\bauthor{\bsnm{McMahon}, \binits{R.G.}},
\bauthor{\bsnm{Melandri}, \binits{A.}},
\bauthor{\bsnm{O'Brien}, \binits{P.T.}},
\bauthor{\bsnm{Osborne}, \binits{J.P.}},
\bauthor{\bsnm{Palazzi}, \binits{E.}},
\bauthor{\bsnm{Perley}, \binits{D.A.}},
\bauthor{\bsnm{Pian}, \binits{E.}},
\bauthor{\bsnm{Piranomonte}, \binits{S.}},
\bauthor{\bsnm{Rabus}, \binits{M.}},
\bauthor{\bsnm{Rol}, \binits{E.}},
\bauthor{\bsnm{Rowlinson}, \binits{A.}},
\bauthor{\bsnm{Schulze}, \binits{S.}},
\bauthor{\bsnm{Sutton}, \binits{P.}},
\bauthor{\bsnm{Th\"one}, \binits{C.C.}},
\bauthor{\bsnm{Ulaczyk}, \binits{K.}},
\bauthor{\bsnm{Watson}, \binits{D.}},
\bauthor{\bsnm{Wiersema}, \binits{K.}},
\bauthor{\bsnm{Wijers}, \binits{R.A.M.J.}}:
\bjtitle{\apj}
\bvolume{848},
\bfpage{27}
(\byear{2017}).
doi:\doiurl{10.3847/2041-8213/aa90b6}
\end{barticle}
\endbibitem

\bibitem[\protect\citeauthoryear{Thoene and
  de~Ugarte~Postigo}{2014}]{Thoene2014_111225A}
\begin{barticle}
\bauthor{\bsnm{Thoene}, \binits{C.C.}},
\bauthor{\bparticle{de} \bsnm{Ugarte~Postigo}, \binits{A.}}:
\bjtitle{GRB Coordinates Network}
\bvolume{16079},
\bfpage{1}
(\byear{2014})
\end{barticle}
\endbibitem

\bibitem[\protect\citeauthoryear{Thoene et~al.}{2006}]{Thoene2006_060505}
\begin{barticle}
\bauthor{\bsnm{Thoene}, \binits{C.C.}},
\bauthor{\bsnm{Fynbo}, \binits{J.P.U.}},
\bauthor{\bsnm{Sollerman}, \binits{J.}},
\bauthor{\bsnm{Jensen}, \binits{B.L.}},
\bauthor{\bsnm{Hjorth}, \binits{J.}},
\bauthor{\bsnm{Jakobsson}, \binits{P.}},
\bauthor{\bsnm{Klose}, \binits{S.}},
\bauthor{\bsnm{Thoene}, \binits{C.C.}},
\bauthor{\bsnm{Fynbo}, \binits{J.P.U.}},
\bauthor{\bsnm{Sollerman}, \binits{J.}},
\bauthor{\bsnm{Jensen}, \binits{B.L.}},
\bauthor{\bsnm{Hjorth}, \binits{J.}},
\bauthor{\bsnm{Jakobsson}, \binits{P.}},
\bauthor{\bsnm{Klose}, \binits{S.}}:
\bjtitle{GCN}
\bvolume{5161},
\bfpage{1}
(\byear{2006})
\end{barticle}
\endbibitem

\bibitem[\protect\citeauthoryear{Tinney et~al.}{1998}]{Tinney1998_980425}
\begin{barticle}
\bauthor{\bsnm{Tinney}, \binits{C.}},
\bauthor{\bsnm{Stathakis}, \binits{R.}},
\bauthor{\bsnm{Cannon}, \binits{R.}},
\bauthor{\bsnm{Galama}, \binits{T.}},
\bauthor{\bsnm{Wieringa}, \binits{M.}},
\bauthor{\bsnm{Frail}, \binits{D.s.}},
\bauthor{\bsnm{Kulkarni}, \binits{S.s.}},
\bauthor{\bsnm{Higdon}, \binits{J.s.}},
\bauthor{\bsnm{Wark}, \binits{R.}},
\bauthor{\bsnm{Bloom}, \binits{J.s.}},
\bauthor{\bsnm{{Bepposax GRB Team}}}:
\bjtitle{\iaucirc}
\bvolume{6896},
\bfpage{3}
(\byear{1998})
\end{barticle}
\endbibitem

\bibitem[\protect\citeauthoryear{{Tominaga} et~al.}{2007}]{Tominaga2007}
\begin{barticle}
\bauthor{\bsnm{{Tominaga}}, \binits{N.}},
\bauthor{\bsnm{{Maeda}}, \binits{K.}},
\bauthor{\bsnm{{Umeda}}, \binits{H.}},
\bauthor{\bsnm{{Nomoto}}, \binits{K.}},
\bauthor{\bsnm{{Tanaka}}, \binits{M.}},
\bauthor{\bsnm{{Iwamoto}}, \binits{N.}},
\bauthor{\bsnm{{Suzuki}}, \binits{T.}},
\bauthor{\bsnm{{Mazzali}}, \binits{P.A.}}:
\bjtitle{\apjl}
\bvolume{657}(\bissue{2}),
\bfpage{77}
(\byear{2007}).
\arxivurl{astro-ph/0702471}.
doi:\doiurl{10.1086/513193}
\end{barticle}
\endbibitem

\bibitem[\protect\citeauthoryear{{Toy} et~al.}{2016}]{Toy2016_130702A}
\begin{barticle}
\bauthor{\bsnm{{Toy}}, \binits{V.L.}},
\bauthor{\bsnm{{Cenko}}, \binits{S.B.}},
\bauthor{\bsnm{{Silverman}}, \binits{J.M.}},
\bauthor{\bsnm{{Butler}}, \binits{N.R.}},
\bauthor{\bsnm{{Cucchiara}}, \binits{A.}},
\bauthor{\bsnm{{Watson}}, \binits{A.M.}},
\bauthor{\bsnm{{Bersier}}, \binits{D.}},
\bauthor{\bsnm{{Perley}}, \binits{D.A.}},
\bauthor{\bsnm{{Margutti}}, \binits{R.}},
\bauthor{\bsnm{{Bellm}}, \binits{E.}},
\bauthor{\bsnm{{Bloom}}, \binits{J.S.}},
\bauthor{\bsnm{{Cao}}, \binits{Y.}},
\bauthor{\bsnm{{Capone}}, \binits{J.I.}},
\bauthor{\bsnm{{Clubb}}, \binits{K.}},
\bauthor{\bsnm{{Corsi}}, \binits{A.}},
\bauthor{\bsnm{{De Cia}}, \binits{A.}},
\bauthor{\bsnm{{de Diego}}, \binits{J.A.}},
\bauthor{\bsnm{{Filippenko}}, \binits{A.V.}},
\bauthor{\bsnm{{Fox}}, \binits{O.D.}},
\bauthor{\bsnm{{Gal-Yam}}, \binits{A.}},
\bauthor{\bsnm{{Gehrels}}, \binits{N.}},
\bauthor{\bsnm{{Georgiev}}, \binits{L.}},
\bauthor{\bsnm{{Gonz{\'a}lez}}, \binits{J.J.}},
\bauthor{\bsnm{{Kasliwal}}, \binits{M.M.}},
\bauthor{\bsnm{{Kelly}}, \binits{P.L.}},
\bauthor{\bsnm{{Kulkarni}}, \binits{S.R.}},
\bauthor{\bsnm{{Kutyrev}}, \binits{A.S.}},
\bauthor{\bsnm{{Lee}}, \binits{W.H.}},
\bauthor{\bsnm{{Prochaska}}, \binits{J.X.}},
\bauthor{\bsnm{{Ramirez-Ruiz}}, \binits{E.}},
\bauthor{\bsnm{{Richer}}, \binits{M.G.}},
\bauthor{\bsnm{{Rom{\'a}n-Z{\'u}{\~n}iga}}, \binits{C.}},
\bauthor{\bsnm{{Singer}}, \binits{L.}},
\bauthor{\bsnm{{Stern}}, \binits{D.}},
\bauthor{\bsnm{{Troja}}, \binits{E.}},
\bauthor{\bsnm{{Veilleux}}, \binits{S.}}:
\bjtitle{\apj}
\bvolume{818}(\bissue{1}),
\bfpage{79}
(\byear{2016}).
\arxivurl{1508.00575}.
doi:\doiurl{10.3847/0004-637X/818/1/79}
\end{barticle}
\endbibitem

\bibitem[\protect\citeauthoryear{Troja et~al.}{2006a}]{Troja2006_060502B}
\begin{barticle}
\bauthor{\bsnm{Troja}, \binits{E.}},
\bauthor{\bsnm{Barthelmy}, \binits{S.D.}},
\bauthor{\bsnm{Boyd}, \binits{P.T.}},
\bauthor{\bsnm{Burrows}, \binits{D.N.}},
\bauthor{\bsnm{Cummings}, \binits{J.R.}},
\bauthor{\bsnm{Gehrels}, \binits{N.}},
\bauthor{\bsnm{Holland}, \binits{S.T.}},
\bauthor{\bsnm{Hunsberger}, \binits{S.D.}},
\bauthor{\bsnm{Kennea}, \binits{J.A.}},
\bauthor{\bsnm{Krimm}, \binits{H.A.}},
\bauthor{\bsnm{Parola}, \binits{V.L.}},
\bauthor{\bsnm{Mangano}, \binits{V.}},
\bauthor{\bsnm{Marshall}, \binits{F.E.}},
\bauthor{\bsnm{O'Brien}, \binits{P.T.}},
\bauthor{\bsnm{Page}, \binits{K.L.}},
\bauthor{\bsnm{Palmer}, \binits{D.M.}},
\bauthor{\bsnm{Sakamoto}, \binits{T.}},
\bauthor{\bsnm{Tagliaferri}, \binits{G.}},
\bauthor{\bparticle{vanden} \bsnm{Berk}, \binits{D.E.}}:
\bjtitle{GRB Coordinates Network}
\bvolume{5055},
\bfpage{1}
(\byear{2006}a)
\end{barticle}
\endbibitem

\bibitem[\protect\citeauthoryear{Troja et~al.}{2006b}]{Troja2006_051109B}
\begin{barticle}
\bauthor{\bsnm{Troja}, \binits{E.}},
\bauthor{\bsnm{Cusumano}, \binits{G.}},
\bauthor{\bsnm{Laparola}, \binits{V.}},
\bauthor{\bsnm{Mangano}, \binits{V.}},
\bauthor{\bsnm{Mineo}, \binits{T.}}:
\bjtitle{Nuovo Cimento B Serie}
\bvolume{121},
\bfpage{1599}
(\byear{2006}b).
doi:\doiurl{10.1393/ncb/i2007-10324-8}
\end{barticle}
\endbibitem

\bibitem[\protect\citeauthoryear{{Troja} et~al.}{2008}]{Troja2008}
\begin{barticle}
\bauthor{\bsnm{{Troja}}, \binits{E.}},
\bauthor{\bsnm{{King}}, \binits{A.R.}},
\bauthor{\bsnm{{O'Brien}}, \binits{P.T.}},
\bauthor{\bsnm{{Lyons}}, \binits{N.}},
\bauthor{\bsnm{{Cusumano}}, \binits{G.}}:
\bjtitle{\mnras}
\bvolume{385}(\bissue{1}),
\bfpage{10}
(\byear{2008}).
\arxivurl{0711.3034}.
doi:\doiurl{10.1111/j.1745-3933.2007.00421.x}
\end{barticle}
\endbibitem

\bibitem[\protect\citeauthoryear{Troja et~al.}{2017}]{Troja2017_170817A}
\begin{barticle}
\bauthor{\bsnm{Troja}, \binits{E.}},
\bauthor{\bsnm{Piro}, \binits{L.}},
\bauthor{\bparticle{van} \bsnm{Eerten}, \binits{H.}},
\bauthor{\bsnm{Wollaeger}, \binits{R.T.}},
\bauthor{\bsnm{Im}, \binits{M.}},
\bauthor{\bsnm{Fox}, \binits{O.D.}},
\bauthor{\bsnm{Butler}, \binits{N.R.}},
\bauthor{\bsnm{Cenko}, \binits{S.B.}},
\bauthor{\bsnm{Sakamoto}, \binits{T.}},
\bauthor{\bsnm{Fryer}, \binits{C.L.}},
\bauthor{\bsnm{Ricci}, \binits{R.}},
\bauthor{\bsnm{Lien}, \binits{A.}},
\bauthor{\bsnm{Ryan}, \binits{R.E.}},
\bauthor{\bsnm{Korobkin}, \binits{O.}},
\bauthor{\bsnm{Lee}, \binits{S.-K.}},
\bauthor{\bsnm{Burgess}, \binits{J.M.}},
\bauthor{\bsnm{Lee}, \binits{W.H.}},
\bauthor{\bsnm{Watson}, \binits{A.M.}},
\bauthor{\bsnm{Choi}, \binits{C.}},
\bauthor{\bsnm{Covino}, \binits{S.}},
\bauthor{\bsnm{D'Avanzo}, \binits{P.}},
\bauthor{\bsnm{Fontes}, \binits{C.J.}},
\bauthor{\bsnm{Gonzalez}, \binits{J.B.}},
\bauthor{\bsnm{Khandrika}, \binits{H.G.}},
\bauthor{\bsnm{Kim}, \binits{J.}},
\bauthor{\bsnm{Kim}, \binits{S.-L.}},
\bauthor{\bsnm{Lee}, \binits{C.-U.}},
\bauthor{\bsnm{Lee}, \binits{H.M.}},
\bauthor{\bsnm{Kutyrev}, \binits{A.}},
\bauthor{\bsnm{Lim}, \binits{G.}},
\bauthor{\bsnm{Sanchez-Ramirez}, \binits{R.}},
\bauthor{\bsnm{Veilleux}, \binits{S.}},
\bauthor{\bsnm{Wieringa}, \binits{M.H.}},
\bauthor{\bsnm{Yoon}, \binits{Y.}}:
\bjtitle{\nat}
\bvolume{551},
\bfpage{71}
(\byear{2017}).
doi:\doiurl{10.1038/nature24290}
\end{barticle}
\endbibitem

\bibitem[\protect\citeauthoryear{Troja et~al.}{2018}]{Troja2018_150101B}
\begin{barticle}
\bauthor{\bsnm{Troja}, \binits{E.}},
\bauthor{\bsnm{Ryan}, \binits{G.}},
\bauthor{\bsnm{Piro}, \binits{L.}},
\bauthor{\bparticle{van} \bsnm{Eerten}, \binits{H.}},
\bauthor{\bsnm{Cenko}, \binits{S.B.}},
\bauthor{\bsnm{Yoon}, \binits{Y.}},
\bauthor{\bsnm{Lee}, \binits{S.-K.}},
\bauthor{\bsnm{Im}, \binits{M.}},
\bauthor{\bsnm{Sakamoto}, \binits{T.}},
\bauthor{\bsnm{Gatkine}, \binits{P.}},
\bauthor{\bsnm{Kutyrev}, \binits{A.}},
\bauthor{\bsnm{Veilleux}, \binits{S.}}:
\bjtitle{Nature Communications}
\bvolume{9},
\bfpage{4089}
(\byear{2018}).
doi:\doiurl{10.1038/s41467-018-06558-7}
\end{barticle}
\endbibitem

\bibitem[\protect\citeauthoryear{Troja et~al.}{2019}]{Troja2019_160821B}
\begin{barticle}
\bauthor{\bsnm{Troja}, \binits{E.}},
\bauthor{\bsnm{Castro-Tirado}, \binits{A.J.}},
\bauthor{\bsnm{González}, \binits{J.B.}},
\bauthor{\bsnm{Hu}, \binits{Y.}},
\bauthor{\bsnm{Ryan}, \binits{G.S.}},
\bauthor{\bsnm{Cenko}, \binits{S.B.}},
\bauthor{\bsnm{Ricci}, \binits{R.}},
\bauthor{\bsnm{Novara}, \binits{G.}},
\bauthor{\bsnm{Sanchez-Ramirez}, \binits{R.}},
\bauthor{\bsnm{Acosta-Pulido}, \binits{J.A.}},
\bauthor{\bsnm{Ackley}, \binits{K.D.}},
\bauthor{\bsnm{Garcia}, \binits{M.D.C.}},
\bauthor{\bsnm{Eikenberry}, \binits{S.S.}},
\bauthor{\bsnm{Guziy}, \binits{S.}},
\bauthor{\bsnm{Jeong}, \binits{S.}},
\bauthor{\bsnm{Lien}, \binits{A.Y.}},
\bauthor{\bsnm{Marquez}, \binits{I.}},
\bauthor{\bsnm{ey}, \binits{S.B.P.}},
\bauthor{\bsnm{Park}, \binits{I.H.}},
\bauthor{\bsnm{Sakamoto}, \binits{T.}},
\bauthor{\bsnm{Tello}, \binits{J.C.}},
\bauthor{\bsnm{Sokolov}, \binits{I.V.}},
\bauthor{\bsnm{Sokolov}, \binits{V.V.}},
\bauthor{\bsnm{Tiengo}, \binits{A.}},
\bauthor{\bsnm{Valeev}, \binits{A.F.}},
\bauthor{\bsnm{Zhang}, \binits{B.B.}},
\bauthor{\bsnm{Veilleux}, \binits{S.}}:
\bjtitle{\mnras}
\bvolume{489},
\bfpage{2104}
(\byear{2019}).
doi:\doiurl{10.1093/mnras/stz2255}
\end{barticle}
\endbibitem

\bibitem[\protect\citeauthoryear{{Turpin} et~al.}{2020}]{Turpin2020}
\begin{barticle}
\bauthor{\bsnm{{Turpin}}, \binits{D.}},
\bauthor{\bsnm{{Wu}}, \binits{C.}},
\bauthor{\bsnm{{Han}}, \binits{X.-H.}},
\bauthor{\bsnm{{Xin}}, \binits{L.-P.}},
\bauthor{\bsnm{{Antier}}, \binits{S.}},
\bauthor{\bsnm{{Leroy}}, \binits{N.}},
\bauthor{\bsnm{{Cao}}, \binits{L.}},
\bauthor{\bsnm{{Cai}}, \binits{H.-B.}},
\bauthor{\bsnm{{Cordier}}, \binits{B.}},
\bauthor{\bsnm{{Deng}}, \binits{J.-S.}},
\bauthor{\bsnm{{Dong}}, \binits{W.-L.}},
\bauthor{\bsnm{{Feng}}, \binits{Q.-C.}},
\bauthor{\bsnm{{Huang}}, \binits{L.}},
\bauthor{\bsnm{{Jia}}, \binits{L.}},
\bauthor{\bsnm{{Klotz}}, \binits{A.}},
\bauthor{\bsnm{{Lachaud}}, \binits{C.}},
\bauthor{\bsnm{{Li}}, \binits{H.-L.}},
\bauthor{\bsnm{{Liang}}, \binits{E.-W.}},
\bauthor{\bsnm{{Liu}}, \binits{S.-F.}},
\bauthor{\bsnm{{Lu}}, \binits{X.-M.}},
\bauthor{\bsnm{{Meng}}, \binits{X.-M.}},
\bauthor{\bsnm{{Qiu}}, \binits{Y.-L.}},
\bauthor{\bsnm{{Wang}}, \binits{H.-J.}},
\bauthor{\bsnm{{Wang}}, \binits{J.}},
\bauthor{\bsnm{{Wang}}, \binits{S.}},
\bauthor{\bsnm{{Wang}}, \binits{X.-G.}},
\bauthor{\bsnm{{Wei}}, \binits{J.-Y.}},
\bauthor{\bsnm{{Wu}}, \binits{B.-B.}},
\bauthor{\bsnm{{Xiao}}, \binits{Y.-J.}},
\bauthor{\bsnm{{Xu}}, \binits{D.-W.}},
\bauthor{\bsnm{{Xu}}, \binits{Y.}},
\bauthor{\bsnm{{Yang}}, \binits{Y.-G.}},
\bauthor{\bsnm{{Zhang}}, \binits{P.-P.}},
\bauthor{\bsnm{{Zhang}}, \binits{R.-S.}},
\bauthor{\bsnm{{Zhang}}, \binits{S.-N.}},
\bauthor{\bsnm{{Zheng}}, \binits{Y.-T.}},
\bauthor{\bsnm{{Zou}}, \binits{S.-C.}}:
\bjtitle{Research in Astronomy and Astrophysics}
\bvolume{20}(\bissue{1}),
\bfpage{013}
(\byear{2020}).
\arxivurl{1902.08476}.
doi:\doiurl{10.1088/1674-4527/20/1/13}
\end{barticle}
\endbibitem

\bibitem[\protect\citeauthoryear{{Ukwatta} et~al.}{2008}]{Ukawatta2008}
\begin{bchapter}
\bauthor{\bsnm{{Ukwatta}}, \binits{T.N.}},
\bauthor{\bsnm{{Sakamoto}}, \binits{T.}},
\bauthor{\bsnm{{Stamatikos}}, \binits{M.}},
\bauthor{\bsnm{{Gehrels}}, \binits{N.}},
\bauthor{\bsnm{{Dhuga}}, \binits{K.S.}}:
In: \beditor{\bsnm{{Galassi}}, \binits{M.}},
\beditor{\bsnm{{Palmer}}, \binits{D.}},
\beditor{\bsnm{{Fenimore}}, \binits{E.}} (eds.)
\bbtitle{American Institute of Physics Conference Series}.
\bsertitle{American Institute of Physics Conference Series},
vol. \bseriesno{1000},
p. \bfpage{166}
(\byear{2008}).
\arxivurl{0802.3815}.
doi:\doiurl{10.1063/1.2943435}.
\burl{https://ui.adsabs.harvard.edu/abs/2008AIPC.1000..166U}
\end{bchapter}
\endbibitem

\bibitem[\protect\citeauthoryear{{Ulanov} et~al.}{2005}]{Ulanov2005_031203}
\begin{barticle}
\bauthor{\bsnm{{Ulanov}}, \binits{M.V.}},
\bauthor{\bsnm{{Golenetskii}}, \binits{S.V.}},
\bauthor{\bsnm{{Frederiks}}, \binits{D.D.}},
\bauthor{\bsnm{{Mazets}}, \binits{R.L.A.E.P.}},
\bauthor{\bsnm{{Kokomov}}, \binits{A.A.}},
\bauthor{\bsnm{{Palshin}}, \binits{V.D.}}:
\bjtitle{Nuovo Cimento C Geophysics Space Physics C}
\bvolume{28}(\bissue{3}),
\bfpage{351}
(\byear{2005}).
doi:\doiurl{10.1393/ncc/i2005-10058-8}
\end{barticle}
\endbibitem

\bibitem[\protect\citeauthoryear{Valeev et~al.}{2019}]{Valeev2019_190829A}
\begin{barticle}
\bauthor{\bsnm{Valeev}, \binits{A.F.}},
\bauthor{\bsnm{Castro-Tirado}, \binits{A.J.}},
\bauthor{\bsnm{Hu}, \binits{Y.-D.}},
\bauthor{\bsnm{Fernandez-Garcia}, \binits{E.}},
\bauthor{\bsnm{Sokolov}, \binits{V.V.}},
\bauthor{\bsnm{Carrasco}, \binits{I.}},
\bauthor{\bsnm{Castellon}, \binits{A.}},
\bauthor{\bsnm{{Garcia Alvarez}}, \binits{D.}},
\bauthor{\bsnm{Rivero}, \binits{M.}},
\bauthor{\bparticle{et} \bsnm{Al.}},
\bauthor{\bsnm{Valeev}, \binits{A.F.}},
\bauthor{\bsnm{Castro-Tirado}, \binits{A.J.}},
\bauthor{\bsnm{Hu}, \binits{Y.-D.}},
\bauthor{\bsnm{Fernandez-Garcia}, \binits{E.}},
\bauthor{\bsnm{Sokolov}, \binits{V.V.}},
\bauthor{\bsnm{Carrasco}, \binits{I.}},
\bauthor{\bsnm{Castellon}, \binits{A.}},
\bauthor{\bsnm{{Garcia Alvarez}}, \binits{D.}},
\bauthor{\bsnm{Rivero}, \binits{M.}},
\bauthor{\bparticle{et} \bsnm{Al.}}:
\bjtitle{GCN}
\bvolume{25565},
\bfpage{1}
(\byear{2019})
\end{barticle}
\endbibitem

\bibitem[\protect\citeauthoryear{Valle et~al.}{2006}]{DellaValle2006_060614}
\begin{barticle}
\bauthor{\bsnm{Valle}, \binits{M.D.}},
\bauthor{\bsnm{Chincarini}, \binits{G.}},
\bauthor{\bsnm{Panagia}, \binits{N.}},
\bauthor{\bsnm{Tagliaferri}, \binits{G.}},
\bauthor{\bsnm{Malesani}, \binits{D.}},
\bauthor{\bsnm{Testa}, \binits{V.}},
\bauthor{\bsnm{Fugazza}, \binits{D.}},
\bauthor{\bsnm{Campana}, \binits{S.}},
\bauthor{\bsnm{Covino}, \binits{S.}},
\bauthor{\bsnm{Mangano}, \binits{V.}},
\bauthor{\bsnm{Antonelli}, \binits{L.A.}},
\bauthor{\bsnm{D'Avanzo}, \binits{P.}},
\bauthor{\bsnm{Hurley}, \binits{K.}},
\bauthor{\bsnm{Mirabel}, \binits{I.F.}},
\bauthor{\bsnm{Pellizza}, \binits{L.J.}},
\bauthor{\bsnm{Piranomonte}, \binits{S.}},
\bauthor{\bsnm{Stella}, \binits{L.}}:
\bjtitle{\nat}
\bvolume{444},
\bfpage{1050}
(\byear{2006}).
doi:\doiurl{10.1038/nature05374}
\end{barticle}
\endbibitem

\bibitem[\protect\citeauthoryear{Valle et~al.}{2018}]{DellaValle2018_170817}
\begin{botherref}
\oauthor{\bsnm{Valle}, \binits{M.D.}},
\oauthor{\bsnm{Guetta}, \binits{D.}},
\oauthor{\bsnm{Cappellaro}, \binits{E.}},
\oauthor{\bsnm{Amati}, \binits{L.}},
\oauthor{\bsnm{Botticella}, \binits{M.T.}},
\oauthor{\bsnm{Branchesi}, \binits{M.}},
\oauthor{\bsnm{Brocato}, \binits{E.}},
\oauthor{\bsnm{Izzo}, \binits{L.}},
\oauthor{\bsnm{Perez-Torres}, \binits{M.A.}},
\oauthor{\bsnm{Stratta}, \binits{G.}}:
Gw170817: implications for the local kilonova rate and for surveys from
  ground-based facilities
\textbf{481},
4355
(2018).
doi:\doiurl{10.1093/mnras/sty2541}
\end{botherref}
\endbibitem

\bibitem[\protect\citeauthoryear{van~der Horst}{2013}]{vanderHorst2013_130702A}
\begin{barticle}
\bauthor{\bparticle{van~der} \bsnm{Horst}, \binits{A.J.}}:
\bjtitle{GRB Coordinates Network}
\bvolume{14987},
\bfpage{1}
(\byear{2013})
\end{barticle}
\endbibitem

\bibitem[\protect\citeauthoryear{van~der Horst
  et~al.}{2006}]{vanderHorst2006_030329}
\begin{barticle}
\bauthor{\bparticle{van~der} \bsnm{Horst}, \binits{A.J.}},
\bauthor{\bsnm{Kamble}, \binits{A.}},
\bauthor{\bsnm{Wijers}, \binits{R.A.M.J.}},
\bauthor{\bsnm{Resmi}, \binits{L.}},
\bauthor{\bsnm{Bhattacharya}, \binits{D.}},
\bauthor{\bsnm{Rol}, \binits{E.}},
\bauthor{\bsnm{Strom}, \binits{R.}},
\bauthor{\bsnm{Kouveliotou}, \binits{C.}},
\bauthor{\bsnm{Oosterloo}, \binits{T.}},
\bauthor{\bsnm{Ishwara-Chandra}, \binits{C.H.}}:
\bjtitle{Nuovo Cimento B Serie}
\bvolume{121},
\bfpage{1605}
(\byear{2006}).
doi:\doiurl{10.1393/ncb/i2007-10327-5}
\end{barticle}
\endbibitem

\bibitem[\protect\citeauthoryear{van Eerten et~al.}{2010}]{vanEerten2010}
\begin{botherref}
\oauthor{\bparticle{van} \bsnm{Eerten}, \binits{H.J.}},
\oauthor{\bsnm{Leventis}, \binits{K.}},
\oauthor{\bsnm{Meliani}, \binits{Z.}},
\oauthor{\bsnm{Wijers}, \binits{R.A.M.J.}},
\oauthor{\bsnm{Keppens}, \binits{R.}}:
Gamma-ray burst afterglows from transrelativistic blast wave simulations
\textbf{403},
300
(2010).
doi:\doiurl{10.1111/j.1365-2966.2009.16109.x}
\end{botherref}
\endbibitem

\bibitem[\protect\citeauthoryear{van Putten}{2004}]{vanPutten2004}
\begin{botherref}
\oauthor{\bparticle{van} \bsnm{Putten}, \binits{M.H.P.M.}}:
The branching ratio of type ib/c supernovae to gamma-ray burst supernovae
\textbf{611},
81
(2004).
doi:\doiurl{10.1086/423934}
\end{botherref}
\endbibitem

\bibitem[\protect\citeauthoryear{Vanderspek
  et~al.}{2004}]{Vanderspek2004_030329}
\begin{barticle}
\bauthor{\bsnm{Vanderspek}, \binits{R.}},
\bauthor{\bsnm{Sakamoto}, \binits{T.}},
\bauthor{\bsnm{Barraud}, \binits{C.}},
\bauthor{\bsnm{Tamagawa}, \binits{T.}},
\bauthor{\bsnm{Graziani}, \binits{C.}},
\bauthor{\bsnm{Suzuki}, \binits{M.}},
\bauthor{\bsnm{Shirasaki}, \binits{Y.}},
\bauthor{\bsnm{Prigozhin}, \binits{G.}},
\bauthor{\bsnm{Villasenor}, \binits{J.}},
\bauthor{\bsnm{Jernigan}, \binits{J.G.}},
\bauthor{\bsnm{Crew}, \binits{G.B.}},
\bauthor{\bsnm{Atteia}, \binits{J.-L.}},
\bauthor{\bsnm{Hurley}, \binits{K.}},
\bauthor{\bsnm{Kawai}, \binits{N.}},
\bauthor{\bsnm{Lamb}, \binits{D.Q.}},
\bauthor{\bsnm{Ricker}, \binits{G.R.}},
\bauthor{\bsnm{Woosley}, \binits{S.E.}},
\bauthor{\bsnm{Butler}, \binits{N.}},
\bauthor{\bsnm{Doty}, \binits{J.P.}},
\bauthor{\bsnm{Dullighan}, \binits{A.}},
\bauthor{\bsnm{Donaghy}, \binits{T.Q.}},
\bauthor{\bsnm{Fenimore}, \binits{E.E.}},
\bauthor{\bsnm{Galassi}, \binits{M.}},
\bauthor{\bsnm{Pizzichini}, \binits{G.}},
\bauthor{\bsnm{Matsuoka}, \binits{M..}},
\bauthor{\bsnm{Takagishi}, \binits{K.}},
\bauthor{\bsnm{Torii}, \binits{K.}},
\bauthor{\bsnm{Yoshida}, \binits{A.}},
\bauthor{\bsnm{Boer}, \binits{M.}},
\bauthor{\bsnm{Dezalay}, \binits{J.-P.}},
\bauthor{\bsnm{Olive}, \binits{J.-F.}},
\bauthor{\bsnm{Braga}, \binits{J.}},
\bauthor{\bsnm{Manchanda}, \binits{R.}}:
\bjtitle{Astrophys. J.}
\bvolume{617}(\bissue{2}),
\bfpage{1251}
(\byear{2004}).
\arxivurl{0401311}.
doi:\doiurl{10.1086/423923}
\end{barticle}
\endbibitem

\bibitem[\protect\citeauthoryear{Vergani et~al.}{2010}]{Vergani2010_100316D}
\begin{barticle}
\bauthor{\bsnm{Vergani}, \binits{S.D.}},
\bauthor{\bsnm{D'Avanzo}, \binits{P.}},
\bauthor{\bsnm{Levan}, \binits{A.J.}},
\bauthor{\bsnm{Covino}, \binits{S.}},
\bauthor{\bsnm{Malesani}, \binits{D.}},
\bauthor{\bsnm{Hjorth}, \binits{J.}},
\bauthor{\bsnm{Antonelli}, \binits{L.A.}}:
\bjtitle{GRB Coordinates Network}
\bvolume{10512},
\bfpage{1}
(\byear{2010})
\end{barticle}
\endbibitem

\bibitem[\protect\citeauthoryear{Villasenor
  et~al.}{2005}]{Villasenor2005_050709}
\begin{barticle}
\bauthor{\bsnm{Villasenor}, \binits{J.S.}},
\bauthor{\bsnm{Lamb}, \binits{D.Q.}},
\bauthor{\bsnm{Ricker}, \binits{G.R.}},
\bauthor{\bsnm{Atteia}, \binits{J.-L.}},
\bauthor{\bsnm{Kawai}, \binits{N.}},
\bauthor{\bsnm{Butler}, \binits{N.}},
\bauthor{\bsnm{Nakagawa}, \binits{Y.}},
\bauthor{\bsnm{Jernigan}, \binits{J.G.}},
\bauthor{\bsnm{Boer}, \binits{M.}},
\bauthor{\bsnm{Crew}, \binits{G.B.}},
\bauthor{\bsnm{Donaghy}, \binits{T.Q.}},
\bauthor{\bsnm{Doty}, \binits{J.}},
\bauthor{\bsnm{Fenimore}, \binits{E.E.}},
\bauthor{\bsnm{Galassi}, \binits{M.}},
\bauthor{\bsnm{Graziani}, \binits{C.}},
\bauthor{\bsnm{Hurley}, \binits{K.}},
\bauthor{\bsnm{Levine}, \binits{A.}},
\bauthor{\bsnm{Martel}, \binits{F.}},
\bauthor{\bsnm{Matsuoka}, \binits{M.}},
\bauthor{\bsnm{Olive}, \binits{J.-F.}},
\bauthor{\bsnm{Prigozhin}, \binits{G.}},
\bauthor{\bsnm{Sakamoto}, \binits{T.}},
\bauthor{\bsnm{Shirasaki}, \binits{Y.}},
\bauthor{\bsnm{Suzuki}, \binits{M.}},
\bauthor{\bsnm{Tamagawa}, \binits{T.}},
\bauthor{\bsnm{Vanderspek}, \binits{R.}},
\bauthor{\bsnm{Woosley}, \binits{S.E.}},
\bauthor{\bsnm{Yoshida}, \binits{A.}},
\bauthor{\bsnm{Braga}, \binits{J.}},
\bauthor{\bsnm{Manchanda}, \binits{R.}},
\bauthor{\bsnm{Pizzichini}, \binits{G.}},
\bauthor{\bsnm{Takagishi}, \binits{K.}},
\bauthor{\bsnm{Yamauchi}, \binits{M.}}:
\bjtitle{Nature}
\bvolume{437}(\bissue{7060}),
\bfpage{855}
(\byear{2005}).
\arxivurl{0510190}.
doi:\doiurl{10.1038/nature04213}
\end{barticle}
\endbibitem

\bibitem[\protect\citeauthoryear{Vrba et~al.}{2000}]{Vrba2000_980827}
\begin{barticle}
\bauthor{\bsnm{Vrba}, \binits{F.J.}},
\bauthor{\bsnm{Henden}, \binits{A.A.}},
\bauthor{\bsnm{Luginbuhl}, \binits{C.B.}},
\bauthor{\bsnm{Guetter}, \binits{H.H.}},
\bauthor{\bsnm{Hartmann}, \binits{D.H.}},
\bauthor{\bsnm{Klose}, \binits{S.}}:
\bjtitle{\apj}
\bvolume{533},
\bfpage{17}
(\byear{2000}).
doi:\doiurl{10.1086/312602}
\end{barticle}
\endbibitem

\bibitem[\protect\citeauthoryear{Wanderman and Piran}{2015}]{Wanderman2015}
\begin{botherref}
\oauthor{\bsnm{Wanderman}, \binits{D.}},
\oauthor{\bsnm{Piran}, \binits{T.}}:
The rate, luminosity function and time delay of non-collapsar short grbs
\textbf{448},
3026
(2015).
doi:\doiurl{10.1093/mnras/stv123}
\end{botherref}
\endbibitem

\bibitem[\protect\citeauthoryear{Wang et~al.}{2019}]{Wang2019}
\begin{barticle}
\bauthor{\bsnm{Wang}, \binits{Y.}},
\bauthor{\bsnm{Rueda}, \binits{J.A.}},
\bauthor{\bsnm{Ruffini}, \binits{R.}},
\bauthor{\bsnm{Becerra}, \binits{L.}},
\bauthor{\bsnm{Bianco}, \binits{C.}},
\bauthor{\bsnm{Becerra}, \binits{L.}},
\bauthor{\bsnm{Li}, \binits{L.}},
\bauthor{\bsnm{Karlica}, \binits{M.}}:
\bjtitle{\apj}
\bvolume{874},
\bfpage{39}
(\byear{2019}).
doi:\doiurl{10.3847/1538-4357/ab04f8}
\end{barticle}
\endbibitem

\bibitem[\protect\citeauthoryear{Watson et~al.}{2004}]{Watson2004_031203}
\begin{barticle}
\bauthor{\bsnm{Watson}, \binits{D.}},
\bauthor{\bsnm{Hjorth}, \binits{J.}},
\bauthor{\bsnm{Levan}, \binits{A.}},
\bauthor{\bsnm{Jakobsson}, \binits{P.}},
\bauthor{\bsnm{O'Brien}, \binits{P.T.}},
\bauthor{\bsnm{Osborne}, \binits{J.P.}},
\bauthor{\bsnm{Pedersen}, \binits{K.}},
\bauthor{\bsnm{Reeves}, \binits{J.N.}},
\bauthor{\bsnm{Tedds}, \binits{J.A.}},
\bauthor{\bsnm{Vaughan}, \binits{S.A.}},
\bauthor{\bsnm{Ward}, \binits{M.J.}},
\bauthor{\bsnm{Willingale}, \binits{R.}}:
\bjtitle{Astrophys. J.}
\bvolume{605}(\bissue{2}),
\bfpage{101}
(\byear{2004}).
\arxivurl{0401225}.
doi:\doiurl{10.1086/420844}
\end{barticle}
\endbibitem

\bibitem[\protect\citeauthoryear{{Watson} et~al.}{2006}]{Watson2006_031203}
\begin{barticle}
\bauthor{\bsnm{{Watson}}, \binits{D.}},
\bauthor{\bsnm{{Vaughan}}, \binits{S.A.}},
\bauthor{\bsnm{{Willingale}}, \binits{R.}},
\bauthor{\bsnm{{Hjorth}}, \binits{J.}},
\bauthor{\bsnm{{Foley}}, \binits{S.}},
\bauthor{\bsnm{{Fynbo}}, \binits{J.P.U.}},
\bauthor{\bsnm{{Jakobsson}}, \binits{P.}},
\bauthor{\bsnm{{Levan}}, \binits{A.}},
\bauthor{\bsnm{{O'Brien}}, \binits{P.T.}},
\bauthor{\bsnm{{Osborne}}, \binits{J.P.}},
\bauthor{\bsnm{{Pedersen}}, \binits{K.}},
\bauthor{\bsnm{{Reeves}}, \binits{J.N.}},
\bauthor{\bsnm{{Tedds}}, \binits{J.A.}},
\bauthor{\bsnm{{Watson}}, \binits{M.G.}}:
\bjtitle{\apj}
\bvolume{636}(\bissue{2}),
\bfpage{967}
(\byear{2006}).
\arxivurl{astro-ph/0509477}.
doi:\doiurl{10.1086/498089}
\end{barticle}
\endbibitem

\bibitem[\protect\citeauthoryear{Waxman}{2004}]{Waxman2004_980425}
\begin{botherref}
\oauthor{\bsnm{Waxman}, \binits{E.}}:
The nature of grb 980425 and the search for off-axis gamma-ray burst signatures
  in nearby type ib/c supernova emission
\textbf{602},
886
(2004).
doi:\doiurl{10.1086/381230}
\end{botherref}
\endbibitem

\bibitem[\protect\citeauthoryear{{Wei} et~al.}{2016}]{Wei2016}
\begin{botherref}
\oauthor{\bsnm{{Wei}}, \binits{J.}},
\oauthor{\bsnm{{Cordier}}, \binits{B.}},
\oauthor{\bsnm{{Antier}}, \binits{S.}},
\oauthor{\bsnm{{Antilogus}}, \binits{P.}},
\oauthor{\bsnm{{Atteia}}, \binits{J.-L.}},
\oauthor{\bsnm{{Bajat}}, \binits{A.}},
\oauthor{\bsnm{{Basa}}, \binits{S.}},
\oauthor{\bsnm{{Beckmann}}, \binits{V.}},
\oauthor{\bsnm{{Bernardini}}, \binits{M.G.}},
\oauthor{\bsnm{{Boissier}}, \binits{S.}},
\oauthor{\bsnm{{Bouchet}}, \binits{L.}},
\oauthor{\bsnm{{Burwitz}}, \binits{V.}},
\oauthor{\bsnm{{Claret}}, \binits{A.}},
\oauthor{\bsnm{{Dai}}, \binits{Z.-G.}},
\oauthor{\bsnm{{Daigne}}, \binits{F.}},
\oauthor{\bsnm{{Deng}}, \binits{J.}},
\oauthor{\bsnm{{Dornic}}, \binits{D.}},
\oauthor{\bsnm{{Feng}}, \binits{H.}},
\oauthor{\bsnm{{Foglizzo}}, \binits{T.}},
\oauthor{\bsnm{{Gao}}, \binits{H.}},
\oauthor{\bsnm{{Gehrels}}, \binits{N.}},
\oauthor{\bsnm{{Godet}}, \binits{O.}},
\oauthor{\bsnm{{Goldwurm}}, \binits{A.}},
\oauthor{\bsnm{{Gonzalez}}, \binits{F.}},
\oauthor{\bsnm{{Gosset}}, \binits{L.}},
\oauthor{\bsnm{{G{\"o}tz}}, \binits{D.}},
\oauthor{\bsnm{{Gouiffes}}, \binits{C.}},
\oauthor{\bsnm{{Grise}}, \binits{F.}},
\oauthor{\bsnm{{Gros}}, \binits{A.}},
\oauthor{\bsnm{{Guilet}}, \binits{J.}},
\oauthor{\bsnm{{Han}}, \binits{X.}},
\oauthor{\bsnm{{Huang}}, \binits{M.}},
\oauthor{\bsnm{{Huang}}, \binits{Y.-F.}},
\oauthor{\bsnm{{Jouret}}, \binits{M.}},
\oauthor{\bsnm{{Klotz}}, \binits{A.}},
\oauthor{\bsnm{{La Marle}}, \binits{O.}},
\oauthor{\bsnm{{Lachaud}}, \binits{C.}},
\oauthor{\bsnm{{Le Floch}}, \binits{E.}},
\oauthor{\bsnm{{Lee}}, \binits{W.}},
\oauthor{\bsnm{{Leroy}}, \binits{N.}},
\oauthor{\bsnm{{Li}}, \binits{L.-X.}},
\oauthor{\bsnm{{Li}}, \binits{S.C.}},
\oauthor{\bsnm{{Li}}, \binits{Z.}},
\oauthor{\bsnm{{Liang}}, \binits{E.-W.}},
\oauthor{\bsnm{{Lyu}}, \binits{H.}},
\oauthor{\bsnm{{Mercier}}, \binits{K.}},
\oauthor{\bsnm{{Migliori}}, \binits{G.}},
\oauthor{\bsnm{{Mochkovitch}}, \binits{R.}},
\oauthor{\bsnm{{O'Brien}}, \binits{P.}},
\oauthor{\bsnm{{Osborne}}, \binits{J.}},
\oauthor{\bsnm{{Paul}}, \binits{J.}},
\oauthor{\bsnm{{Perinati}}, \binits{E.}},
\oauthor{\bsnm{{Petitjean}}, \binits{P.}},
\oauthor{\bsnm{{Piron}}, \binits{F.}},
\oauthor{\bsnm{{Qiu}}, \binits{Y.}},
\oauthor{\bsnm{{Rau}}, \binits{A.}},
\oauthor{\bsnm{{Rodriguez}}, \binits{J.}},
\oauthor{\bsnm{{Schanne}}, \binits{S.}},
\oauthor{\bsnm{{Tanvir}}, \binits{N.}},
\oauthor{\bsnm{{Vangioni}}, \binits{E.}},
\oauthor{\bsnm{{Vergani}}, \binits{S.}},
\oauthor{\bsnm{{Wang}}, \binits{F.-Y.}},
\oauthor{\bsnm{{Wang}}, \binits{J.}},
\oauthor{\bsnm{{Wang}}, \binits{X.-G.}},
\oauthor{\bsnm{{Wang}}, \binits{X.-Y.}},
\oauthor{\bsnm{{Watson}}, \binits{A.}},
\oauthor{\bsnm{{Webb}}, \binits{N.}},
\oauthor{\bsnm{{Wei}}, \binits{J.J.}},
\oauthor{\bsnm{{Willingale}}, \binits{R.}},
\oauthor{\bsnm{{Wu}}, \binits{C.}},
\oauthor{\bsnm{{Wu}}, \binits{X.-F.}},
\oauthor{\bsnm{{Xin}}, \binits{L.-P.}},
\oauthor{\bsnm{{Xu}}, \binits{D.}},
\oauthor{\bsnm{{Yu}}, \binits{S.}},
\oauthor{\bsnm{{Yu}}, \binits{W.-F.}},
\oauthor{\bsnm{{Yu}}, \binits{Y.-W.}},
\oauthor{\bsnm{{Zhang}}, \binits{B.}},
\oauthor{\bsnm{{Zhang}}, \binits{S.-N.}},
\oauthor{\bsnm{{Zhang}}, \binits{Y.}},
\oauthor{\bsnm{{Zhou}}, \binits{X.L.}}:
arXiv e-prints,
1610
(2016).
\arxivurl{1610.06892}
\end{botherref}
\endbibitem

\bibitem[\protect\citeauthoryear{Weiler et~al.}{2004}]{Weiler2004}
\begin{botherref}
\oauthor{\bsnm{Weiler}, \binits{K.W.}},
\oauthor{\bsnm{Dyk}, \binits{S.D.V.}},
\oauthor{\bsnm{Sramek}, \binits{R.A.}},
\oauthor{\bsnm{Panagia}, \binits{N.}}:
Radio emission from supernovae and gamma-ray bursters and the need for the ska
\textbf{48},
1377
(2004).
doi:\doiurl{10.1016/j.newar.2004.09.017}
\end{botherref}
\endbibitem

\bibitem[\protect\citeauthoryear{Wieringa et~al.}{2010}]{Wieringa2010_100316D}
\begin{barticle}
\bauthor{\bsnm{Wieringa}, \binits{M.}},
\bauthor{\bsnm{Soderberg}, \binits{A.}},
\bauthor{\bsnm{Edwards}, \binits{P.}}:
\bjtitle{GRB Coordinates Network}
\bvolume{10533},
\bfpage{1}
(\byear{2010})
\end{barticle}
\endbibitem

\bibitem[\protect\citeauthoryear{Wiersema et~al.}{2010}]{Wiersema2010_100316D}
\begin{barticle}
\bauthor{\bsnm{Wiersema}, \binits{K.}},
\bauthor{\bsnm{D'Avanzo}, \binits{P.}},
\bauthor{\bsnm{Levan}, \binits{A.J.}},
\bauthor{\bsnm{Tanvir}, \binits{N.R.}},
\bauthor{\bsnm{Malesani}, \binits{D.}},
\bauthor{\bsnm{Covino}, \binits{S.}}:
\bjtitle{GRB Coordinates Network}
\bvolume{10525},
\bfpage{1}
(\byear{2010})
\end{barticle}
\endbibitem

\bibitem[\protect\citeauthoryear{{Winkler} et~al.}{2003}]{Winkler2003}
\begin{barticle}
\bauthor{\bsnm{{Winkler}}, \binits{C.}},
\bauthor{\bsnm{{Courvoisier}}, \binits{T.J.-L.}},
\bauthor{\bsnm{{Di Cocco}}, \binits{G.}},
\bauthor{\bsnm{{Gehrels}}, \binits{N.}},
\bauthor{\bsnm{{Gim{\'e}nez}}, \binits{A.}},
\bauthor{\bsnm{{Grebenev}}, \binits{S.}},
\bauthor{\bsnm{{Hermsen}}, \binits{W.}},
\bauthor{\bsnm{{Mas-Hesse}}, \binits{J.M.}},
\bauthor{\bsnm{{Lebrun}}, \binits{F.}},
\bauthor{\bsnm{{Lund}}, \binits{N.}},
\bauthor{\bsnm{{Palumbo}}, \binits{G.G.C.}},
\bauthor{\bsnm{{Paul}}, \binits{J.}},
\bauthor{\bsnm{{Roques}}, \binits{J.-P.}},
\bauthor{\bsnm{{Schnopper}}, \binits{H.}},
\bauthor{\bsnm{{Sch{\"o}nfelder}}, \binits{V.}},
\bauthor{\bsnm{{Sunyaev}}, \binits{R.}},
\bauthor{\bsnm{{Teegarden}}, \binits{B.}},
\bauthor{\bsnm{{Ubertini}}, \binits{P.}},
\bauthor{\bsnm{{Vedrenne}}, \binits{G.}},
\bauthor{\bsnm{{Dean}}, \binits{A.J.}}:
\bjtitle{\aap}
\bvolume{411},
\bfpage{1}
(\byear{2003}).
doi:\doiurl{10.1051/0004-6361:20031288}
\end{barticle}
\endbibitem

\bibitem[\protect\citeauthoryear{Woosley et~al.}{1999}]{Woosley1999}
\begin{botherref}
\oauthor{\bsnm{Woosley}, \binits{S.E.}},
\oauthor{\bsnm{Eastman}, \binits{R.G.}},
\oauthor{\bsnm{Schmidt}, \binits{B.P.}}:
Gamma-ray bursts and type ic supernova sn 1998bw
\textbf{516},
788
(1999).
doi:\doiurl{10.1086/307131}
\end{botherref}
\endbibitem

\bibitem[\protect\citeauthoryear{{Wu} et~al.}{2012}]{Wu2012_VT}
\begin{bchapter}
\bauthor{\bsnm{{Wu}}, \binits{C.}},
\bauthor{\bsnm{{Qiu}}, \binits{Y.L.}},
\bauthor{\bsnm{{Cai}}, \binits{H.B.}}:
In: \beditor{\bsnm{{Roming}}, \binits{P.}},
\beditor{\bsnm{{Kawai}}, \binits{N.}},
\beditor{\bsnm{{Pian}}, \binits{E.}} (eds.)
\bbtitle{Death of Massive Stars: Supernovae and Gamma-Ray Bursts}.
\bsertitle{IAU Symposium},
vol. \bseriesno{279},
p. \bfpage{421}
(\byear{2012}).
doi:\doiurl{10.1017/S1743921312013646}.
\burl{https://ui.adsabs.harvard.edu/abs/2012IAUS..279..421W}
\end{bchapter}
\endbibitem

\bibitem[\protect\citeauthoryear{Xie et~al.}{2012}]{Xie2012}
\begin{barticle}
\bauthor{\bsnm{Xie}, \binits{W.}},
\bauthor{\bsnm{Lei}, \binits{W.-H.}},
\bauthor{\bsnm{Zou}, \binits{Y.-C.}},
\bauthor{\bsnm{Wang}, \binits{D.-X.}},
\bauthor{\bsnm{Wu}, \binits{Q.}},
\bauthor{\bsnm{Wang}, \binits{J.-Z.}}:
\bjtitle{Research in Astronomy and Astrophysics}
\bvolume{12},
\bfpage{817}
(\byear{2012}).
doi:\doiurl{10.1088/1674-4527/12/7/010}
\end{barticle}
\endbibitem

\bibitem[\protect\citeauthoryear{Xu et~al.}{2015}]{Xu2015_150518Aafter}
\begin{barticle}
\bauthor{\bsnm{Xu}, \binits{D.}},
\bauthor{\bsnm{Wu}, \binits{X.-B.}},
\bauthor{\bsnm{Yang}, \binits{Q.}}:
\bjtitle{GRB Coordinates Network}
\bvolume{17829},
\bfpage{1}
(\byear{2015})
\end{barticle}
\endbibitem

\bibitem[\protect\citeauthoryear{Xu et~al.}{2009}]{Xu2009_060505}
\begin{barticle}
\bauthor{\bsnm{Xu}, \binits{D.}},
\bauthor{\bsnm{Starling}, \binits{R.L.C.}},
\bauthor{\bsnm{Fynbo}, \binits{J.P.U.}},
\bauthor{\bsnm{Sollerman}, \binits{J.}},
\bauthor{\bsnm{Yost}, \binits{S.}},
\bauthor{\bsnm{Watson}, \binits{D.}},
\bauthor{\bsnm{Foley}, \binits{S.}},
\bauthor{\bsnm{O'Brien}, \binits{P.T.}},
\bauthor{\bsnm{Hjorth}, \binits{J.}}:
\bjtitle{\apj}
\bvolume{696},
\bfpage{971}
(\byear{2009}).
doi:\doiurl{10.1088/0004-637X/696/1/971}
\end{barticle}
\endbibitem

\bibitem[\protect\citeauthoryear{{Xu} et~al.}{2011}]{Xu2011_111005A}
\begin{barticle}
\bauthor{\bsnm{{Xu}}, \binits{D.}},
\bauthor{\bsnm{{Michalowski}}, \binits{M.}},
\bauthor{\bsnm{{Stevens}}, \binits{J.}},
\bauthor{\bsnm{{Edwards}}, \binits{P.}}:
\bjtitle{GRB Coordinates Network}
\bvolume{12435},
\bfpage{1}
(\byear{2011})
\end{barticle}
\endbibitem

\bibitem[\protect\citeauthoryear{Xu et~al.}{2015a}]{Xu2015_150518A}
\begin{barticle}
\bauthor{\bsnm{Xu}, \binits{D.}},
\bauthor{\bsnm{Levan}, \binits{A.J.}},
\bauthor{\bsnm{{de Ugarte Postigo}}, \binits{A.}},
\bauthor{\bsnm{Tanvir}, \binits{N.R.}},
\bauthor{\bsnm{Malesani}, \binits{D.}},
\bauthor{\bsnm{Fynbo}, \binits{J.P.U.}},
\bauthor{\bsnm{Xu}, \binits{D.}},
\bauthor{\bsnm{Levan}, \binits{A.J.}},
\bauthor{\bsnm{{de Ugarte Postigo}}, \binits{A.}},
\bauthor{\bsnm{Tanvir}, \binits{N.R.}},
\bauthor{\bsnm{Malesani}, \binits{D.}},
\bauthor{\bsnm{Fynbo}, \binits{J.P.U.}}:
\bjtitle{GCN}
\bvolume{17832},
\bfpage{1}
(\byear{2015}a)
\end{barticle}
\endbibitem

\bibitem[\protect\citeauthoryear{Xu et~al.}{2015b}]{Xu2015_150518Aredshift}
\begin{barticle}
\bauthor{\bsnm{Xu}, \binits{D.}},
\bauthor{\bsnm{Levan}, \binits{A.J.}},
\bauthor{\bparticle{de} \bsnm{Ugarte~Postigo}, \binits{A.}},
\bauthor{\bsnm{Tanvir}, \binits{N.R.}},
\bauthor{\bsnm{Malesani}, \binits{D.}},
\bauthor{\bsnm{Fynbo}, \binits{J.P.U.}}:
\bjtitle{GRB Coordinates Network}
\bvolume{17832},
\bfpage{1}
(\byear{2015}b)
\end{barticle}
\endbibitem

\bibitem[\protect\citeauthoryear{Xu et~al.}{2016}]{Xu2016_160821B}
\begin{barticle}
\bauthor{\bsnm{Xu}, \binits{D.}},
\bauthor{\bsnm{Malesani}, \binits{D.}},
\bauthor{\bparticle{de} \bsnm{Ugarte~Postigo}, \binits{A.}},
\bauthor{\bsnm{Gafton}, \binits{E.}},
\bauthor{\bsnm{Losada}, \binits{I.R.}}:
\bjtitle{GRB Coordinates Network}
\bvolume{19834},
\bfpage{1}
(\byear{2016})
\end{barticle}
\endbibitem

\bibitem[\protect\citeauthoryear{Xue et~al.}{2009}]{Xue2009}
\begin{botherref}
\oauthor{\bsnm{Xue}, \binits{R.R.}},
\oauthor{\bsnm{Tam}, \binits{P.H.}},
\oauthor{\bsnm{Wagner}, \binits{S.J.}},
\oauthor{\bsnm{Behera}, \binits{B.}},
\oauthor{\bsnm{Fan}, \binits{Y.Z.}},
\oauthor{\bsnm{Wei}, \binits{D.M.}}:
Very high energy gamma-ray afterglow emission of nearby gamma-ray bursts
\textbf{703},
60
(2009).
doi:\doiurl{10.1088/0004-637X/703/1/60}
\end{botherref}
\endbibitem

\bibitem[\protect\citeauthoryear{{Yamazaki} et~al.}{2004}]{Yamazaki2004}
\begin{barticle}
\bauthor{\bsnm{{Yamazaki}}, \binits{R.}},
\bauthor{\bsnm{{Ioka}}, \binits{K.}},
\bauthor{\bsnm{{Nakamura}}, \binits{T.}}:
\bjtitle{\apjl}
\bvolume{607}(\bissue{2}),
\bfpage{103}
(\byear{2004}).
\arxivurl{astro-ph/0401142}.
doi:\doiurl{10.1086/421872}
\end{barticle}
\endbibitem

\bibitem[\protect\citeauthoryear{Yamazaki et~al.}{2003}]{Yamazaki2003_980425}
\begin{barticle}
\bauthor{\bsnm{Yamazaki}, \binits{R.}},
\bauthor{\bsnm{Yonetoku}, \binits{D.}},
\bauthor{\bsnm{Nakamura}, \binits{T.}}:
\bjtitle{Astrophys. J.}
\bvolume{594}(\bissue{2}),
\bfpage{79}
(\byear{2003}).
\arxivurl{0306615}.
doi:\doiurl{10.1086/378736}
\end{barticle}
\endbibitem

\bibitem[\protect\citeauthoryear{Yang et~al.}{2015}]{Yang2015_060614}
\begin{barticle}
\bauthor{\bsnm{Yang}, \binits{B.}},
\bauthor{\bsnm{Jin}, \binits{Z.-P.}},
\bauthor{\bsnm{Li}, \binits{X.}},
\bauthor{\bsnm{Covino}, \binits{S.}},
\bauthor{\bsnm{Zheng}, \binits{X.-Z.}},
\bauthor{\bsnm{Hotokezaka}, \binits{K.}},
\bauthor{\bsnm{Fan}, \binits{Y.-Z.}},
\bauthor{\bsnm{Piran}, \binits{T.}},
\bauthor{\bsnm{Wei}, \binits{D.-M.}}:
\bjtitle{Nature Communications}
\bvolume{6},
\bfpage{7323}
(\byear{2015}).
doi:\doiurl{10.1038/ncomms8323}
\end{barticle}
\endbibitem

\bibitem[\protect\citeauthoryear{Zeh et~al.}{2005}]{Zeh2005}
\begin{barticle}
\bauthor{\bsnm{Zeh}, \binits{A.}},
\bauthor{\bsnm{Kann}, \binits{D.A.}},
\bauthor{\bsnm{Klose}, \binits{S.}},
\bauthor{\bsnm{Hartmann}, \binits{D.H.}}:
\bjtitle{Nuovo Cimento C Geophysics Space Physics C}
\bvolume{28},
\bfpage{617}
(\byear{2005}).
doi:\doiurl{10.1393/ncc/i2005-10114-5}
\end{barticle}
\endbibitem

\bibitem[\protect\citeauthoryear{{Zhang} et~al.}{2004}]{Zhang2004}
\begin{barticle}
\bauthor{\bsnm{{Zhang}}, \binits{B.}},
\bauthor{\bsnm{{Dai}}, \binits{X.}},
\bauthor{\bsnm{{Lloyd-Ronning}}, \binits{N.M.}},
\bauthor{\bsnm{{M{\'e}sz{\'a}ros}}, \binits{P.}}:
\bjtitle{\apjl}
\bvolume{601}(\bissue{2}),
\bfpage{119}
(\byear{2004}).
\arxivurl{astro-ph/0311190}.
doi:\doiurl{10.1086/382132}
\end{barticle}
\endbibitem

\bibitem[\protect\citeauthoryear{Zhang et~al.}{2006}]{Zhang2006_060614}
\begin{barticle}
\bauthor{\bsnm{Zhang}, \binits{B.}},
\bauthor{\bsnm{Zhang}, \binits{B.-B.}},
\bauthor{\bsnm{Liang}, \binits{E.-W.}},
\bauthor{\bsnm{Gehrels}, \binits{N.}},
\bauthor{\bsnm{Burrows}, \binits{D.N.}},
\bauthor{\bsnm{Meszaros}, \binits{P.}}:
\bjtitle{Astrophys. J.}
\bvolume{655}(\bissue{1}),
\bfpage{25}
(\byear{2006}).
\arxivurl{0612238}.
doi:\doiurl{10.1086/511781}
\end{barticle}
\endbibitem

\bibitem[\protect\citeauthoryear{{Zhang}
  et~al.}{2020a}]{Zhang2020_amati_gflare}
\begin{botherref}
\oauthor{\bsnm{{Zhang}}, \binits{H.-M.}},
\oauthor{\bsnm{{Liu}}, \binits{R.-Y.}},
\oauthor{\bsnm{{Zhong}}, \binits{S.-Q.}},
\oauthor{\bsnm{{Wang}}, \binits{X.-Y.}}:
arXiv e-prints,
2008
(2020a).
\arxivurl{2008.05097}
\end{botherref}
\endbibitem

\bibitem[\protect\citeauthoryear{{Zhang} et~al.}{2020b}]{Zhang2020}
\begin{barticle}
\bauthor{\bsnm{{Zhang}}, \binits{S.-N.}},
\bauthor{\bsnm{{Li}}, \binits{T.}},
\bauthor{\bsnm{{Lu}}, \binits{F.}},
\bauthor{\bsnm{{Song}}, \binits{L.}},
\bauthor{\bsnm{{Xu}}, \binits{Y.}},
\bauthor{\bsnm{{Liu}}, \binits{C.}},
\bauthor{\bsnm{{Chen}}, \binits{Y.}},
\bauthor{\bsnm{{Cao}}, \binits{X.}},
\bauthor{\bsnm{{Bu}}, \binits{Q.}},
\bauthor{\bsnm{{Chang}}, \binits{Z.}},
\bauthor{\bsnm{{Chen}}, \binits{G.}},
\bauthor{\bsnm{{Chen}}, \binits{L.}},
\bauthor{\bsnm{{Chen}}, \binits{T.}},
\bauthor{\bsnm{{Chen}}, \binits{Y.}},
\bauthor{\bsnm{{Chen}}, \binits{Y.}},
\bauthor{\bsnm{{Cui}}, \binits{W.}},
\bauthor{\bsnm{{Cui}}, \binits{W.}},
\bauthor{\bsnm{{Deng}}, \binits{J.}},
\bauthor{\bsnm{{Dong}}, \binits{Y.}},
\bauthor{\bsnm{{Du}}, \binits{Y.}},
\bauthor{\bsnm{{Fu}}, \binits{M.}},
\bauthor{\bsnm{{Gao}}, \binits{G.}},
\bauthor{\bsnm{{Gao}}, \binits{H.}},
\bauthor{\bsnm{{Gao}}, \binits{M.}},
\bauthor{\bsnm{{Ge}}, \binits{M.}},
\bauthor{\bsnm{{Gu}}, \binits{Y.}},
\bauthor{\bsnm{{Guan}}, \binits{J.}},
\bauthor{\bsnm{{Gungor}}, \binits{C.}},
\bauthor{\bsnm{{Guo}}, \binits{C.}},
\bauthor{\bsnm{{Han}}, \binits{D.}},
\bauthor{\bsnm{{Hu}}, \binits{W.}},
\bauthor{\bsnm{{Huang}}, \binits{Y.}},
\bauthor{\bsnm{{Huo}}, \binits{J.}},
\bauthor{\bsnm{{Jia}}, \binits{S.}},
\bauthor{\bsnm{{Jiang}}, \binits{L.}},
\bauthor{\bsnm{{Jiang}}, \binits{W.}},
\bauthor{\bsnm{{Jin}}, \binits{J.}},
\bauthor{\bsnm{{Jin}}, \binits{Y.}},
\bauthor{\bsnm{{Li}}, \binits{B.}},
\bauthor{\bsnm{{Li}}, \binits{C.}},
\bauthor{\bsnm{{Li}}, \binits{G.}},
\bauthor{\bsnm{{Li}}, \binits{M.}},
\bauthor{\bsnm{{Li}}, \binits{W.}},
\bauthor{\bsnm{{Li}}, \binits{X.}},
\bauthor{\bsnm{{Li}}, \binits{X.}},
\bauthor{\bsnm{{Li}}, \binits{X.}},
\bauthor{\bsnm{{Li}}, \binits{Y.}},
\bauthor{\bsnm{{Li}}, \binits{Z.}},
\bauthor{\bsnm{{Li}}, \binits{Z.}},
\bauthor{\bsnm{{Liang}}, \binits{X.}},
\bauthor{\bsnm{{Liao}}, \binits{J.}},
\bauthor{\bsnm{{Liu}}, \binits{G.}},
\bauthor{\bsnm{{Liu}}, \binits{H.}},
\bauthor{\bsnm{{Liu}}, \binits{S.}},
\bauthor{\bsnm{{Liu}}, \binits{X.}},
\bauthor{\bsnm{{Liu}}, \binits{Y.}},
\bauthor{\bsnm{{Liu}}, \binits{Y.}},
\bauthor{\bsnm{{Lu}}, \binits{B.}},
\bauthor{\bsnm{{Lu}}, \binits{X.}},
\bauthor{\bsnm{{Luo}}, \binits{T.}},
\bauthor{\bsnm{{Ma}}, \binits{X.}},
\bauthor{\bsnm{{Meng}}, \binits{B.}},
\bauthor{\bsnm{{Nang}}, \binits{Y.}},
\bauthor{\bsnm{{Nie}}, \binits{J.}},
\bauthor{\bsnm{{Ou}}, \binits{G.}},
\bauthor{\bsnm{{Qu}}, \binits{J.}},
\bauthor{\bsnm{{Sai}}, \binits{N.}},
\bauthor{\bsnm{{Shang}}, \binits{R.}},
\bauthor{\bsnm{{Shen}}, \binits{G.}},
\bauthor{\bsnm{{Sun}}, \binits{L.}},
\bauthor{\bsnm{{Tan}}, \binits{Y.}},
\bauthor{\bsnm{{Tao}}, \binits{L.}},
\bauthor{\bsnm{{Tuo}}, \binits{Y.}},
\bauthor{\bsnm{{Wang}}, \binits{C.}},
\bauthor{\bsnm{{Wang}}, \binits{C.}},
\bauthor{\bsnm{{Wang}}, \binits{G.}},
\bauthor{\bsnm{{Wang}}, \binits{H.}},
\bauthor{\bsnm{{Wang}}, \binits{J.}},
\bauthor{\bsnm{{Wang}}, \binits{W.}},
\bauthor{\bsnm{{Wang}}, \binits{Y.}},
\bauthor{\bsnm{{Wen}}, \binits{X.}},
\bauthor{\bsnm{{Wu}}, \binits{B.}},
\bauthor{\bsnm{{Wu}}, \binits{B.}},
\bauthor{\bsnm{{Wu}}, \binits{M.}},
\bauthor{\bsnm{{Xiao}}, \binits{G.}},
\bauthor{\bsnm{{Xiong}}, \binits{S.}},
\bauthor{\bsnm{{Yan}}, \binits{L.}},
\bauthor{\bsnm{{Yang}}, \binits{J.}},
\bauthor{\bsnm{{Yang}}, \binits{S.}},
\bauthor{\bsnm{{Yang}}, \binits{Y.}},
\bauthor{\bsnm{{Yi}}, \binits{Q.}},
\bauthor{\bsnm{{Yuan}}, \binits{B.}},
\bauthor{\bsnm{{Zhang}}, \binits{A.}},
\bauthor{\bsnm{{Zhang}}, \binits{C.}},
\bauthor{\bsnm{{Zhang}}, \binits{C.}},
\bauthor{\bsnm{{Zhang}}, \binits{F.}},
\bauthor{\bsnm{{Zhang}}, \binits{H.}},
\bauthor{\bsnm{{Zhang}}, \binits{J.}},
\bauthor{\bsnm{{Zhang}}, \binits{Q.}},
\bauthor{\bsnm{{Zhang}}, \binits{S.}},
\bauthor{\bsnm{{Zhang}}, \binits{S.}},
\bauthor{\bsnm{{Zhang}}, \binits{T.}},
\bauthor{\bsnm{{Zhang}}, \binits{W.}},
\bauthor{\bsnm{{Zhang}}, \binits{W.}},
\bauthor{\bsnm{{Zhang}}, \binits{W.}},
\bauthor{\bsnm{{Zhang}}, \binits{Y.}},
\bauthor{\bsnm{{Zhang}}, \binits{Y.}},
\bauthor{\bsnm{{Zhang}}, \binits{Y.}},
\bauthor{\bsnm{{Zhang}}, \binits{Y.}},
\bauthor{\bsnm{{Zhang}}, \binits{Z.}},
\bauthor{\bsnm{{Zhang}}, \binits{Z.}},
\bauthor{\bsnm{{Zhang}}, \binits{Z.}},
\bauthor{\bsnm{{Zhao}}, \binits{H.}},
\bauthor{\bsnm{{Zhao}}, \binits{X.}},
\bauthor{\bsnm{{Zheng}}, \binits{S.}},
\bauthor{\bsnm{{Zhou}}, \binits{J.}},
\bauthor{\bsnm{{Zhu}}, \binits{Y.}},
\bauthor{\bsnm{{Zhu}}, \binits{Y.}},
\bauthor{\bsnm{{Zhuang}}, \binits{R.}},
\bauthor{\bsnm{{Insight-HXMT team}}}:
\bjtitle{Science China Physics, Mechanics, and Astronomy}
\bvolume{63}(\bissue{4}),
\bfpage{249502}
(\byear{2020}b).
\arxivurl{1910.09613}.
doi:\doiurl{10.1007/s11433-019-1432-6}
\end{barticle}
\endbibitem

\bibitem[\protect\citeauthoryear{{Zhao} et~al.}{2012}]{Zhao2012}
\begin{barticle}
\bauthor{\bsnm{{Zhao}}, \binits{D.}},
\bauthor{\bsnm{{Cordier}}, \binits{B.}},
\bauthor{\bsnm{{Sizun}}, \binits{P.}},
\bauthor{\bsnm{{Wu}}, \binits{B.}},
\bauthor{\bsnm{{Dong}}, \binits{Y.}},
\bauthor{\bsnm{{Schanne}}, \binits{S.}},
\bauthor{\bsnm{{Song}}, \binits{L.}},
\bauthor{\bsnm{{Liu}}, \binits{J.}}:
\bjtitle{Experimental Astronomy}
\bvolume{34}(\bissue{3}),
\bfpage{705}
(\byear{2012}).
\arxivurl{1208.2493}.
doi:\doiurl{10.1007/s10686-012-9313-2}
\end{barticle}
\endbibitem

\end{thebibliography}

\begin{appendix} 

\section{Long GRBs}
\label{sub:lgrbs}

This section discusses the properties of the \Nlong\ long GRBs in our sample, starting with some comments about peculiar GRBs.

\begin{itemize}
\item GRB~150518A has only been detected by MAXI/GSC located on the ISS \citep{Sakamoto2015_150518A} and partially by Konus-Wind \citep{Golenetskii2015_150518A}. Its energy has not been measured by Konus-Wind because of a data gap during the main emitting episode. The light curve has been partially recorded by MAXI/GSC, but Konus-Wind observations seem to indicate that this burst was already emitting $\sim 250\sec$ before MAXI/GSC observations, implying that the light curve and the GRB properties recovered by MAXI/GSC are only a fraction of an ultra-long GRB. For this reason, the simulations that have been made on this burst are incomplete and only account for the part visible by MAXI/GSC.
\item GRB~190829A has two distinct peaks separated by $\sim\ 45\ \sec$, with completely different spectra. The case is similar for GRB~180728A, with a precursor and then the main emission of the burst. Thus, the emission properties of those two GRBs have been divided into two parts.
\end{itemize}
Tables \ref{tab:detection_long}, Table \ref{tab:spectrum_long} and Table \ref{tab:intrinsic_long} summarize the features detected for the long GRBs.

The population observed is diverse in terms of observed characteristics:

\begin{itemize}
    \item Concerning the fluence, the minimum fluence observed is $1.0\times 10^{-8}\ \fluence$ for GRB~150818A while the maximum one is $1.63\times 10^{-4}\ \fluence$ for GRB~030329, with the median equal to $3.55\times 10^{-6}\ \fluence$.
    \item For the 1s peak flux, the minimum is $0.1^{+0.06}_{-0.03}\ \phflux$ for GRB~100316D and the maximum is $451^{+25}_{-25}\ \phflux$ for   GRB~030329, with the median equal to $2.65 \pm 0.63\ \phflux$ for GRB~060505.
    \item For the peak energy of the fluence spectrum, $13$ out of the $30$ long GRBs do not have any peak energy information. 
    This is mainly because they have been observed with instruments such as \textit{Swift}/BAT whose energy band is often too narrow to measure properly the spectrum peak energy. GRB~020903, GRB~120422A and GRB~040701 only have a lower estimation. For the remaining $14$ long GRBs with peak energy, the minimum $E_\mathrm{peak}$ for long GRBs is equal to $4.9 \pm 0.4\ \mathrm{keV}$ for GRB~060218 and the maximum equal to $302^{+214}_{-85}\ \mathrm{keV}$ for the first part of GRB~060614, with a median of $68\ \mathrm{keV}$ from GRB~030329.
    \item For \Tqvd-values, the minimum value is $4.0\sec$ for GRB~060505, the maximum $2100\pm 100\sec$ for GRB~060218 and the median is $34\sec$.
\end{itemize}
There are also some differences in the intrinsic properties:
\begin{itemize}
    \item The closest long GRB that has been detected is GRB~980425, at a redshift of $0.0085$ ($\approx 38\ \mathrm{Mpc}$). The furthest one in our local GRB sample is at a redshift of $\mathrm{z} = 0.297$ ($\approx 1590\ \mathrm{Mpc}$), GRB~111225A. 
    the median redshift of the sample is $\mathrm{z} = 0.135$ ($\approx 660\ \mathrm{Mpc}$).
    \item The minimum isotropic energy detected in the long GRB sample is $E_\mathrm{iso} = 5\times 10^{47}\ \mathrm{erg}$ for GRB~111005A \citep{Tanga2018_111005A}, while the maximum isotropic energy is equal to $1.86\times 10^{52}\ \mathrm{erg}$ for GRB~030329. The median value for the long GRBs is $E_\mathrm{iso} = 1.65\times 10^{50}\ \mathrm{erg}$, significantly lower than the median \eiso\ of classical long GRBs observed at larger distances.

\end{itemize}

Based on the observations on Fig. \ref{fig:T90_Pflux}, the 1s peak flux in the energy range 15--150 keV for the 2 closest long GRBs (GRB~980425 and GRB~111005A) is of the order of $1\ \phflux$, which is slightly lower than for classical GRBs (the 1s peak flux median in the \textit{Swift}/BAT catalog is $\sim 1.4\ \phflux$, see \citealt{BATcat2016}), while these low-energy GRBs are one order of magnitude closer. This suggests that these two GRBs are the tip of the iceberg of a much larger population \citep{Soderberg2004_031203}. Two long GRBs of the sample,  GRB~040701 and GRB~150518A, have no peak flux mentioned in the literature and therefore do not appear in Fig. \ref{fig:T90_Pflux}.

\begin{table}

\centering
\caption{Detected features for Long GRBs}
\label{tab:detection_long}
\begin{tabular}{lllllll}
\toprule
    Name & Instrument(s)\ \tablenotemark{a} &                              Class &         Afterglow\ \tablenotemark{b} &                                  SN/KN &                           Host &           T\textsubscript {90} (s)\ \tablenotemark{c} \\
\midrule
  980425 &                             C B  &                            llLGRB &       X(O)R \textsuperscript {{(1)}} &      SN1998bw \textsuperscript {{(2)}} &       \textsuperscript {{(3)}} &         $34.9^{+3.8}_{-3.8}$ \textsuperscript {{(4)}} \\
  020903 &                               H  &      XRF  \textsuperscript {{(5)}} &        OR \textsuperscript {{(6,7)}} &            SN \textsuperscript {{(6)}} &       \textsuperscript {{(6)}} &         $10.0^{+0.7}_{-0.7}$ \textsuperscript {{(5)}} \\
  030329 &                             K H  &                               LGRB &    XOR \textsuperscript {{(8,9,10)}} &  SN2003dh \textsuperscript {{(11,12)}} &      \textsuperscript {{(13)}} &         $33.1^{+0.5}_{-0.5}$ \textsuperscript {{(5)}} \\
  031203 &                               I  &    LGRB  \textsuperscript {{(14)}} &  XOR \textsuperscript {{(15,16,17)}} &     SN2003lw \textsuperscript {{(18)}} &      \textsuperscript {{(19)}} &                      $30.0$ \textsuperscript {{(20)}} \\
  040701 &                               H  &     XRF  \textsuperscript {{(21)}} &          X \textsuperscript {{(21)}} &           NO \textsuperscript {{(21)}} &      \textsuperscript {{(22)}} &        $11.7^{+5.7}_{-5.7}$ \textsuperscript {{(23)}} \\
 050219A &                               S  &                               LGRB &    X \textsuperscript {{(24,25,26)}} &                                    ... &      \textsuperscript {{(27)}} &                                  $23.8^{+2.3}_{-2.3}$ \\
  050826 &                               S  &                               LGRB &      XO \textsuperscript {{(28,29)}} &                                    ... &      \textsuperscript {{(29)}} &        $35.0^{+8.0}_{-8.0}$ \textsuperscript {{(30)}} \\
 051109B &                               S  &                               LGRB &       X \textsuperscript {{(31,32)}} &                                    ... &      \textsuperscript {{(33)}} &                                  $15.7^{+4.1}_{-4.1}$ \\
  060218 &                               S  &  ulGRB  \textsuperscript {{(34)}} &     XOR \textsuperscript {{(35,36)}} &     SN2006aj \textsuperscript {{(37)}} &      \textsuperscript {{(34)}} &  $2100^{+100}_{-100}$ \textsuperscript {{(34)}} \\
  060505 &                           S SUZ  &                               LGRB &      XO \textsuperscript {{(38,39)}} &           NO \textsuperscript {{(40)}} &      \textsuperscript {{(41)}} &                                                 $4$ \\
  060614 &                             S K  &                               LGRB &         XO \textsuperscript {{(42)}} &     NO \textsuperscript {{(43,44,45)}} &   \textsuperscript {{(44,45)}} &                                 $109.1^{+3.4}_{-3.4}$ \\
  080517 &                               S  &                               LGRB &          X \textsuperscript {{(46)}} &                                    ... &      \textsuperscript {{(46)}} &                                $64.5^{+22.3}_{-22.3}$ \\
 100316D &                               S  &                             ulGRB &      XO \textsuperscript {{(47,48)}} &     SN2010bh \textsuperscript {{(49)}} &      \textsuperscript {{(50)}} &  $1300^{+500}_{-500}$ \textsuperscript {{(51)}} \\
 111005A &                               S  &                            llLGRB &          R \textsuperscript {{(52)}} &        NO \textsuperscript {{(53,54)}} &      \textsuperscript {{(55)}} &        $27.0^{+8.0}_{-8.0}$ \textsuperscript {{(56)}} \\
 111225A &                               S  &                               LGRB &         XO \textsuperscript {{(57)}} &                                      - &      \textsuperscript {{(58)}} &                               $105.7^{+26.2}_{-26.2}$ \\
 120422A &                               S  &                               LGRB &   XO(R) \textsuperscript {{(59,60)}} &     SN2012bz \textsuperscript {{(61)}} &      \textsuperscript {{(62)}} &                                  $60.4^{+5.7}_{-5.7}$ \\
 130702A &                             F K  &                               LGRB &  XOR \textsuperscript {{(63,64,65)}} &     SN2013dx \textsuperscript {{(66)}} &      \textsuperscript {{(67)}} &                                  $58.9^{+6.2}_{-6.2}$ \\
 150518A &                             K M  &  ulGRB  \textsuperscript {{(68)}} &  XOR \textsuperscript {{(69,70,71)}} &           SN \textsuperscript {{(72)}} &      \textsuperscript {{(70)}} &                    $1000.0$ \textsuperscript {{(68)}} \\
 150818A &                             S K  &    LGRB  \textsuperscript {{(73)}} &      XO \textsuperscript {{(74,75)}} &           SN \textsuperscript {{(76)}} &      \textsuperscript {{(77)}} &                               $143.3^{+21.8}_{-21.8}$ \\
 161219B &                             S K  &                               LGRB &  XOR \textsuperscript {{(78,79,80)}} &    SN2016jca \textsuperscript {{(81)}} &      \textsuperscript {{(82)}} &                                   $6.9^{+0.8}_{-0.8}$ \\
 171205A &                             S K  &                               LGRB &  XOR \textsuperscript {{(83,84,85)}} &    SN2017iuk \textsuperscript {{(86)}} &      \textsuperscript {{(87)}} &                               $190.5^{+33.9}_{-33.9}$ \\
 180728A &                        S F K AS  &    LGRB  \textsuperscript {{(88)}} &      XO \textsuperscript {{(89,90)}} &    SN2018flp \textsuperscript {{(91)}} &      \textsuperscript {{(92)}} &                      $24.0$ \textsuperscript {{(88)}} \\
 190829A &                         S F K A  &                               LGRB &    HXOR \textsuperscript {{(93,94)}} &     SN2010bh \textsuperscript {{(95)}} &      \textsuperscript {{(93)}} &        $59.4^{+0.6}_{-0.6}$ \textsuperscript {{(96)}} \\
 191019A &                               S  &                               LGRB &   XO \textsuperscript {{(97,98,99)}} &                                    ... &   \textsuperscript {{(99,98)}} &                                  $64.3^{+4.5}_{-4.5}$ \\
\bottomrule
\end{tabular}

\tablenotetext{a}{A for AGILE/GRID and/or AGILE/SA, AS for Astrosat/CZTI, B for BeppoSax/WFC, C for CGRO/BATSE, F for \textit{Fermi}/GBM, H for HETE-2/FREGATE and/or WXM, I for INTEGRAL/IBIS and/or INTEGRAL/SPI, K for Wind/KONUS, M for MAXI/GSC, S for \textit{Swift}/BAT, SUZ for \textit{Suzaku}/WAM}
\tablenotetext{b}{The letters X, O and R stands respectively for X-ray, Optical and Radio afterglow. H represents the High-Energy detection by H.E.S.S.}
\tablenotetext{c}{Unless quoted differently, T\textsubscript{90}-values are measured by the \textit{Swift}/BAT instrument in the $15-350\ \mathrm{keV}$ band}

\tablerefs{
(1) \cite{Pian1998_980425}; (2) \cite{Tinney1998_980425}; (3) \cite{Galama1998_980425}; (4) \cite{Soffitta1998_980425}; (5) \cite{Sakamoto2005}; (6) \cite{Soderberg2002_020903}; (7) \cite{Soderberg2004_020903}; (8) \cite{Marshall2003_030329}; (9) \cite{Peterson2003_030329}; (10) \cite{vanderHorst2006_030329}; (11) \cite{Stanek2003_030329}; (12) \cite{Hjorth2003_030329}; (13) \cite{Greiner2003_030329}; (14) \cite{Watson2004_031203}; (15) \cite{SantosLleo2003_031203}; (16) \cite{Reichart2003_031203}; (17) \cite{Frail2003_031203}; (18) \cite{Bersier2004_031203}; (19) \cite{Prochaska2003_031203}; (20) \cite{Mereghetti2003_031203}; (21) \cite{Soderberg2005_040701}; (22) \cite{Kelson2004_040701}; (23) \cite{Pelangeon2008_040701}; (24) \cite{Romano2005_050219A}; (25) \cite{Berger2005_050219A}; (26) \cite{UgartePostigo2005_050219A}; (27) \cite{Rossi2014_050219A}; (28) \cite{Mangano2005_050826}; (29) \cite{Halpern2006_050826}; (30) \cite{Markwardt2005_050826}; (31) \cite{Campana2005_051109B}; (32) \cite{DePasquale2005_051109B}; (33) \cite{Perley2006_051109B}; (34) \cite{Campana2006_060218}; (35) \cite{Cusumano2006_060218}; (36) \cite{Soderberg2006_060218}; (37) \cite{Masetti2006_060218}; (38) \cite{Conciatore2006_060505}; (39) \cite{Ofek2006_060505}; (40) \cite{Xu2009_060505}; (41) \cite{Thoene2006_060505}; (42) \cite{Parsons2006_060614}; (43) \cite{Fynbo2006_060614}; (44) \cite{GalYam2006_060614}; (45) \cite{DellaValle2006_060614}; (46) \cite{Parsons2008_080517}; (47) \cite{Stamatikos2010_100316D}; (48) \cite{Wieringa2010_100316D}; (49) \cite{Wiersema2010_100316D}; (50) \cite{Vergani2010_100316D}; (51) \cite{Starling2011_100316D}; (52) \cite{Xu2011_111005A}; (53) \cite{Levan2011_111005A}; (54) \cite{Michalowski2018_111005A}; (55) \cite{Malesani2011_111005A}; (56) \cite{Tanga2018_111005A}; (57) \cite{Siegel2011_111225A}; (58) \cite{Thoene2014_111225A}; (59) \cite{Beardmore2012_120422A}; (60) \cite{Kuin2012_120422A}; (61) \cite{Malesani2012_120422A}; (62) \cite{Tanvir2012_120422A}; (63) \cite{Davanzo2013_130702A}; (64) \cite{Guidorzi2013_130702A}; (65) \cite{vanderHorst2013_130702A}; (66) \cite{Schulze2013_130702A}; (67) \cite{Singer2013_130702A}; (68) \cite{Sakamoto2015_150518A}; (69) \cite{Sbarufatti2015_150518A}; (70) \cite{Xu2015_150518A}; (71) \cite{Kamble2015_150518A}; (72) \cite{Pozanenko2015_150518A}; (73) \cite{Palmer2015_150818A}; (74) \cite{Delia2015_150818A}; (75) \cite{Marshall2015_150818A}; (76) \cite{Mazaeva2015_150818A}; (77) \cite{SanchezRamirez2015_150818A}; (78) \cite{Beardmore2016_161219B}; (79) \cite{DAi2016_161219B}; (80) \cite{Alexander2016_161219B}; (81) \cite{UgartePostigo2016_161219B}; (82) \cite{Kruehler2016_161219B}; (83) \cite{Kennea2017_171205A}; (84) \cite{Osborne2017_171205A}; (85) \cite{Chandra2017_171205A}; (86) \cite{UgartePostigo2017_171205A}; (87) \cite{Izzo2017_171205A}; (88) \cite{Wang2019}; (89) \cite{Perri2018_180728A}; (90) \cite{Laporte2018_180728A}; (91) \cite{Izzo2018_180728A}; (92) \cite{Rossi2018_180728A}; (93) \cite{Dichiara2019_190829A}; (94) \cite{Rhodes2020_190829A}; (95) \cite{Perley2019_190829A}; (96) \cite{GBMcat2016}; (97) \cite{Sbarufatti2019_191019A}; (98) \cite{Reva2019_191019A}; (99) \cite{Perley2019_191019A}; 
}

\end{table}

\begin{table}

\centering
\caption{Spectral properties of Long GRBs}
\label{tab:spectrum_long}
\resizebox{\textwidth}{!}{
\begin{tabular}{lllllll}
\toprule
       Name &                                            Fluence &                                       1-s P.-flux & Energy Band\ \tablenotemark{a} & P.-flux (norm)\ \tablenotemark{b} &                            $\mathrm{E_{peak}}$ &                  Spectral Model\ \tablenotemark{c} \\
 & ($10^{-6}$ erg cm$^{-2}$) & (ph cm$^{-2}$ s$^{-1}$) & (keV - Inst.) & (ph cm$^{-2}$ s$^{-1}$) & (keV) & \\ 
\midrule
     980425 &    $4.0^{+0.74}_{-0.74}$ \textsuperscript {{(1)}} &      $1.0^{+0.05}_{-0.05}$ \textsuperscript {{(1)}} &                 20 - 2000 - C  &               $1.1^{+0.1}_{-0.1}$ &      $55^{+21}_{-21}$ \textsuperscript {{(1)}} &    BAND: 55, -1.00, -2.10 \textsuperscript {{(1)}} \\
     020903 &    $0.10^{+0.06}_{-0.06}$ \textsuperscript {{(2)}} &      $2.8^{+0.7}_{-0.7}$ \textsuperscript {{(2)}} &                   2 - 400 - H  &               $0.1^{+0.0}_{-0.0}$ &                $ < 5$ \textsuperscript {{(2)}} &                 PL: -2.60 \textsuperscript {{(2)}} \\
     030329 &                  $186$ \textsuperscript {{(3)}} &  $451^{+25}_{-25}$ \textsuperscript {{(2)}} &                   2 - 400 - H  &             $124.5^{+6.9}_{-6.9}$ &        $82^{+3}_{-3}$ \textsuperscript {{(3)}} &    BAND: 70, -1.32, -2.44 \textsuperscript {{(3)}} \\
     031203 &    $2.0^{+0.4}_{-0.4}$ \textsuperscript {{(4)}} &                    $1.2$ \textsuperscript {{(5)}} &                  20 - 200 - I  &                             $1.4$ &              $ > 190$ \textsuperscript {{(4)}} &                 PL: -1.63 \textsuperscript {{(4)}} \\
     040701 &    $0.54^{+0.05}_{-0.05}$ \textsuperscript {{(6)}} &                                               ... &                    2 - 30 - H  &                               ... &                $ < 3$ \textsuperscript {{(6)}} &                 PL: -2.30 \textsuperscript {{(6)}} \\
    050219A &    $5.2^{+0.4}_{-0.4}$ \textsuperscript {{(7)}} &                               $3.5^{+0.3}_{-0.3}$ &                  15 - 350 - S  &               $3.3^{+0.3}_{-0.3}$ &        $90^{+9}_{-9}$ \textsuperscript {{(7)}} &            CPL: 90, -0.75 \textsuperscript {{(7)}} \\
     050826 &                             $0.41^{+0.07}_{-0.07}$ &                               $0.4^{+0.1}_{-0.1}$ &                  15 - 150 - S  &               $0.4^{+0.1}_{-0.1}$ &                                            ... &                                          PL: -1.23 \\
    051109B &                             $0.27^{+0.04}_{-0.04}$ &                               $0.6^{+0.1}_{-0.1}$ &                  15 - 150 - S  &               $0.6^{+0.1}_{-0.1}$ &                                            ... &                                          PL: -1.98 \\
     060218 &                             $1.57^{+0.15}_{-0.15}$ &                               $0.2^{+0.1}_{-0.1}$ &                  15 - 150 - S  &               $0.2^{+0.1}_{-0.1}$ &         $5^{+0}_{-0}$ \textsuperscript {{(8)}} &                                          PL: -2.26 \\
     060505 &    $2.30^{+1.08}_{-1.08}$ \textsuperscript {{(9)}} &                               $2.6^{+0.6}_{-0.6}$ &                 15 - 2000 - S  &               $1.2^{+0.3}_{-0.3}$ &   $397^{+485}_{-485}$ \textsuperscript {{(9)}} &   BAND: 397, -1.19, -2.39 \textsuperscript {{(9)}} \\
  060614-p1 &   $8.19^{+0.56}_{-0.56}$ \textsuperscript {{(10)}} &                              $11.5^{+0.7}_{-0.7}$ &                20 - 2000 - K*  &              $11.5^{+0.7}_{-0.7}$ &  $302^{+214}_{-214}$ \textsuperscript {{(10)}} &          CPL: 302, -1.57 \textsuperscript {{(10)}} \\
  060614-p2 &  $32.7^{+1.7}_{-1.7}$ \textsuperscript {{(10)}} &                                               ... &                 20 - 2000 - K  &                               ... &                                            ... &                PL: -2.13 \textsuperscript {{(10)}} \\
     080517 &                             $0.56^{+0.12}_{-0.12}$ &                               $0.6^{+0.2}_{-0.2}$ &                  15 - 150 - S  &               $0.6^{+0.2}_{-0.2}$ &              $ > 55$ \textsuperscript {{(11)}} &                                          PL: -1.54 \\
    100316D &   $5.10^{+0.39}_{-0.39}$ \textsuperscript {{(12)}} &                                             $0.1$ &                  15 - 150 - S  &                             $0.1$ &       $33^{+7}_{-7}$ \textsuperscript {{(12)}} &           CPL: 33, -1.33 \textsuperscript {{(12)}} \\
    111005A &                   $0.48$ \textsuperscript {{(13)}} &                               $1.1^{+0.3}_{-0.3}$ &                  15 - 150 - S  &               $1.1^{+0.3}_{-0.3}$ &                                            ... &                PL: -2.40 \textsuperscript {{(13)}} \\
    111225A &                             $1.30^{+0.12}_{-0.12}$ &                               $0.7^{+0.1}_{-0.1}$ &                  15 - 150 - S  &               $0.7^{+0.1}_{-0.1}$ &                                            ... &                                          PL: -1.70 \\
    120422A &                                             $0.23$ &                                             $0.6$ &                  15 - 150 - S  &                             $0.6$ &              $ < 56$ \textsuperscript {{(14)}} &                PL: -1.91 \textsuperscript {{(14)}} \\
    130702A &                             $5.72^{+0.12}_{-0.12}$ &                               $7.0^{+0.9}_{-0.9}$ &                 10 - 1000 - F  &               $4.4^{+0.5}_{-0.5}$ &                                            ... &                                          PL: -2.44 \\
    150518A &   $0.01^{+0.003}_{-0.003}$ \textsuperscript {{(15)}} &                                               ... &                    2 - 20 - M  &                               ... &                                            ... &                PL: -1.30 \textsuperscript {{(15)}} \\
    150818A &   $5.3^{+0.7}_{-0.7}$ \textsuperscript {{(16)}} &     $2.4^{+0.3}_{-0.3}$ \textsuperscript {{(16)}} &                20 - 1000 - K*  &               $2.4^{+0.3}_{-0.3}$ &    $100^{+29}_{-29}$ \textsuperscript {{(16)}} &          CPL: 100, -1.40 \textsuperscript {{(16)}} \\
    161219B &   $3.1^{+0.8}_{-0.8}$ \textsuperscript {{(17)}} &                               $5.3^{+0.4}_{-0.4}$ &                20 - 1000 - K*  &               $5.3^{+0.4}_{-0.4}$ &     $91^{+21}_{-21}$ \textsuperscript {{(17)}} &           CPL: 91, -1.59 \textsuperscript {{(17)}} \\
    171205A &   $6.0^{+1.8}_{-1.8}$ \textsuperscript {{(18)}} &                               $1.0^{+0.3}_{-0.3}$ &                15 - 1500 - K*  &               $1.0^{+0.3}_{-0.3}$ &  $122^{+111}_{-111}$ \textsuperscript {{(18)}} &          CPL: 122, -0.85 \textsuperscript {{(18)}} \\
 180728A-p1 &                                             $1.59$ &                                               ... &                 10 - 1000 - F  &                               ... &                                            ... &                PL: -2.31 \textsuperscript {{(19)}} \\
 180728A-p2 &                                            $54.3$ &                             $231.0^{+1.2}_{-1.2}$ &                 10 - 1000 - F  &             $155.3^{+0.8}_{-0.8}$ &                $129$ \textsuperscript {{(19)}} &  BAND: 129, -1.55, -3.48 \textsuperscript {{(19)}} \\
 190829A-p1 &   $2.4^{+0.7}_{-0.7}$ \textsuperscript {{(20)}} &                                               ... &                 10 - 1000 - F  &                               ... &    $130^{+20}_{-20}$ \textsuperscript {{(20)}} &          CPL: 130, -1.40 \textsuperscript {{(20)}} \\
 190829A-p2 &  $13.8^{+0.4}_{-0.4}$ \textsuperscript {{(20)}} &                              $33.5^{+0.5}_{-0.5}$ &                 10 - 1000 - F  &              $18.2^{+0.3}_{-0.3}$ &       $11^{+1}_{-1}$ \textsuperscript {{(20)}} &   BAND: 11, -0.92, -2.51 \textsuperscript {{(20)}} \\
    191019A &                            $10.40^{+0.26}_{-0.26}$ &                               $5.7^{+0.4}_{-0.4}$ &                  15 - 150 - S  &               $5.7^{+0.4}_{-0.4}$ &                               $54^{+12}_{-12}$ &                                          PL: -2.26 \\
\bottomrule
\end{tabular}
}

\tablecomments{For the \textit{Swift}/BAT and \textit{Fermi}/GBM instruments, unless specified, properties are respectively from the \textit{Swift}/BAT  \citep{BATcat2016} and the \textit{Fermi}/GBM \citep{GBMcat2016} catalogs.}

\tablenotetext{a}{Energy range and Instrument (abbreviation from Table \ref{tab:detection_long}) used to obtain the fluence, peak flux and spectral model determination.}
\tablenotetext{b}{Peak Flux translated in the $15-150\ \mathrm{keV}$ band, used with \Tqvd to plot Fig. \ref{fig:T90_Pflux}. Units are in $\phflux$}
\tablenotetext{c}{Parameters for the spectral model: BAND: $E_\mathrm{peak}\ (\mathrm{keV}), \alpha, \beta$ ; CPL: $E_\mathrm{peak}\ (\mathrm{keV}), \alpha$ ; PL: $\Gamma$}
\tablenotetext{*}{Fluence and Spectral Model are from Wind/Konus, Peak-Flux is from \textit{Swift}/BAT in the $15-150\ \mathrm{keV}$ energy range.}

\tablerefs{
(1) \cite{Yamazaki2003_980425}; (2) \cite{Sakamoto2005}; (3) \cite{Vanderspek2004_030329}; (4) \cite{Sazonov2004_031203}; (5) \cite{Gotz2003_031203}; (6) \cite{Pelangeon2008_040701}; (7) \cite{Tagliaferri2005_050219A}; (8) \cite{Campana2006_060218}; (9) \cite{Krimm2009_060505}; (10) \cite{Golenetskii2006_060614}; (11) \cite{Stanway2015_080517}; (12) \cite{Starling2011_100316D}; (13) \cite{Tanga2018_111005A}; (14) \cite{Melandri2012_120422A}; (15) \cite{Sakamoto2015_150518A}; (16) \cite{Golenetskii2015_150818A}; (17) \cite{Frederiks2016_161219B}; (18) \cite{Delia2018_171205A}; (19) \cite{Wang2019}; (20) \cite{Chand2020_190829A}; 
}

\end{table}

\begin{table}

\centering
\caption{Intrinsic properties of Long GRBs}
\label{tab:intrinsic_long}
\begin{tabular}{llll}
\toprule
    Name &                              Redshift &                            $E_\mathrm{peak,i}$ & $E_\mathrm{iso}$\ \tablenotemark{a} \\
 & & (keV) &($\times 10^{50}\ \mathrm{erg}$) \\ 
\midrule
  980425 &      $0.0085$ \textsuperscript {{(1)}} &      $55^{+21}_{-21}$ \textsuperscript {{(2)}} &                                0.01 \\
  020903 &      $0.25$ \textsuperscript {{(3)}} &                $ < 6$ \textsuperscript {{(4)}} &       0.24 \textsuperscript {{(5)}} \\
  030329 &      $0.168$ \textsuperscript {{(6)}} &        $96^{+3}_{-3}$ \textsuperscript {{(7)}} &     186 \textsuperscript {{(7)}} \\
  031203 &      $0.105$ \textsuperscript {{(8)}} &              $ > 210$ \textsuperscript {{(9)}} &                                1.4 \\
  040701 &     $0.215$ \textsuperscript {{(10)}} &               $ < 4$ \textsuperscript {{(11)}} &      0.8 \textsuperscript {{(11)}} \\
 050219A &     $0.211$ \textsuperscript {{(12)}} &    $109^{+11}_{-11}$ \textsuperscript {{(13)}} &                                6.8 \\
  050826 &     $0.297$ \textsuperscript {{(14)}} &                                            ... &                                 ... \\
 051109B &     $0.080$ \textsuperscript {{(16)}} &                                            ... &                                0.17 \\
  060218 &     $0.033$ \textsuperscript {{(18)}} &        $5^{+0}_{-0}$ \textsuperscript {{(19)}} &      0.62 \textsuperscript {{(19)}} \\
  060505 &     $0.089$ \textsuperscript {{(21)}} &  $432^{+528}_{-528}$ \textsuperscript {{(22)}} &                                0.57 \\
  060614 &     $0.125$ \textsuperscript {{(26)}} &  $340^{+241}_{-241}$ \textsuperscript {{(27)}} &                               20.13 \\
  080517 &     $0.089$ \textsuperscript {{(29)}} &              $ > 60$ \textsuperscript {{(29)}} &      0.10 \textsuperscript {{(29)}} \\
 100316D &     $0.059$ \textsuperscript {{(30)}} &       $35^{+7}_{-7}$ \textsuperscript {{(30)}} &      > 0.59 \textsuperscript {{(30)}} \\
 111005A &     $0.013$ \textsuperscript {{(31)}} &                                            ... &      0.005 \textsuperscript {{(32)}} \\
 111225A &     $0.297$ \textsuperscript {{(33)}} &                                            ... &                               18.5 \\
 120422A &  $0.283$ \textsuperscript {{(34,35)}} &              $ < 72$ \textsuperscript {{(36)}} &                                2.1 \\
 130702A &     $0.145$ \textsuperscript {{(37)}} &                                            ... &                               10.0 \\
 150518A &     $0.256$ \textsuperscript {{(39)}} &                                            ... &                                 ... \\
 150818A &     $0.282$ \textsuperscript {{(40)}} &    $128^{+37}_{-37}$ \textsuperscript {{(41)}} &     10 \textsuperscript {{(41)}} \\
 161219B &     $0.147$ \textsuperscript {{(43)}} &    $104^{+24}_{-24}$ \textsuperscript {{(44)}} &      1.6 \textsuperscript {{(44)}} \\
 171205A &     $0.037$ \textsuperscript {{(46)}} &  $126^{+115}_{-115}$ \textsuperscript {{(45)}} &      0.22 \textsuperscript {{(45)}} \\
 180728A &     $0.117$ \textsuperscript {{(48)}} &                                            ... &     28.1 \textsuperscript {{(47)}} \\
 190829A &     $0.079$ \textsuperscript {{(49)}} &                                            ... &      2.22 \textsuperscript {{(50)}} \\
 191019A &     $0.248$ \textsuperscript {{(51)}} &                               $67^{+15}_{-15}$ &                               72.1 \\
\bottomrule
\end{tabular}

\tablenotetext{a}{(erg) Isotropic energy released between $1\ \mathrm{keV}$ and $10\ \mathrm{MeV}$. The values given are either extracted from the literature if mentioned or calculated with the corresponding spectrum and fluence that can be found in Table \ref{tab:spectrum_long}}

\tablerefs{
(1) \cite{Tinney1998_980425}; (2) \cite{Yamazaki2003_980425}; (3) \cite{Soderberg2002_020903}; (4) \cite{Sakamoto2005}; (5) \cite{Sakamoto2004_020903}; (6) \cite{Greiner2003_030329}; (7) \cite{Vanderspek2004_030329}; (8) \cite{Prochaska2003_031203}; (9) \cite{Sazonov2004_031203}; (10) \cite{Kelson2004_040701}; (11) \cite{Pelangeon2008_040701}; (12) \cite{Rossi2014_050219A}; (13) \cite{Tagliaferri2005_050219A}; (14) \cite{Halpern2006_050826}; (15) \cite{Mirabal2007_050826}; (16) \cite{Perley2006_051109B}; (17) \cite{Bromberg2011}; (18) \cite{Mirabal2006_060218}; (19) \cite{Campana2006_060218}; (20) \cite{Fynbo2006_060505}; (21) \cite{Thoene2006_060505}; (22) \cite{Krimm2009_060505}; (23) \cite{Ofek2007_060505}; (24) \cite{GalYam2006_060614}; (25) \cite{Fynbo2006_060614}; (26) \cite{Price2006_060614}; (27) \cite{Golenetskii2006_060614}; (28) \cite{Zhang2006_060614}; (29) \cite{Stanway2015_080517}; (30) \cite{Starling2011_100316D}; (31) \cite{Levan2011_111005A}; (32) \cite{Tanga2018_111005A}; (33) \cite{Thoene2014_111225A}; (34) \cite{Schulze2012_120422A}; (35) \cite{Tanvir2012_120422A}; (36) \cite{Melandri2012_120422A}; (37) \cite{Leloudas2013_130702A}; (38) \cite{Toy2016_130702A}; (39) \cite{Xu2015_150518A}; (40) \cite{SanchezRamirez2015_150818A}; (41) \cite{Golenetskii2015_150818A}; (42) \cite{Cano2017_161219B}; (43) \cite{Tanvir2016_161219B}; (44) \cite{Frederiks2016_161219B}; (45) \cite{Delia2018_171205A}; (46) \cite{Izzo2017_171205A}; (47) \cite{Wang2019}; (48) \cite{Rossi2018_180728A}; (49) \cite{Valeev2019_190829A}; (50) \cite{Chand2020_190829A}; (51) \cite{Fynbo2019_191019A}; 
}

\end{table}

\clearpage

\section{Short GRBs} 
\label{sub:sgrbs}

Tables \ref{tab:detection_short}, \ref{tab:spectrum_short} and \ref{tab:intrinsic_short} summarize the information regarding the \Nshort\ short GRBs in our sample.

The detection of the kilonova AT2017gfo associated with GRB~170817A \citep{Tanvir2017_170817A} was the first and only detection of a kilonova in real-time. However, by looking into the archive data, some short GRBs have been tentatively associated with a kilonova.
\begin{itemize}
\item GRB~050709 has an optical and infrared signal at $\mathrm{t} > 2.5$ days after the trigger, which could be dominated by a kilonova \citep{Jin2016_050709}.
\item For GRB~070809, \cite{Jin2020_070809} has found a possible optical kilonova at $\mathrm{t}\sim0.47$ d after the trigger. 
\item The work of \cite{Lamb2019_160821B} has enabled to find for GRB~160821B the best-sampled kilonova light curve, without having a gravitational wave trigger.
\item \cite{Troja2018_150101B} associate GRB~150101B to an off-axis short GRB with a kilonova, similar to the well-known GRB~170817A.
\end{itemize}

The short GRBs in our sample have diverse prompt emission properties:

\begin{itemize}
    \item Concerning the fluence, the minimum fluence observed is $9\times 10^{-9}\ \fluence$ for GRB~050509B while the maximum one is $5.33\times 10^{-6}\ \fluence$ for GRB~061201, with the median equal to $3.6\times 10^{-7}\ \fluence$.
    \item For the 64ms peak flux, the minimum is $3.21\ \phflux$ for GRB~070809 and the maximum is $92.1\ \phflux$ for GRB~050709, with the median equal to $7.74\ \phflux$.
    \item For the peak energy of the fluence spectrum, $6$ out of the $10$ short GRBs do not have any peak energy information. GRB~050709 only has a peak energy for the first part of the GRB, the second part of the burst being best fitted by a power-law \citep{Villasenor2005_050709}. For the remaining $4$ short GRBs with a peak energy, the minimum $E_\mathrm{peak}$ is equal to $84 \pm 19\ \mathrm{keV}$ for GRB~160821B and the maximum equal to $302^{+214}_{-85}\ \mathrm{keV}$ for the first part of GRB~060614, with a median of $68\ \mathrm{keV}$ from GRB~030329.
    \item For \Tqvd-values, the minimum value is $0.024 \sec$ for GRB~050509B, the maximum $160 \sec$ for GRB~050709 and the median is $0.90\sec$.
\end{itemize}

There are also some differences in the intrinsic properties:
\begin{itemize}
    \item The closest short GRB that has been detected is GRB~170817A, at a redshift of $0.0093$ ($\approx 42\ \mathrm{Mpc}$). The furthest one in our local GRB sample is at a redshift of $\mathrm{z} = 0.287$ ($\approx 1530\ \mathrm{Mpc}$), GRB~050724. the median redshift of the sample is $\mathrm{z} = 0.160$ ($\approx 790\ \mathrm{Mpc}$).
    \item The minimum isotropic energy detected in the short GRB sample is $E_\mathrm{iso} = 4\pm 1\times 10^{46}\ \mathrm{erg}$ for GRB~170817A \citep{Abbott2017b}, while the maximum isotropic energy is equal to $7.65\times 10^{50}\ \mathrm{erg}$ for GRB~050724. The median value for the short GRBs is $E_\mathrm{iso} = 8.6\times 10^{49}\ \mathrm{erg}$: there are two orders of magnitude between GRB~170817A and the second dimmest short GRB, GRB~050509B.
\end{itemize}

\begin{table}

\centering
\caption{Detected features for Short GRBs}
\label{tab:detection_short}
\begin{tabular}{lllllll}
\toprule
    Name & Instrument(s)\ \tablenotemark{a} &                               Class &         Afterglow\ \tablenotemark{b} &                                  SN/KN &                              Host &       T\textsubscript {90} (s)\ \tablenotemark{c} \\
\midrule
 050509B &                               S  &                                SGRB &           X \textsuperscript {{(1)}} &                                    ... &          \textsuperscript {{(2)}} &                            $0.024^{+0.0089}_{-0.0089}$ \\
  050709 &                               H  &                             eeSGRB &        XO \textsuperscript {{(3,4)}} &          (KN) \textsuperscript {{(5)}} &          \textsuperscript {{(4)}} &                 $160$ \textsuperscript {{(6)}} \\
  050724 &                               S  &   eeSGRB  \textsuperscript {{(7)}} &       XOR \textsuperscript {{(8,9)}} &                                    ... &          \textsuperscript {{(9)}} &   $3^{+1}_{-1}$ \textsuperscript {{(8)}} \\
 060502B &                               S  &                                SGRB &       X \textsuperscript {{(10,11)}} &                                    ... &         \textsuperscript {{(12)}} &                            $0.14^{+0.05}_{-0.05}$ \\
  061201 &                             S K  &                                SGRB &      XO \textsuperscript {{(13,14)}} &                                    ... &         \textsuperscript {{(14)}} &                            $0.78^{+0.10}_{-0.10}$ \\
  070809 &                               S  &                                SGRB &         XO \textsuperscript {{(15)}} &         (KN) \textsuperscript {{(16)}} &         \textsuperscript {{(15)}} &                            $1.28^{+0.37}_{-0.37}$ \\
 080905A &                       S F I SUZ  &                                SGRB &      XO \textsuperscript {{(17,18)}} &                                    ... &   \textsuperscript {{(19,20,18)}} &                            $1.02^{+0.08}_{-0.08}$ \\
 150101B &                           S F I  &  eeSGRB  \textsuperscript {{(21)}} &         XO \textsuperscript {{(22)}} &         (KN) \textsuperscript {{(23)}} &         \textsuperscript {{(22)}} &                            $0.23^{+0.06}_{-0.06}$ \\
 160821B &                             S F  &                                SGRB &  XOR \textsuperscript {{(24,25,26)}} &      (KN) \textsuperscript {{(27,28)}} &         \textsuperscript {{(25)}} &                            $0.48^{+0.07}_{-0.07}$ \\
 170817A &                           F K I  &                             llSGRB &     XOR \textsuperscript {{(29,30)}} &  KNAT2017gfo \textsuperscript {{(31)}} &                                   &  $2.05^{+0.47}_{-0.47}$ \textsuperscript {{(32)}} \\
\bottomrule
\end{tabular}

\tablenotetext{a}{F for \textit{Fermi}/GBM, H for HETE-2/FREGATE and/or WXM, I for INTEGRAL/IBIS and/or INTEGRAL/SPI, K for Wind/KONUS, S for \textit{Swift}/BAT, SUZ for \textit{Suzaku}/WAM}
\tablenotetext{b}{The letters X, O and R stands respectively for X-ray, Optical and Radio afterglow.}
\tablenotetext{c}{Unless quoted differently, the T\textsubscript{90} are measured by the \textit{Swift}/BAT instrument in the $15-350\ \mathrm{keV}$ band}

\tablerefs{
(1) \cite{Gehrels2005_050509B}; (2) \cite{Prochaska2005_050509B}; (3) \cite{Fox2005_050709}; (4) \cite{Hjorth2005_050709}; (5) \cite{Jin2016_050709}; (6) \cite{Villasenor2005_050709}; (7) \cite{Barthelmy2005_050724}; (8) \cite{Berger2005_050724}; (9) \cite{Prochaska2005_050724}; (10) \cite{Troja2006_060502B}; (11) \cite{Poole2006_060502B}; (12) \cite{Bloom2006_060502B}; (13) \cite{Marshall2006_061201}; (14) \cite{Holland2006_061201}; (15) \cite{Perley2008_070809}; (16) \cite{Jin2020_070809}; (17) \cite{Pagani2008_080905A}; (18) \cite{Rowlinson2010_080905A}; (19) \cite{Malesani2008_080905A}; (20) \cite{UgartePostigo2008_080905A}; (21) \cite{Burns2018_150101B}; (22) \cite{Fong2016_150101B}; (23) \cite{Troja2018_150101B}; (24) \cite{Levan2016_160821B}; (25) \cite{Xu2016_160821B}; (26) \cite{Fong2016_160821B}; (27) \cite{Troja2019_160821B}; (28) \cite{Lamb2019_160821B}; (29) \cite{Troja2017_170817A}; (30) \cite{Hallinan2017_170817A}; (31) \cite{Tanvir2017_170817A}; (32) \cite{GBMcat2016}; 
}

\end{table}

\begin{table}

\centering
\caption{Spectral properties of Short GRBs}
\label{tab:spectrum_short}
\resizebox{\textwidth}{!}{
\begin{tabular}{lllllll}
\toprule
      Name &                                          Fluence &                                     64-ms P.-flux & Energy Band\ \tablenotemark{a} & P.-flux (norm)\ \tablenotemark{b} &                           $\mathrm{E_{peak}}$ &         Spectral Model\ \tablenotemark{c} \\
 & ($10^{-6}$ erg cm$^{-2}$) & (ph cm$^{-2}$ s$^{-1}$) & (keV - Inst.) & (ph cm$^{-2}$ s$^{-1}$) & (keV) & \\ 
\midrule
   050509B &  $0.007^{+0.002}_{-0.002}$ \textsuperscript {{(1)}} &                   $1.3$ \textsuperscript {{(1)}} &                  15 - 150 - S  &                            $1.3$ &                                           ... &                                 PL: -1.57 \\
 050709-p1 &  $0.40^{+0.04}_{-0.04}$ \textsuperscript {{(2)}} &  $92.1^{+7.6}_{-7.6}$ \textsuperscript {{(2)}} &                   2 - 400 - H  &           $50.9^{+4.2}_{-4.2}$ &     $84^{+11}_{-11}$ \textsuperscript {{(2)}} &   CPL: 84, -0.53 \textsuperscript {{(2)}} \\
 050709-p2 &  $1.10^{+0.14}_{-0.14}$ \textsuperscript {{(2)}} &   $2.72^{+0.47}_{-0.47}$ \textsuperscript {{(2)}} &                    2 - 25 - H  &            $0.37^{+0.06}_{-0.06}$ &                                           ... &        PL: -1.98 \textsuperscript {{(2)}} \\
    050724 &                           $1.01^{+0.12}_{-0.12}$ &                                           $12.05$ &                  15 - 150 - S  &                           $12.05$ &                                           ... &                                 PL: -1.93 \\
   060502B &                           $0.05^{+0.01}_{-0.01}$ &                   $3.4$ \textsuperscript {{(1)}} &                  15 - 150 - S  &                            $3.4$ &                                           ... &                                 PL: -0.99 \\
    061201 &                  $5.33$ \textsuperscript {{(3)}} &                                           $13.2$ &                20 - 3000 - K*  &                           $13.2$ &  $873^{+458}_{-458}$ \textsuperscript {{(3)}} &  CPL: 873, -0.36 \textsuperscript {{(3)}} \\
    070809 &                           $0.102^{+0.015}_{-0.015}$ &                   $1.9$ \textsuperscript {{(1)}} &                  15 - 150 - S  &                            $1.9$ &                                           ... &                                 PL: -1.66 \\
   080905A &                           $0.85^{+0.05}_{-0.05}$ &                   $3.7$ \textsuperscript {{(1)}} &                 10 - 1000 - F  &                            $2.2$ &                                           ... &                                 PL: -1.33 \\
   150101B &                           $0.238^{+0.015}_{-0.015}$ &                           $10.48^{+1.35}_{-1.35}$ &                 10 - 1000 - F  &            $5.66^{+0.73}_{-0.73}$ &  $550^{+190}_{-190}$ \textsuperscript {{(4)}} &                                 PL: -1.76 \\
   160821B &                  $0.168$ \textsuperscript {{(5)}} &   $9.16^{+1.19}_{-1.19}$ \textsuperscript {{(5)}} &                 10 - 1000 - F  &            $5.75^{+0.75}_{-0.75}$ &     $84^{+19}_{-19}$ \textsuperscript {{(5)}} &   CPL: 84, -1.37 \textsuperscript {{(5)}} \\
   170817A &  $0.31^{+0.07}_{-0.07}$ \textsuperscript {{(6)}} &                            $3.73^{+0.93}_{-0.93}$ &                 10 - 1000 - F  &            $2.60^{+0.65}_{-0.65}$ &    $185^{+62}_{-62}$ \textsuperscript {{(6)}} &  CPL: 185, -0.62 \textsuperscript {{(6)}} \\
\bottomrule
\end{tabular}
}

\tablecomments{For the \textit{Swift}/BAT and \textit{Fermi}/GBM instruments, unless specified, properties are respectively from the \textit{Swift}/BAT  \citep{BATcat2016} and the \textit{Fermi}/GBM \citep{GBMcat2016} catalogs.}

\tablenotetext{a}{Energy range and Instrument (abbreviation from Table \ref{tab:detection_short}) used to obtain the fluence, peak flux and spectral model determination.}
\tablenotetext{b}{Peak Flux translated in the $15-150 \mathrm{keV}$ band, used with \Tqvd to plot Fig. \ref{fig:T90_Pflux}}
\tablenotetext{c}{Parameters for the spectral model: BAND: $E_\mathrm{peak}\ (\mathrm{keV}), \alpha, \beta$ ; CPL: $E_\mathrm{peak}\ (\mathrm{keV}), \alpha$ ; PL: $\Gamma$}
\tablenotetext{*}{Fluence and Spectral Model are from Wind/Konus, Peak-Flux is from \textit{Swift}/BAT in the $15-150\ \mathrm{keV}$ energy range.}

\tablerefs{
(1) \cite{DAvanzo2014}; (2) \cite{Villasenor2005_050709}; (3) \cite{Golenetskii2006_061201}; (4) \cite{Burns2018_150101B}; (5) \cite{Stanbro2016_160821B}; (6) \cite{Goldstein2017_170817A}; 
}
\end{table}

\begin{table}

\centering
\caption{Intrinsic properties of Short GRBs}
\label{tab:intrinsic_short}
\begin{tabular}{llll}
\toprule
    Name &                           Redshift &                            $E_\mathrm{peak,i}$ & $E_\mathrm{iso}$\ \tablenotemark{a} \\
 & & (keV) &($\times 10^{50}\ \mathrm{erg}$) \\ 
\midrule
 050509B &   $0.225$ \textsuperscript {{(1)}} &                                            ... &                                 ... \\
  050709 &   $0.161$ \textsuperscript {{(3)}} &                                            ... &       1 \textsuperscript {{(4)}} \\
  050724 &   $0.258$ \textsuperscript {{(5)}} &                                            ... &                                7.41 \\
 060502B &   $0.287$ \textsuperscript {{(6)}} &                                            ... &                                 ... \\
  061201 &   $0.111$ \textsuperscript {{(7)}} &   $970^{+509}_{-509}$ \textsuperscript {{(8)}} &       1.4 \textsuperscript {{(9)}} \\
  070809 &  $0.219$ \textsuperscript {{(11)}} &                                            ... &      > 0.1 \textsuperscript {{(10)}} \\
 080905A &  $0.122$ \textsuperscript {{(12)}} &                                            ... &                                 ... \\
 150101B &  $0.134$ \textsuperscript {{(14)}} &  $624^{+215}_{-215}$ \textsuperscript {{(15)}} &      0.23 \textsuperscript {{(16)}} \\
 160821B &  $0.160$ \textsuperscript {{(18)}} &     $97^{+22}_{-22}$ \textsuperscript {{(19)}} &                                0.13 \\
 170817A &                            $0.0093$ &    $187^{+63}_{-63}$ \textsuperscript {{(22)}} &      0.00053 \textsuperscript {{(16)}} \\
\bottomrule
\end{tabular}

\tablenotetext{a}{(erg) Isotropic energy released between $1\ \mathrm{keV}$ and $10\ \mathrm{MeV}$. The values given are either extracted from the literature if mentioned or calculated with the corresponding spectrum and fluence that can be found in Table \ref{tab:spectrum_short}}

\tablerefs{
(1) \cite{Bloom2006_050509B}; (2) \cite{Jin2016_050709}; (3) \cite{Price2005_050709}; (4) \cite{Villasenor2005_050709}; (5) \cite{Prochaska2005_050724}; (6) \cite{Bloom2006_060502B}; (7) \cite{Berger2006_061201}; (8) \cite{Golenetskii2006_061201}; (9) \cite{Stratta2007_061201}; (10) \cite{Jin2020_070809}; (11) \cite{Perley2008_070809}; (12) \cite{Rowlinson2010_080905A}; (13) \cite{Troja2018_150101B}; (14) \cite{Levan2015_150101B}; (15) \cite{Burns2018_150101B}; (16) \cite{Abbott2017b}; (17) \cite{Lamb2019_160821B}; (18) \cite{Levan2016_160821B}; (19) \cite{Stanbro2016_160821B}; (20) \cite{Lu2017_160821B}; (21) \cite{Abbott2017a}; (22) \cite{Goldstein2017_170817A}; 
}
\end{table}

\clearpage

\section{SGR Giant-Flares}
\label{sub:sgr_gflares}

Tables \ref{tab:detection_gflare}, \ref{tab:spectrum_gflare} and \ref{tab:intrinsic_gflare} summarize the information detected for the SGR giant flares.

\begin{itemize}
    \item Concerning the fluence, the minimum fluence observed is $5.6\times 10^{-9}\ \fluence$ for GF~050906 while the maximum one is $0.61\ \fluence$
    for GF~041227, with the median equal to $2.7\times 10^{-5}\ \fluence$. 
    \item For the 64ms peak flux, the minimum is $1.95\ \phflux$ for GF~050906 and the maximum is $11.9\times 10^6\ \phflux$ for GF~041227, with the median equal to $990\ \phflux$ for GF~070201.
    \item For the peak energy of the fluence spectrum, $6$ out of the $7$ SGR giant flares have spectra showing a peak energy. This is the consequence of the extremely high-luminosity of such events, in addition to the large diversity of instruments that have observed some of them (see Table \ref{tab:detection_gflare}). GF~050906 has a fluence model that is best fitted by a power law, mainly because it was a faint burst ($5.6\times 10^{-9}\ \fluence$) only detected by \textit{Swift}/BAT. The minimum $E_\mathrm{peak}$-value is equal to $240\ \mathrm{keV}$ for GF~980827 with an OTTB spectrum, the maximum equal to $2080\ \mathrm{keV}$ for GF~051103 with a Band spectrum, with a median of $548\ \mathrm{keV}$.
    \item For \Tqvd-values, the minimum value is $0.1 \sec$ for GF~051103, the maximum $0.310 \sec$ for GF~980827 and the median is $0.2 \sec$ for GF~200415A.
\end{itemize}

There are also some differences in the intrinsic properties:
\begin{itemize}
    \item The closest Giant Flare detected is GF~041227. At an assumed distance of $8.7\ \mathrm{kpc}$, it comes from one of the two galactic SGRs that have emitted a giant flare (the other being \mbox{GF~980827}). GF~050906 is the furthest one at a distance of $130\ \mathrm{Mpc}$. The median distance of the sample is $0.780\ \mathrm{Mpc}$.
    \item The minimum isotropic energy detected for the giant flares is $E_\mathrm{iso} = 4.3\pm 1\times 10^{44}\ \mathrm{erg}$ for GF~980827, while the maximum isotropic energy is equal to $8.0\times 10^{46}\ \mathrm{erg}$ for GF~050906. The median value is $E_\mathrm{iso} = 1.2\times 10^{46}\ \mathrm{erg}$ for GF~041227.
\end{itemize}

As it can be seen in Table \ref{tab:intrinsic_gflare}, GF~050906 is an outlier among the giant flares, being detected at a distance that is forty times larger than the second most distant Giant Flare GF~051103. This will be discussed further in Section \ref{sub:sgr_gflares}.

\begin{table}

\centering
\caption{Detected features for SGR Giant Flares}
\label{tab:detection_gflare}
\begin{tabular}{lllll}
\toprule
    Name & Instrument(s)\ \tablenotemark{a} &                        Host &                          T\textsubscript {90} (s) & Pulsating Tail \\
\midrule
  790305 &                K ICE HEL IS V P  &    \textsuperscript {{(1)}} &                   $0.2$ \textsuperscript {{(2)}} &            Yes \\
  980827 &                           K B U  &    \textsuperscript {{(3)}} &                   $0.31$ \textsuperscript {{(4)}} &            Yes \\
  041227 &                 S K I R COR SOP  &    \textsuperscript {{(5)}} &                   $0.16$ \textsuperscript {{(5)}} &            Yes \\
  050906 &                               S  &    \textsuperscript {{(6)}} &   $0.13^{+0.02}_{-0.02}$ \textsuperscript {{(7)}} &            No \\
  051103 &                               S  &   \textsuperscript {{(10)}} &  $0.1^{+0.004}_{-0.004}$ \textsuperscript {{(10)}} &            No \\
  070201 &                           S K I  &   \textsuperscript {{(11)}} &                  $0.28$ \textsuperscript {{(11)}} &            No \\
 200415A &                         S F K I  &   \textsuperscript {{(12)}} &  $0.124^{+0.005}_{-0.005}$ \textsuperscript {{(13)}} &            No \\
\bottomrule
\end{tabular}

\tablenotetext{a}{B for BeppoSax/WFC, C for CGRO/BATSE, COR for CORONAS-F, F for \textit{Fermi}/GBM, H for HETE-2/FREGATE and/or WXM, HEL for Helios B, I for INTEGRAL/IBIS and/or INTEGRAL/SPI, ICE for ICE, IS for ISEE 3, K for Wind/KONUS, P for PVO and Prognoz 7, R for RHESSI, S for \textit{Swift}/BAT, SOP for SOPA, U for Ulysses, V for Vela 5A 5B 6A and Venera 11 12}

\tablerefs{
(1) \cite{Evans1980_790305}; (2) \cite{Fenimore1996_790305}; (3) \cite{Vrba2000_980827}; (4) \cite{Tanaka2007_980827}; (5) \cite{Mazets2005_041227}; (6) \cite{Levan2007_050906}; (7) \cite{Parsons2005_050906}; (8) \cite{Lipunov2005_051103}; (9) \cite{Cameron2005_051103}; (10) \cite{Hurley2010_051103}; (11) \cite{Ofek2008_070201}; (12) \cite{Svinkin2020_200415A}; (13) \cite{Minaev2020_200415A}; 
}
\end{table}

\begin{table}

\centering
\caption{Spectral properties of SGR Giant Flare}
\label{tab:spectrum_gflare}
\resizebox{\textwidth}{!}{
\begin{tabular}{lllllll}
\toprule
    Name &                                                      Fluence &                                   64-ms P.-flux & Energy Band\ \tablenotemark{a} & P.-flux (norm)\ \tablenotemark{b} &                 $\mathrm{E_{peak}}/\mathrm{kT}$ &                 Spectral Model\ \tablenotemark{c} \\
 & ($10^{-6}$ erg cm$^{-2}$) & (ph cm$^{-2}$ s$^{-1}$) & (keV - Inst.) & (ph cm$^{-2}$ s$^{-1}$) & (keV) & \\ 
\midrule
  790305 &                             $4.5\times10^2$ \textsuperscript {{(1)}} &              $1.8\times 10^4$ \textsuperscript {{(1)}} &                 30 - 2000 - K  &                         $2.0\times 10^4$ &                  $246$ \textsuperscript {{(1)}} &                OTTB: 246 \textsuperscript {{(1)}} \\
  980827 &     $1.6^{+2.0}_{-2.0}\times 10^4$ \textsuperscript {{(2)}} &             $5.5\times 10^5$ \textsuperscript {{(2)}} &                15 - 20000 - C  &                        $4.4\times 10^5$ &                  $240$ \textsuperscript {{(3)}} &                OTTB: 240 \textsuperscript {{(3)}} \\
  041227 &  $6.1^{+3.5}_{-3.5} \times 10^5$ \textsuperscript {{(4)}} &           $11.9\times 10^6$ \textsuperscript {{(4)}} &              20 - 10000 - COR  &                       $6.5\times 10^6$ &  $850^{+1259}_{-1259}$ \textsuperscript {{(5)}} &          CPL: 800, -0.70 \textsuperscript {{(4)}} \\
  050906 &                                          $5.6^{+2.4}_{-2.4}\times 10^{-3}$ &    $2.0^{+1.0}_{-1.0}$ \textsuperscript {{(6)}} &                  15 - 150 - S  &               $2.0^{+1.0}_{-1.0}$ &                                             ... &                                         PL: -1.66 \\
  051103 &                $33.3^{+2.1}_{-2.1}$ \textsuperscript {{(7)}} &                $245.7$ \textsuperscript {{(7)}} &                20 - 10000 - K  &                            $23.3$ &   $2080^{+180}_{-180}$ \textsuperscript {{(7)}} &  BAND: 2080, 0.13, -2.78 \textsuperscript {{(7)}} \\
  070201 &                $20^{+1}_{-2.6}$ \textsuperscript {{(8)}} &                $990.0$ \textsuperscript {{(8)}} &                 20 - 1200 - K  &                           $674.2$ &      $296^{+38}_{-38}$ \textsuperscript {{(8)}} &          CPL: 296, -0.98 \textsuperscript {{(8)}} \\
 200415A &                 $5.2^{+0.1}_{-0.1}$ \textsuperscript {{(9)}} &  $74.0^{+2.0}_{-2.0}$ \textsuperscript {{(10)}} &                20 - 10000 - K  &              $14.6^{+0.4}_{-0.4}$ &      $950^{+50}_{-50}$ \textsuperscript {{(9)}} &           CPL: 950, 0.07 \textsuperscript {{(9)}} \\
\bottomrule
\end{tabular}
}

\tablecomments{For the \textit{Swift}/BAT and \textit{Fermi}/GBM instruments, unless specified, properties are respectively from the \textit{Swift}/BAT  \citep{BATcat2016} and the \textit{Fermi}/GBM \citep{GBMcat2016} catalogs.}

\tablenotetext{a}{Energy range and Instrument (abbreviation from Table \ref{tab:detection_gflare}) used to obtain the fluence, peak flux and spectral model determination.}
\tablenotetext{b}{Peak Flux translated in the $15-150 \mathrm{keV}$ band, used with \Tqvd to plot Fig. \ref{fig:T90_Pflux}}
\tablenotetext{c}{Parameters for the spectral model: BAND: $E_\mathrm{peak}\ (\mathrm{keV}), \alpha, \beta$ ; CPL: $E_\mathrm{peak}\ (\mathrm{keV}), \alpha$ ; PL: $\Gamma$ ; OTTB: $\mathrm{kT}\ (\mathrm{keV})$}
\tablenotetext{*}{Fluence and Spectral Model are from Wind/Konus, Peak-Flux is from \textit{Swift}/BAT in the $15-150\ \mathrm{keV}$ energy range.}

\tablerefs{
(1) \cite{Fenimore1996_790305}; (2) \cite{Tanaka2007_980827}; (3) \cite{Hurley1999_980827}; (4) \cite{Mazets2005_041227}; (5) \cite{Frederiks2007_041227}; (6) \cite{Parsons2005_050906}; (7) \cite{Hurley2010_051103}; (8) \cite{Mazets2008}; (9) \cite{Bissaldi2020_200415A}; (10) \cite{Frederiks2020_200415A}; 
}
\end{table}

\begin{table}

\centering
\caption{Intrinsic properties of SGR Giant Flares}
\label{tab:intrinsic_gflare}
\begin{tabular}{lllll}
\toprule
    Name &         Origin &                         Distance &                             $E_\mathrm{peak,i}$ & $E_\mathrm{iso}$\ \tablenotemark{a} \\
 & & (kpc) & (keV) & ($\times 10^{45}\ \mathrm{erg}$) \\ 
\midrule
  790305* &    SGR 0526-66 &      55 \textsuperscript {{(1)}} &                  $246$ \textsuperscript {{(2)}} &       0.7 \textsuperscript {{(1)}} \\
  980827 &    SGR 1900+14 &      15 \textsuperscript {{(3)}} &                  $240$ \textsuperscript {{(4)}} &       0.43 \textsuperscript {{(5)}} \\
  041227 &    SGR 1806-20 &       8.7 \textsuperscript {{(6)}} &  $850^{+1259}_{-1259}$ \textsuperscript {{(7)}} &      23 \textsuperscript {{(7)}} \\
  050906 &  SGR 0331-1439 &  130000 \textsuperscript {{(8)}} &                                             ... &                               80 \\
  051103 &    SGR 0952+69 &    3600 \textsuperscript {{(9)}} &   $2082^{+180}_{-180}$ \textsuperscript {{(9)}} &      75 \textsuperscript {{(9)}} \\
  070201 &    SGR 0044+42 &    780 \textsuperscript {{(10)}} &      $296^{+38}_{-38}$ \textsuperscript {{(1)}} &       1.5 \textsuperscript {{(1)}} \\
 200415A &            ... &   3500 \textsuperscript {{(11)}} &     $951^{+50}_{-50}$ \textsuperscript {{(12)}} &     12.2 \textsuperscript {{(12)}} \\
\bottomrule
\end{tabular}

\tablenotetext{a}{Isotropic energy released between $1\ \mathrm{keV}$ and $10\ \mathrm{MeV}$. The values given are either extracted from the literature if mentioned or calculated with the corresponding spectrum and fluence that can be found in \ref{tab:spectrum_gflare}}
\tablenotetext{*}{The fluence and spectral properties from \cite{Fenimore1996_790305} given in Table \ref{tab:spectrum_gflare} have a different dead time correction than the $E_\mathrm{iso}$ value calculated by \cite{Mazets2008}.}

\tablerefs{
(1) \cite{Mazets2008}; (2) \cite{Fenimore1996_790305}; (3) \cite{Vrba2000_980827}; (4) \cite{Hurley1999_980827}; (5) \cite{Tanaka2007_980827}; (6) \cite{Hurley2005_041227}; (7) \cite{Frederiks2007_041227}; (8) \cite{Levan2007_050906}; (9) \cite{Hurley2010_051103}; (10) \cite{Ofek2008_070201}; (11) \cite{Svinkin2020_200415A}; (12) \cite{Bissaldi2020_200415A}; 
}
\end{table}

\clearpage

\section{Summary of \textit{SVOM}/ECLAIRs characteristics}

\begin{table*}
\centering
\caption{Main characteristics and expected performance of \textit{SVOM}/ECLAIRs} 
\begin{tabular}{ll}
\toprule
  Measured property 					& Value 						\\
\midrule
  Energy range 							&  4 -- 150~keV  					\\
  Mode of operation						& Photon counting					\\
  Time resolution						& 20 $\mathrm{\mu s}$				\\
  Detecting area 						& $\sim 1000$~cm$^{2}$  			\\
  Detectors							& CdTe							\\
  Number of detectors					& 6400							\\
  Size of detectors						& 4~mm $\times$ 4~mm $\times$ 1~mm			\\
  Energy resolution (median)			& $<1.6\ \mathrm{keV}$ @ $60\ \mathrm{keV}$						\\
  Field of view (half-coded / full)				& 0.9 / 2.0 sr  					\\
  Coded mask dimensions					& 54~cm $\times$ 54~cm			\\
  Coded mask open fraction				& 40 \%						\\
  Coded mask element size				& 10.4~mm					\\
  Detector-Mask distance					& 46~cm						\\
  Telescope PSF 						& 52~arcminute FWHM			\\
\midrule
\midrule
  Expected performance 					& Value 						\\
\midrule
  Effective area at 20 keV					& $\sim 400$~cm$^{2}$  			\\
  Background level (empty sky)				& $\sim4000$~cts s$^{-1}$			 \\
  Point source localization			&  13~arcminute @ SNR = 7 \\
  (90\% confidence) &  	\\
  Limiting flux for a 	&  $1.5\times10^{-8}\ \mathrm{erg\ cm^{-2}\ s^{-1}}$ 			\\
  20~s long on-axis GRB & \\
  Dead time						&  $\leq 5\%$ for 10$^{5}$	~cts s$^{-1}$ 	\\
\bottomrule
\end{tabular}
\label{tab:eclairs}
\end{table*}

\clearpage

\section{Signal-to-Noise Ratio for local high-energy transients}

\begin{table}
\centering
\caption{SVOM/ECLAIRs Signal To Noise Ratio for Long GRBs}
\label{tab:SNR_long}
\begin{tabular}{llllll}
\toprule
       Name & Energy Range & Time Scale &         Count SNR &         Image SNR & Fraction FoV \\
 & (keV) & (s) & & & \\ 
\midrule
     980425 &      4 - 120 &      20.48 &              53.7 &              52.0 &         0.86 \\
     020903 &          ... &        ... &  \textless \  6.5 &  \textless \  6.5 &          ... \\
     030329 &      4 - 120 &      20.48 &            2943.0 &             722.3 &         0.99 \\
     031203 &      4 - 120 &      20.48 &              67.9 &              63.9 &         0.89 \\
     040701 &       4 - 25 &      20.48 &              17.3 &              18.3 &         0.63 \\
    050219A &      4 - 120 &      20.48 &              98.6 &              86.1 &         0.92 \\
     050826 &      4 - 120 &      20.48 &  \textless \  6.5 &               7.6 &          ... \\
    051109B &      4 - 120 &      20.48 &              14.5 &              16.0 &         0.56 \\
     060218 &       4 - 25 &      20.48 &              13.4 &              13.8 &         0.52 \\
     060505 &      4 - 120 &       5.12 &              29.9 &              28.4 &         0.78 \\
  060614-p1 &      4 - 120 &       5.12 &             253.2 &             134.9 &         0.97 \\
  060614-p2 &      4 - 120 &      20.48 &             613.7 &             304.7 &         0.98 \\
     080517 &      4 - 120 &      20.48 &              14.1 &              15.6 &         0.53 \\
    100316D &      4 - 120 &      20.48 &              14.0 &              15.8 &         0.51 \\
    111005A &       4 - 25 &      20.48 &              40.2 &              39.9 &         0.83 \\
    111225A &      4 - 120 &      20.48 &              24.8 &              25.7 &         0.74 \\
    120422A &      4 - 120 &      10.24 &               9.4 &              10.4 &         0.25 \\
    130702A &       4 - 25 &      10.24 &             600.6 &             246.2 &         0.98 \\
    150518A &          ... &        ... &  \textless \  6.5 &  \textless \  6.5 &          ... \\
    150818A &      4 - 120 &      20.48 &              73.6 &              68.2 &         0.90 \\
    161219B &      4 - 120 &       5.12 &             168.0 &             104.2 &         0.95 \\
    171205A &      4 - 120 &      20.48 &              15.7 &              16.6 &         0.61 \\
 180728A-p1 &       4 - 25 &       2.56 &             265.7 &             115.6 &         0.97 \\
 180728A-p2 &      4 - 120 &       5.12 &            2414.1 &             470.0 &         0.98 \\
 190829A-p1 &      4 - 120 &       5.12 &              98.2 &              71.4 &         0.91 \\
 190829A-p2 &       4 - 25 &      10.24 &            1171.6 &             356.3 &         0.98 \\
    191019A &       4 - 25 &      20.48 &             402.2 &             222.6 &         0.98 \\
\bottomrule
\end{tabular}
\end{table}

\begin{table}
\centering
\caption{SVOM/ECLAIRs Signal To Noise Ratio for Short GRBs}
\label{tab:SNR_short}
\begin{tabular}{llllll}
\toprule
      Name & Energy Range & Time Scale &         Count SNR &         Image SNR & Fraction FoV \\
 & (keV) & (s) & & & \\ 
\midrule
   050509B &          ... &        ... &  \textless \  6.5 &  \textless \  6.5 &          ... \\
 050709-p1 &      15 - 50 &       0.01 &              70.0 &               8.5 &         0.35 \\
 050709-p2 &      4 - 120 &      20.48 &              15.4 &              15.9 &         0.60 \\
    050724 &      4 - 120 &       0.16 &             282.3 &              66.8 &         0.96 \\
   060502B &          ... &        ... &  \textless \  6.5 &  \textless \  6.5 &          ... \\
    061201 &     25 - 120 &       0.64 &              26.4 &              17.1 &         0.72 \\
    070809 &      4 - 120 &       1.28 &              15.9 &              14.9 &         0.63 \\
   080905A &      4 - 120 &       1.28 &              20.5 &              18.9 &         0.69 \\
   150101B &      4 - 120 &       0.04 &              79.2 &              24.1 &         0.85 \\
   160821B &      4 - 120 &       0.08 &              26.7 &              15.0 &         0.70 \\
   170817A &     25 - 120 &       0.32 &              13.2 &               9.5 &         0.43 \\
\bottomrule
\end{tabular}
\end{table}

\begin{table}
\centering
\caption{SVOM/ECLAIRs Signal To Noise Ratio for SGR Giant Flares}
\label{tab:SNR_gflare}
\begin{tabular}{llllll}
\toprule
    Name & Energy Range & Time Scale &              Count SNR &         Image SNR & Fraction FoV \\
 & (keV) & (s) & & & \\ 
\midrule
  790305 &     25 - 120 &       0.02 &  \textgreater \  10000 &             337.9 &         0.31 \\
  980827 &     25 - 120 &       0.01 &  \textgreater \  10000 &             647.6 &         0.19 \\
  041227 &     25 - 120 &       0.01 &  \textgreater \  10000 &             705.7 &         0.19 \\
  050906 &          ... &        ... &       \textless \  6.5 &  \textless \  6.5 &          ... \\
  051103 &     25 - 120 &       0.01 &                   77.0 &               9.6 &         0.45 \\
  070201 &      15 - 50 &       0.01 &                 3113.1 &              63.3 &         0.95 \\
 200415A &     25 - 120 &       0.01 &                   74.0 &               9.8 &         0.44 \\
\bottomrule
\end{tabular}

\tablecomments{
The energy range and the timescale given in Tables \ref{tab:SNR_long}, \ref{tab:SNR_short} and \ref{tab:SNR_gflare} are the configurations for which the best Signal to Noise Ratio displayed in those tables is obtained. The Fraction FoV is the fraction in which both the count SNR and the resulting image SNR are above the SNR limit 6.5.
}
\end{table}

\clearpage

\section{Light curves used in the simulations}

\begin{figure}[t]
\centering
\includegraphics[width=\linewidth]{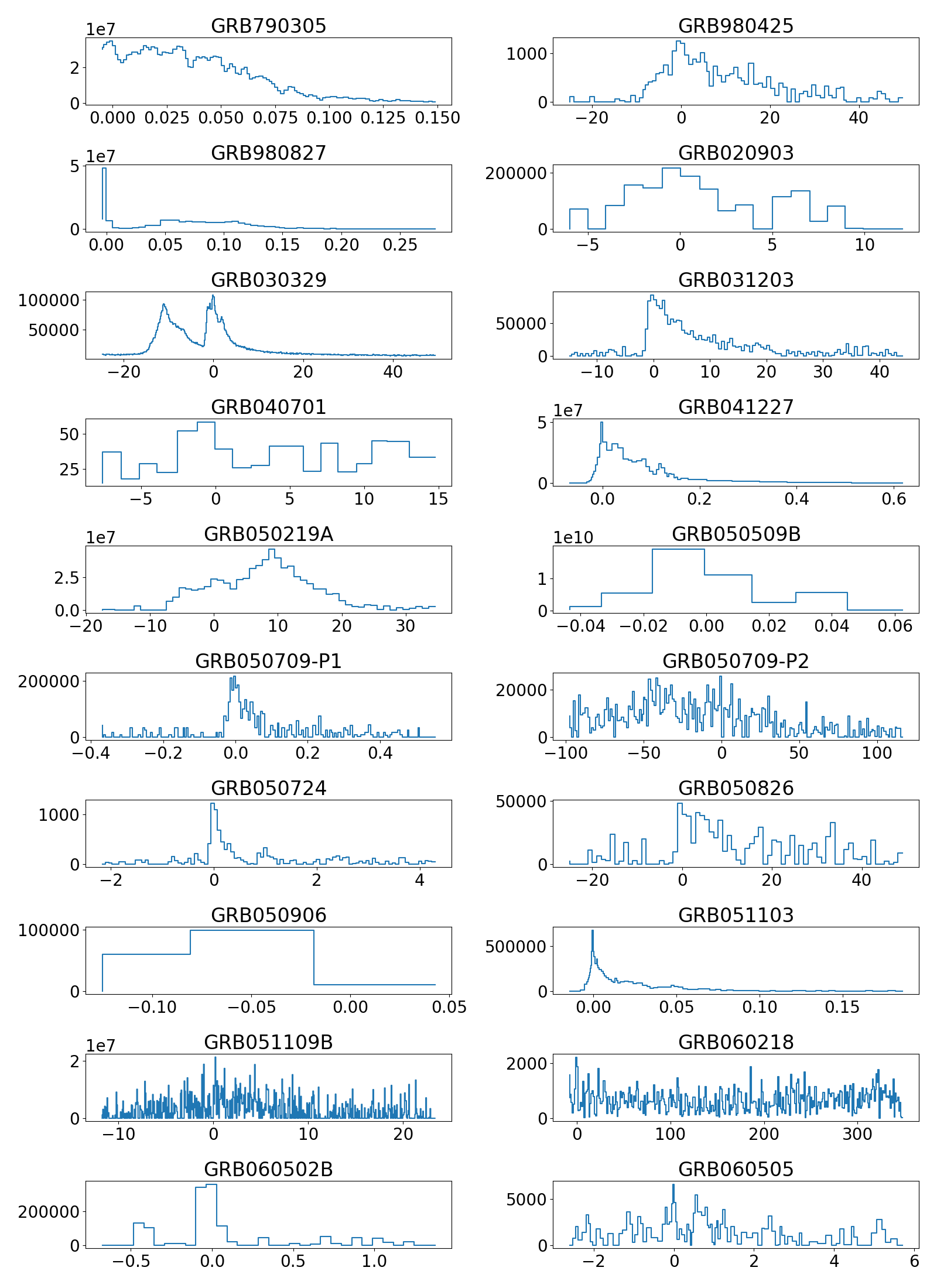}
\label{fig:LC1}
\end{figure}

\begin{figure}[t]
\centering
\includegraphics[width=\linewidth]{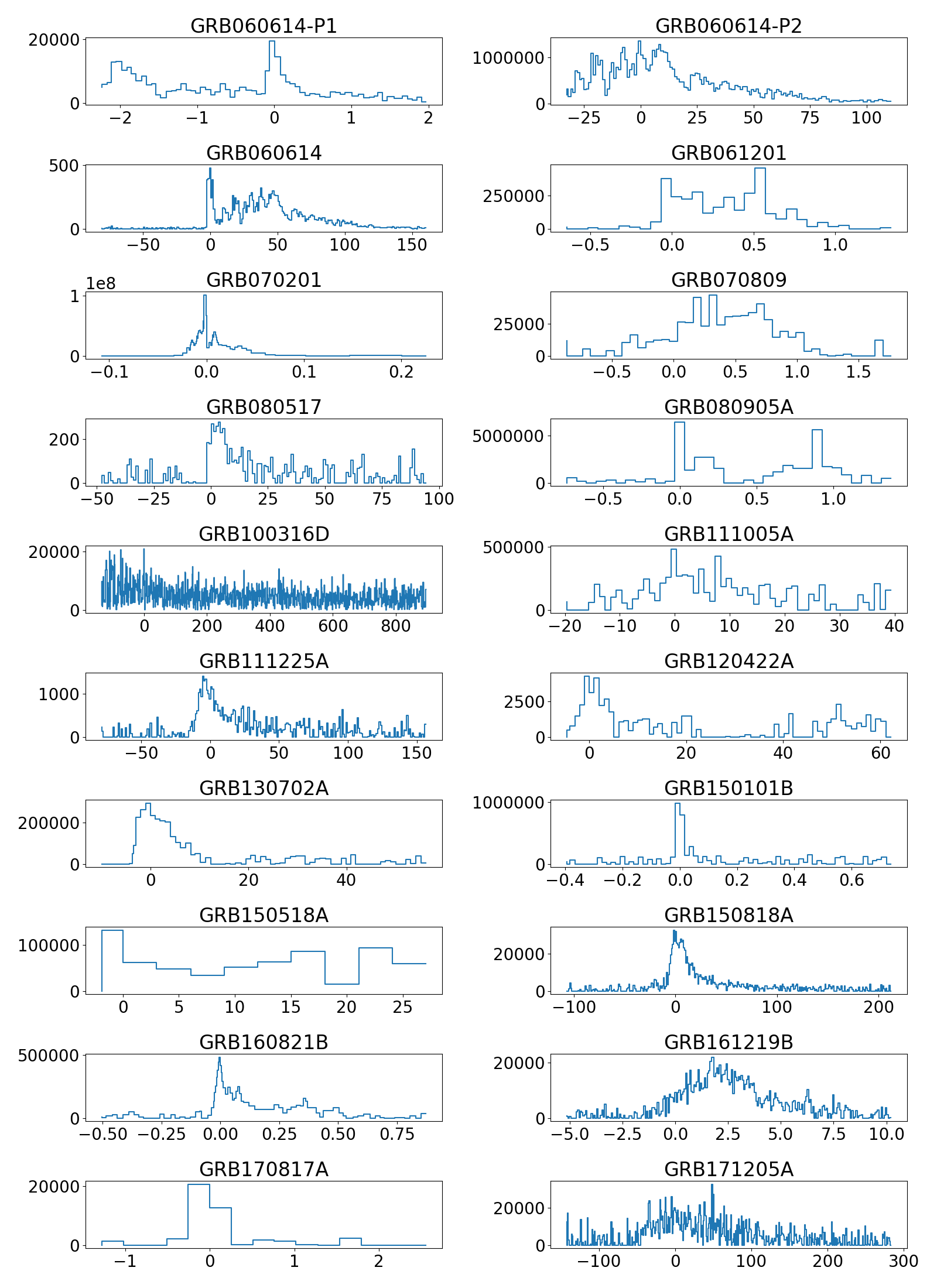}
\label{fig:LC2}
\end{figure}

\begin{figure}[t]
\centering
\includegraphics[width=\linewidth]{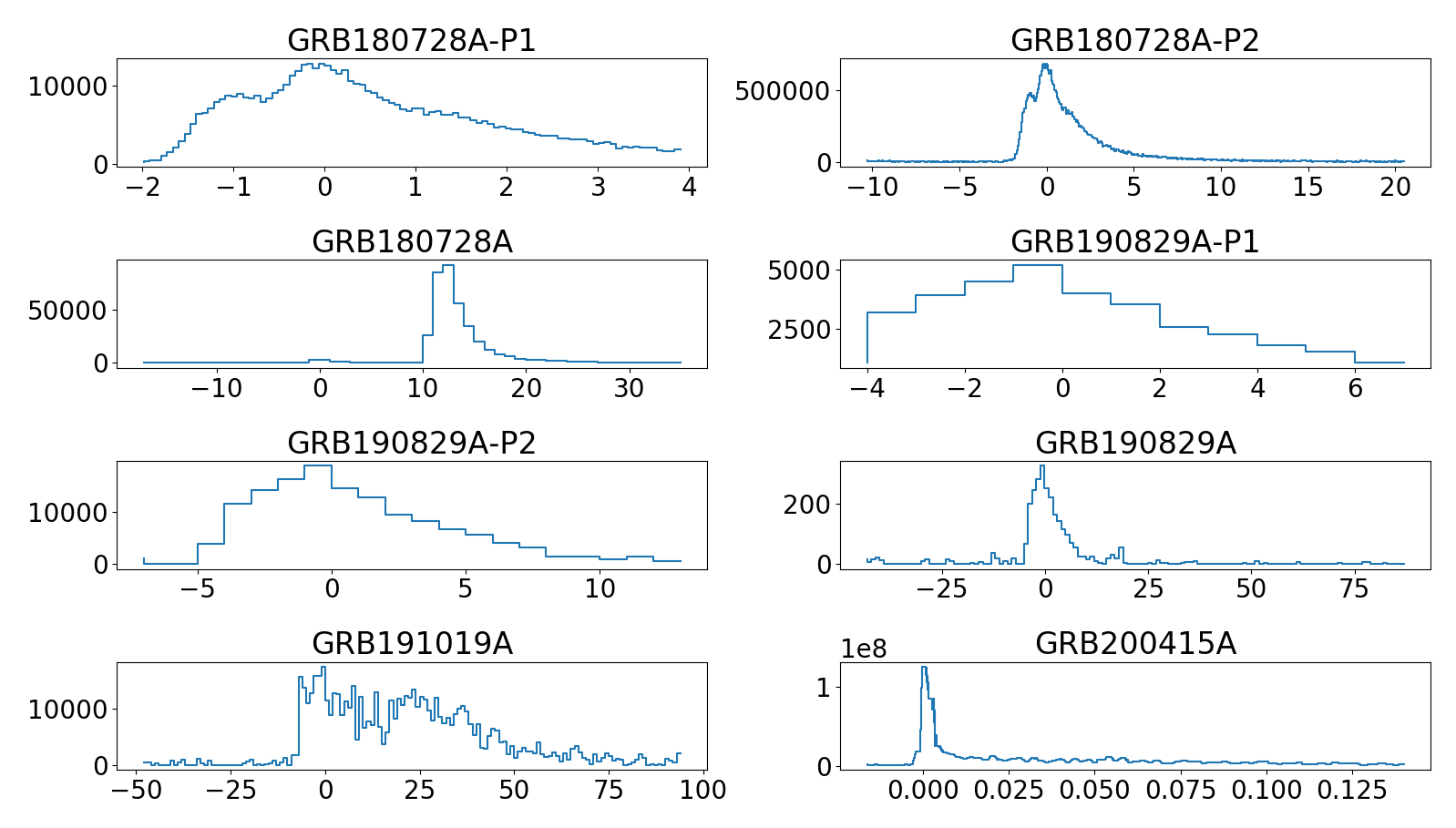}
\label{fig:LC3}
\caption{light curves used to simulate the photon counts from GRBs in the \textit{movegrb-simulator} program. The x-axis is in $\sec$ while the y-axis is the expected $\mathrm{cnt}/\sec$ per bin.}
\end{figure}

\end{appendix}

\end{document}